\documentclass[english,showpacs,aps,multicol]{revtex4}
\usepackage[T1]{fontenc}
\usepackage[latin1]{inputenc}
\usepackage{array}
\usepackage{amsmath}
\usepackage{graphicx}
\usepackage{amssymb}
\usepackage{feynmf}

\makeatletter

\providecommand{\tabularnewline}{\\}

\usepackage{babel}
\makeatother
\begin{document}

\title{Statistical Theory of Magnetohydrodynamic Turbulence: Recent Results}

\author{Mahendra K. Verma}

\email{mkv@iitk.ac.in}

\homepage{http://home.iitk.ac.in/~mkv}

\affiliation{Department of Physics, IIT Kanpur, Kanpur 208016, India}

\begin{abstract}
In this review article we will describe recent developments in statistical
theory of magnetohydrodynamic (MHD) turbulence. Kraichnan and Iroshnikov
first proposed a phenomenology of MHD turbulence where Alfv\'{e}n time-scale
dominates the dynamics, and the energy spectrum $E(k)$ is proportional
to $k^{-3/2}$. In the last decade, many numerical simulations show
that spectral index is closer to 5/3, which is Kolmogorov's index
for fluid turbulence. We review recent theoretical results based on
anisotropy and Renormalization Groups which support Kolmogorov's scaling
for MHD turbulence. 

Energy transfer among Fourier modes, energy flux, and shell-to-shell
energy transfers are important quantities in MHD turbulence. We report
recent numerical and field-theoretic results in this area. Role of
these quantities in magnetic field amplification (dynamo) are also
discussed. There are new insights into the role of magnetic helicity
in turbulence evolution. Recent interesting results in intermittency,
large-eddy simulations, and shell models of magnetohydrodynamics are
also covered.

\end{abstract}





\maketitle

\tableofcontents

\section{Introduction}

Fluid and plasma flows exhibit complex random behaviour at high Reynolds
number; this phenomena is called turbulence. On the Earth this phenomena
is seen in atmosphere, channel and rivers flows etc. In the universe,
most of the astrophysical systems are turbulent. Some of the examples
are solar wind, convective zone in stars, galactic plasma, accretion
disk etc. 

Reynolds number, defined as $UL/\nu$ ($U$ is the large-scale velocity,
$L$ is the large length scale, and $\nu$ is the kinematic viscosity),
has to be large (typically 2000 or more) for turbulence to set in.
At large Reynolds number, there are many active modes which are nonlinearly
coupled. These modes show random behaviour along with rich structures
and long-range correlations. Presence of large number of modes and
long-range correlations makes turbulence a very difficult problem
that remains largely unsolved for more than hundred years. 

Fortunately, random motion and presence of large number of modes make
turbulence amenable to statistical analysis. Notice that the energy
supplied at large-scales $(L)$ gets dissipated at small scales, say
$l_{d}$. Experiments and numerical simulations show that the velocity
difference $u(\mathbf{x+l})-u(\mathbf{x})$ has a universal probability
density function (pdf) for $l_{d}\ll l\ll L$. That is, the pdf is
independent of experimental conditions, forcing and dissipative mechanisms
etc. Because of its universal behaviour, the above quantity has been
of major interest among physicists for last sixty years. Unfortunately,
we do not yet know how to derive the form of this pdf from the first
principle, but some of the moments have been computed analytically.
The range of scales $l$ satisfying $l_{d}\ll l\ll L$ is called inertial
range.

In 1941 Kolmogorov \cite{K41a,K41b,K41c} computed an exact expression
for the third moment of velocity difference. He showed that under
vanishing viscosity, third moment for velocity difference for homogeneous,
isotropic, incompressible, and steady-state fluid turbulence is\[
\left\langle \left(u_{||}(\mathbf{x+l})-u_{||}(\mathbf{x})\right)^{3}\right\rangle =\frac{4}{5}\Pi l\]
where $||$ is the parallel component along $\mathbf{l}$, $\left\langle .\right\rangle $
stands for ensemble average, and $\Pi$ is the energy cascade rate,
which is also equal to the energy supply rate at large scale $L$
and dissipation rate at the small scale $l_{d}$. Assuming fractal
structure for the velocity field, and $\Pi$ to be constant for all
$l$, we can show that the energy spectrum $E(k)$ is\[
E(k)=K_{Ko}\Pi^{2/3}k^{-5/3},\]
where $K_{Ko}$ is a universal constant, called Kolmogorov's constant,
and $L^{-1}\ll k\ll l_{d}^{-1}$. Numerical simulations and experiments
verify the above energy spectrum apart from a small deviation called
intermittency correction. 

Physics of magnetohydrodynamic (MHD) turbulence is more complex than
fluid turbulence. There are two coupled vector fields, velocity $\mathbf{u}$
and magnetic $\mathbf{b}$, and two dissipative parameters, viscosity
and resistivity. In addition, we have mean magnetic field $B_{0}$
which cannot be transformed away (unlike mean velocity field which
can be transformed away using Galilean transformation). The mean magnetic
field makes the turbulence anisotropic, further complicating the problem.
Availability of powerful computers and sophisticated theoretical tools
have helped us understand several aspects of MHD turbulence. In the
last ten years, there has been major advances in the understanding
of energy spectra and fluxes of MHD turbulence. Some of these theories
have been motivated by Kolmogorov's theory for fluid turbulence. Note
that incompressible turbulence is better understood than compressible
turbulence. Therefore, our discussion on MHD turbulence is primarily
for incompressible plasma. \emph{In this paper we focus on the universal
statistical properties of MHD turbulence, which are valid in the inertial
range. In this paper we will review the statistical properties of
following quantities:}

\begin{enumerate}
\item Inertial-range energy spectrum for MHD turbulence.
\item Various energy fluxes in MHD turbulence.
\item Energy transfers between various wavenumber shells.
\item Anisotropic effects of mean magnetic field.
\item Structure functions $<\left(u_{||}(\mathbf{x+l})-u_{||}(\mathbf{x})\right)^{n}>$
and $<\left(b_{||}(\mathbf{x+l})-b_{||}(\mathbf{x})\right)^{n}>$,
where $u_{||}$ and $b_{||}$ are components of velocity and magnetic
fields along vector $\mathbf{l}$. 
\item Growth of magnetic field (dynamo).
\end{enumerate}
Currently energy spectra and fluxes of isotropic MHD turbulence is
quite well established, but anisotropy, intermittency, and dynamo
is not yet fully understood. Therefore, items 1-3 will be discussed
in greater detail. 

Basic modes of incompressible MHD are Alfvén waves, which travel parallel
and antiparallel to the mean magnetic field with speed $B_{0}$. The
nonlinear terms induce interactions among these modes. In mid sixties
Kraichnan \cite{Krai:65} and Iroshnikov \cite{Iros} postulated that
the time-scale for the nonlinear interaction is proportional to $B_{0}^{-1}$,
leading to $E(k)\sim B_{0}^{1/2}k^{-3/2}$. However, research in last
ten years \cite{Srid2,MKV:B0_RG,Maro:Simulation,ChoVish:localB} show
that the energy spectrum of MHD turbulence Kolmogorov-like ($k^{-5/3}$).
Current understanding is that Alfvén waves are scattered by {}``local
mean magnetic field'' $B_{0}(k)\sim k^{-1/3}$, leading to Kolmogorov's
spectrum for MHD turbulence. The above ideas will be discussed in
Sections \ref{sec:Renormalization-Group-Analysis-MHD} and \ref{sec:Field-Theor-of-anisotropic}. 

In MHD turbulence there are exchanges of energies among the velocity-velocity,
velocity-magnetic, and magnetic-magnetic modes. These exchanges lead
to energy fluxes from inside of the velocity/magnetic wavenumber sphere
to the outside of the velocity/magnetic wavenumber sphere. Similarly
we have shell-to-shell energy transfers in MHD turbulence. We have
developed a new formalism called {}``mode-to-mode'' energy transfer
rates, using which we have computed energy fluxes and shell-to-shell
energy transfers numerically and analytically \cite{Dar:flux,MKV:MHD_PRE,MKV:MHD_Flux,MKV:MHD_Helical}.
The analytic calculations are based on field-theoretic techniques.
Note that some of the fluxes and shell-to-shell energy transfers are
possible only using {}``mode-to-mode'' energy transfer, and cannot
be computed using {}``combined energy transfer'' in a triad \cite{Lesi:book}.

Many analytic calculations in fluid and MHD turbulence have been done
using field-theoretic techniques. Even though these methods are plagued
with some inconsistencies, we get many meaningful results using them.
In Sections \ref{sec:Renormalization-Group-Analysis-MHD}, \ref{sec:analytic-energy},
and \ref{sec:Field-Theor-of-anisotropic} we will review the field-theoretic
calculations of energy spectrum, energy fluxes, and shell-to-shell
energy transfers.

Growth of magnetic field in MHD turbulence (dynamo) is of central
importance in MHD turbulence research. Earlier dynamo models (kinematic)
assumed a given form of velocity field and computed the growth of
large-scale magnetic field. These models do not take into account
the back-reaction of magnetic field on the velocity field. In last
ten years, many dynamic dynamo simulations have been done which also
include the above mentioned back reaction. Role of magnetic helicity
($\mathbf{a}\cdot\mathbf{b}$, where $\mathbf{a}$ is the vector potential)
in the growth of large-scale magnetic field is better understood now.
Recently, Field et al. \cite{Fiel:Dynamo}, Chou \cite{Chou:theo},
Schekochihin et al. \cite{Maro:DynamoNlin} and Blackman \cite{Blac:Rev_Dynamo}
have constructed theoretical dynamical models of dynamo, and studied
nonlinear evolution and saturation mechanisms.

As mentioned above, pdf of velocity difference in fluid turbulence
is still unsolved. We know from experiments and simulation that pdf
is close to Gaussian for small $\delta u$, but is nongaussian for
large $\delta u$. This phenomena is called intermittency. Note that
various moments called Structure functions are connected to pdf. It
can be shown that the structure functions are related to the {}``local
energy cascade rate'' $\Pi(k)$. Some phenomenological models, notably
by She and Leveque \cite{SheLeve} based on log-Poisson process, have
been developed to compute $\Pi(k)$; these models quite successfully
capture intermittency in both fluid and MHD turbulence. The predictions
of these models are in good agreement with numerical results. We will
discuss these issues in Section \ref{sec:Intermittency-in-MHD}.

Numerical simulations have provided many important data and clues
for understanding the dynamics of turbulence. They have motivated
new models, and have verified/rejected existing models. In that sense,
they have become another type of experiment, hence commonly termed
as numerical experiments. Modern computers have made reasonably high
resolution simulations possible. The highest resolution simulation
in fluid turbulence is on $4096^{3}$ grid (e.g., by Gotoh \cite{Goto:DNS}),
and in MHD turbulence is on $1024^{3}$ grid (e.g., by Haugen et al.
\cite{Bran:Nonhelical_simulation}). Simulations of Biskamp \cite{Bisk:Kolm1,Bisk:Kolm2},
Cho et al. \cite{ChoVish:localB}, Maron and Goldreich \cite{Maro:Simulation}
have verified 5/3 spectrum for MHD turbulence. Dar et al. \cite{Dar:flux}
have computed various energy fluxes in 2D MHD turbulence. Earlier,
based on energy fluxes Verma et al. \cite{MKV:MHD_Simulation} could
conclude that Kolmogorov-like phenomenology models MHD turbulence
better that Kraichnan and Iroshnikov's phenomenology. Many interesting
simulations have been done to simulate dynamo, e. g., Chou \cite{Chou:num}
and Brandenburg \cite{Bran:Alpha}.

Because of large values of dissipative parameters, MHD turbulence
requires large length and velocity scales. This make terrestrial experiments
on MHD turbulence impossible. However, astrophysical plasmas are typically
turbulent because of large length and velocity scales. Taking advantage
of this fact, large amount of solar-wind in-situ data have been collected
by spacecrafts. These data have been very useful in understanding
the physics of MHD turbulence. In fact, in 1982 Matthaeus and Goldstein
\cite{MattGold} had shown that solar wind data favors Kolmogorov's
$k^{-5/3}$ spectrum over Kraichnan and Iroshnikov's $k^{-3/2}$ spectrum.
Solar wind data also shows that MHD turbulence exhibits intermittency.
some of the observational results of solar wind will discussed in
Section \ref{sec:Solar-Wind}. In addition to the above topics, we
will also state the current results on the absolute equilibrium theories,
decay of global quantities, two-dimensional turbulence, shell model
of MHD turbulence, compressible turbulence etc.

Literature on MHD turbulence is quite extensive. Recent book {}``Magnetohydrodynamic
Turbulence'' by Biskamp \cite{BiskTurb:book} covers most of the
basics. MHD turbulence normally figures as one of the chapters in
many books on Magnetohydrodynamics, namely Biskamp \cite{BiskNonl:book},
Priest \cite{Prie:book}, Raichoudhury \cite{Arna:book}, Shu \cite{Shu:book},
Cowling \cite{Cowl:book}, Vedenov \cite{Vede:book}. The recent developments
are nicely covered by the review articles in an edited volume \cite{Pass:book}.
Some of the important review articles are by Montgomery \cite{Mont:SW},
Pouquet \cite{Pouq:Rev}, Krommes \cite{Krom:Rev_Intermittency,Krom:Rev_Plasma}.
On dynamo, the key references are books by Moffatt \cite{Moff:book}
and Krause and Rädler \cite{Krau:book}, and recent review articles
\cite{Gilb:inbook,Robe:Rev_Geo,Bran:PR}. Relatively, fluid turbulence
has a larger volume of literature. Here we will list only some of
the relevant ones. Leslie \cite{Lesl:book}, McComb \cite{McCo:book,McCo:rev,McCo:RGBook},
Zhou et al. \cite{ZhouMcCo:RGrev}, and Smith and Woodruff \cite{Smit}
have reviewed field-theoretic treatment of fluid turbulence. The recent
books by Frisch \cite{Fris:book} and Lesieur \cite{Lesi:book} cover
recent developments and phenomenological theories. The review articles
by Orszag \cite{Orsz:Rev}, Kraichnan and Montgomery \cite{KraiMont},
and Sreenivasan \cite{Sree:RMP} are quite exhaustive.

In this review paper, we have focussed on statistical theory of MHD
turbulence, specially on energy spectra, energy fluxes, and shell-to-shell
energy transfers. These quantities have been analyzed analytically
and numerically. A significant portion of the paper is devoted to
self-consistent field-theoretic calculations of MHD turbulence and
{}``mode-to-mode'' energy transfer rates because of their power
of analysis as well as our familiarity with these topics. These topics
are new and are of current interest. Hence, this review article complements
the earlier work. Universal laws are observed in the inertial range
of homogeneous and isotropic turbulence. Following the similar approach,
in analytic calculations of MHD turbulence, homogeneity and isotropy
are assumed except in the presence of mean magnetic field. 

To keep our discussion focussed, we have left out many important topics
like coherent structures, astrophysical objects like accretion disks
and Sun, transition to turbulence etc. Our discussion on compressible
turbulence and intermittency is relatively brief because final word
on these topics still awaited. Dynamo theory is only touched upon;
the reader is referred to the above mentioned references for a detailed
discussion. In the discussion on the solar wind, only a small number
of results connected to energy spectra are covered. 

The outline of the paper is as follows: Section \ref{sec:MHD:-Definitions-and}
contains definition of various global and spectral quantities along
with their governing equations. In Section \ref{sec:Mode-to-mode-Energy-Transfer}
we discuss the formalism of {}``mode-to-mode'' energy transfer rates
in fluid and MHD turbulence. Using this formalism, formulas for energy
fluxes and shell-to-shell energy transfer rates have been derived.
Section \ref{sec:MHD-Turbulence-Models} contains the existing MHD
turbulence phenomenologies which include Kraichnan's 3/2 mode; Kolmogorov-like
models of Goldreich and Sridhar. Absolute equilibrium theories and
Selective decay are also discussed here. In Section \ref{sec:Solar-Wind}
we review the observed energy spectra of the solar wind. Section \ref{sec:Numerical-Investigation-MHD}
describes Pseudo-spectral method along with the numerical results
on energy spectra, fluxes, and shell-to-shell energy transfers. In
these discussions we verify which of the turbulence phenomenologies
are in agreement with the solar wind data and numerical results.

Next three sections cover applications of field-theoretic techniques
to MHD turbulence. In Section \ref{sec:Renormalization-Group-Analysis-MHD}
we introduce Renormalization-group analysis of MHD turbulence, with
an emphasis on the renormalization of {}``mean magnetic field''
\cite{MKV:B0_RG}, viscosity and resistivity \cite{MKV:MHD_RG}. In
Section \ref{sec:analytic-energy} we compute various energy fluxes
and shell-to-shell energy transfers in MHD turbulence using field-theoretic
techniques. Here we also review eddy-damped quasi-normal Markovian
(EDQNM) calculations of MHD turbulence. In Section \ref{sec:Field-Theor-of-anisotropic}
we discuss the anisotropic turbulence calculations of Goldreich and
Sridhar \cite{Srid1,Srid2} and Galtiers et al. \cite{Galt:Weak}
in significant details. The variations of turbulence properties with
space dimensions have been discussed. 

In Section \ref{sec:Magnetic-Field-Growth} we briefly mention the
main numerical and analytic results on homogeneous and isotropic dynamo.
We include both kinematic and dynamic dynamo models, with emphasis
being on the later. Section \ref{sec:Intermittency-in-MHD} contains
a brief discussion on intermittency models of fluid and MHD turbulence.
Next section \ref{sec:Miscellaneous-Topics} contains a brief discussion
on the large-eddy simulations, decay of global energy, compressible
turbulence, and shell model of MHD turbulence. Appendix \ref{sec:Fourier-Series}
contains the definitions of Fourier series and transforms of fields
in homogeneous turbulence. Appendix B and C contain the Feynman diagrams
for MHD turbulence; these diagrams are used in the field-theoretic
calculations. In the last Appendix \ref{sec:Digression-to-Fluid}
we briefly mention the main results of spectral theory of fluid turbulence
in 2D and 3D.

\section{MHD: Definitions and Governing equations \label{sec:MHD:-Definitions-and}}

\subsection{MHD Approximations and Equations}

MHD fluid is quasi-neutral, i.e., local charges of ions and electrons
almost balance each other. The conductivity of MHD fluid is very high.
As a consequence, the magnetic field lines are frozen, and the matter
(ions and electrons) moves with the field. A slight imbalance in the
motion creates electric currents, that in turn generates the magnetic
field. The fluid approximation implies that the plasma is collisional,
and the equations are written for the coarse-grained fluid volume
(called fluid element) containing many ions and electrons. In the
MHD picture, the ions (heavier particle) carry momentum, and the electrons
(lighter particle) carry current. In the following discussion we will
make the above arguments quantitative. In this paper we will use CGS
units. For detailed discussions on MHD, refer to Cowling \cite{Cowl:book},
Siscoe \cite{Sisc:book}, and Shu \cite{Shu:book}.

Consider MHD plasma contained in a volume. In the rest frame of the
fluid element, the electric field $\mathbf{E'=J}/\sigma$, where $\mathbf{J}$
is the electric current density, and $\sigma$ is the electrical conductivity.
If $\mathbf{E}$ is the electric field in the laboratory frame, Lorenz
transformation for nonrelativistic flows yields\begin{equation}
\mathbf{E'=E+\frac{u\times B}{\textrm{$c$}}=\frac{J}{\sigma},}\label{eq:E_eq_J/sigma}\end{equation}
where $\mathbf{u}$ is the velocity of the fluid element, $\mathbf{B}$
is the magnetic field, and $c$ is the speed of light. Note that the
current density, which is proportional to the relative velocity of
electrons with relative to ions, remains unchanged under Galilean
transformation. Since MHD fluid is highly conducting $(\sigma\rightarrow\infty)$,\[
E\approx\frac{u}{c}B.\]
This implies that for the nonrelativistic flows, $E\ll B$. Now let
us look at one of the Maxwell's equations\[
\nabla\times\mathbf{B}=\frac{4\pi}{c}\mathbf{J}+\frac{1}{c}\frac{\partial\mathbf{E}}{\partial t}.\]
The last term of the above equation is $(u/c)^{2}$ times smaller
as compared to $\nabla\times\mathbf{B}$, hence it can be ignored.
Therefore,\begin{equation}
\mathbf{J}=\frac{c}{4\pi}\nabla\times\mathbf{B}.\label{eq:J_eq_del_cross_B}\end{equation}
Hence both $\mathbf{E}$ and $\mathbf{J}$ are dependent variables,
and they can be written in terms of $\mathbf{B}$ and $\mathbf{u}$
as discussed above.

In MHD both magnetic and velocity fields are dynamic. To determine
the magnetic field we make use of one of Maxwell's equation\begin{equation}
\frac{\partial\mathbf{B}}{\partial t}=-c\nabla\times\mathbf{E}.\label{eq:Bdot_eq_del_cross_E}\end{equation}
An application of Eqs. (\ref{eq:E_eq_J/sigma},\ref{eq:J_eq_del_cross_B})
yields\begin{equation}
\frac{\partial\mathbf{B}}{\partial t}=\nabla\times\left(\mathbf{u}\times\mathbf{B}\right)+\eta\nabla^{2}\mathbf{B},\end{equation}
or,\begin{equation}
\frac{\partial\mathbf{B}}{\partial t}+(\mathbf{u}\cdot\nabla)\mathbf{B}=(\mathbf{B}\cdot\nabla)\mathbf{u}-\mathbf{B}\nabla\cdot\mathbf{u}+\eta\nabla^{2}\mathbf{B}.\label{eq:MHDBx_compressible}\end{equation}
The parameter $\eta$ is called the resistivity, and is equal to $c^{2}/(4\pi\sigma)$.
The magnetic field obeys the following constraint:

\begin{equation}
\nabla\cdot\mathbf{B}=0.\label{eq:div_B_eq_0}\end{equation}

The time evolution of the velocity field is given by the Navier-Stokes
equation. In this paper, we work in an inertial frame of reference
in which the mean flow speed is zero. This transformation is possible
because of Galilean invariance. The Navier-Stokes equation is \cite{LandFlui:book,Kund:book}

\begin{equation}
\rho\left(\frac{\partial\mathbf{u}}{\partial t}+(\mathbf{u}\cdot\nabla)\mathbf{u}\right)=-\nabla p_{th}+\frac{1}{c}\mathbf{J\times B}+\mu\nabla^{2}\mathbf{u}+\frac{2\mu}{3}\nabla\nabla\cdot\mathbf{u},\end{equation}
where $\rho(\mathbf{x})$ is the density of the fluid, $p_{th}$ is
the thermal pressure, and $\mu$ is the dynamic viscosity. Note that
kinematic viscosity $\nu=\mu/\rho$. Substitution of $\mathbf{J}$
in terms of $\mathbf{B}$ {[}Eq. (\ref{eq:J_eq_del_cross_B}){]} yields

\begin{equation}
\frac{\partial\mathbf{u}}{\partial t}+(\mathbf{u}\cdot\nabla)\mathbf{u}=\frac{1}{\rho}\left[-\nabla\left(p_{th}+\frac{B^{2}}{8\pi}\right)+(\mathbf{B}\cdot\nabla)\mathbf{B}\right]+\nu\nabla^{2}\mathbf{u}+\frac{2\nu}{3}\nabla\nabla\cdot\mathbf{u},\label{eq:MHDux_compressible}\end{equation}
where $p_{th}+\frac{B^{2}}{8\pi}=p$ is called total pressure. The
ratio $p_{th}/(B^{2}/8\pi)$ is called $\beta$, which describes the
strength of the magnetic field with relative to thermal pressure.

Mass conservation yields the following equation for density field
$\rho(\mathbf{x})$

\begin{equation}
\frac{\partial\rho}{\partial t}+\nabla\cdot\left(\rho\mathbf{u}\right)=0\label{eq:mass_conservation}\end{equation}

Pressure can be computed from $\rho$ using equation of state

\begin{equation}
p=f\left(\rho\right)\label{eq:p_rho}\end{equation}
This completes the basic equations of MHD, which are (\ref{eq:MHDBx_compressible},
\ref{eq:MHDux_compressible}, \ref{eq:mass_conservation}, \ref{eq:p_rho}).
Using these equations we can determine the unknowns $(\mathbf{u,B,}\rho,p)$.
Note that the number of equations and unknowns are the same.

When $\beta$ is high, $B^{2}$ is much less than $p_{th}$, and it
can be ignored. On nondimensionalization of the Navier-Stokes equation,
the term $\nabla p$ becomes $\left(d\rho/dx'\right)/\rho$ $\times$$(C_{s}/U)^{2}$,
where $C_{s}$ is the sound speed, $U$ is the typical velocity of
the flow, $x'$ is the position coordinate normalized with relative
to the length scale of the system \cite{Trit:Book}. $C_{s}\rightarrow\infty$
is the incompressible limit, which is widely studied because water,
the most commonly found fluid on earth, is almost incompressible ($\delta\rho/\rho<$0.01)
in most practical situations. The other limit $C_{s}\rightarrow0$
or $U\gg C_{s}$ (supersonic) is the fully compressible limit, and
it is described by Burgers equation. As we will see later, the energy
spectrum for both these extreme limits well known. When $U/C_{s}\ll1$
but nonzero, then we call the fluid to be nearly incompressible; Zank
and Matthaeus \cite{Zank:Compress_PoF,Zank:Compress_PRL} have given
theories for this limit. The energy and density spectra are not well
understood for arbitrary $U/C_{s}$.

When $\beta$ is low, $p_{th}$ can be ignored. Now the Alfvén speed
$C_{A}=B/\sqrt{4\pi\rho}$ plays the role of $C_{s}$. Hence, the
fluid is incompressible if $U\ll C_{A}$ \cite{BiskTurb:book}. For
most part of this paper, we assume the magnetofluid to be incompressible.
In many astrophysical and terrestrial situations (except shocks),
incompressibility is a reasonably good approximation for the MHD plasma
because typical velocity fluctuations are much smaller compared to
the sound speed or the Alfvén speed. This assumption simplifies the
calculations significantly. In Section \ref{sub:Compressible-Turbulence}
we will discuss the compressible MHD.

The incompressibility approximation can also be interpreted as the
limit when volume of a fluid parcel will not change along its path,
that is, $d\rho/dt=0$. From the continuity equation (\ref{eq:mass_conservation}),
the incompressibility condition reduces to

\begin{equation}
\nabla\cdot\mathbf{u}=0\label{eq:div_u_eq_0}\end{equation}
This is a constraint on the velocity field \textbf{u}. Note that incompressibility
does not imply constant density. However, for simplicity we take density
to be constant and equal to 1. Under this condition, Eqs. (\ref{eq:MHDBx_compressible},
\ref{eq:MHDux_compressible}) reduce to\begin{eqnarray}
\frac{\partial\mathbf{u}}{\partial t}+(\mathbf{u}\cdot\nabla)\mathbf{u} & = & -\nabla p+(\mathbf{B}\cdot\nabla)\mathbf{B}+\nu\nabla^{2}\mathbf{u}\label{eq:MHDux}\\
\frac{\partial\mathbf{B}}{\partial t}+(\mathbf{u}\cdot\nabla)\mathbf{B} & = & (\mathbf{B}\cdot\nabla)\mathbf{u}+\eta\nabla^{2}\mathbf{B}\label{eq:MHDbx}\end{eqnarray}

To summarize, the incompressible MHD equations are

\begin{center}\begin{tabular}{|c|}
\hline 
$\frac{\partial\mathbf{u}}{\partial t}+(\mathbf{u}\cdot\nabla)\mathbf{u}=-\nabla p+(\mathbf{B}\cdot\nabla)\mathbf{B}+\nu\nabla^{2}\mathbf{u}$\tabularnewline
$\frac{\partial\mathbf{B}}{\partial t}+(\mathbf{u}\cdot\nabla)\mathbf{B}=(\mathbf{B}\cdot\nabla)\mathbf{u}+\eta\nabla^{2}\mathbf{B}$\tabularnewline
$\nabla\cdot\mathbf{u}=0$\tabularnewline
$\nabla\cdot\mathbf{B}=0$\tabularnewline
\hline
\end{tabular}\end{center}

When we take divergence of the equation Eq. (\ref{eq:MHDux}), we
obtain Poisson's equation \[
-\nabla^{2}p=\nabla\cdot\left[(\mathbf{u}\cdot\nabla)\mathbf{u}-(\mathbf{B}\cdot\nabla)\mathbf{B}.\right]\]
Hence, given \textbf{u} and \textbf{B} fields at any given time, we
can evaluate $p$. Hence $p$ is a dependent variable in the incompressible
limit. 

Incompressible MHD has two unknown vector fields $(\mathbf{u,B})$.
They are determined using Eqs. (\ref{eq:MHDux}, \ref{eq:MHDbx})
under the constraints (\ref{eq:div_B_eq_0}, \ref{eq:div_u_eq_0}).
The fields $\mathbf{E},\mathbf{J}$ and $p$ are dependent variables
that can be obtained in terms of $\mathbf{u}$ and $\mathbf{B}$.

The MHD equations are nonlinear, and that is the crux of the problem.
There are two dissipative terms: viscous $(\nu\nabla^{2}\mathbf{u)}$
and resistive $(\eta\nabla^{2}\mathbf{B})$. The ratio of the nonlinear
vs. viscous dissipative term is called Reynolds number $Re=UL/\nu$,
where $U$ is the velocity scale, and $L$ is the length scale. There
is another parameter called magnetic Reynolds number $Re_{m}=UL/\eta$.
For turbulent flows, Reynolds number should be high, typically more
than 2000 or so. The magnetic Prandtl number $\nu/\eta$ also plays
an important role in MHD turbulence. Typical values of parameters
in commonly studied MHD systems are given in Table \ref{Table:MHDParameters}
\cite{Encl,Mont:SW,MKV:SW_Nonclassical,Kuls3}. %
\begin{table}

\caption{\label{Table:MHDParameters} Typical values of parameters in commonly
studied MHD systems. Viscosity and resistivity of first 4 columns
are rough estimates \cite{Encl,Mont:SW,MKV:SW_Nonclassical,Kuls3}}

\begin{tabular}{|p{1.5in}|c|c|c|c|>{\centering}p{1in}|>{\centering}p{0.5in}|}
\hline 
System&
Earth's Core&
\multicolumn{1}{p{1.1in}|}{Solar Conve-ctive Zone}&
Solar Wind&
Galactic Disk&
Ioniz-ed H $(10^{5}K,42Pa)$&
Hg\tabularnewline
\cline{1-1} \cline{2-2} \cline{4-4} \cline{5-5} \cline{6-6} \cline{7-7} 
\hline 
Length $(cm)$&
$10^{8}$&
$10^{10}$&
$10^{13}$&
$10^{22}$&
$10$&
$10$\tabularnewline
\hline 
Velocity $(cm/s)$&
$10^{-2}$&
$10^{4}$&
$10^{6}$&
$10^{6}$&
$10^{2}$&
$10$\tabularnewline
\hline 
Mean Mag. Field $(G)$ &
$10^{2}$&
$10^{3}$&
$10^{-5}$&
$10^{-5}$&
$10^{3}$&
$10^{4}$\tabularnewline
\hline 
Density $(gm/cc)$&
10&
$10^{-5}$&
$10^{-23}$&
$10^{-24}$&
$10^{-10}$&
$10$\tabularnewline
\hline 
Kinematic viscosity $(cm^{2}/s)$&
$10^{-2}$&
$10^{11}$&
$10^{4}$&
$10^{21}$&
$10^{5}$&
$10^{-3}$\tabularnewline
\hline 
Reynolds Number&
$10^{8}$&
$10^{3}$&
$10^{13}$&
$10^{7}$&
$10^{-2}$&
$10^{5}$\tabularnewline
\hline 
Resistivity $(cm^{2}/s)$&
$10^{4}$&
$10^{11}$&
$10^{4}$&
$10^{7}$&
$1.5\times10^{5}$&
$10^{4}$\tabularnewline
\hline 
Magnetic Reynolds no&
$10^{2}$&
$10^{3}$&
$10^{4}$&
$10^{21}$&
$7\times10^{-3}$&
$10^{-2}$\tabularnewline
\hline 
Magnetic Prandtl no&
$10^{-6}$&
$1$&
(1)?&
$10^{14}$&
0.7&
$10^{-7}$\tabularnewline
\hline
\end{tabular}
\end{table}
The calculation of viscosity and resistivity of MHD plasma is quite
involved because of anisotropy caused by mean magnetic field \cite{Shu:book}.
In Table \ref{Table:MHDParameters} we have provided rough estimates
of these quantities.

\subsection{Energy Equations and Conserved Quantities}

In this subsection we derive energy equations for compressible and
incompressible fluids. For compressible fluids we can construct equations
for energy using Eqs. (\ref{eq:MHDBx_compressible},\ref{eq:MHDux_compressible}).
Following Landau \cite{LandFlui:book} we derive the following energy
equation for the kinetic energy \begin{equation}
\frac{\partial}{\partial t}\left(\frac{1}{2}\rho u^{2}+\rho\epsilon\right)=-\nabla\cdot\left[\left(\frac{1}{2}u^{2}+\epsilon\right)\rho\mathbf{u}\right]-\nabla\cdot p\mathbf{u}+\frac{1}{c}\mathbf{u}\cdot(\mathbf{J}\times\mathbf{B})+\Phi\label{eq:u2_compressible}\end{equation}
where $\epsilon$ is the internal energy function. The first term
in the RHS is the energy flux, and the second term is the work done
by the pressure, which enhances the energy of the system. The third
term in the RHS is work done by magnetic force on the fluid, while
$\Phi$, a complex function of strain tensor, is the energy change
due to surface forces.

For the evolution of magnetic energy we use Eq. (\ref{eq:Bdot_eq_del_cross_E})
and obtain \cite{Kund:book}\begin{eqnarray}
\frac{\partial}{\partial t}\frac{1}{8\pi}B^{2} & = & -\frac{c}{4\pi}\mathbf{B}\cdot\nabla\times\mathbf{E}\nonumber \\
 & = & -\nabla\cdot\frac{c}{4\pi}\mathbf{E}\times\mathbf{B}-\mathbf{J}\cdot\mathbf{E}\label{eq:B2_compressible}\end{eqnarray}
The first term in the RHS is the Poynting flux (energy flux of the
electro-magnetic field), and the second term is the work done by the
electro-magnetic field on fluid. The second term also includes the
Joule dissipation term. Combination of Eqs. (\ref{eq:u2_compressible},
\ref{eq:B2_compressible}) yields the following dynamical equation
for the energy in MHD\begin{eqnarray*}
\frac{\partial}{\partial t}\left(\frac{1}{2}\rho u^{2}+\rho\epsilon+\frac{1}{8\pi}B^{2}\right) & = & -\nabla\cdot\left[\left(\frac{1}{2}u^{2}+\epsilon\right)\rho\mathbf{u}+\frac{c}{4\pi}\mathbf{E}\times\mathbf{B}\right]\\
 &  & -\nabla\cdot p\mathbf{u}+\Phi+\frac{1}{\sigma}J^{2}\end{eqnarray*}
Here $\frac{1}{2}\rho u^{2}+\rho\epsilon+\frac{1}{8\pi}B^{2}$ is
the total energy. Physical interpretation of the above equation is
the following: the rate of change of total energy is the sum of energy
flux, work done by pressure, and the viscous and resistive dissipation.
It is convenient to use a new variable for magnetic field $B=B_{CGS}/\sqrt{4\pi}$.
In terms of new variable, the total energy is $\frac{1}{2}\rho u^{2}+\rho\epsilon+\frac{1}{2}B^{2}$.
From this point onward we use this new variable for magnetic field.

In the above equations we apply isoentropic and incompressibility
conditions. For the incompressible fluids we can choose $\rho=1$.
Landau \cite{LandFlui:book} showed that under this condition $\epsilon$
is a constant. Hence, for incompressible MHD fluid we treat $(u^{2}+B^{2})/2$
as total energy. For ideal incompressible MHD ($\nu=\eta=0)$ the
energy evolution equation is\[
\frac{\partial}{\partial t}\frac{1}{2}\left(u^{2}+B^{2}\right)=-\nabla\cdot\left[\left(\frac{1}{2}u^{2}+\frac{1}{2}B^{2}+p\right)\mathbf{u}\right]-2\nabla\cdot\left[\left(\mathbf{B}\cdot\mathbf{u}\right)\mathbf{B}\right]\]
By applying Gauss law we find that \[
\frac{\partial}{\partial t}\int\frac{1}{2}\left(u^{2}+B^{2}\right)d\mathbf{x}=-\oint\left[\left(\frac{1}{2}u^{2}+\frac{1}{2}B^{2}+p\right)\mathbf{u}+\left(\mathbf{B}\cdot\mathbf{u}\right)\mathbf{B}\right]\cdot d\mathbf{S}\]
For the boundary condition $B_{n}=u_{n}=0$ or periodic boundary condition,
the total energy $\int$$1/2(u^{2}+B^{2})$ is conserved.

There are some more important quantities in MHD turbulence. They are
listed in Table \ref{table:Globals}. Note that $\mathbf{A}$ is the
vector potential and $\mathbf{\omega}$ is the vorticity field. %
\begin{table}

\caption{\label{table:Globals} Global Quantities in MHD}

\begin{tabular}{|c|c|c|c|}
\hline 
Quantity&
Symbol&
Definition&
Conserved in MHD?\tabularnewline
\hline
\hline 
Kinetic Energy&
$E^{u}$ &
$\int d\mathbf{x}u^{2}/2$ &
No\tabularnewline
\hline 
Magnetic Energy&
$E^{B}$ &
$\int d\mathbf{x}B^{2}/2$ &
No\tabularnewline
\hline 
Total Energy&
$E$&
$\int d\mathbf{x}(u^{2}+B^{2})/2$&
Yes (2D,3D)\tabularnewline
\hline 
Cross Helicity&
$H_{c}$&
$\int d\mathbf{x}(\mathbf{u}\cdot\mathbf{B})/2$ &
Yes (2D,3D)\tabularnewline
\hline 
Magnetic Helicity&
$H_{M}$&
$\int d\mathbf{x}(\mathbf{A}\cdot\mathbf{B})$/2 &
Yes (3D)\tabularnewline
\hline 
Kinetic Helicity&
$H_{K}$&
$\int d\mathbf{x}(\mathbf{u}\cdot\mathbf{\omega})/2$&
No\tabularnewline
\hline 
Mean-square Vector Potential&
$A2$&
$\int d\mathbf{x}$$A^{2}/2$&
Yes (2D)\tabularnewline
\hline 
Enstrophy&
$\Omega$&
$\int d\mathbf{x}\omega^{2}/2$&
No\tabularnewline
\hline
\end{tabular}
\end{table}
By following the same procedure described above, we can show that
$E,H_{c},$ and $H_{M}$ are conserved in 3D MHD, while $E,H_{c}$
and $A2$ are conserved in 2D MHD \cite{MattGold,BiskNonl:book}.
Note that in 3D fluids, $E^{u}$ and $H_{K}$ are conserved, while
in 2D fluids, $E^{u}$ and $\Omega$ are conserved \cite{Lesl:book,Lesi:book}.

Magnetic helicity is a tricky quantity. Because of the choice of gauge
it can be shown that magnetic helicity is not unique unless $B_{n}=0$
at the boundary. Magnetic helicity is connected with flux tubes, and
plays important role in magnetic field generation. For details refer
to Biskamp \cite{BiskNonl:book}.

In addition to the above global quantities, there are infinite many
conserved quantities. In the following we will show that the magnetic
flux defined as \[
\Phi=\int\mathbf{B}\cdot d\mathbf{S},\]
where $d\mathbf{S}$ is the area enclosed by any closed contour moving
with the plasma, is conserved. Since infinitely many closed curves
are possible in any given volume, we have infinitely many conserved
quantities. To prove the above conservation law, we use vector potential
\textbf{A}, whose dynamical evolution is given by\[
\frac{\partial}{\partial t}\mathbf{A=u}\times\mathbf{B}+\nabla\phi,\]
where $\phi$ is the scalar potential \cite{MattGold}. The above
equation can be rewritten as\[
\frac{dA_{i}}{dt}=u_{k}\partial_{i}A_{k}+\partial_{i}\phi.\]
Now we write magnetic flux $\Phi$ in terms of vector potential \[
\Phi=\oint\mathbf{A}\cdot\mathbf{dl}.\]
The time derivative of $\Phi$ will be\begin{eqnarray*}
\frac{d\Phi}{dt} & = & \oint\frac{dA_{i}}{dt}dl_{i}+A_{i}\frac{d}{dt}dl_{i}\\
 & = & \oint d\phi+dl_{i}u_{k}\partial_{i}A_{k}+dl_{i}A_{k}\partial_{i}u_{k}\\
 & = & 0\end{eqnarray*}
Hence, magnetic flux over any surface moving with the plasma is conserved.

The conserved quantities play very important role in turbulence. These
aspects will be discussed later in Sections \ref{sec:Renormalization-Group-Analysis-MHD}
and \ref{sec:analytic-energy} of this review. Now we turn to the
linear solutions of MHD equations.

\subsection{Linearized MHD Equations and their Solutions; MHD Waves}

The fields can be separated into their mean and fluctuating parts:
$\mathbf{B=B}_{0}+\mathbf{b}\textrm{ and }\rho=\rho_{0}+\delta\rho.$
Here $\mathbf{B}_{0}$ and $\rho_{0}$ denote the mean, and $\mathbf{b}$
and $\delta\rho$ denote the fluctuating fields. Note that the velocity
field $\mathbf{u}$ is purely fluctuating field; its mean can be eliminated
by Galilean transformation.

\begin{figure}
\includegraphics[%
  scale=0.7,bb=0 420 350 700 ]{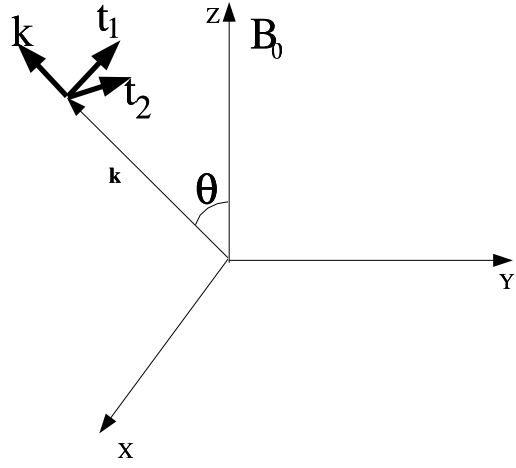}

\caption{\label{Fig:Alfven} Basis vectors for MHD waves. Compressible waves
have components along $\mathbf{k}$, while Alfvén waves have components
along $\mathbf{t}_{1}$and $\mathbf{t}_{2}$. }
\end{figure}

The linearized MHD equations are (cf. Eqs. {[}\ref{eq:MHDBx_compressible},
\ref{eq:MHDux_compressible}, \ref{eq:mass_conservation}{]})\begin{eqnarray*}
\frac{\partial\mathbf{u}}{\partial t}-\left(\mathbf{B}_{0}\cdot\nabla\right)\mathbf{b} & = & -\frac{1}{\rho_{0}}\nabla p-\nabla\mathbf{B}_{0}\cdot\mathbf{b},\\
\frac{\partial\mathbf{b}}{\partial t}-\left(\mathbf{B}_{0}\cdot\nabla\right)\mathbf{u} & = & -\mathbf{B}_{0}\nabla\cdot\mathbf{u},\\
\frac{\partial\delta\rho}{\partial t}+\nabla\cdot\left(\rho_{0}\mathbf{u}\right) & = & 0.\end{eqnarray*}
We attempt a plane-wave solution for the above equations: \[
(\mathbf{u,b,}p,\delta\rho)=(\mathbf{u(k),b(k),}p(\mathbf{k)},\rho(\mathbf{k}))\exp\left(i\mathbf{k}\cdot\mathbf{x}-i\omega t\right).\]
Substitutions of these waves in the linearized equations yield\begin{eqnarray*}
\omega\mathbf{u\left(\mathbf{k}\right)}+\left(\mathbf{B}_{0}.\mathbf{k}\right)\mathbf{b\left(\mathbf{k}\right)} & = & \frac{1}{\rho_{0}}\mathbf{k}p\left(\mathbf{k}\right)+\mathbf{k}\left(\mathbf{B}_{0}\cdot\mathbf{b}\right)\\
\omega\mathbf{b\left(\mathbf{k}\right)}+\left(\mathbf{B}_{0}.\mathbf{k}\right)\mathbf{u\left(\mathbf{k}\right)} & = & \mathbf{B}_{0}\left(\mathbf{k}\cdot\mathbf{u\left(\mathbf{k}\right)}\right)\\
\omega\rho\left(\mathbf{k}\right)-\rho_{0}\mathbf{k}\cdot\mathbf{u\left(\mathbf{k}\right)} & = & 0\end{eqnarray*}
Let us solve the above equations in coordinate system $(\mathbf{k},\mathbf{t}_{1},\mathbf{t_{2}})$
shown in Fig. \ref{Fig:Alfven}. Here $\mathbf{t}_{1,2}$ are transverse
to $\mathbf{k}$, with $\mathbf{t}_{1}$ in $\mathbf{B_{0}}$-$\mathbf{k}$
plane, and $\mathbf{t}_{2}$ perpendicular to this plane. %
 The components of velocity and magnetic field along $\mathbf{t}_{1,2}$
are denoted by $u_{\perp}^{(1)},u_{\perp}^{(2)},b_{\perp}^{(1)},b_{\perp}^{(2)}$,
and along $\mathbf{k}$ are $u_{||}$ and $b_{||}$. The angle between
$\mathbf{B}_{0}$ and $\mathbf{k}$ is $\theta$. The equations along
the new basis vectors are\begin{eqnarray}
\omega u_{||}-\frac{C_{s}^{2}}{\omega}k^{2}u_{||} & = & B_{0}b_{\perp}^{(1)}k\sin{\theta}\label{eq:Alf1}\\
b_{||} & = & 0\label{eq:Alf2}\\
\omega u_{\perp}^{(1)}+B_{0}k\cos{\theta}b_{\perp}^{(1)} & = & 0\label{eq:Alf3}\\
\omega b_{\perp}^{(1)}+B_{0}k\cos{\theta}u_{\perp}^{(1)} & = & B_{0}k\sin{\theta}u_{||}\label{eq:Alf4}\\
\omega u_{\perp}^{(2)}+B_{0}k\cos{\theta}b_{\perp}^{(2)} & = & 0\label{eq:Alf5}\\
\omega b_{\perp}^{(2)}+B_{0}k\cos{\theta}u_{\perp}^{(2)} & = & 0\label{eq:Alf6}\end{eqnarray}
using $C_{s}=\sqrt{{p(k)/\rho(k)}}$. Note that $b_{||}=0$, which
also follows from $\nabla\cdot\mathbf{b}=0$. From the above equations
we can infer the following basic wave modes:

\begin{enumerate}
\item Alfvén wave (Incompressible Mode): Here $u_{||}=u_{\perp}^{(1)}=b_{\perp}^{(1)}=0$,
and $u_{\perp}^{(2)}\ne0,b_{\perp}^{(2)}\ne0$, and the relevant equations
are (\ref{eq:Alf5}, \ref{eq:Alf6}). There are two solutions, which
correspond to waves travelling antiparallel and parallel to the mean
magnetic field with phase velocity $\pm C_{A}\cos{\theta}$ ($C_{A}=B_{0}$).
For these waves thermal and magnetic pressures are constants. These
waves are also called shear Alfvén waves.
\item Pseudo Alfvén wave (Incompressible Mode): Here $u_{||}=u_{\perp}^{(2)}=b_{\perp}^{(2)}=0$,
and $u_{\perp}^{(1)}\ne0,b_{\perp}^{(1)}\ne0$, and the relevant equations
are (\ref{eq:Alf3}, \ref{eq:Alf4}). The two solutions correspond
to waves moving antiparallel and parallel to the mean magnetic field
with velocity $\pm C_{A}\cos{\theta}$.
\item Compressible Mode (Purely Fluid): Here $u_{\perp}^{(1)}=b_{\perp}^{(1)}=u_{\perp}^{(2)}=b_{\perp}^{(2)}=0$,
and $u_{||}\ne0$, and the relevant equation is (\ref{eq:Alf1}).
This is the sound wave in fluid arising due to the fluctuations of
thermal pressure only.
\item MHD compressible Mode: Here $u_{\perp}^{(2)}=b_{\perp}^{(2)}=0$,
and $u_{\perp}^{(1)}\ne0,b_{\perp}^{(1)}\ne0$, $u_{||}\ne0$. Clearly,
$u_{||}$ is coupled to $b_{\perp}^{(1)}$ as evident from Eqs. (\ref{eq:Alf1},
\ref{eq:Alf4}). Solving Eqs. (\ref{eq:Alf1}, \ref{eq:Alf3}, \ref{eq:Alf4})
yields\[
\omega^{4}-\omega^{2}k^{2}\left(C_{s}^{2}+C_{A}^{2}\right)+C_{A}^{2}C_{S}^{2}k^{4}\cos^{2}{\theta}=0.\]
Hence, the two compressible modes move with velocities\begin{equation}
V_{ph}^{2}=\frac{1}{2}\left[\left(C_{A}^{2}+C_{s}^{2}\right)\pm\sqrt{\left(C_{A}^{2}+C_{s}^{2}\right)^{2}-4C_{s}^{2}C_{A}^{2}\cos\theta}\right].\label{eq:Fast_slow_vel}\end{equation}
 The faster between the two is called fast wave, and the other one
is called slow waves. The pressure variation for these waves are provided
by both thermal and magnetic pressure. For details on these waves,
refer to Sisco \cite{Sisc:book} and Priest \cite{Prie:book}.
\end{enumerate}
Turbulent flow contains many interacting waves, and the solution cannot
be written in a simple way. A popular approach to analyze the turbulent
flows is to use statistical tools. We will describe below the application
of statistical methods to turbulence.

\subsection{Necessity for Statistical Theory of Turbulence}

In turbulent fluid the field variables are typically random both in
space and time. Hence the exact solutions given initial and boundary
conditions will not be very useful even when they were available (they
are not!). However statistical averages and probability distribution
functions are reproducible in experiments under steady state, and
they shed important light on the dynamics of turbulence. For this
reason many researchers study turbulence statistically. The idea is
to use the tools of statistical physics for understanding turbulence.
Unfortunately, only systems at equilibrium or near equilibrium have
been understood reasonably well, and a good understanding of nonequilibrium
systems (turbulence being one of them) is still lacking. 

The statistical description of turbulent flow starts by dividing the
field variables into mean and fluctuating parts. Then we compute averages
of various functions of fluctuating fields. There are three types
are averages: ensemble, temporal, and spatial averages. Ensemble averages
are computed by considering a large number of identical systems and
taking averages at corresponding instants over all these systems.
Clearly, ensemble averaging demands heavily in experiments and numerical
simulations. So, we resort to temporal and/or spatial averaging. Temporal
averages are computed by measuring the quantity of interest at a point
over a long period and then averaging. Temporal averages make sense
for steady flows. Spatial averages are computed by measuring the quantity
of interest at various spatial points at a given time, and then averaging.
Clearly, spatial averages are meaningful for homogeneous systems.
Steady-state turbulent systems are generally assumed to be ergodic,
for which the temporal average is equal to the ensemble average \cite{Fris:book}. 

As discussed above, certain symmetries like homogeneity help us in
statistical description. Formally, homogeneity \emph{}indicates that
the average properties do not vary with absolute position in a particular
direction, but depends only on the separation between points. For
example, a homogeneous two-point correlation function is\[
\left\langle u_{i}(\mathbf{x},t)u_{j}(\mathbf{x'},t)\right\rangle =C_{ij}(\mathbf{x-x'},t)=C_{ij}(\mathbf{r},t).\]
Similarly, \emph{s}tationarity or \emph{}steady-state implies that
average properties depend on time difference, not on the absolute
time. That is, \[
\left\langle u_{i}(\mathbf{x},t)u_{j}(\mathbf{x},t')\right\rangle =C_{ij}(\mathbf{x},t-t').\]
 Another important symmetry is isotropy. A system is said to be isotropic
if its average properties are invariant under rotation. For isotropic
systems\[
\left\langle u_{i}(\mathbf{x},t)u_{j}(\mathbf{x'},t)\right\rangle =C_{ij}(\left|\mathbf{x-x'}\right|,t)=C_{ij}(\left|\mathbf{r}\right|,t).\]
Isotropy reduces the number of independent correlation functions.
Batchelor \cite{BatcTurb:book} showed that the isotropic two-point
correlation could be written as\[
C_{ij}(\mathbf{r})=C^{(1)}(r)r_{i}r_{j}+C^{(2)}(r)\delta_{ij}\]
where $C^{(1)}$ and $C^{(2)}$ are even functions of $r=|\mathbf{r}|.$
Hence we have reduced the independent correlation functions to two.
For incompressible flows, there is only one independent correlation
function \cite{BatcTurb:book}.

In the previous subsection we studied the global conserved quantities.
We revisit those quantities in presence of mean magnetic field. Note
that mean flow velocity can be set to zero because of Galilean invariance,
but the same trick cannot be used for the mean magnetic field. Matthaeus
and Goldstein \cite{MattGold} showed that the total energy and cross
helicity formed using the fluctuating fields are conserved. We denote
the fluctuating magnetic energy by $E^{b}$, in contrast to total
magnetic energy $E^{B}$. The magnetic helicity $\int\mathbf{a}\cdot\mathbf{b}/2$
is not conserved, but $\mathbf{B}_{0}\cdot<\mathbf{A}>+\int\mathbf{a}\cdot\mathbf{b}/2$
instead is conserved.

In turbulent fluid, fluctuations of all scales exist. Therefore, it
is quite convenient to use Fourier basis for the representation of
turbulent fluid velocity and magnetic field. Note that in recent times
another promising basis called wavelet is becoming popular. In this
paper we focus our attention on Fourier expansion, which is the topic
of the next subsection.

\subsection{Turbulence Equations in Spectral Space}

Turbulent fluid velocity $\mathbf{u}(\mathbf{x},t)$ is represented
in Fourier space as\begin{eqnarray*}
\mathbf{u}\left(\mathbf{x},t\right) & = & \int\frac{d\mathbf{k}}{(2\pi)^{d}}\mathbf{u}\left(\mathbf{k},t\right)\exp\left(i\mathbf{k\cdot x}\right)\\
\mathbf{u(}\mathbf{k},t) & = & \int d\mathbf{x}\mathbf{u}\left(\mathbf{x},t\right)\exp\left(-i\mathbf{k\cdot x}\right)\end{eqnarray*}
where $d$ is the space dimensionality. 

In Fourier space, the equations for \emph{incompressible} MHD are
\cite{BiskTurb:book}\begin{eqnarray}
\left(\frac{\partial}{\partial t}-i\left(\mathbf{B}_{0}\cdot\mathbf{k}\right)+\nu k^{2}\right)u_{i}(\mathbf{k},t) & = & -ik_{i}p_{tot}\left(\mathbf{k},t\right)-ik_{j}\int\frac{d\mathbf{p}}{(2\pi)^{d}}u_{j}(\mathbf{k-p},t)u_{i}(\mathbf{p},t)\nonumber \\
 &  & +ik_{j}\int\frac{d\mathbf{p}}{(2\pi)^{d}}b_{j}(k-\mathbf{p},t)b_{i}(\mathbf{p},t),\label{eq:MHDuk}\\
\left(\frac{\partial}{\partial t}-i\left(\mathbf{B}_{0}\cdot\mathbf{k}\right)+\eta k^{2}\right)b_{i}(\mathbf{k},t) & = & -ik_{j}\int\frac{d\mathbf{p}}{(2\pi)^{d}}u_{j}(\mathbf{k-p},t)b_{i}(\mathbf{p},t)\nonumber \\
 &  & +ik_{j}\int\frac{d\mathbf{p}}{(2\pi)^{d}}b_{j}(\mathbf{k-p},t)u_{i}(\mathbf{p},t),\label{eq:MHDbk}\end{eqnarray}
with the following constrains \begin{eqnarray*}
\mathbf{k}\cdot\mathbf{u\left(\mathbf{k}\right)} & = & 0,\\
\mathbf{k}\cdot\mathbf{b\left(\mathbf{k}\right)} & = & 0\end{eqnarray*}
The substitution of the incompressibility condition $\mathbf{k}\cdot\mathbf{u\left(\mathbf{k}\right)}=0$
into Eq. (\ref{eq:MHDuk}) yields the following expression for the
pressure field \[
p\left(\mathbf{k}\right)=-\frac{k_{i}k_{j}}{k^{2}}\int\frac{d\mathbf{p}}{(2\pi)^{d}}\left[u_{j}(\mathbf{k-p},t)u_{i}(\mathbf{p},t)-b(\mathbf{k-p},t)b_{i}(\mathbf{p},t)\right].\]
Note that the density field has been taken to be a constant, and has
been set equal to 1. 

It is also customary to write the evolution equations symmetrically
in terms of $\mathbf{p}$ and \textbf{$\mathbf{k-p}$} variables.
The symmetrized equations are

\begin{eqnarray}
\left(\frac{\partial}{\partial t}-i\left(\mathbf{B}_{0}\cdot\mathbf{k}\right)+\nu k^{2}\right)u_{i}(\mathbf{k},t) & = & -\frac{i}{2}P_{ijm}^{+}(\mathbf{k})\int\frac{d\mathbf{p}}{(2\pi)^{d}}[u_{j}(\mathbf{p},t)u_{m}(\mathbf{k-p},t)\nonumber \\
 &  & \hspace{3cm}-b_{j}(\mathbf{p},t)b_{m}(\mathbf{k-p},t)],\label{eq:MHDuk_P}\\
\left(\frac{\partial}{\partial t}-i\left(\mathbf{B}_{0}\cdot\mathbf{k}\right)+\eta k^{2}\right)b_{i}(\mathbf{k},t) & = & -iP_{ijm}^{-}(\mathbf{k})\int\frac{d\mathbf{p}}{(2\pi)^{d}}[u_{j}(\mathbf{p},t)b_{m}(\mathbf{k-p},t)],\label{eq:MHDbk_P}\end{eqnarray}
where \begin{eqnarray*}
P_{ijm}^{+}(\mathbf{k}) & = & k_{j}P_{im}(\mathbf{k})+k_{m}P_{ij}(\mathbf{k});\\
P_{im}(\mathbf{k}) & = & \delta_{im}-\frac{k_{i}k_{m}}{k^{2}};\\
P_{ijm}^{-}(\mathbf{k}) & = & k_{j}\delta_{im}-k_{m}\delta_{ij}.\end{eqnarray*}

Alfvén waves are fundamental modes of incompressible MHD. It turns
out that the equations become more transparent when they are written
in terms of Elsässer variables $\mathbf{z^{\pm}=u\pm b}$, which {}``represent''
the amplitudes of Alfvénic fluctuations with positive and negative
correlations. Note that no pure wave exist in turbulent medium, but
the interactions can be conveniently written in terms of these variables.
The MHD equations in terms of $\mathbf{z^{\pm}}$ are\begin{eqnarray}
\left(\frac{\partial}{\partial t}\mp i\left(\mathbf{B}_{0}\cdot\mathbf{k}\right)+\nu_{+}k^{2}\right)z_{i}^{\pm}(\mathbf{k})+\nu_{-}k^{2}z_{i}^{\mp}\left(\mathbf{k}\right) & = & -iM_{ijm}(\mathbf{k})\int d\mathbf{p}z_{j}^{\mp}(\mathbf{p})z_{m}^{\pm}(\mathbf{k-p}),\label{eq:MHDzk}\\
k_{i}z_{i}^{\pm}(\mathbf{k}) & = & 0,\nonumber \end{eqnarray}
where $\nu_{\pm}=(\nu\pm\eta)/2$ and

\[
M_{ijm}(\mathbf{k})=k_{j}P_{im}(\mathbf{k}).\]
From the Eq. (\ref{eq:MHDzk}) it is clear that the interactions are
between $\mathbf{z^{+}}$ and $\mathbf{z^{-}}$ modes. 

Energy and other second-order quantities play important roles in MHD
turbulence. For a homogeneous system these quantities are defined
as \[
\left\langle X_{i}(\mathbf{k},t)Y_{j}(\mathbf{k'},t)\right\rangle =C_{ij}^{XY}(\mathbf{k},t)(2\pi)^{d}\delta(\mathbf{k+k'}),\]
where $\mathbf{X,Y}$ are vector fields representing $\mathbf{u}$,
$\mathbf{b},$ or $\mathbf{z}^{\pm}$. The spectrum is also related
to correlation function in real space\[
C_{ij}^{XY}(\mathbf{r})=\int\frac{d\mathbf{k}}{(2\pi)^{d}}C_{ij}^{XY}\left(\mathbf{k}\right)\exp\left(i\mathbf{k\cdot r}\right).\]

When mean magnetic field is absent, or its effects are ignored, then
we can take $C_{ij}^{XY}(\mathbf{k})$ to be an isotropic tensor,
and it can be written as \cite{BatcTurb:book}\begin{equation}
C_{ij}^{XY}(\mathbf{k})=P_{ij}(\mathbf{k})C^{XY}(k).\label{eq:Cij_isotropic}\end{equation}
When turbulence is isotropic and $\mathbf{X=Y}$, then a quantity
called 1D spectrum or reduced spectrum $E^{X}(k)$ defined below is
very useful.\begin{eqnarray*}
E^{X}=\frac{1}{2}\left\langle X^{2}\right\rangle  & =\frac{1}{2} & \int\frac{d\mathbf{k}}{(2\pi)^{d}}C_{ii}^{XX}\left(\mathbf{k}\right)\\
\int E^{X}(k)dk & = & \frac{1}{2}\int dk\frac{S_{d}k^{d-1}}{(2\pi)^{d}}P_{ii}(\mathbf{k})C^{XX}\left(\mathbf{k}\right)\\
 & = & \int dk\frac{S_{d}k^{d-1}(d-1)}{2(2\pi)^{d}}C^{XX}\left(\mathbf{k}\right),\end{eqnarray*}
where $S_{d}=2\pi^{d/2}/\Gamma{(d/2)}$ is the area of $d-$dimensional
unit sphere. Therefore,\begin{equation}
E^{X}(k)=C^{XX}(\mathbf{k})k^{d-1}\frac{S_{d}(d-1)}{2(2\pi)^{d}}.\label{eq:E(k)_eq_C(k)}\end{equation}
Note that the above formula is valid only for isotropic turbulence.
We have tabulated all the important spectra of MHD turbulence in Table
\ref{table:Spectra}. The vector potential $\mathbf{A=A_{0}+a}$,
where $\mathbf{A_{0}}$ is the mean field, and $\mathbf{a}$ is the
fluctuation. %
\begin{table}

\caption{\label{table:Spectra} Various Spectra of MHD Turbulence }

\begin{tabular}{|c|c|c|c|}
\hline 
Quantity&
Symbol&
Derived from&
Symbol for 1D\tabularnewline
\hline
\hline 
Kinetic energy spectrum&
$C^{uu}\left(\mathbf{k}\right)$&
$\left\langle u_{i}(\mathbf{k})u_{j}(\mathbf{k'})\right\rangle $&
$E^{u}$$\left(k\right)$\tabularnewline
\hline 
Magnetic energy spectrum&
$C^{bb}$$\left(\mathbf{k}\right)$&
$\left\langle b_{i}(\mathbf{k})b_{j}(\mathbf{k'})\right\rangle $&
$E^{b}\left(k\right)$\tabularnewline
\hline 
Total energy spectrum &
$C$$\left(\mathbf{k}\right)$&
$C^{uu}+C^{bb}$&
$E$$\left(k\right)$\tabularnewline
\hline 
Cross helicity spectrum&
$C^{ub}$$\left(\mathbf{k}\right)$&
$\left\langle u_{i}(\mathbf{k})b_{j}(\mathbf{k'})\right\rangle $&
$H_{c}\left(k\right)$\tabularnewline
\hline 
Elsässer variable spectrum &
$C^{\pm\pm}$$\left(\mathbf{k}\right)$&
$\left\langle z_{i}^{\pm}(\mathbf{k})z_{j}^{\pm}(\mathbf{k'})\right\rangle $&
$E^{\pm}$$\left(k\right)$\tabularnewline
\hline 
Elsässer variable spectrum&
$C^{\pm\mp}\left(\mathbf{k}\right)$&
$\left\langle z_{i}^{\pm}(\mathbf{k})z_{j}^{\mp}(\mathbf{k'})\right\rangle $&
$E^{R}\left(k\right)$\tabularnewline
\hline 
Enstrophy spectrum&
$\Omega\left(\mathbf{k}\right)$&
$\left\langle \omega_{i}(\mathbf{k})\omega_{j}(\mathbf{k'})\right\rangle $&
$\Omega\left(k\right)$\tabularnewline
\hline 
Mean-square vector pot. spectrum&
$A2$$\left(\mathbf{k}\right)$&
$\left\langle a_{i}(\mathbf{k})a_{j}(\mathbf{k'})\right\rangle $&
$A2$$\left(k\right)$\tabularnewline
\hline
\end{tabular}
\end{table}

The global quantities defined in Table \ref{table:Globals} are related
to the spectra defined in Table \ref{table:Spectra} by Perceval's
theorem \cite{BatcTurb:book}. Since the fields are homogeneous, Fourier
integrals are not well defined. In Appendix \ref{sec:Fourier-Series}
we show that energy spectra defined using correlation functions are
still meaningful because correlation functions vanish at large distances.
We consider energy per unit volume, which are finite for homogeneous
turbulence. As an example, the kinetic energy per unit volume is related
to energy spectrum in the following manner:

\[
\frac{1}{L^{d}}\int d\mathbf{x}\frac{1}{2}\left\langle u^{2}\right\rangle =\frac{1}{2}\int\frac{d\mathbf{k}}{(2\pi)^{d}}C_{ii}(\mathbf{k})=\int E^{u}(k)dk\]
Similar identities can be written for other fields. 

In three dimensions we have two more important quantities, magnetic
and kinetic helicities. In Fourier space magnetic helicity $H_{M}(\mathbf{k})$
is defined using\[
\left\langle a_{i}\left(\mathbf{k},t\right)b_{j}\left(\mathbf{k'},t\right)\right\rangle =P_{ij}\left(\mathbf{k}\right)H_{M}(\mathbf{k})(2\pi)^{d}\delta(\mathbf{k+k'})\]
The total magnetic helicity $H_{M}$ can be written in terms of\begin{eqnarray*}
H_{M} & = & \frac{1}{2}\left\langle \mathbf{a}(\mathbf{x})\cdot\mathbf{b}(\mathbf{x})\right\rangle \\
 & =\frac{1}{2} & \int\frac{d\mathbf{k}}{(2\pi)^{d}}\frac{d\mathbf{k'}}{(2\pi)^{d}}\left\langle \mathbf{a}(\mathbf{k})\cdot\mathbf{b}(\mathbf{k'})\right\rangle \\
 & = & \int\frac{d\mathbf{k}}{(2\pi)^{d}}H_{M}(\mathbf{k})\\
 & = & \int dkH_{M}(k)\end{eqnarray*}
Therefore, one dimensional magnetic helicity $H_{M}$ is\[
H_{M}(k)=\frac{4\pi k^{2}}{(2\pi)^{3}}H_{M}(\mathbf{k}).\]
Using the definition $\mathbf{b}(\mathbf{k})=i\mathbf{k}\times\mathbf{a}(\mathbf{k})$,
we obtain\[
\left\langle b_{i}\left(\mathbf{k},t\right)b_{j}\left(\mathbf{k'},t\right)\right\rangle =\left[P_{ij}\left(\mathbf{k}\right)C^{bb}(\mathbf{k})-i\epsilon_{ijl}k_{l}H_{M}(\mathbf{k})\right](2\pi)^{d}\delta(\mathbf{k+k'})\]
The first term is the usual tensor described in Eq. (\ref{eq:Cij_isotropic}),
but the second term involving magnetic helicity is new. We illustrate
the second term with an example. If $\mathbf{k}$ is along $z$ axis,
then \[
b_{x}(\mathbf{k})b_{y}(\mathbf{k})=-ikH_{M}(\mathbf{k}).\]
 This is a circularly polarized field where $b_{x}$ and $b_{y}$
differ by a phase shift of $\pi/2$. Note that the magnetic helicity
breaks mirror symmetry.

A similar analysis for kinetic helicity shows that \[
\left\langle u_{i}\left(\mathbf{k},t\right)\Omega_{j}\left(\mathbf{k'},t\right)\right\rangle =P_{ij}\left(\mathbf{k}\right)H_{K}(\mathbf{k})(2\pi)^{d}\delta(\mathbf{k+k'})\]

\[
H_{K}=\frac{1}{2}\left\langle \mathbf{u}\cdot\mathbf{\Omega}\right\rangle =\int\frac{d\mathbf{k}}{(2\pi)^{d}}H_{K}(\mathbf{k})\]
and\[
\left\langle u_{i}\left(\mathbf{k},t\right)u_{j}\left(\mathbf{k'},t\right)\right\rangle =\left[P_{ij}\left(\mathbf{k}\right)C^{uu}(\mathbf{k})-i\epsilon_{ijl}k_{l}\frac{H_{M}(\mathbf{k})}{k^{2}}\right](2\pi)^{d}\delta(\mathbf{k+k'}).\]

We can Fourier transform time as well using \begin{eqnarray*}
\mathbf{u}\left(\mathbf{x},t\right) & = & \int d\hat{{k}}\mathbf{u}\left(\mathbf{k},\omega\right)\exp\left(i\mathbf{k\cdot x}-i\omega t\right)\\
\mathbf{u(}\mathbf{k},\omega) & = & \int d\mathbf{x}dt\mathbf{u}\left(\mathbf{x},t\right)\exp\left(-i\mathbf{k\cdot x}+i\omega t\right)\end{eqnarray*}
where $d\hat{{k}}=d\mathbf{k}d\omega/(2\pi)^{d+1}$. The resulting
MHD equations in $\hat{{k}}=(\mathbf{k},\omega)$ space are\begin{eqnarray}
\left(-i\omega+\nu k^{2}\right)u_{i}(\hat{{k}}) & = & -\frac{i}{2}P_{ijm}^{+}({\textbf{k}})\int_{\hat{{p}}+\hat{{q}}=\hat{{k}}}d\hat{{p}}\left[u_{j}(\hat{{p}})u_{m}(\hat{{q}})-b_{j}(\hat{{p}})b_{m}(\hat{{q}})\right],\label{eq:MHDukw}\\
\left(-i\omega+\eta k^{2}\right)b_{i}(\hat{{k}}) & = & -iP_{ijm}^{-}({\textbf{k}})\int_{\hat{{p}}+\hat{{q}}=\hat{{k}}}d\hat{{p}}\left[u_{j}(\hat{{p}})b_{m}(\hat{{q}})\right],\label{eq:MHDbkw}\end{eqnarray}
or,\begin{equation}
\left(-i\omega\mp i\left({\textbf{B}}_{0}\cdot{\textbf{k}}\right)+\nu_{+}k^{2}\right)z_{i}^{\pm}(\hat{k})+\nu_{-}k^{2}z_{i}^{\mp}(\hat{{k}})=-iM_{ijm}({\textbf{k}})\int d\hat{{p}}z_{j}^{\mp}(\hat{{p}})z_{m}^{\pm}(\hat{{k}}-\hat{{p}})\label{eq:MHDzkw}\end{equation}

After we have introduced the energy spectra and other second-order
correlation functions, we move on to discuss their evolution.

\subsection{Energy Equations }

The energy equation for general (compressible) Navier-Stokes is quite
complex. However, incompressible Navier-Stokes and MHD equations are
relatively simpler, and are discussed below.

From the evolution equations of fields, we can derive the following
spectral evolution equations for incompressible MHD\begin{eqnarray}
\left(\frac{\partial}{\partial t}+2\nu k^{2}\right)C^{uu}\left(\mathbf{k},t\right) & = & \frac{2}{\left(d-1\right)\delta\left(\mathbf{k+k'}\right)}\int_{\mathbf{k'+p+q=0}}\frac{d\mathbf{p}}{(2\pi)^{2d}}[-\Im\left\langle \left(\mathbf{k'}\cdot\mathbf{u(q)}\right)\left(\mathbf{u(p)}\cdot\mathbf{u(k')}\right)\right\rangle \nonumber \\
 &  & \hspace{3cm}+\Im\left\langle \left(\mathbf{k'}\cdot\mathbf{b(q)}\right)\left(\mathbf{b(p)}\cdot\mathbf{u(k')}\right)\right\rangle ],\label{eq:Cuu(k)vst}\\
\left(\frac{\partial}{\partial t}+2\eta k^{2}\right)C^{bb}\left(\mathbf{k},t\right) & = & \frac{2}{\left(d-1\right)\delta\left(\mathbf{k+k'}\right)}\int_{\mathbf{k'+p+q=0}}\frac{d\mathbf{p}}{(2\pi)^{2d}}[-\Im\left\langle \left(\mathbf{k'}\cdot\mathbf{u(q)}\right)\left(\mathbf{b(p)}\cdot\mathbf{b(k')}\right)\right\rangle \nonumber \\
 &  & \hspace{3cm}+\Im\left\langle \left(\mathbf{k'}\cdot\mathbf{b(q)}\right)\left(\mathbf{u(p)}\cdot\mathbf{b(k')}\right)\right\rangle ],\label{eq:Cbb(k)vst}\end{eqnarray}
where $\Im$ stands for the imaginary part. Note that $\mathbf{k'+p+q=0}$
and $\mathbf{k'=-k}$. In Eq. (\ref{eq:Cuu(k)vst}) the first term
in the RHS provides the energy transfer from the velocity modes to
$\mathbf{u(k)}$ mode, and the second term provides the energy transfer
from the magnetic modes to $\mathbf{u(k)}$ mode. While in Eq. (\ref{eq:Cbb(k)vst})
the first term in the RHS provides the energy transfer from the magnetic
modes to $\mathbf{b(k)}$ mode, and the second term provides the energy
transfer from the velocity modes to $\mathbf{b(k)}$ mode. Note that
pressure couples with compressible modes, hence it is absent in the
incompressible equations. Simple algebraic manipulation shows that
the mean magnetic field also disappears in the energy equation. In
a finite box, using $\left\langle \left|\mathbf{u}(\mathbf{k})\right|^{2}\right\rangle =C(\mathbf{k})/((d-1)L^{d})$,
and $\delta(\mathbf{k})(2\pi)^{d}=L^{d}$ (see Appendix A), we can
show that\begin{eqnarray*}
\left(\frac{\partial}{\partial t}+2\nu k^{2}\right)\frac{1}{2}\left\langle \left|\mathbf{u}(\mathbf{k})\right|^{2}\right\rangle  & = & \sum[-\Im\left\langle \left(\mathbf{k'}\cdot\mathbf{u(q)}\right)\left(\mathbf{u(p)}\cdot\mathbf{u(k')}\right)\right\rangle +\Im\left\langle \left(\mathbf{k'}\cdot\mathbf{b(q)}\right)\left(\mathbf{b(p)}\cdot\mathbf{u(k')}\right)\right\rangle ],\\
\left(\frac{\partial}{\partial t}+2\nu k^{2}\right)\frac{1}{2}\left\langle \left|\mathbf{b}(\mathbf{k})\right|^{2}\right\rangle  & = & \sum[-\Im\left\langle \left(\mathbf{k'}\cdot\mathbf{u(q)}\right)\left(\mathbf{b(p)}\cdot\mathbf{b(k')}\right)\right\rangle +\Im\left\langle \left(\mathbf{k'}\cdot\mathbf{b(q)}\right)\left(\mathbf{u(p)}\cdot\mathbf{b(k')}\right)\right\rangle ].\end{eqnarray*}

Many important quantities, e.g. energy fluxes, can be derived from
the energy equations. We will discuss these quantities in the next
section.

\section{Mode-to-mode Energy Transfers and Fluxes in MHD Turbulence \label{sec:Mode-to-mode-Energy-Transfer}}

In turbulence energy exchange takes place between various Fourier
modes because of nonlinear interactions. Basic interactions in turbulence
involves a wavenumber triad $(\mathbf{k',p,q)}$ satisfying $\mathbf{k'+p+q=0}$.
Usually, energy gained by a mode in the triad is computed using the
\emph{combined energy transfer} from the other two modes \cite{Lesi:book}.
Recently Dar et al. \cite{Dar:flux} devised a new scheme to compute
the energy transfer rate between two modes in a triad\emph{,} and
called it \emph{{}``mode-to-mode energy transfer}''. They computed
energy cascade rates and energy transfer rates between two wavenumber
shells using this scheme. We will review these ideas in this section.
Note that we are considering only the interactions of incompressible
modes.

\subsection{{}``Mode-to-Mode'' Energy Transfer in Fluid Turbulence \label{sub:'Mode-to-Mode'-fluid}}

In this subsection we discuss evolution of energy in turbulent fluid
\emph{in a periodic bo}x. The equation for MHD will be discussed subsequently.
We consider ideal case where viscous dissipation is zero $(\nu=0).$
The equations are given in Lesieur \cite{Lesi:book}\begin{eqnarray}
\frac{\partial}{\partial t}\frac{1}{2}\left|u\left(\mathbf{k'}\right)\right|^{2} & =\sum_{\mathbf{k'+p+q}=0} & -\frac{1}{2}\Im\left[\left(\mathbf{k'\cdot u(q)}\right)\left(\mathbf{u(k')}\cdot\mathbf{u(p)}\right)+\left(\mathbf{k'\cdot u(p)}\right)\left(\mathbf{u(k')}\cdot\mathbf{u(q)}\right)\right],\end{eqnarray}
where $\Im$ denotes the imaginary part. Note that the pressure does
not appear in the energy equation because of the incompressibility
condition.

Consider a case in which only three modes $\mathbf{u(k'),u(p),u(q)}$,
and their conjugates are nonzero. Then the above equation yields \begin{eqnarray}
\frac{\partial}{\partial t}\frac{1}{2}\left|u\left(\mathbf{k'}\right)\right|^{2} & = & \frac{1}{2}S(\mathbf{k'|p,q)},\label{eq:Fluid_triad}\end{eqnarray}
where \begin{eqnarray}
S(\mathbf{k'|p,q)} & = & -\Im\left[\left(\mathbf{k'\cdot u(q)}\right)\left(\mathbf{u(k')}\cdot\mathbf{u(p)}\right)+\left(\mathbf{k'\cdot u(p)}\right)\left(\mathbf{u(k')}\cdot\mathbf{u(q)}\right)\right].\label{eq:Fluid_Sk,pq}\end{eqnarray}
Lesieur and other researchers physically interpreted $S(\mathbf{k'|p,q)}$
as the \emph{combined energy transfer rate} from modes $\mathbf{p}$
and $\mathbf{q}$ to mode $\mathbf{k'}$. The evolution equations
for $\left|u\left(\mathbf{p}\right)\right|^{2}$ and $\left|u\left(\mathbf{q}\right)\right|^{2}$
are similar to that for $\left|u\left(\mathbf{k'}\right)\right|^{2}$.
By adding the energy equations for all three modes, we obtain

\begin{eqnarray*}
\frac{\partial}{\partial t}\left[\left|u\left(\mathbf{k'}\right)\right|^{2}+\left|u\left(\mathbf{p}\right)\right|^{2}+\left|u\left(\mathbf{q}\right)\right|^{2}\right]/2 & = & S(\mathbf{k'|p,q)}+S(\mathbf{p|q,k')}+S(\mathbf{q|k',p)}\\
 & = & \Im[\left(\mathbf{q\cdot u(q)}\right)\left(\mathbf{u(k')}\cdot\mathbf{u(p)}\right)\\
 &  & +\left(\mathbf{p\cdot u(p)}\right)\left(\mathbf{u(k')}\cdot\mathbf{u(q)}\right)\\
 &  & +\left(\mathbf{k'\cdot u(k')}\right)\left(\mathbf{u(p)}\cdot\mathbf{u(q)}\right)]\end{eqnarray*}
For incompressible fluid the right-hand-side is identically zero because
$\mathbf{k'\cdot u(k')}=0$. Hence the energy in each interacting
triad is conserved , i.e., \[
\left|u\left(\mathbf{k'}\right)\right|^{2}+\left|u\left(\mathbf{p}\right)\right|^{2}+\left|u\left(\mathbf{q}\right)\right|^{2}=const.\]

The question is whether we can derive an expression for mode-to-mode
energy transfer rates from mode $\mathbf{p}$ to mode \emph{$\mathbf{k'}$,}
and from mode \emph{$\mathbf{q}$} to mode \emph{$\mathbf{k'}$} separately.
Dar et al. \cite{Dar:flux} showed that it is meaningful to talk about
energy transfer rate between two modes. They derived an expression
for the mode-to-mode energy transfer, and showed it to be unique apart
from an irrelevant arbitrary constant. They referred to this quantity
as {}``mode-to-mode energy transfer''. Even though they talk about
mode-to-mode transfer, they are still within the framework of triad
interaction, i.e., a triad is still the fundamental entity of interaction.

\subsubsection{Definition of Mode-to-Mode Transfer in a Triad}

Consider a triad ($\mathbf{k'|p,q}$). Let the quantity $R^{uu}(\mathbf{k'|p|q})$
denote the energy transferred from mode \textbf{$\mathbf{p}$} to
mode \textbf{$\mathbf{k'}$} with mode \textbf{$\mathbf{q}$} playing
the role of a mediator. We wish to obtain an expression for $R$.

The $R$'s should satisfy the following relationships : 

\begin{enumerate}
\item The sum of energy transfer from mode \textbf{$\mathbf{p}$} to mode
\textbf{$\mathbf{k'}$ $(R^{uu}(\mathbf{k'|p|q})$),} and from mode
\textbf{$\mathbf{q}$} to mode \textbf{$\mathbf{k'}$} $(R^{uu}(\mathbf{k'|p|q}))$
should be equal to the total energy transferred to mode \textbf{$\mathbf{k'}$}
from modes \textbf{$\mathbf{p}$} and $\mathbf{q}$, i.e., $S^{uu}(\mathbf{k'|p,q})$
{[}see Eq.~(\ref{eq:Fluid_triad}){]}. That is, \begin{equation}
R^{uu}(\mathbf{k'|p|q})+R^{uu}(\mathbf{k'|q|p})=S^{uu}(\mathbf{k'|p,q}),\label{eq:Skpq}\end{equation}
\begin{equation}
R^{uu}(\mathbf{p|k',q})+R^{uu}(\mathbf{p|q|k'})=S^{uu}(\mathbf{p|k',q}),\end{equation}
\begin{equation}
R^{uu}(\mathbf{q|k'|p})+R^{uu}(\mathbf{q|p|k'})=S^{uu}(\mathbf{q|k',p}).\end{equation}

\item By definition, the energy transferred from mode \textbf{$\mathbf{p}$}
to mode \textbf{$\mathbf{k}'$}, $R^{uu}(\mathbf{k'|p|q})$, will
be equal and opposite to the energy transferred from mode $\mathbf{k'}$
to mode $\mathbf{p}$, $R^{uu}(\mathbf{p|k'|q})$. Thus, \begin{equation}
R^{uu}(\mathbf{k'|p|q})+R^{uu}(\mathbf{p|k'|q})=0,\label{eq:Rkpq}\end{equation}
\begin{equation}
R^{uu}(\mathbf{k'|q|p})+R^{uu}(\mathbf{q|k'|p})=0,\label{eq:Rkqp}\end{equation}
\begin{equation}
R^{uu}(\mathbf{p|q|k'})+R^{uu}(\mathbf{q|p|k'})=0.\label{eq:Rpqk}\end{equation}

\end{enumerate}
These are six equations with six unknowns. However, the value of the
determinant formed from the Eqs.~(\ref{eq:Skpq}-\ref{eq:Rpqk})
is zero. Therefore we cannot find unique $R$'s given just these equations.
In the following discussion we will study the set of solutions of
the above equations.

\subsubsection{Solutions of equations of mode-to-mode transfer \label{sub:Solutions-mode-to-mode}}

Consider a function\begin{equation}
S^{uu}(\mathbf{k'|p|q})=-\Im\left(\left[\mathbf{k'}\cdot\mathbf{u}(\mathbf{q})\right]\left[\mathbf{u}(\mathbf{k}')\cdot\mathbf{u}(\mathbf{q})\right]\right)\label{eq:Suu_Dar}\end{equation}
Note that $S^{uu}(\mathbf{k'|p|q})$ is altogether different function
compared to $S(\mathbf{k'|p,q})$. In the expression for $S^{uu}(\mathbf{k'|p|q})$,
the field variables with the first and second arguments are dotted
together, while the field variable with the third argument is dotted
with the first argument. 

It is very easy to check that $S^{uu}(\mathbf{k'|p|q})$ satisfy the
Eqs. (\ref{eq:Skpq}-\ref{eq:Rpqk}). Note that $S^{uu}(\mathbf{k'|p|q})$
satisfy the Eqs. (\ref{eq:Rkpq}-\ref{eq:Rpqk}) because of incompressibility
condition. The above results implies that the set of $S^{uu}(||)$'s
is \textit{one instance} of the $R^{uu}(||)$'s. However, $S^{uu}(\mathbf{k'|p|q})$
is not a unique solution. If another solution $R^{uu}(\mathbf{k'|p|q})$
differs from $S(\mathbf{k'|p|q})$ by an arbitrary function $X_{\Delta}$,
i.e., $R^{uu}(\mathbf{k'|p|q})=S^{uu}(\mathbf{k'|p|q})+X_{\Delta}$,
then by inspection we can easily see that the solution of Eqs.~(\ref{eq:Skpq}-\ref{eq:Rpqk})
must be of the form \begin{equation}
R^{uu}(\mathbf{k'|p|q})=S^{uu}(\mathbf{k'|p|q})+X_{\Delta}\end{equation}
\begin{equation}
R^{uu}(\mathbf{k'|q|p})=S^{uu}(\mathbf{k'|q|p})-X_{\Delta}\end{equation}
\begin{equation}
R^{uu}(\mathbf{p|k'|q})=S^{uu}(\mathbf{p|k'|q})-X_{\Delta}\end{equation}
\begin{equation}
R^{uu}(\mathbf{p|q|k'})=S(\mathbf{p|q|k'})+X_{\Delta}\end{equation}
\begin{equation}
R^{uu}(\mathbf{q|k'|p})=S(\mathbf{q|k'|p})+X_{\Delta}\end{equation}
\begin{equation}
R^{uu}(\mathbf{q|p|k'})=S(\mathbf{q|p|k'})-X_{\Delta}\end{equation}
 Hence, the solution differs from $S^{uu}(\mathbf{k'|p|q})$ by a
constant. %
\begin{figure}
\includegraphics[bb=0 500 380 755] {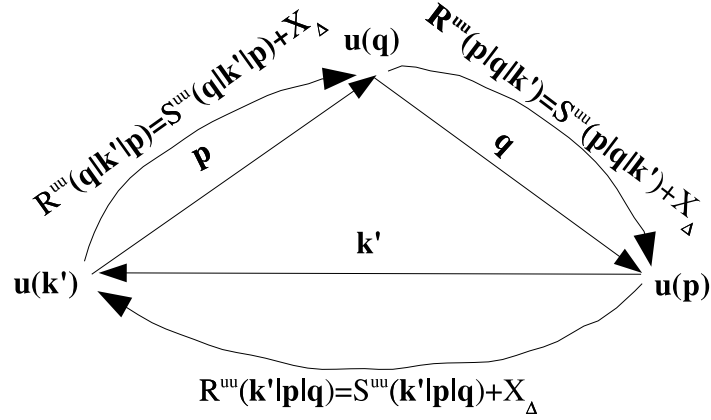}

\caption{\label{Fig:fluid-mode} Mode-to-mode energy transfer in fluid turbulence.
$S^{uu}({\textbf{k'|p|q}})$ represents energy transfer rate from
mode $\mathbf{u}(\mathbf{p})$ to mode $\mathbf{u}(\mathbf{k'})$
with the mediation of mode $\mathbf{u}(\mathbf{q})$. $X_{\Delta}$
is the arbitrary circulating transfer.}
\end{figure}
See Fig. \ref{Fig:fluid-mode} for illustration. A careful observation
of the figure indicates that the quantity $X_{\Delta}$ flows along
${\textbf{p}}\rightarrow{\mathbf{k'}}\rightarrow{\textbf{q}}\rightarrow{\textbf{p}}$,
circulating around the entire triad without changing the energy of
any of the modes. Therefore we will call it the \textit{Circulating
transfer}. Of the total energy transfer between two modes, $S^{uu}+X_{\Delta}$,
only $S^{uu}$ can bring about a change in modal energy. $X_{\Delta}$
transferred from mode \textbf{p} to mode \textbf{$\mathbf{k'}$} is
transferred back to mode \textbf{p} via mode \textbf{q}. Thus the
energy that is effectively transferred from mode \textbf{p} to mode
\textbf{$\mathbf{k'}$} is just $S^{uu}(\mathbf{k'|p|q})$. Therefore
$S^{uu}(\mathbf{k'|p|q})$ can be termed as the \textit{effective
mode-to-mode energy transfer} from mode \textbf{p} to mode $\mathbf{k'}$. 

Note that $X_{\Delta}$ can be a function of wavenumbers \textbf{$\mathbf{k',p,q}$,}
and the Fourier components $\mathbf{u(k')},\mathbf{u(p)},\mathbf{u(q)}$.
It may be possible to determine $X_{\Delta}$ using constraints based
on invariance, symmetries, etc. Dar et al. \cite{Dar:Modetomode}
attempted to obtain $X_{\Delta}$ using this approach, but could show
that $X_{\Delta}$ is zero to linear order in the expansion. However,
a general solution for $X_{\Delta}$ could not be found. Unfortunately,
$X_{\Delta}$ cannot be calculated even by simulation or experiment,
because we can experimentally compute only the energy transfer rate
to a mode, which is a sum of two mode-to-mode energy transfers. The
mode-to-mode energy transfer rate is really an abstract quantity,
somewhat similar to {}``gauges'' in electrodynamics. 

The terms $u_{j}\partial_{j}u_{i}$ and $u_{i}u_{j}\partial_{j}u_{i}$
are nonlinear terms in the Navier-Stokes equation and the energy equation
respectively. When we look at the formula (\ref{eq:Suu_Dar}) carefully,
we find that the $u_{j}(\mathbf{q})$ term is contracted with $k_{j}$
in the formula. Hence, $u_{j}$ field is the mediator in the energy
exchange between first $(u_{i})$ and third field $(u_{i})$ of $u_{i}u_{j}\partial_{j}u_{i}$.

In this following discussion we will compute the energy fluxes and
the shell-to-shell energy transfer rates using $S^{uu}(\mathbf{k'|p|q})$.

\subsection{Shell-to-Shell Energy Transfer in Fluid Turbulence Using Mode-to-mode
Formalism \label{sub:Shell-to-Shell-fluid}}

In turbulence energy transfer takes place from one region of the wavenumber
space to another region. Domaradzki and Rogallo \cite{Doma:Local2}
have discussed the energy transfer between two shells using the combined
energy transfer $S^{uu}({\textbf{k'|p,q}})$. They interpret the quantity
\begin{equation}
T_{nm}^{uu}=\frac{1}{2}\sum_{\mathbf{k'}\in n}\sum_{\mathbf{p}\in m}S^{uu}(\mathbf{k'|p,q}).\label{eq:shell_old_defn}\end{equation}
 as the rate of energy transfer from shell \textit{$m$} to shell
\textit{$n$} . Note that \textbf{$\mathbf{k'}$}-sum is over shell
$n$, $\mathbf{p}$-sum over shell $m$, and $\mathbf{k'+p+q}=0$.
However, Domaradzki and Rogallo \cite{Doma:Local2} themselves points
out that it may not be entirely correct to interpret the formula (\ref{eq:shell_old_defn})
as the shell-to-shell energy transfer. The reason for this is as follows.

\begin{figure}
\includegraphics[%
  scale=0.7,bb = 100 200 500 600]{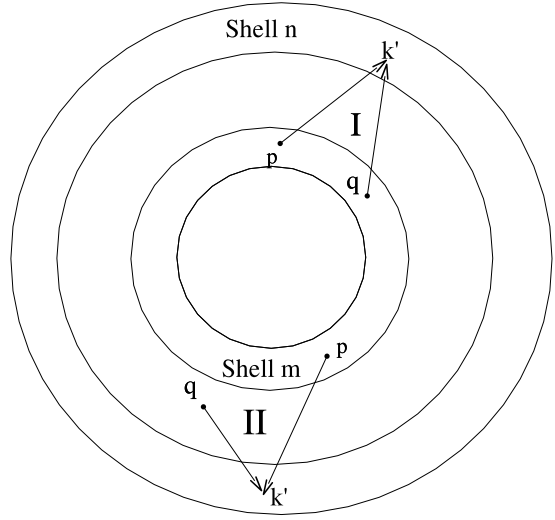}

\caption{\label{Fig:shell-to-shell} Shell-to-shell energy transfer from wavenumber-shell
$m$ to wavenumber-shell $n$. The triads involved in this process
fall in two categories: Type I, where both $\mathbf{p}$ and $\mathbf{q}$
are inside shell $m$, and Type II, where only $\mathbf{p}$ is inside
shell $m$. }
\end{figure}

In the energy transfer between two shells \textit{m} and \textit{n},
two types of wavenumber triads are involved, as shown in Fig. \ref{Fig:shell-to-shell}.
The real energy transfer from the shell \textit{$m$} to the shell
\textit{$n$} takes place through both $\mathbf{k'}$-$\mathbf{p}$
and $\mathbf{k'}$-$\mathbf{q}$ legs of triad I, but only through
$\mathbf{k'}$-$\mathbf{p}$ leg of triad II. But in Eq. (\ref{eq:shell_old_defn})
summation erroneously includes $\mathbf{k'}$-$\mathbf{q}$ leg of
triad II also along with the three legs given above. Hence  Domaradzki
and Ragallo's formalism \cite{Doma:Local2} do not yield totally correct
shell-to-shell energy transfers, as was pointed out by Domaradzki
and Rogallo themselves. We will show below how Dar et al.'s formalism
\cite{Dar:flux} overcomes this difficulty.

By definition of the the mode-to-mode transfer function $R^{uu}(\mathbf{k'|p|q})$,
the energy transfer from shell \textit{$m$} to shell \textit{$n$}
can be defined as \begin{equation}
T_{nm}^{uu}=\sum_{\mathbf{k'}\in n}\sum_{\mathbf{p}\in m}R^{uu}(\mathbf{k'|p|q})\label{eq:shell_new_defn}\end{equation}
 where the $\mathbf{k'}$-sum is over the shell $n$, and $\mathbf{p}$-sum
is over the shell $m$. The quantity $R^{uu}$ can be written as a
sum of an effective transfer $S^{uu}(\mathbf{k'|p|q})$ and a circulating
transfer $X_{\Delta}$. As discussed in the last section, the circulating
transfer $X_{\Delta}$ does not contribute to the energy change of
modes. From Figs. \ref{Fig:fluid-mode} and \ref{Fig:shell-to-shell}
we can see that $X_{\Delta}$ flows from the shell \textit{$m$} to
the shell \textit{$n$} and then flows back to \textit{$m$} indirectly
through the mode $\mathbf{q}$. Therefore the \textit{effective} energy
transfer from the shell \textit{m} to the shell \textit{n} is just
$S^{uu}(\mathbf{k'|p|q})$ summed over all the \textbf{$\mathbf{k'}$-}modes
in the shell \textit{$n$} and all the \textbf{$\mathbf{p}$-}modes
in the shell $m$, i.e., \begin{equation}
T_{nm}^{uu}=\sum_{\mathbf{k'}\in n}\sum_{\mathbf{p}\in m}S^{uu}(\mathbf{k'|p|q}).\label{eq:shell_eff_defn}\end{equation}
 Clearly, the energy transfer through $\mathbf{k'-q}$ of the triad
II of Fig. \ref{Fig:shell-to-shell} is not present in $T_{nm}^{uu}$
In Dar et al.'s formalism because $\mathbf{q}\notin m$. Hence, the
formalism of the mode-to-mode energy transfer rates provides us a
correct and convenient method to compute the shell-to-shell energy
transfer rates in fluid turbulence.

\subsection{Energy Cascade Rates in Fluid Turbulence Using Mode-to-mode Formalism
\label{sub:Energy-Cascade-Rates-fluid}}

The kinetic energy cascade rate (or flux) $\Pi$ in fluid turbulence
is defined as the rate of loss of kinetic energy by the modes inside
a sphere to the modes outside the sphere. Let $k_{0}$ be the radius
of the sphere under consideration. Kraichnan \cite{Krai:59}, Leslie
\cite{Lesl:book}, and others have computed the energy flux in fluid
turbulence using $S^{uu}(\mathbf{k'|p|q})$\begin{equation}
\Pi(k_{0})=-\sum_{|\mathbf{k}|<k_{0}}\sum_{|\mathbf{p}|>k_{0}}\frac{1}{2}S^{uu}(\mathbf{k'|p|q}).\label{eq:flux_Krai59}\end{equation}
 Although the energy cascade rate in fluid turbulence can be found
by the above formula, the mode-to-mode approach of Dar et al. \cite{Dar:flux}
provides a more natural way of looking at the energy flux. Since $R^{uu}(\mathbf{k'|p|q})$
represents energy transfer from $\mathbf{p}$ to $\mathbf{k'}$ with
\textbf{$\mathbf{q}$} as a mediator, we may alternatively write the
energy flux as \begin{equation}
\Pi(k_{0})=\sum_{|\mathbf{k'}|>k_{0}}\sum_{|\mathbf{p}|<k_{0}}R^{uu}(\mathbf{k'|p|q}).\label{eq:fluid_flux_Dar_R}\end{equation}
 However, $R^{uu}(\mathbf{k'|p|q})=S^{uu}(\mathbf{k'|p|q})+X_{\Delta}$,
and the circulating transfer $X_{\Delta}$ makes no contribution to
the the energy flux from the sphere because the energy lost from the
sphere through $X_{\Delta}$ returns to the sphere. Hence,\begin{equation}
\Pi(k_{0})=\sum_{|\mathbf{k}'|>k_{0}}\sum_{|\mathbf{p}|<k_{0}}S^{uu}(\mathbf{k'|p|q}).\label{eq:fluid_flux_Dar}\end{equation}
 Both the formulas given above, Eqs. (\ref{eq:flux_Krai59}) and (\ref{eq:fluid_flux_Dar}),
are equivalent as shown by Dar et al. \cite{Dar:Modetomode}.

Frisch \cite{Fris:book} has derived a formula for energy flux as\[
\Pi(k_{0})=\left\langle \mathbf{u_{k_{0}}^{<}}\cdot\left(\mathbf{u_{k_{0}}^{<}}\cdot\nabla\mathbf{u_{k_{0}}^{>}}\right)\right\rangle +\left\langle \mathbf{u_{k_{0}}^{<}}\cdot\left(\mathbf{u_{k_{0}}^{>}}\cdot\nabla\mathbf{u_{k_{0}}^{>}}\right)\right\rangle .\]
It is easy to see that the above formula is consistent with mode-to-mode
formalism. As discussed in the Subsection \ref{sub:Solutions-mode-to-mode},
the second field of both the terms are mediators in the energy transfer.
Hence in mode-to-mode formalism, the above formula will translate
to \[
\Pi(k_{0})=\sum_{k>k_{0}}\sum_{p<k_{0}}-\Im\left[\left(\mathbf{k'}\cdot\mathbf{u^{<}}(\mathbf{q})\right)\left(\mathbf{u^{<}}(\mathbf{p})\cdot\mathbf{u^{>}}(\mathbf{k'})\right)+\left(\mathbf{k}'\cdot\mathbf{u^{>}}(\mathbf{q})\right)\left(\mathbf{u^{<}}(\mathbf{p})\cdot\mathbf{u^{>}}(\mathbf{k'})\right)\right],\]
which is same as mode-to-mode formula (\ref{eq:fluid_flux_Dar}) of
Dar et al. \cite{Dar:flux}.

After discussion on energy transfers in fluid turbulence, we move
on to MHD turbulence.

\subsection{Mode-to-Mode Energy Transfer in MHD Turbulence \label{sub:Mode-to-Mode-Energy-Transfer-MHD}}

In Fourier space, the kinetic energy and magnetic energy evolution
equations are \cite{Stan:book}

\begin{equation}
\frac{\partial E^{u}(\mathbf{k})}{\partial t}+2\nu k^{2}E^{u}(\mathbf{k})=\sum_{\mathbf{k'+p+q}=0}\frac{1}{2}S^{uu}(\mathbf{k'|p,q})+\sum_{\mathbf{k'+p+q}=0}\frac{1}{2}S^{ub}(\mathbf{k'|p,q}),\label{eq:eu_mhd}\end{equation}

\begin{equation}
\frac{\partial E^{b}(\mathbf{k})}{\partial t}+2\mu k^{2}E^{b}(\mathbf{k})=\sum_{\mathbf{k'+p+q}=0}\frac{1}{2}S^{bb}(\mathbf{k'|p,q})+\sum_{\mathbf{k'+p+q}=0}\frac{1}{2}S^{bu}(\mathbf{k'|p,q}),\label{eq:eb_mhd}\end{equation}
 where $E^{u}(\mathbf{k})=|\mathbf{u}(\mathbf{k})|^{2}/2$ is the
kinetic energy, and $E^{b}(\mathbf{k})=|\mathbf{b}(\mathbf{k})|^{2}/2$
is the magnetic energy. The four nonlinear terms $S^{uu}(\mathbf{k'|p,q})$,
$S^{ub}(\mathbf{k'|p,q})$, $S^{bb}(\mathbf{k'|p,q})$ and $S^{bu}(\mathbf{k'|p,q})$
are

\begin{equation}
S^{uu}(\mathbf{k'|p,q})=-\Im\left(\mathbf{\left[\mathbf{k'\mathbf{.u(q)}}\right]\left[u(k').u(p)\right]+\left[k'.u(p)\right]\left[u(k').u(q)\right]}\right),\end{equation}
\begin{equation}
S^{bb}(\mathbf{k'|p,q})=-\Im\left(\mathbf{\left[\mathbf{k\mathbf{'.u(q)}}\right]\left[b(k').b(p)\right]+\left[k'.u(p)\right]\left[b(k').b(q)\right]}\right),\end{equation}
\begin{equation}
S^{ub}(\mathbf{k'|p,q})=\Im\left(\mathbf{\left[\mathbf{k\mathbf{'.b(q)}}\right]\left[u(k').b(p)\right]+\left[k'.b(p)\right]\left[u(k').b(q)\right]}\right),\end{equation}
\begin{equation}
S^{bu}(\mathbf{k'|p,q})=\Im\left(\mathbf{\left[\mathbf{k\mathbf{'.b(q)}}\right]\left[b(k').u(p)\right]+\left[k'.b(p)\right]\left[b(k').u(q)\right]}\right).\end{equation}
These terms are conventionally taken to represent the nonlinear transfer
from modes $\mathbf{p}$ and $\mathbf{q}$ to mode $\mathbf{k}'$
of a triad \cite{Stan:book,Lesi:book}. The term $S^{uu}(\mathbf{k'|p,q})$
represents the net transfer of kinetic energy from modes $\mathbf{p}$
and $\mathbf{q}$ to mode $\mathbf{k'}$. Likewise the term $S^{ub}(\mathbf{k'|p,q})$
is the net magnetic energy transferred from modes $\mathbf{p}$ and
$\mathbf{q}$ to the kinetic energy in mode $\mathbf{k'}$, whereas
$S^{bu}(\mathbf{k'|p,q})$ is the net kinetic energy transferred from
modes $\mathbf{p}$ and $\mathbf{q}$ to the magnetic energy in mode
$\mathbf{k'}$. The term $S^{bb}(\mathbf{k'|p,q})$ represents the
transfer of magnetic energy from modes $\mathbf{p}$ and \textbf{$\mathbf{q}$}
to mode $\mathbf{k}'$. 

Stani\u{s}i\'{c} \cite{Stan:book} showed that the nonlinear terms
satisfy the following detailed conservation properties: 

\begin{equation}
S^{uu}(\mathbf{k'|p,q})+S^{uu}(\mathbf{p|k',q})+S^{uu}(\mathbf{q}|\mathbf{k',p})=0,\label{eq:Suu_k,pq_conserve}\end{equation}

\begin{equation}
S^{bb}(\mathbf{k'|p,q})+S^{bb}(\mathbf{p|k',q})+S^{bb}(\mathbf{q}|\mathbf{k',p})=0,\label{eq:Sbb_k,pq_conserve}\end{equation}
 and \begin{equation}
S^{ub}(\mathbf{k'|p,q})+S^{ub}(\mathbf{p|k',q})+S^{ub}(\mathbf{q}|\mathbf{k',p})+S^{bu}(\mathbf{k'|p,q})+S^{bu}(\mathbf{p|k',q})+S^{bu}(\mathbf{q}|\mathbf{k',p})=0.\label{eq:Sub_k,pq_conserve}\end{equation}
 The Eqs. (\ref{eq:Suu_k,pq_conserve}, \ref{eq:Sbb_k,pq_conserve})
implies that kinetic/magnetic energy are transferred conservatively
between the velocity/magnetic modes of a wavenumber triad. The Eq.
(\ref{eq:Sub_k,pq_conserve}) implies that the cross-transfers of
kinetic and magnetic energy, $S^{ub}(\mathbf{k'|p,q})$ and $S^{bu}(\mathbf{k'|p,q})$,
within a triad are also energy conserving.

Dar et al. \cite{Dar:flux,Dar:Modetomode} provided an alternative
formalism called \emph{mode-to-mode energy transfer} for MHD turbulence.
This is a generalization fluid's mode-to-mode formalism described
in the previous subsection. We consider ideal MHD fluid $(\nu=\eta=0)$.
The basic unit of nonlinear interaction in MHD is a triad involving
modes $\mathbf{u(k'),u(p),u(q),b(k'),b(p)},\mathbf{b(q)}$ with $\mathbf{k'+p+q=0}$,
and the {}``mode-to-mode energy transfer'' is from velocity to velocity,
from magnetic to magnetic, from velocity to magnetic, and from magnetic
to velocity mode. We will discuss these transfers below.

\subsubsection{Velocity mode to velocity mode energy transfer}

\begin{figure}
\includegraphics[%
  scale=0.7,bb=0 290 470 755 ]{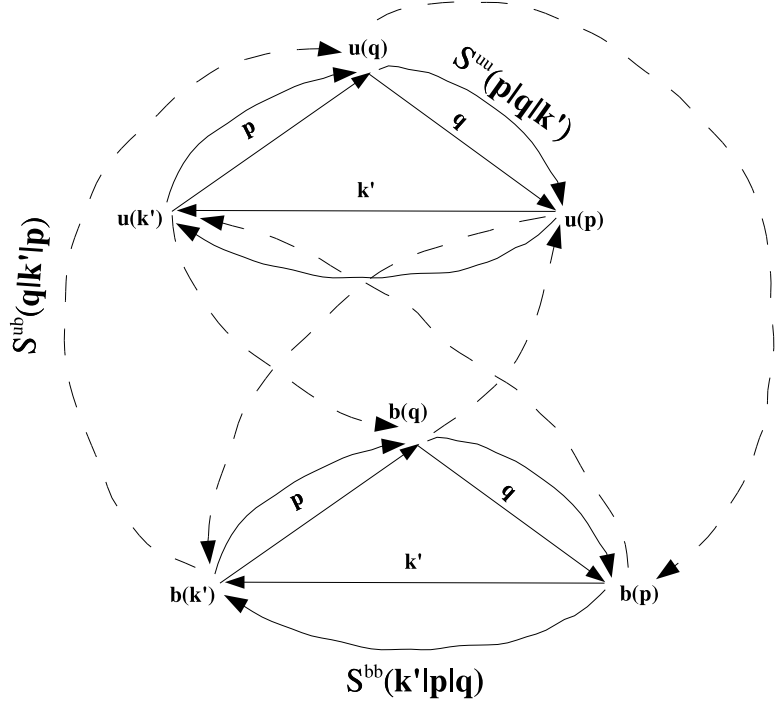}

\caption{\label{Fig:MHD-mode} Mode-to-mode energy transfers in MHD turbulence.
$S^{bb}(\mathbf{k'|p|q})$ represents energy transfer rate from mode
$\mathbf{b}(\mathbf{p})$ to mode $\mathbf{b}(\mathbf{k'})$ with
the mediation of mode $\mathbf{u}(\mathbf{q})$, while $S^{ub}(\mathbf{q|k'|p})$
represents energy transfer rate from mode $\mathbf{b}(\mathbf{k'})$
to mode $\mathbf{u}(\mathbf{q})$ with the mediation of mode $\mathbf{b}(\mathbf{p})$.}
\end{figure}

In Section \ref{sub:'Mode-to-Mode'-fluid} we discussed the mode-to-mode
transfer, $R^{uu}$, between velocity modes in fluid flows. In this
section we will find $R^{uu}$ for MHD flows. Let $R^{uu}(\mathbf{k}'|\mathbf{p}|\mathbf{q})$
be the energy transfer rate from the mode $\mathbf{u}(\mathbf{p)}$
to the mode $\mathbf{\mathbf{u}(\mathbf{k')}}$ in mediation of the
mode $\mathbf{u}(\mathbf{q)}$. The transfer of kinetic energy between
the velocity modes is brought about by the term $\mathbf{u}\cdot\nabla\mathbf{u}$,
both in the Navier-Stokes and MHD equations. Therefore, the expression
for the combined kinetic energy transfer in MHD will be same as that
in fluid. Consequently, $R^{uu}$ for MHD will satisfy the constraints
given in Eqs. (\ref{eq:Skpq}-\ref{eq:Rpqk}). As a result, $R^{uu}(\mathbf{k'|p|q})$
in MHD can be expressed as a sum of a circulating transfer $X_{\Delta}$
and the effective transfer $S^{uu}(\mathbf{k'|p|q})$ given by Eq.
(\ref{eq:Suu_Dar}), i.e., 

\begin{equation}
R^{uu}(\mathbf{k'|p|q})=S^{uu}(\mathbf{k'|p|q})+X_{\Delta}\label{eq:Ruu=Suu+X}\end{equation}
As discussed in Subsection \ref{sub:'Mode-to-Mode'-fluid}, the circulating
transfer $X_{\Delta}$ is irrelevant for the energy flux or the shell-to-shell
energy transfer. Therefore, we we use $S^{uu}(\mathbf{k'|p|q})$ as
the energy transfer rate from the mode $\mathbf{u}(\mathbf{p)}$ to
the mode $\mathbf{\mathbf{u}(\mathbf{k')}}$ with the mediation of
the mode $\mathbf{u}(\mathbf{q)}$. $S^{uu}(\mathbf{k'|p|q})$ and
other transfers in MHD turbulence are shown in Fig. \ref{Fig:MHD-mode}.

\subsubsection{Magnetic mode to Magnetic mode energy transfers}

Now we consider the magnetic energy transfer $R^{bb}(\mathbf{k'|p|q})$
from \textbf{$\mathbf{b(p)}$} to \textbf{$\mathbf{b(k')}$} in the
triad $\mathbf{(k',p,q})$ (see Fig. \ref{Fig:MHD-mode}). This transfer
is due to the term \textbf{$\mathbf{u}\cdot\nabla\mathbf{b}$} of
induction equation (Eq. {[}\ref{eq:MHDbx}{]}). The function $R^{bb}(\mathbf{k'|p|q})$
should satisfy the same relationships as (\ref{eq:Skpq}-\ref{eq:Rpqk})
with $R^{uu}$ and $S^{uu}$ replaced by $R^{bb}$ and $S^{bb}$ respectively.
The solution of $R^{uu}$'s are not unique. Following arguments of
Subsection \ref{sub:'Mode-to-Mode'-fluid} we can show that \begin{equation}
R^{bb}(\mathbf{k'|p|q})=S^{bb}(\mathbf{k'|p|q})+Y_{\Delta}\end{equation}
where\begin{equation}
S^{bb}(\mathbf{k'|p|q})=-\Im\left(\left[\mathbf{k'}\cdot\mathbf{u}(\mathbf{q})\right]\left[\mathbf{b}(\mathbf{k'})\cdot\mathbf{b}(\mathbf{p})\right]\right),\label{eq:Sbb_Dar}\end{equation}
and $Y_{\Delta}$ is the circulating energy transfer that is transferred
from $\mathbf{b(p)}\rightarrow\mathbf{b(k')}\rightarrow\mathbf{b(q)}$
and back to $\mathbf{b(p)}$. $Y_{\Delta}$ does not cause any change
in modal energy. Hence, the magnetic energy \textit{effectively} transferred
from \textbf{$\mathbf{b(p)}$} to \textbf{$\mathbf{b(k')}$} is just
$S^{bb}(\mathbf{k'|p|q})$, i.e., \begin{equation}
R_{eff}^{bb}(\mathbf{k'|p|q})=S^{bb}(\mathbf{k'|p|q}).\end{equation}

\subsubsection{Energy Transfer Between a Velocity Mode to a Magnetic mode}

We now consider the energy transfer $R^{ub}(\mathbf{k'|p|q})$ (from
$\mathbf{b}(\mathbf{p})$ to $\mathbf{u}(\mathbf{k}')$) and $R^{bu}(\mathbf{k'|p|q})$
(from $\mathbf{u}(\mathbf{p})$ to $\mathbf{b}(\mathbf{k'})$) as
illustrated in Fig. \ref{Fig:MHD-mode}. These functions satisfy properties
similar to Eqs. (\ref{eq:Skpq}-\ref{eq:Rpqk}). For example, for
energies coming to $\mathbf{u}(\mathbf{k'})$, we have\begin{equation}
R^{ub}(\mathbf{k'|p|q})+R^{ub}(\mathbf{k'|q|p})=S^{ub}(\mathbf{k'|p,q}),\end{equation}

\begin{equation}
R^{ub}(\mathbf{k'|p|q})+R^{bu}(\mathbf{p|k'|q})=0.\end{equation}

The solutions of these equations are not unique. Using arguments similar
to those in Subsection \ref{sub:'Mode-to-Mode'-fluid}, we can show
that the general solution of $R'$s are 

\begin{equation}
S^{bu}(\mathbf{k'|p|q})=S^{bu}(\mathbf{k'|p|q})+Z_{\Delta},\end{equation}
\begin{equation}
S^{ub}(\mathbf{k'|q|p})=S^{ub}(\mathbf{k'|q|p})-Z_{\Delta},\end{equation}
where\begin{equation}
S^{bu}(\mathbf{k'|p|q})=\Im\left(\left[\mathbf{k}'\cdot\mathbf{b}(\mathbf{q})\right]\left[\mathbf{b(k'})\cdot\mathbf{u}(\mathbf{p})\right]\right),\label{eq:Sbu_Dar}\end{equation}
\begin{equation}
S^{ub}(\mathbf{k'|p|q})=\Im\left(\left[\mathbf{k}'\cdot\mathbf{b}(\mathbf{q})\right]\left[\mathbf{u}(\mathbf{k'})\cdot\mathbf{b}(\mathbf{p})\right]\right),\label{eq:Sub_Dar}\end{equation}
and $Z_{\Delta}$ is the circulating transfer, transferring energy
from $\mathbf{u(p)}\rightarrow\mathbf{b(k')}\rightarrow\mathbf{u(q)}\rightarrow\mathbf{b(p)}\rightarrow\mathbf{u(k')}\rightarrow\mathbf{b(q)}$
 and back to $\mathbf{u(p)}$ without resulting in any change in modal
energy. See Fig. \ref{Fig:MHD-mode} for illustration. Since the circulating
transfer does not affect the net energy transfer, we interpret $S^{bu}$
and $S^{ub}$ as the effective mode-to-mode energy transfer rates.
For example, $S^{bu}(\mathbf{k'|p|q})$ is the effective energy transfer
rate from $\mathbf{u(p)}$ to \textbf{$\mathbf{b(k')}$} with the
mediation of $\mathbf{b}(\mathbf{q})$, i.e, \begin{equation}
R_{eff}^{bu}(\mathbf{k'|p|q})=S^{bu}(\mathbf{k'|p|q}).\end{equation}

To summarize, the energy evolution equations for a triad $(\mathbf{k,p,q})$
are

\begin{eqnarray}
\frac{\partial}{\partial t}\frac{1}{2}\left|u\left(\mathbf{k'}\right)\right|^{2} & = & S^{uu}(\mathbf{k'|p|q)}+S^{uu}(\mathbf{k'|q|p)}+S^{ub}(\mathbf{k'|p|q)+}S^{ub}(\mathbf{k'|q|p)}\\
\frac{\partial}{\partial t}\frac{1}{2}\left|b\left(\mathbf{k'}\right)\right|^{2} & = & S^{bb}(\mathbf{k'|p|q)}+S^{bb}(\mathbf{k'|q|p)}+S^{bu}(\mathbf{k'|p|q)+}S^{bu}(\mathbf{k'|q|p)}\end{eqnarray}
As discussed above $S^{YX}(\mathbf{k'|p|q)}$ $(X,Y=u$ or $b)$ is
the mode-to-mode energy transfer rate from mode $\mathbf{p}$ of field
$X$ to mode $\mathbf{k'}$ of field $Y$ with mode $\mathbf{q}$
acting as a mediator. These transfers have been schematically shown
in Fig. \ref{Fig:MHD-mode}.

The triads interactions can are also be written in terms of Elsässer
variables. Here the participating modes are $\mathbf{z^{\pm}(k')},\mathbf{z^{\pm}(p)}$
and $\mathbf{z^{\pm}(q)}.$ The energy equations for these modes are\begin{equation}
\frac{\partial}{\partial t}\frac{1}{2}\left|z^{\pm}\left(\mathbf{k'}\right)\right|^{2}=S^{\pm}(\mathbf{k'|p|q)}+S^{\pm}(\mathbf{k'|q|p)},\end{equation}
where\begin{eqnarray}
S^{\pm}(\mathbf{k'|p|q)} & = & -\Im\left(\left[\mathbf{k'}\cdot\mathbf{z^{\mp}}(\mathbf{q})\right]\left[\mathbf{z^{\pm}}(\mathbf{k}')\cdot\mathbf{z^{\pm}}(\mathbf{p})\right]\right).\label{eq:Spm_Dar}\end{eqnarray}
From Eq. (\ref{eq:Spm_Dar}) we deduce that the $\mathbf{z^{+}}$
modes transfer energy only to $\mathbf{z^{+}}$ modes, and $\mathbf{z^{-}}$
modes transfer energy only to $\mathbf{z^{-}}$ modes. This is in
spite of the fact that nonlinear interaction involves both $\mathbf{z}^{+}$
and $\mathbf{z}^{-}$ modes. These deductions became possible only
because of mode-to-mode energy transfers proposed by Dar et al. 

The evolution equation of magnetic helicity in a triad interaction
is given by\begin{eqnarray}
\frac{\partial}{\partial t}H_{M}(\mathbf{k)} & = & \frac{1}{2}\left[\mathbf{b^{*}(k)}\cdot\frac{\partial\mathbf{a(k)}}{\partial t}+\mathbf{a^{*}(k)}\cdot\frac{\partial\mathbf{b(k)}}{\partial t}\right]\\
 & = & S^{H_{M}}(\mathbf{k'|p|q)}+S^{H_{M}}(\mathbf{k'|q|p)},\label{eq:MHD_Hm_triad}\end{eqnarray}
where\begin{eqnarray}
S^{H_{M}}(\mathbf{k'|p|q}) & = & \frac{1}{4}\Re\left[\mathbf{b(k')}\cdot\left(\mathbf{u(p)}\times\mathbf{b(q)}\right)\right]\nonumber \\
 &  & +\frac{1}{4}\Im\left[\left(\mathbf{k'}\cdot\mathbf{b(q)}\right)\left(\mathbf{a(k')}\cdot\mathbf{u(p)}\right)-\left(\mathbf{k'}\cdot\mathbf{u(q)}\right)\left(\mathbf{a(k')}\cdot\mathbf{b(p)}\right)\right]\label{eq:Skpq_HM}\end{eqnarray}

In ideal MHD, the functions $S^{YX}(\mathbf{k'|p|q)}$ and energy
functions have the following interesting properties:

\begin{enumerate}
\item Energy transfer rate from $\mathbf{X(p)}$ to $\mathbf{Y(k')}$ is
equal and opposite to that from $\mathbf{Y(k')}$ to $\mathbf{X(p)}$,
i. e.,\[
S^{YX}(\mathbf{k'|p|q)}=-S^{XY}(\mathbf{p|k'|q).}\]

\item Sum of all energy transfer rates along $u$-$u$, $b$-$b$, $z^{+}$-$z^{+}$,
and $z^{-}$-$z^{-}$ channels are zero, i.e.,\begin{eqnarray*}
S^{XX}(\mathbf{k'|p|q)}+S^{XX}(\mathbf{k'|q|p)}+S^{XX}(\mathbf{p|k'|q)}\\
+S^{XX}(\mathbf{p|q|k')}+S^{XX}(\mathbf{q|k'|p)}+S^{XX}(\mathbf{q|p|k')} & = & 0,\end{eqnarray*}
where $X$ could be a vector field ($\mathbf{u,b,z^{+},z^{-})}$.
\item Sum of all energy transfer rates along $u$-$b$ channel is zero,
i.e., \begin{eqnarray*}
S^{bu}(\mathbf{k'|p|q)}+S^{bu}(\mathbf{k'|q|p)}+S^{bu}(\mathbf{p|k'|q)}+S^{bu}(\mathbf{p|q|k')}+S^{bu}(\mathbf{q|k'|p)}+S^{bu}(\mathbf{q|p|k')}\\
+S^{ub}(\mathbf{k'|p|q)}+S^{ub}(\mathbf{k'|q|p)}+S^{ub}(\mathbf{p|k'|q)}+S^{ub}(\mathbf{p|q|k')}+S^{ub}(\mathbf{q|k'|p)}+S^{ub}(\mathbf{q|p|k')} & = & 0.\end{eqnarray*}

\item Using the above identities we can show that total energy in a triad
interaction is conserved, i. e., \begin{eqnarray*}
E^{u}(\mathbf{k')}+E^{u}(\mathbf{p)}+E^{u}(\mathbf{q)}+E^{b}(\mathbf{k')}+E^{b}(\mathbf{p)}+E^{b}(\mathbf{q)} & = & const.\end{eqnarray*}
 Kinetic energy and magnetic energies are \emph{not} conserved individually.
\item Sum of all $E^{+}$ energies of in a triad are conserved. Similarly,
sum of all $E^{-}$ energies are conserved, i. e.,\begin{eqnarray*}
E^{\pm}(\mathbf{k')}+E^{\pm}(\mathbf{p)}+E^{\pm}(\mathbf{q)} & = & const.\end{eqnarray*}
Since cross helicity $H_{c}=(E^{+}-E^{-})/4$, we find the cross helicity
is also conserved in a triad interaction. 
\item Sum of transfer rates of magnetic helicity in a triad is zero, i.
e.,\begin{eqnarray*}
S^{H_{M}}(\mathbf{k'|p|q)}+S^{H_{M}}(\mathbf{k'|q|p)}+S^{H_{M}}(\mathbf{p|k'|q)}\\
+S^{H_{M}}(\mathbf{k'|p|q)}+S^{H_{M}}(\mathbf{k'|q|p)}+S^{H_{M}}(\mathbf{p|k'|q)} & = & 0.\end{eqnarray*}

\item Sum of $H_{M}$ in a triad is conserved, i.e.,\[
H_{M}(\mathbf{k'})+H_{\mathbf{M}}(\mathbf{p})+H_{M}(\mathbf{q})=const.\]

\item In incompressible flows, $i\mathbf{k}p(\mathbf{k})$ is perpendicular
to both the transverse components (transverse to $\mathbf{k}$), and
it does not couple with them. That is why pressure is absent in the
energy transfer formulas for incompressible flows. Pressure does not
isotropize energy in the transverse direction, contrary to Orszag's
conjecture \cite{Orsz:Rev}. In compressive flows pressure couples
with the compressive component of velocity field and internal energy.
\item Mean magnetic field only convects the waves; it does not participate
in energy exchange. Hence, it is absent in the energy transfer formulas.
\end{enumerate}
In the above subsections we derived formulas for mode-to-mode energy
transfer rates in MHD turbulence. In the next subsections, we will
use these formulas to define (a) shell-to-shell transfers and (b)
cascade rates in MHD turbulence.

\subsection{Shell-to-Shell Energy Transfer Rates in MHD Turbulence}

Using the definition of the mode-to-mode energy transfer function
$S^{YX}(\mathbf{k'|p|q})$, the energy transfer rate from $m$-th
shell of field $X$ to  $n$-th shell of field $Y$ is\begin{equation}
T_{nm}^{YX}=\sum_{\mathbf{k'}\in n}\sum_{\mathbf{p}\in m}S^{YX}(\mathbf{k'|p|q}).\label{eq:T_MHD}\end{equation}
The $\mathbf{p}$-sum is over $m$-th shell, and the $\mathbf{k'}$-sum
is over $n$-th shell. As discussed in Subsection \ref{sub:Shell-to-Shell-fluid},
the circulating transfer rates $X_{\Delta},Y_{\Delta}$, and $Z_{\Delta}$
do not appear in the expressions for shell-to-shell energy transfer
rates. Also, as discussed in Section \ref{sub:Shell-to-Shell-fluid},
shell-to-shell energy transfer can be reliably computed only by mode-to-mode
transfer $S(\mathbf{k'}|\mathbf{p}|\mathbf{q})$. 

The numerical and analytical computation of shell-to-shell energy
transfer rates will be discussed in the later part of the paper.

\subsection{Energy Cascade Rates in MHD Turbulence \label{sub:Energy-Cascade-Rates-formula-MHD}}

\begin{figure}
\includegraphics[bb=0 330 440 655 ] {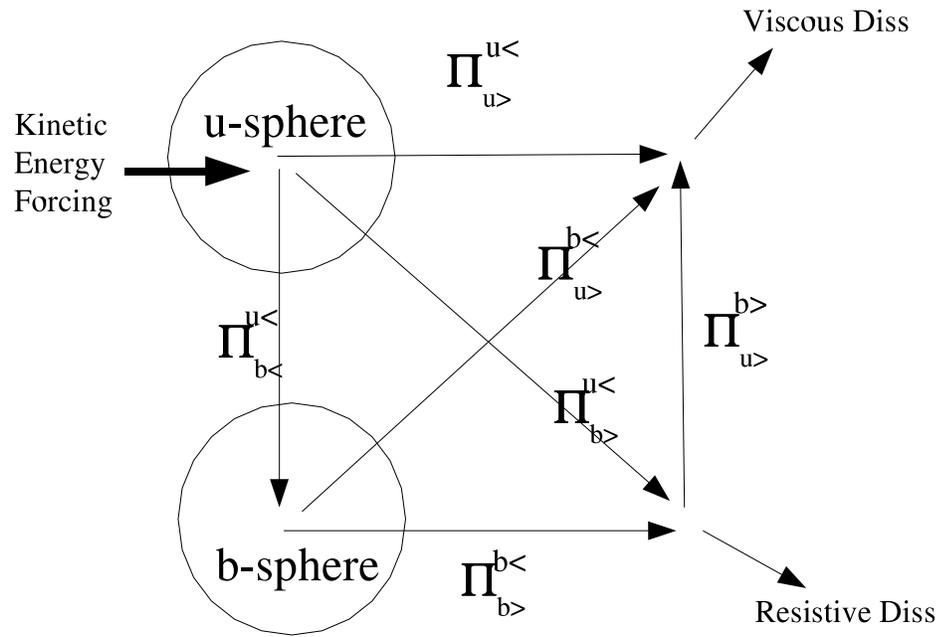}

\caption{\label{Fig:MHD-flux-ub} Various energy fluxes in MHD turbulence.
$\Pi_{Y>}^{X<}$ represents energy flux from the inside of $X$-sphere
to the outside of $Y$-sphere. }
\end{figure}

The energy cascade rate (or flux) is defined as the rate of loss of
energy from a sphere in the wavenumber space to the modes outside
the sphere. There are various types of cascade rates in MHD turbulence.
We have shown them schematically in Fig. \ref{Fig:MHD-flux-ub}. %
 For flux studies, we split the wavenumber space into two regions:
$k<k_{0}$ (inside ``$k_{0}$ sphere'') and $k>k_{0}$ (outside ``$k_{0}$
sphere''). The energy transfer could take place from the inside/outside
of the $u/b$-sphere to the inside/outside of the $u/b$-sphere. In
terms of $S^{YX}(\mathbf{k'|p|q})$ the energy transfer rate from
region $A$ of $X$ field to region $B$ of $Y$field is\begin{equation}
\Pi_{Y,B}^{X,A}=\sum_{\mathbf{k'}\in B}\sum_{\mathbf{p}\in A}S^{YX}(\mathbf{k'|p|q}).\label{eq:Pi_MHD}\end{equation}
 For example, energy flux from $u$-sphere of radius $k_{0}$ to $b$-sphere
of the same radius is\[
\Pi_{b<}^{u<}(k_{0})=\sum_{\mathbf{\left|k'\right|<k_{0}}}\sum_{\left|\mathbf{p}\right|<k_{0}}S^{bu}(\mathbf{k'|p|q}).\]
In this paper we denote inside of a sphere by $<$ sign and outside
of a sphere by $>$ sign. Note that the energy flux is independent
of circulatory energy transfer. The total energy flux is defined as
the total energy (kinetic+magnetic) lost by the $k_{0}$-sphere to
the modes outside of the sphere, i. e.,\[
\Pi_{tot}(k_{0})=\Pi_{u>}^{u<}(k_{0})+\Pi_{b>}^{u<}(k_{0})+\Pi_{b>}^{b<}(k_{0})+\Pi_{u>}^{b<}(k_{0}).\]

Using arguments of Subsection \ref{sub:Shell-to-Shell-fluid}, it
can be easily seen that the fluxes $\Pi_{u>}^{u<}(k_{0}),\Pi_{b>}^{u<}(k_{0}),\Pi_{b>}^{b<}(k_{0}),\Pi_{u>}^{b<}(k_{0})$
can all be computed using the combined energy transfer $S(\mathbf{k'|p|q})$,
and mode-to-mode energy transfer $S(\mathbf{k'|p|q})$. However, $\Pi_{b<}^{u<}(k_{0})$
and $\Pi_{b>}^{u>}(k_{0})$ can be computed only using $S(\mathbf{k'|p|q})$,
not by $S(\mathbf{k'|p,q})$.

\begin{figure}
\includegraphics[bb=0 380 370 650 ]{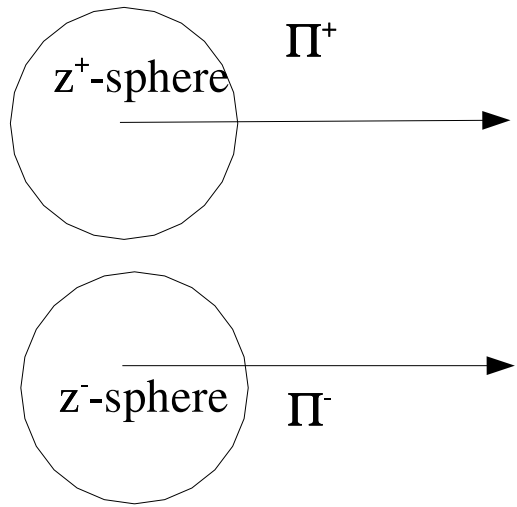}

\caption{\label{Fig:MHD-flux-z} Energy flux $\Pi^{+}(\Pi^{-})$ from inside
the $z^{+}-$sphere ($z^{-}-$sphere) to outside of $z^{+}$-sphere
($z^{-}$-sphere). }
\end{figure}

We also define the energy flux $\Pi^{+}(\Pi^{-})$ from inside the
$z^{+}$-sphere ($z^{-}$-sphere) to outside of $z^{+}$-sphere ($z^{-}$-sphere)
\[
\Pi^{\pm}(k_{0})=\sum_{\mathbf{\left|k'\right|>k_{0}}}\sum_{\left|\mathbf{p}\right|<k_{0}}S^{\pm}(\mathbf{k'|p|q}).\]
as shown in Fig. \ref{Fig:MHD-flux-z}. Note that there is no cross
transfer between $z^{+}$-sphere and $z^{-}$-sphere. %

The energy fluxes have been computed analytically and numerically
by researchers. These results will be described in later part of the
paper.

\subsection{Digression to Infinite Box \label{sub:Digression-to-Infinite-box:flux}}

In the above discussion we assumed that the fluid is contained in
a finite volume. In simulations, box size is typically taken to $2\pi$.
However, most analytic calculations assume infinite box. It is quite
easy to transform the equations given above to those for infinite
box using the method described in Appendix. Here, the evolution of
energy spectrum is given by (see Section \ref{sec:MHD:-Definitions-and})\begin{eqnarray}
\left(\frac{\partial}{\partial t}+2\nu k^{2}\right)C^{uu}\left(\mathbf{k},t\right) & = & \frac{2}{\left(d-1\right)\delta\left(\mathbf{k+k'}\right)}\int_{\mathbf{k'+p+q=0}}\frac{d\mathbf{p}}{(2\pi)^{2d}}\nonumber \\
 &  & \left[S^{uu}(\mathbf{k'}|\mathbf{p}|\mathbf{q})+S^{ub}(\mathbf{k'}|\mathbf{p}|\mathbf{q})\right]\label{eq:Cuu(k)-evolve}\\
\left(\frac{\partial}{\partial t}+2\eta k^{2}\right)C^{bb}\left(\mathbf{k},t\right) & = & \frac{2}{\left(d-1\right)\delta\left(\mathbf{k+k'}\right)}\int_{\mathbf{k'+p+q=0}}\frac{d\mathbf{p}}{(2\pi)^{2d}}\nonumber \\
 &  & \left[S^{bu}(\mathbf{k'}|\mathbf{p}|\mathbf{q})+S^{bb}(\mathbf{k'}|\mathbf{p}|\mathbf{q})\right]\label{eq:Cbb(k)-evolve}\end{eqnarray}
The shell-to-shell energy transfer rate $T_{nm}^{YX}$ from the $m$-th
shell of field $X$ to the $n$-th shell of field $Y$ is\begin{equation}
T_{nm}^{YX}=\frac{1}{(2\pi)^{d}\delta(\mathbf{k'+p+q)}}\int_{k'\in n}\frac{d\mathbf{k'}}{(2\pi)^{d}}\int_{p\in m}\frac{d\mathbf{p}}{(2\pi)^{d}}\left\langle S^{YX}(\mathbf{k'|p|q})\right\rangle ,\label{eq:T_MHD_FT}\end{equation}
In terms of Fourier transform, the energy cascade rate from region
$A$ of field $X$ to region $B$ of field $Y$ is\begin{equation}
\Pi_{Y,B}^{X,A}=\frac{1}{(2\pi)^{d}\delta(\mathbf{k'+p+q)}}\int_{B}\frac{d\mathbf{k'}}{(2\pi)^{d}}\int_{A}\frac{d\mathbf{p}}{(2\pi)^{d}}\left\langle S^{YX}(\mathbf{k'|p|q})\right\rangle .\label{eq:Pi_MHD_FT}\end{equation}
In $\mathbf{z}^{\pm}$ variables, the energy evolution equations are\[
\left(\frac{\partial}{\partial t}+2\nu_{+}k^{2}\right)C^{\pm\pm}\left(\mathbf{k},t\right)+2\nu_{+}k^{2}C^{\pm\mp}\left(\mathbf{k},t\right)=\frac{2}{\left(d-1\right)\delta\left(\mathbf{k+k'}\right)}\int_{\mathbf{k'+p+q=0}}\frac{d\mathbf{p}}{(2\pi)^{2d}}S^{\pm}(\mathbf{k'}|\mathbf{p}|\mathbf{q}),\]
and the energy fluxes $\Pi^{\pm}(k_{0})$ coming out of a wavenumber
sphere of radius $k_{0}$ is \begin{equation}
\Pi^{\pm}(k_{0})=\frac{1}{(2\pi)^{d}\delta(\mathbf{k'+p+q)}}\int_{|\mathbf{k'}|>k_{0}}\frac{d\mathbf{k'}}{(2\pi)^{d}}\int_{|\mathbf{p}|<k_{0}}\frac{d\mathbf{p}}{(2\pi)^{d}}\left\langle S^{\pm\pm}(\mathbf{k'|p|q})\right\rangle .\label{eq:Pipm_FT}\end{equation}

For isotropic flows, after some manipulation and using Eq. (\ref{eq:E(k)_eq_C(k)}),
we obtain \cite{Lesi:book}\begin{equation}
\left(\frac{\partial}{\partial t}+2\nu k^{2}\right)E^{u,b,\pm}(k,t)=T^{u,b,\pm}(k,t),\label{eq:dtE(k,t)_eq_T(k,t)}\end{equation}
where $T(k,t)$, called \emph{transfer function}, can be written in
terms of $S^{YX}(\mathbf{k'}|\mathbf{p}|\mathbf{q})$. The above formulas
will be used in analytic calculations.

The mode-to-mode formalism discussed here is quite general, and it
can be applied to scalar turbulence \cite{MKV:scalar}, Rayleigh-Benard
convection, enstrophy, Electron MHD etc. One key assumption however
is incompressibility. With this remark we close our  formal discussion
on energy transfers in MHD turbulence. In the next section we will
discuss various turbulence phenomenologies and models of MHD turbulence.

\section{MHD Turbulence Phenomenological Models \label{sec:MHD-Turbulence-Models}}

In the last two sections we introduced Navier-Stokes and MHD equations,
and spectral quantities like the energy spectra and fluxes. These
quantities will be analyzed in most part of this paper using (a) phenomenological
(b) numerical (c) analytical (d) observational or experimental methods.
In the present section we will present some of the existing phenomenological
models of MHD turbulence.

Many MHD turbulence models are motivated by fluid turbulence models.
Therefore, we present a brief review of fluid turbulence models before
going to MHD turbulence. The most notable theory in fluid turbulence
is due to Kolmogorov, which will be presented below.

\subsection{Kolmogorov's 1941 Theory for Fluid Turbulence \label{sub:Kolmogorov's-1941-Theory}}

For homogeneous, isotropic, incompressible, and steady fluid turbulence
with vanishing viscosity (large $Re$), Kolmogorov \cite{K41a,K41b,K41c,LandFlui:book}
derived an exact relation that 

\begin{equation}
\left\langle \left(\bigtriangleup u\right)_{\parallel}^{3}\right\rangle =-\frac{4}{5}\epsilon l\label{eq:K41}\end{equation}
where $(\bigtriangleup u)_{||}$ is component of $\mathbf{u}(\mathbf{x}+\mathbf{l})-\mathbf{u}(\mathbf{x})$
along $\mathbf{l}$, $\epsilon$ is the dissipation rate, and $l$
lies between forcing scale $(L)$ and dissipative scales $(l_{d})$,
i.e., $l_{d}\ll l\ll L$. This intermediate range of scales is called
inertial range. Note that the above relationship is universal, which
holds independent of forcing and dissipative mechanisms, properties
of fluid (viscosity), and initial conditions. Therefore it finds applications
in wide spectrum of phenomena, e. g., atmosphere, ocean, channels,
pipes, and astrophysical objects like stars, accretion disks etc. 

More popular than Eq. (\ref{eq:K41}) is its equivalent statement
on energy spectrum. If we assume $\bigtriangleup u$ to be fractal,
and $\epsilon$ to be independent of scale, then\[
\left\langle \left(\bigtriangleup u\right)^{2}\right\rangle \propto\epsilon^{2/3}l^{2/3}\]
Fourier transform of the above equation yields\begin{equation}
E(k)=K_{Ko}\epsilon^{2/3}k^{-5/3}\label{eq:Kolm_fluid}\end{equation}
where $K_{Ko}$ is a universal constant, commonly known as Kolmogorov's
constant. 

Kolmogorov's derivation of Eq. (\ref{eq:K41}) is quite involved.
However, Eqs. (\ref{eq:K41}, \ref{eq:Kolm_fluid}) can be derived
using scaling arguments (dimensional analysis) under the assumption
that 

\begin{enumerate}
\item The energy spectrum in the inertial range does not depend on the large-scaling
forcing processes and the small-scale dissipative processes, hence
it must be a power law in the local wavenumber.
\item The energy transfer in fluid turbulence is local in the wavenumber
space. The energy supplied to the fluid at the forcing scale cascades
to smaller scales, and so on. Under steady-state the energy cascade
rate is constant in the wavenumber space, i. e., $\Pi(k)=constant=\epsilon$. 
\end{enumerate}
Eq. (\ref{eq:Kolm_fluid}) has been supported by numerous experiments
and numerical simulations. Kolmogorov's constant $K_{Ko}$ has been
found to lie between 1.4-1.6 or so. It is quite amazing that complex
interactions among fluid eddies in various different situations can
be quite well approximated by Eq. (\ref{eq:Kolm_fluid}). 

In the framework of Kolmogorov's theory, several interesting deductions
can be made.

\begin{enumerate}
\item Kolmogorov's theory assumes homogeneity and isotropy. In real flows,
large-scales (forcing) as well as dissipative scales do not satisfy
these properties. However, experiments and numerical simulations show
that in the inertial range ($l_{d}\ll l\ll L$), the fluid flows are
typically homogeneous and isotropic.
\item The velocity fluctuations at any scale $l$ goes as\[
u_{l}\approx\epsilon^{1/3}l^{1/3}.\]
Therefore, the effective time-scale for the interaction among eddies
of size $l$ is \[
\tau_{l}\approx\frac{l}{u_{l}}\approx\epsilon^{-1/3}l^{2/3}.\]

\item An extrapolation of Kolmogorov's scaling to the forcing and the dissipative
scales yields\[
\epsilon\approx\frac{u_{L}^{3}}{L}\approx\frac{u_{l_{d}}^{3}}{l_{d}}.\]
Taking $\nu\approx u_{l_{d}}l_{d}$, one gets \[
l_{d}\approx\left(\frac{\nu^{3}}{\epsilon}\right)^{1/4}.\]
Note that the dissipation scale, also known as Kolmogorov's scale,
depends on the large-scale quantity $\epsilon$ apart from kinematic
viscosity.
\item From the definition of Reynolds number\[
Re=\frac{U_{L}L}{\nu}\approx\frac{U_{L}L}{u_{l_{d}}l_{d}}\approx\left(\frac{L}{l_{d}}\right)^{4/3}\]
Therefore, \[
\frac{L}{l_{d}}\approx Re^{3/4}.\]
Onset of turbulence depends on geometry, initial conditions, noise
etc. Still, in most experiments turbulences sets in after $Re$ of
2000 or more. Therefore, in three dimensions, number of active modes
$(L/l_{d})^{3}$ is larger than 26 million. These large number of
modes make the problem quite complex and intractable.
\item Space dimension does not appear in the scaling arguments. Hence, one
may expect Kolmogorov's scaling to hold in all dimensions. It is however
found that the above scaling law is applicable in three dimension
only. In two dimension (2D), conservation of enstrophy changes the
behaviour significantly (see Appendix \ref{sec:Digression-to-Fluid}).
The solution for one-dimensional incompressible Navier-Stokes is $\mathbf{u}(\mathbf{x},t)=const$,
which is a trivial solution.
\item Mode-to-mode energy transfer term $S(k|p|q)$ measures the strength
of nonlinear interaction. Kolmogorov's theory implicitly assumes that
energy cascades from larger to smaller scales. It is called local
energy transfer in Fourier space. These issues will be discussed in
Section \ref{sec:analytic-energy} and Appendix \ref{sec:Digression-to-Fluid}.
\item Careful experiments show that the spectral index is close to 1.71
instead of 1.67. This correction of $\approx0.04$ is universal and
is due to the small-scale structures. This phenomena is known as intermittency,
and will be discussed in Section \ref{sec:Intermittency-in-MHD}.
\item Kolmogorov's model for turbulence works only for incompressible flow.
It is connected to the fact that incompressible flow has local energy
transfer in wavenumber space. Note that Burgers equation, which represents
compressible flow $(U\gg C_{s})$, has $k^{-2}$ energy spectrum,
very different from Kolmogorov's spectrum.
\end{enumerate}
Kolmogorov's theory of turbulence had a major impact on turbulence
research because of its universality. Properties of scalar, MHD, Burgers,
Electron MHD, wave turbulence have been studied using similar arguments.
In the next subsection we will investigate the properties of MHD flows.

\subsection{MHD Turbulence Models for Energy Spectra and Fluxes \label{sub:MHD-Turbulence-Models}}

Alfv\'{e}n waves are the basic modes of incompressible MHD equations.
In absence of the nonlinear term $({\textbf{z}}^{\mp}\cdot\bigtriangledown){\textbf{z}}^{\pm}$,
${\textbf{z}}^{\pm}$ are the two independent modes travelling antiparallel
and parallel to the mean magnetic field. However, when the nonlinear
term is present, new modes are generated, and they interact with each
other, resulting in a turbulent behaviour. In the following we will
discuss various phenomenologies of MHD turbulence.

\subsubsection{Kraichnan , Iroshnikov, and Dobrowolny et al.'s (KID) Phenomenology
- $E(k)\propto k^{-3/2}$}

In the mid-sixties, Kraichnan \cite{Krai:65} and Iroshnikov \cite{Iros}
gave the first phenomenological theory of MHD turbulence. For MHD
plasma with mean magnetic field $B_{0}$, Kraichnan and Iroshnikov
argued that localized $z^{+}$ and $z^{-}$ modes travel in apposite
directions with phase velocity of $B_{0}$. When the mean magnetic
field $B_{0}$ is much stronger than the fluctuations ($B_{0}\gg u_{k}$),
the fluctuations (oppositely moving waves) will interact weakly. They
suggested that Alfvén time-scale $\tau_{A}(k)=(B_{0}k)^{-1}$ is the
effective time-scale for the relaxation of the locally built-up phase
correlations, thereby concluding that triple correlation and the energy
flux $\Pi$ will be proportional to $(B_{0}k)^{-1}$. Note that $(B_{0}k)^{-1}\ll(u_{k}k)^{-1}$.
Using dimensional arguments they concluded \begin{equation}
\Pi=A^{2}\tau_{A}(k)\left(E^{b}(k)\right)^{2}k^{4}=A^{2}B_{0}^{-1}\left(E^{b}(k)\right)^{2}k^{3}\label{eq:Pi_eq_tau*}\end{equation}
or\begin{equation}
E^{b}(k)=A\left(\Pi B_{0}\right)^{1/2}k^{-3/2},\label{eq:Kraich3/2}\end{equation}
where $A$ is a nondimensional constant of order 1.

The above approximation yields {}``weak turbulence''. In absence
of any $B_{0}$, the magnetic field of the large eddies was assumed
to play the role of $B_{0}$. Kraichnan \cite{Krai:65} and Iroshnikov
\cite{Iros} also argued that the Alfvén waves are not strongly affected
by the weak interaction among themselves, hence kinetic and magnetic
energy remain equipartitioned. This phenomenon is called {}``Alfvén
effect''. Note that Kraichnan's spectral index is 3/2 as compared
to Kolmogorov's index of 5/3.

In 1980 Dobrowolny et al. \cite{Dobr} derived Kraichnan's 3/2 spectrum
based on random interactions of $z^{+}$ and $z^{-}$ modes. Dobrowolny
et al.'s argument is however more general, and provide us energy spectrum
even when $u_{k}$ is comparable to $B_{0}$. They assumed that the
interaction between the fluctuations are local in wavenumber space,
and that in one interaction, the eddies $z_{k}^{\pm}$ interact with
the other eddies of similar sizes for time interval $\tau_{k}^{\pm}$.
Then from Eq. (\ref{eq:MHDzk}), the variation in the amplitudes of
these eddies, $\delta z_{k}^{\pm}$, during this interval is given
by \begin{equation}
\delta z_{k}^{\pm}\approx\tau_{k}^{\pm}z_{k}^{+}z_{k}^{-}k.\end{equation}
 In $N$ such interactions, because of their stochastic nature, the
amplitude variation will be $\Delta z_{k}^{\pm}\approx\sqrt{N}(\delta z_{k}^{\pm})$.
Therefore, the number of interactions $N_{k}^{\pm}$ required to obtain
a variation equal to its initial amplitude $z_{k}^{\pm}$ is \begin{equation}
N_{k}^{\pm}\approx\frac{1}{k^{2}\left(z_{k}^{\mp}\right)^{2}\left(\tau_{k}^{\pm}\right)^{2}}\end{equation}
 and the corresponding time $T_{k}^{\pm}=N_{k}\tau_{k}^{\pm}$ is
\begin{equation}
T_{k}^{\pm}\approx\frac{1}{k^{2}\left(z_{k}^{\mp}\right)^{2}\tau_{k}^{\pm}}.\end{equation}
 The time scale of the energy transfer at wavenumber $k$ is assumed
to be $T_{k}^{\pm}$. Therefore, the fluxes $\Pi^{\pm}$ of the fluctuations
$z_{k}^{\pm}$ can be estimated to be\begin{equation}
\Pi^{\pm}\approx\frac{\left(z_{k}^{\mp}\right)^{2}}{T_{k}^{\pm}}\approx\tau_{k}^{\pm}\left(z_{k}^{\pm}\right)^{2}\left(z_{k}^{\mp}\right)^{2}k^{2}.\label{eq:MHD_tau}\end{equation}

By choosing different interaction time-scales, one can obtain different
energy spectra. Using the same argument as Kraichnan \cite{Krai:65},
Dobrowolny et al. \cite{Dobr} chose Alfv\'{e}n time scale $\tau_{A}=(kB_{0})^{-1}$
as the relevant time-scale, and found that \begin{equation}
\Pi^{+}\approx\Pi^{-}\approx\frac{1}{B_{0}}E^{+}(k)E^{-}(k)k^{3}=\Pi.\label{eq:Dobro}\end{equation}
 If $E^{+}(k)\approx E^{-}(k),$ then \begin{equation}
E^{+}(k)\approx E^{-}(k)\approx\left(B_{0}\Pi\right)^{1/2}k^{-3/2}\end{equation}
This result of Dobrowolny et al. is the same as that of Kraichnan
\cite{Krai:65}. We refer to the above as KID's (Kraichnan, Iroshnikov,
Dobrowolny et al.) phenomenology.

\subsubsection{Marsch, Matthaeus and Zhou's Kolmogorov-like Phenomenology - $E(k)\propto k^{-5/3}$}

In 1990 Marsch \cite{Mars:Kolm} chose the nonlinear time-scale $\tau_{NL}^{\pm}\approx(kz_{k}^{\mp})^{-1}$
as the interaction time-scale for the eddies $z_{k}^{\pm}$, and substituted
those in Eq. (\ref{eq:MHD_tau}) to obtain \begin{equation}
\Pi^{\pm}\approx\left(z_{k}^{\pm}\right)^{2}\left(z_{k}^{\mp}\right)k,\end{equation}
 which in turn led to \begin{equation}
E^{\pm}(k)=K^{\pm}(\Pi^{\pm})^{4/3}(\Pi^{\mp})^{-2/3}k^{-5/3},\label{eq:MHD_Kolm_zpm}\end{equation}
 where $K^{\pm}$ are constants, referred to as Kolmogorov's constants
for MHD turbulence. Because of its similarity with Kolmogorov's fluid
turbulence phenomenology, we refer to this phenomenology as Kolmogorov-like
MHD turbulence phenomenology. 

During the same time, Matthaeus and Zhou \cite{MattZhou}, and Zhou
and Matthaeus \cite{ZhouMatt} attempted to combine 3/2 and 5/3 spectrum
for an arbitrary ratio of $u_{k}$ and $B_{0}$. They postulated that
the relevant time-scales $\tau^{\pm}(k)$ for MHD turbulence are given
by\begin{eqnarray*}
\frac{1}{\tau^{\pm}(k)} & = & \frac{1}{\tau_{A}(k)}+\frac{1}{\tau_{NL}^{\pm}(k)}\\
 & = & kB_{0}+kz_{k}^{\mp}.\end{eqnarray*}
Substitution of $\tau^{\pm}(k)$ in Eq. (\ref{eq:MHD_tau}) yields
\begin{equation}
\Pi^{\pm}=\frac{A^{2}E^{+}(k)E^{-}(k)k^{3}}{B_{0}+\sqrt{kE^{\pm}(k)}}\label{eq:MHD_E(k)_ZM}\end{equation}
 where $A$ is a constant. If Matthaeus and Zhou's phenomenology (Eq.
{[}\ref{eq:MHD_E(k)_ZM}{]}) were correct, the small wavenumbers ($\sqrt{kE^{\pm}(k)}\gg B_{0}$)
would follow 5/3 spectrum, whereas the large wavenumbers ($\sqrt{kE^{\pm}(k)}\ll B_{0}$)
would follow 3/2 spectrum.

\subsubsection{Grappin et al. - Alfvénic Turbulence \label{sub:Grappin-et-al.-Phenomenology}}

Grappin et al. \cite{Grap83} analyzed MHD turbulence for nonzero
cross helicity; this is also referred to as Alfvén\emph{ic MHD}. They
used Alfvén time-scale as relaxation time-scale for triple correlations,
and derived the transfer function (Eq. {[}\ref{eq:dtE(k,t)_eq_T(k,t)}{]})
to be\[
T^{\pm}(k,t)=\int dpdq(k+p+q)^{-1}(m_{kpq}/p)\left[k^{2}E^{\pm}(p)E^{\mp}(q)-p^{2}E^{\mp}(q)E^{\pm}(k)\right]\]
They postulated that in the inertial range, energy spectra $E^{\pm}(k)=K^{\pm}k^{-m^{\pm}}$.
Using $\Pi^{\pm}(k_{0})=-\int_{0}^{k_{0}}dkT_{k}^{\pm}$, and demanding
that fluxes are independent of $k_{0}$, they derived\begin{equation}
m^{+}+m^{-}=3.\label{eq:m+m-=3}\end{equation}
In addition, using\[
\epsilon^{\pm}=2\nu\int_{k_{0}}^{k_{D}^{\pm}}dpp^{2}E^{\pm}(p),\]
and assuming $K^{+}=K^{-}$, and $k_{D}^{+}\approx k_{D}^{-}\approx k_{D}$,
they concluded that\begin{equation}
\frac{\epsilon^{+}}{\epsilon^{-}}=\frac{m^{+}}{m^{-}}.\label{eq:m+/m-}\end{equation}
Later we will show that the solar wind observations and numerical
results are inconsistent with the above predictions. We will show
later that Grappin et al.'s key assumptions (1) Alfvén time-scale
to be the relevant time scale, and (2) $K^{+}=K^{-}$ are incorrect.

\subsubsection{Goldreich and Sridhar - $E(k_{\perp})\propto k_{\perp}^{-5/3}$}

\begin{figure}
\includegraphics[%
  scale=0.7,bb=0 390 475 755 ]{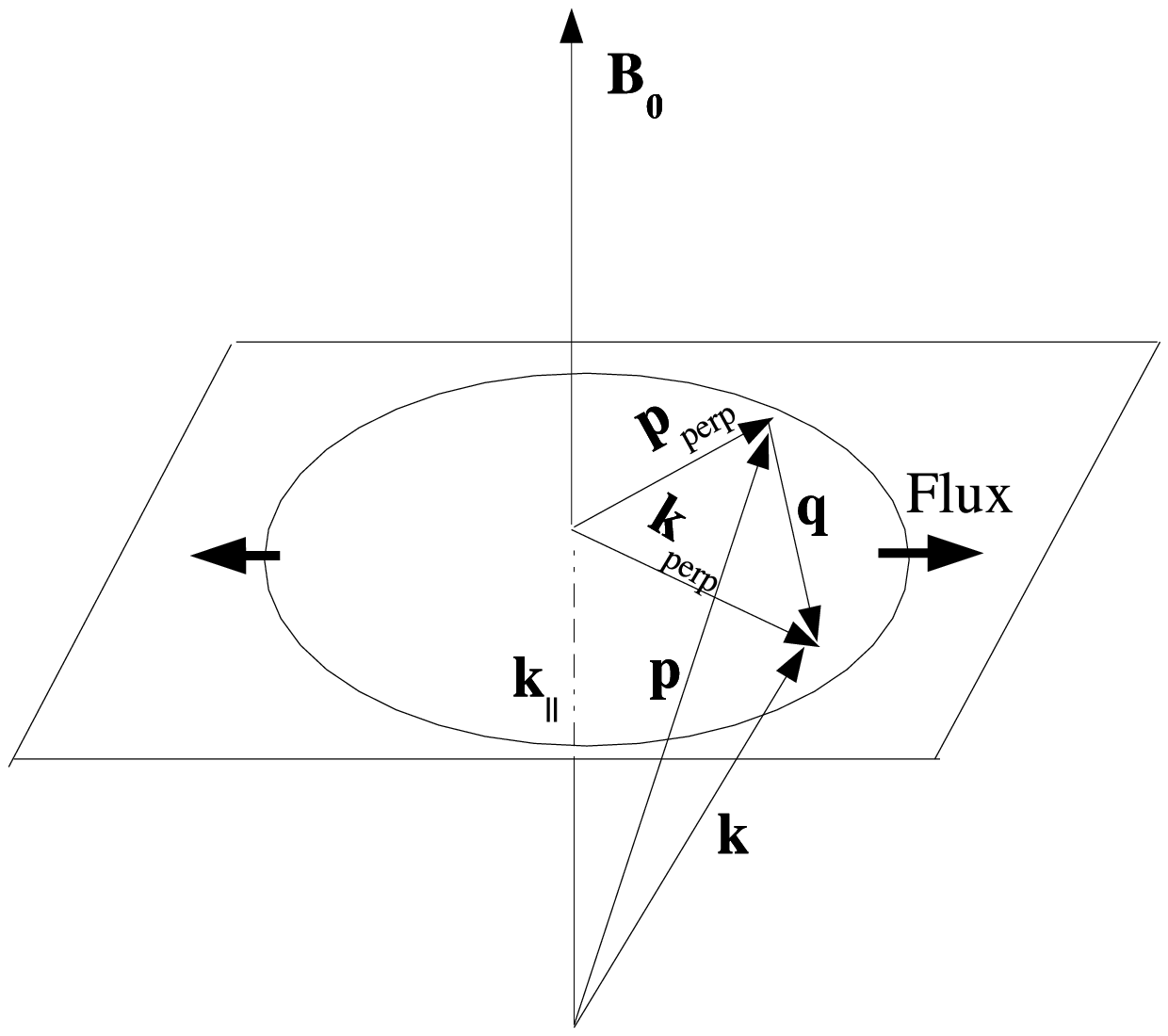}

\caption{\label{Fig:anisotropy} An illustration of an interacting MHD triad
in the presence of strong mean magnetic field.}
\end{figure}

When the mean magnetic field is strong, the oppositely moving Alfvén
waves interact weakly. Suppose three Alfvén waves under discussion
are $\mathbf{z^{+}(p},\omega_{p}),\mathbf{z^{-}(q,}\omega_{q})$ and
$\mathbf{z^{+}(k},\omega_{k})$. The wavenumbers and frequency of
the triads must satisfy the following relationships:\begin{eqnarray*}
\mathbf{p+q} & = & \mathbf{k},\\
\omega_{p}^{+}+\omega_{q}^{-} & = & \omega_{k}^{+},\end{eqnarray*}
 where $\omega_{k}^{\pm}=\mp B_{0}k_{||}$, and $||$ and $\perp$
represent parallel and perpendicular components respectively to the
mean magnetic field (Shebalin et al. \cite{Sheb}). Above relationships
immediately imply that $q_{||}=0.$ Hence, energy transfer could take
place from $\mathbf{p}$ to $\mathbf{k}$ in a plane perpendicular
to the mean magnetic field, as shown in Fig. \ref{Fig:anisotropy}.

Under a strong mean magnetic field, the turbulence is termed as weak.
In 1994 Sridhar and Goldreich \cite{Srid1} argued that the three-wave
resonant interaction is absent in MHD turbulence. They constructed
a kinetic theory based on four-wave interaction and showed that\[
E(k_{||},k_{\perp})\sim\epsilon^{1/3}V_{A}k_{\perp}^{-10/3}.\]
Later, Galtier et al. \cite{Galt:Weak} showed that three-wave interactions
are present in MHD, and modified the above arguments (to be discussed
in Section \ref{sub:Galtiers-et-al.}).

In a subsequent paper, Goldreich and Sridhar \cite{Srid2} constructed
a phenomenology for the strong turbulence. According to them, the
strong turbulence occurs when the time $\tau_{cascade}$ for eddies
of width $\lambda_{\perp}$ and length $\lambda_{\parallel}$ to pass
their energy to the smaller eddies is approximately $\lambda_{||}/C_{A}\sim\lambda_{perp}/z_{\lambda_{\perp}}^{\pm}$.
Assuming local interactions in the wavenumber-space, the turbulence
cascade rate $\Pi$ will be $(z_{\lambda_{\perp}}^{\pm})^{2}/\tau_{cascade}\sim(z_{\lambda_{\perp}}^{\pm})^{3}/\lambda_{\perp}$.
Since steady-state $\Pi$ is independent of $\lambda_{\perp}$, \begin{equation}
z_{\lambda_{\perp}}^{\pm}\propto\lambda_{\perp}^{1/3},\label{eq:GSz}\end{equation}
that immediately implies that \begin{equation}
E(k_{\perp})\propto k_{\perp}^{-5/3}.\end{equation}
The condition $\lambda_{||}/C_{A}\sim\lambda_{perp}/z_{\lambda_{\perp}}^{\pm}$
along with Eq. (\ref{eq:GSz}) yields \[
\lambda_{||}\propto\lambda_{\perp}^{2/3}.\]
The above results were expressed in the combined form as\begin{equation}
E(k_{\perp},k_{||})\sim\Pi^{2/3}k_{\perp}^{-10/3}g\left(k_{||}/k_{\perp}^{2/3}\right),\label{eq:GS_Ek}\end{equation}
from which we can derive \[
E(k_{\perp})\sim\int E(k_{\perp},k_{||})dk_{||}\sim k_{\perp}^{-8/3},\]
and \[
E(k_{||})\sim\int E(k_{\perp},k_{||})k_{\perp}dk_{\perp}\sim k_{||}^{-2}.\]

Thus Goldreich and Sridhar exploited anisotropy in MHD turbulence
and obtained Kolmogorov-like spectrum for energy. The above argument
is phenomenological. In Section \ref{sub:Goldreich-and-Sridhar's analytic}
we will present Goldreich and Sridhar's analytic argument \cite{Srid2}.
As will be discussed later, 5/3 exponent matches better with solar
wind observations and numerical simulation results.

\subsubsection{Verma- Effective Mean Magnetic Field and $E(k)\propto k^{-5/3}$}

In 1999, Verma \cite{MKV:B0_RG} argued that the scattering of Alfvén
waves at a wavenumber $k$ is caused by the combined effect of the
magnetic field with wavenumbers smaller than $k$. Hence, $B_{0}$
of Kraichnan and Iroshnikov theory should be replaced by an {}``effective
mean magnetic field''. Using renormalization group procedure Verma
could construct this effective field, and showed that $B_{0}$ is
scale dependent:\[
B_{0}(k)\propto k^{-1/3}.\]
By substituting the above expression in Eq. (\ref{eq:Kraich3/2}),
Verma  \cite{MKV:B0_RG} obtained Kolmogorov's spectrum for MHD turbulence.
The {}``effective'' mean magnetic field is the same as {}``local''
mean magnetic field of Cho et al. \cite{ChoVish:localB}.

\subsubsection{Galtier et al.- Weak turbulence and $E(k_{\perp})\propto k_{\perp}^{-2}$\label{sub:Galtiers-et-al.}}

Galtier et al. \cite{Galt:Weak} showed that the three-wave interaction
in weak MHD turbulence is not null, contrary to theory of Sridhar
and Goldreich \cite{Srid1}. Their careful field-theoretic calculation
essentially modified Eq. (\ref{eq:Dobro}) to \[
\Pi\sim\frac{1}{k_{||}B_{0}}E^{+}(k_{perp})E^{-}(k_{perp})k_{\perp}^{4}.\]
Hence, Galtier et al. effectively replaced $(kB_{0})^{-1}$ of KID's
model with more appropriate expression for Alfvén time-scale $(k_{||}B_{0})^{-1}$.
From the above equation, it can be immediately deduced that \[
E(k_{\perp})\propto k_{\perp}^{-2}.\]
In Section \ref{sub:Galtiers-et-al.'s analytic} we will present Galtier
et al.'s \cite{Galt:Weak} analytic arguments.

In the later part of the paper we will compare the predictions of
the above phenomenological theories with the solar wind observations
and numerical results. We find that Kolmogorov-like scaling models
MHD turbulence better than KID's phenomenology. We will apply analytic
techniques to understand the dynamics of MHD turbulence.

As discussed in earlier sections, apart from energy spectra, there
are many other quantities of interest in MHD turbulence. Some of them
are cross helicity, magnetic helicity, kinetic helicity, enstrophy
etc. The statistical properties of these quantities are quite interesting,
and they are addressed using (a) Absolute Equilibrium State (b) Selective
Decays (c) Dynamic Alignment, which are discussed below.

\subsection{Absolute Equilibrium States \label{sub:Absolute-Equilibrium-States}}

In fluid turbulence when viscosity is identically zero (inviscid limit),
kinetic energy is conserved in the incompressible limit. Now consider
independent Fourier modes (transverse to wavenumbers) as state variables
$y_{a}(t)$. Lesieur \cite{Lesi:book} has shown that these variables
move in a constant energy surface, and the motion is area preserving
like in Liouville's theorem. Now we look for equilibrium probability-distribution
function $P(\{ y_{a}\})$ for these state variables. Once we assume
ergodicity, the ideal incompressible fluid turbulence can be mapped
to equilibrium statistical mechanics \cite{Lesi:book}.

By applying the usual arguments of equilibrium statistical mechanics
we can deduce that at equilibrium, the probability distribution function
will be \[
P(y_{1},...,y_{m})=\frac{1}{Z}\exp{\left(-\frac{1}{2}\sigma\sum_{a=1}^{m}y_{a}^{2}\right)},\]
where $\sigma$ is a positive constant. The parameter $\sigma$ corresponds
to inverse temperature in the Boltzmann distribution. Clearly \[
\left\langle y_{a}^{2}\right\rangle =\int\Pi_{i}dy_{i}y_{a}^{2}P(\{ y_{i}\})=\frac{1}{\sigma},\]
 independent of $a$. Hence energy spectrum $C(\mathbf{k})$ is constant,
and 1-d spectrum will be proportional to $k^{d-1}$ \cite{Lesi:book}.
This is very different from Kolmogorov's spectrum for large $Re$
turbulence. Hence, the physics of turbulence at $\nu=0$ (inviscid)
differs greatly from the physics at $\nu\rightarrow0$. This is not
surprising because (a) turbulence is a nonequilibrium process, and
(b) Navier-Stokes equation is singular in $\nu$.

The equilibrium properties of inviscid MHD equations too has been
obtained by mapping it to statistical equilibrium system (Frisch et
al. \cite{Fris:HM}, Stribling and Matthaeus \cite{Stri:Abso}). Here
additional complications arise due to the conservation of cross helicity
and magnetic helicity along with energy. Stribling and Matthaeus \cite{Stri:Abso}
provide us with the analytic and numerical energy spectra for the
inviscid MHD turbulence. The algebra is straight forward, but somewhat
involved. In Fig. \ref{Fig:MHD-Spectrum} we illustrate their analytic
prediction for the spectrum \cite{Stri:Abso}. %
\begin{figure}
\includegraphics[%
  scale=0.5, bb=14 14 550 530 ]{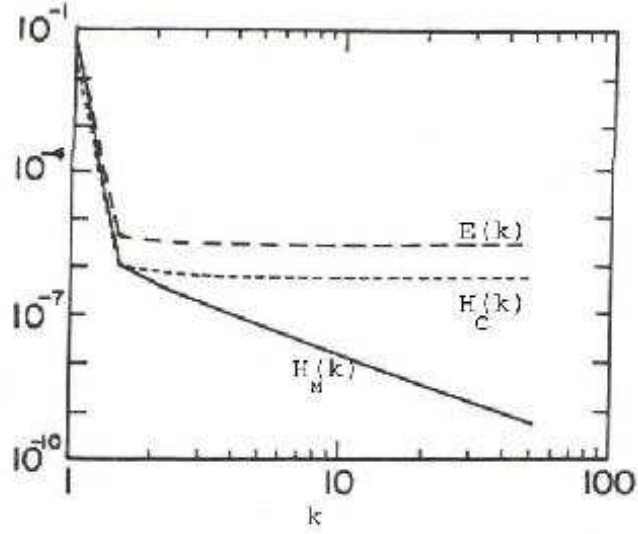}

\caption{\label{Fig:Troy} Spectrum of energy, cross helicity, and magnetic
helicity of absolute equilibrium state. Adopted from Stribling et
al. \cite{Stri:Abso}.}
\end{figure}
Clearly total energy and cross helicity appear to cascade to larger
wavenumbers, and magnetic helicity is peaking at smaller wavenumbers.

Even though nature of inviscid flow is very different from turbulent
flow, Kraichnan and Chen \cite{KraiChen} suggested that the tendency
of the energy cascade in turbulent flow could be anticipated from
the absolute equilibrium states. Suppose energy or helicity is injected
in some intermediate range, and if the inviscid spectrum peaks at
high wavenumber, then one may expect a direct cascade. On the contrary,
if the inviscid spectrum peaks at smaller wavenumber, then we expect
an inverse cascade. Frisch \cite{Fris:HM} and Stribling and Matthaeus
\cite{Stri:Abso} have done detailed analysis, and shown that the
energy and cross helicity may have forward cascade, and magnetic helicity
may have an inverse cascade.

Ting et al. \cite{Ting} studied the absolute equilibrium states for
2D inviscid MHD. They concluded that energy peaks at larger wavenumbers
compared to cross helicity and mean-square vector potential. Hence,
energy is expected to have a forward cascade. This is a very interesting
property because we can get reasonable information about 3D energy
spectra and fluxes by doing 2D numerical simulation, which are much
cheaper compared to 3D simulations.

\subsection{Spectrum of Magnetic Helicity and Cross Helicity}

As discussed in the previous subsection, absolute equilibrium states
of MHD suggest a forward energy cascade for energy and cross helicity,
and an inverse cascade for magnetic helicity (3D) or mean-square vector
potential (2D). The forward energy cascade has already been discussed
in subsection \ref{sub:MHD-Turbulence-Models}. Here we will discuss
the phenomenologies for the inverse cascade regime. 

\begin{figure}
\includegraphics[%
  scale=0.5,bb= 0 400 575 755]{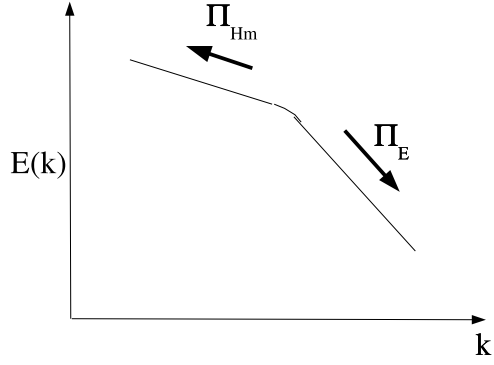}

\caption{\label{Fig:MHD-Spectrum}Cascade direction of energy $\Pi_{E}$ and
magnetic helicity $\Pi_{H_{M}}$in MHD turbulence.}
\end{figure}

The arguments are similar to the derivation of Kolmogorov's spectrum
for fluid turbulence (Sec. \ref{sub:Kolmogorov's-1941-Theory}). We
postulate a constant negative flux of magnetic helicity $\Pi_{H_{M}}$
at low wavenumbers (see Fig. \ref{Fig:MHD-Spectrum}). %
Hence, the energy spectrum in this range will have the form\[
E(k)\sim\left|\Pi_{H_{M}}\right|^{\chi}k^{\psi}.\]
Simple dimensional matching yields $\chi=2/3$ and $\psi=-1$. Hence\[
E(k)\sim\left|\Pi_{H_{M}}\right|^{2/3}k^{-1}.\]
We will show later that the inverse cascade of magnetic helicity assists
the growth of magnetic energy at large-scales, a process known as
{}``dynamo''.

Using similar analysis for 2D MHD, Biskamp showed that \[
E(k)\sim\left|\Pi_{A2}\right|^{2/3}k^{-1/3},\]
where $\Pi_{A2}$ is the flux of mean-square vector potential. Note
however that in 2D fluid turbulence, energy has \emph{inverse cascade},
but enstrophy ($\Omega=\int|\nabla\times\mathbf{u}|^{2}/2$) has \emph{forward
cascade} (Kraichnan \cite{Krai:71}), and the energy spectrum is \begin{eqnarray*}
E(k) & \sim & \Pi^{2/3}k^{-5/3}\,\,\,\,\, k\ll k_{f}\\
E(k) & \sim & \Pi_{\Omega}^{2/3}k^{-3}\,\,\,\,\,\, k\gg k_{f},\end{eqnarray*}
where $k_{f}$ is the forcing wavenumber, and $\Pi_{\Omega}$ is the
enstrophy flux.

\subsection{Dynamic Alignment}

In a decaying turbulence, energy decreases with time. Researchers
found that the evolution of other global quantities also have very
interesting properties. Matthaeus et al. \cite{Matt:DynaicAlign}
studied the evolution of normalized cross helicity $2H_{c}/E$ using
numerical simulations and observed that it increases with time. In
other words, cross helicity decays slower than energy. Matthaeus et
al. termed this phenomena as \emph{dynamic alignment} because higher
normalized cross helicity corresponds to higher alignment of velocity
and magnetic field. Pouquet et al. \cite{Pouq:HcGrowth} also observed
growth of normalized cross helicity in their simulation. The argument
of Matthaeus et al. \cite{Matt:DynaicAlign} to explain this phenomena
is as follows:

In KID's model of MHD turbulence, the energy fluxes $\Pi^{+}$ and
$\Pi^{-}$ are equal (see Eq. {[}\ref{eq:Dobro}{]}). Hence both $E^{+}$
and $E^{-}$ will get depleted at the same rate. If initial condition
were such that $E^{+}>E^{-}$, then $E^{+}/E^{-}$ ratio will increase
with time. Consequently $\sigma_{c}=(E^{+}-E^{-})/(E^{+}+E^{-})$
will also increase with time.

However, recent development in the field show that Kolmogorov-like
phenomenology (Marsch \cite{Mars:Kolm}, Goldreich and Sridhar \cite{Srid1,Srid2},
Verma \cite{MKV:B0_RG}) models the dynamics of MHD turbulence better
than KID's phenomenology. Keeping this in mind, we generalize the
arguments of Matthaeus et al. The rate of change of $E^{+}/E^{-}$is\[
\frac{d}{dt}\frac{E^{+}}{E^{-}}=\frac{1}{\left(E^{-}\right)^{2}}\left[E^{-}\dot{E^{+}}-E^{+}\dot{E^{-}}\right].\]
Clearly, $E^{+}/E^{-}$ will increase with time if \begin{equation}
\frac{\dot{E^{+}}}{\dot{E^{-}}}>\frac{E^{+}}{E^{-}}\,\,\, or\,\,\,\frac{\epsilon^{+}}{\epsilon^{-}}<\frac{E^{+}}{E^{-}}\label{eq:dynamic-align}\end{equation}
using $-\dot{{E^{\pm}}}=\epsilon^{\pm}$. If we assume $E^{+}/E^{-}\sim E^{+}(k)/E^{-}(k)$,
then Eq. (\ref{eq:MHD_Kolm_zpm}) yields\[
\frac{E^{+}}{E^{-}}\sim\frac{K^{+}}{K^{-}}\left(\frac{\epsilon^{+}}{\epsilon^{-}}\right)^{2}.\]
When $E^{+}/E^{-}$ is not much greater than 1, $K^{+}$and $K^{-}$
are probably very close. Hence, \[
\frac{E^{+}}{E^{-}}\sim\left(\frac{\epsilon^{+}}{\epsilon^{-}}\right)^{2}>\left(\frac{\epsilon^{+}}{\epsilon^{-}}\right).\]
Therefore, according to Eq. (\ref{eq:dynamic-align}) $E^{+}/E^{-}$
will increase with time in this limit. For the case $E^{+}/E^{-}\gg1$,
Verma \cite{MKV:MHD_Flux} showed that \[
\frac{\epsilon^{+}}{\epsilon^{-}}\approx\frac{1}{0.4}.\]
 Since $E^{+}/E^{-}\gg1$, $E^{+}/E^{-}>\epsilon^{+}/\epsilon^{-}$.
Hence, growth of normalized cross helicity $\sigma_{c}$ is consistent
with Kolmogorov-like model of MHD turbulence.

The above arguments are not applicable when the initial $\sigma_{c}=0$.
Numerically simulations show that $\sigma_{c}$ typically could deviate
up to 0.1-0.15. Also, cross helicity is quite sensitive to phases
of Fourier modes; we will discuss this phenomena in Section \ref{sub:Phase-Sensitivity}.
It would be interesting to study the evolution of cross helicity in
the language of symmetry-breaking and its possible generalization
to nonequilibrium situations.

\subsection{Selective Decay}

We saw in the previous section that the cross helicity $(E^{+}-E^{-})$
decays slower than energy $(E^{+}+E^{-})$. Let us look at it from
the decay equation of global quantities: \begin{eqnarray*}
\frac{dE}{dt} & = & \frac{d}{dt}\int d\tau\frac{1}{2}(u^{2}+b^{2})=-\nu\int d\tau\left|\bigtriangledown\times\mathbf{u}\right|^{2}-\eta\int d\tau j^{2},\\
\frac{dH_{c}}{dt} & = & \frac{d}{dt}\int d\tau\mathbf{u}\cdot\mathbf{B}=-(\nu+\eta)\int d\tau\mathbf{j}\cdot\bigtriangledown\times\mathbf{u}\\
\frac{dH_{M}}{dt} & = & \frac{d}{dt}\int d\tau\frac{1}{2}\left(\mathbf{A}\cdot\mathbf{B}\right)=-\frac{1}{2}\eta\int d\tau\mathbf{j}\cdot\mathbf{B}.\end{eqnarray*}
where $\mathbf{j}$ represents the current density. Since the dissipation
terms of $H_{M}$ has lower power of spatial derivatives as compared
to $E$, $H_{M}$ will decay slower than $E$. The decay rate of $H_{c}$
is slower because $H_{C}$ can take both positive and negative values.
Hence, $H_{c}$ and $H_{M}$ decay slower than $E$. This phenomena
is called \emph{selective decay,} first proposed by Matthaeus and
Montgomery \cite{Matt:Selective-decay}.

Several researchers argued that turbulence may relax to minimum energy
state under the constraint of constant magnetic helicity $H_{M}$.
This condition can be written as\[
\delta\left(\int d\tau\frac{1}{2}(u^{2}+b^{2})-\lambda\int d\tau\frac{1}{2}\left(\mathbf{A}\cdot\mathbf{B}\right)\right)=0.\]
 Variation with respect to $\mathbf{A}$ yields\[
\bigtriangledown\times\mathbf{B}-\lambda\mathbf{B}=0.\]
Variation with relative to $\mathbf{u}$ yields $\mathbf{u=0}$. The
above equation imply that current $\mathbf{j}=\bigtriangledown\times\mathbf{B}$
is parallel to $\mathbf{B}$, therefore force $\mathbf{j\times B}=0$.
Hence, the minimum-energy state is a static force-free field. This
result finds application in reversed-field-pinch plasma.

Slow decay or growth due to inverse cascade could produce coherent
structure. This process is referred to as self-organization process.
See Yanase et al. \cite{Yana:Coherent} for more detailed study of
this phenomena.

\subsection{{}``Phase'' Sensitivity of Global Quantities \label{sub:Phase-Sensitivity}}

Let us consider the complex Fourier mode $\mathbf{z}^{\pm}\left(\mathbf{k}\right)=\left|\mathbf{z}^{\pm}\left(\mathbf{k}\right)\right|\exp\left(i\theta^{\pm}\right)$.
Clearly there are four independent variables. The $\left|\mathbf{z}^{+}\left(\mathbf{k}\right)\right|$
and $\left|\mathbf{z}^{-}\left(\mathbf{k}\right)\right|$ fix $E^{+}\left(\mathbf{k}\right)$
and $E^{-}\left(\mathbf{k}\right)$ respectively. Since $E^{R}=Re(\mathbf{z}^{+}\left(\mathbf{k}\right)\cdot\mathbf{z}^{-*}\left(\mathbf{k}\right))\propto(E^{u}-E^{b})$,
$\theta^{+}-\theta^{-}$ together with $\left|\mathbf{z}^{\pm}\left(\mathbf{k}\right)\right|$
fix $E^{u}/E^{b}$. Hence, three global quantities $(E^{\pm},E^{R})$
or $(E^{u},E^{b},H_{c})$ are fixed by $\left|\mathbf{z}^{\pm}\left(\mathbf{k}\right)\right|$
and $\theta^{+}-\theta^{-}$, leaving the absolute value of $\theta^{+}$
free. Dar et al. \cite{Dar:Phase} studied the evolution of global
quantities by varying the absolute value of initial phase $\theta^{+}$
while keeping $\theta^{+}-\theta^{-}$ fixed. We term this as {}``phase''.

\begin{figure}
\includegraphics[%
  scale=0.5,bb= 14 14 550 430 ]{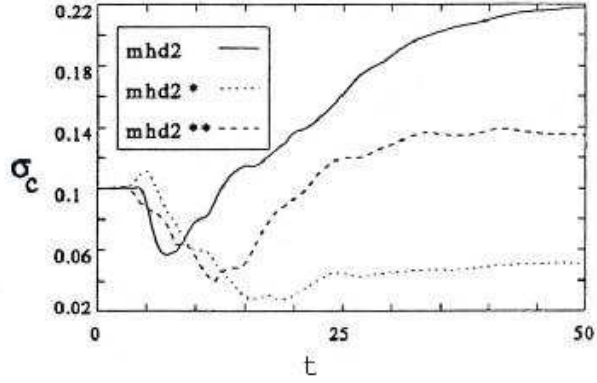}

\caption{\label{Fig:Dar-phase} Evolution of normalized cross helicity $\sigma_{c}$
for initial $\sigma_{c}=0.1,$ $r_{A}=0.5$. The curves correspond
to three different set of {}``phases'' . Adopted from Dar et al.
\cite{Dar:Phase}. }
\end{figure}

Dar et al. \cite{Dar:Phase} performed DNS on $512^{2}$ grid. They
performed one set of run (mhd) for random values of $\theta^{+}$
keeping $\theta^{+}-\theta^{-}$ fixed (by choosing appropriate $r_{A})$.
In the second run (mhd{*}) they changed $\theta^{+}$ uniformly for
all the modes by an amount $\Delta$, and the third run (mhd{*}{*})
the phase $\theta^{+}$ were shifted by a random amount. Dar et al.
found that total energy and Alfvén ratio do not depend on the shift
of $\theta^{+}$, however cross helicity depends quite sensitively
on the shift, specially when $\sigma_{c}$ is small. This result is
illustrated in Fig. \ref{Fig:Dar-phase}.%
Dar et al.'s result is very surprising, and its consequences have
not been studied in detail. This result raises the question on randomness
of initial conditions, ergodicity etc.

In this section we studied some of the basic phenomenological models
of MHD turbulence. We will compare their predictions with the numerical
results and solar wind observations. These are the topics of discussion
of the next two sections.

\section{Solar Wind: A Testbed for MHD Turbulence \label{sec:Solar-Wind}}

Analytical results are very rare in turbulence research because of
complex nature of turbulence. Therefore, experiments and numerical
simulations play very important role in turbulence research. In fluid
turbulence, engineers have been able to obtain necessary information
from experiments (e.g., wind tunnels), and successfully design complex
machines like aeroplanes, spacecrafts etc. Unfortunately, terrestrial
experiments exhibiting MHD turbulence are typically impossible because
of large value of resistivity and viscosity of plasmas. For example,
hydrogen plasma at temperature $10^{4}$K has resistivity approximately
$10^{5}cm^{2}/s$ (see Table \ref{Table:MHDParameters}). For a typical
laboratory setup of size $10cm$ and velocity scale of $10cm/s$,
magnetic Reynolds number will be $10^{-3}$, which is far from turbulent
regime. On the other hand, astrophysical plasmas have large length
and velocity scales, and are typically turbulent. They are a natural
testbed for MHD turbulence theories. We have been able to make large
set of measurements on nearest astrophysical plasma, the solar wind,
using spacecrafts. The data obtained from these measurements have
provided many interesting clues in understanding the physics of MHD
turbulence. Direct or indirect measurements on planetary and solar
atmosphere, galaxies etc. also provide us with useful data, and MHD
turbulence is applied to understand these astrophysical objects; due
to lack of space, we will not cover these astrophysical objects in
our review.

The Sun (or any typical star) spews out plasma, called solar wind.
This was first predicted by Parker in 1958, and later observed by
spacecrafts. The flow starts at the corona base and extends radially
outward beyond the planetary system, and terminates at around 100
AU (1 AU = Earth's orbital radius$\approx1.5\times10^{8}km$). Typical
observational data for the solar wind and the corona base is given
in Table \ref{table:SW-data} \cite{BiskTurb:book}. %
\begin{table}

\caption{\label{table:SW-data} Typical Observational Data on the Solar Wind }

\begin{tabular}{|c|c|c|}
\hline 
Quantity&
Corona Base&
1 AU\tabularnewline
\hline
\hline 
Ion Density&
$10^{9}$$cm^{-3}$&
3-20 $cm^{-3}$\tabularnewline
\hline 
Mean Velocity field&
300-800 $km/s$&
300-800 $km/s$\tabularnewline
\hline 
Velocity fluctuations&
?&
10-20 $km/s$\tabularnewline
\hline 
Mean magnetic field&
$10Gauss$&
$(3-20)\times10^{-5}Gauss$\tabularnewline
\hline 
Magnetic field fluctuations&
?&
$(1-3)\times10^{-5}Gauss$\tabularnewline
\hline 
Temperature&
$10^{6}$&
$10^{4}-10^{6}$\tabularnewline
\hline
\end{tabular}
\end{table}
The density of the wind decreases approximately as $r^{-2}$. The
mean magnetic field is largely polar in north-south direction, but
spirals out in the equatorial plane. Typical Sound speed ($C_{s}\approx\sqrt{{k_{B}T/m_{p}}}$)
is of the order of several hundred km/s. The density fluctuation $\delta\rho/\rho\approx(u/C_{s})^{2}\approx0.01$,
hence solar wind can be treated as incompressible fluid. 

\begin{figure}
\includegraphics[bb=86 14 447 441]{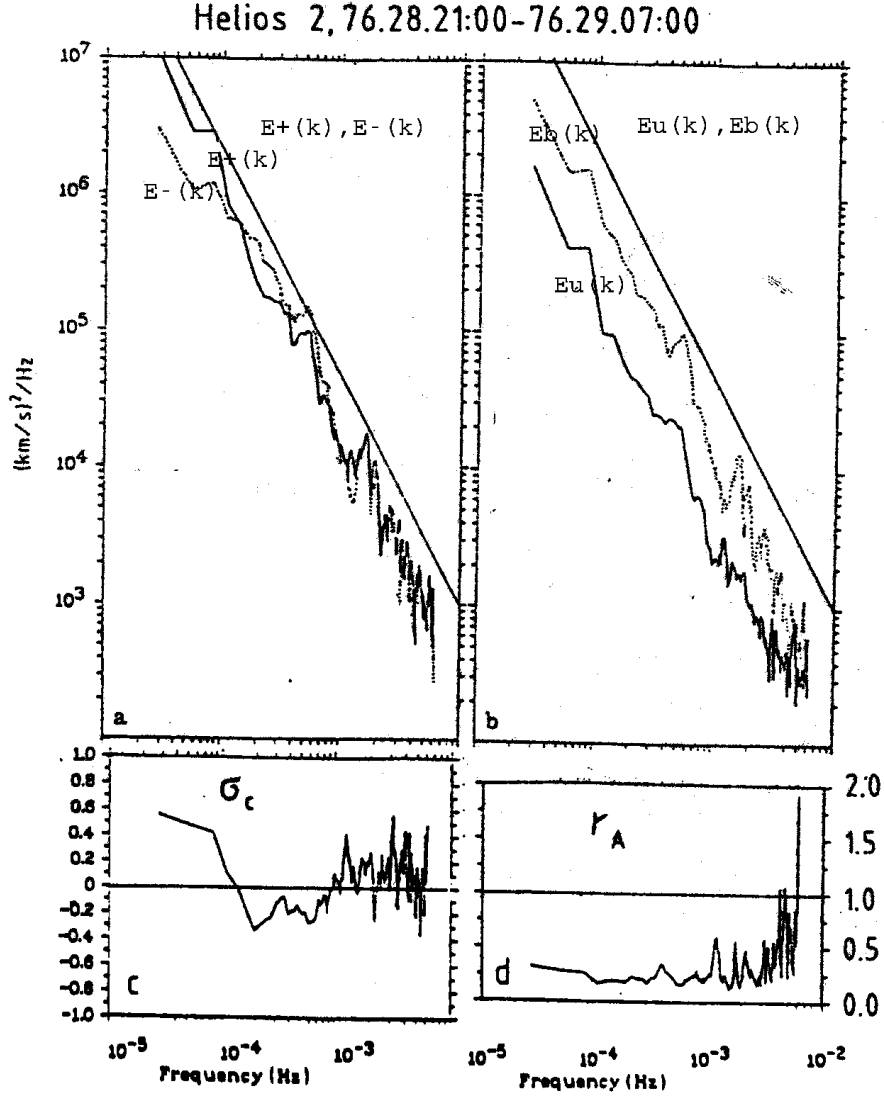}

\caption{\label{Fig:Ek-SW} Energy spectra $E^{\pm,u,b}(k)$ of a typical
solar wind stream. The normalized cross helicity $\sigma_{c}$ and
Alfvén ratio $r_{A}$ are also shown in the figure. Adopted from Tu
and Marsch \cite{TuMars91}}
\end{figure}

The solar wind data has been analyzed by many scientists. For details
the reader is referred to reviews by Goldstein et al. \cite{Gold:RevAnnual}
and Tu and Marsch \cite{TuMars:book}. The Alfvén ratio $r_{A}$,
which is the ratio of kinetic to magnetic energy, is dependent on
heliocentric distance and length-scale. The average value of $r_{A}$
in the inertial range decreases from near 5 at 0.3 AU to near 0.5
at 1AU and beyond \cite{MattGold,Robe:87a,Robe:87b}. The normalized
cross helicity $\sigma_{c}$, in general, decreases with increasing
heliocentric distance, with asymptotic values near $+1$ (purely outward
propagating Alfvén waves) near 0.3 AU, and near 0 by 8 AU or so \cite{MattGold,Robe:87a,Robe:87b}.
See Fig. \ref{Fig:Ek-SW} for an illustration.

Now let is focus on energy spectrum and turbulent dissipation rates
in the solar wind. Matthaeus and Goldstein \cite{MarsTu90} computed
the exponent of the total energy and magnetic energy. They found the
exponents to be $1.69\pm0.08$ and $1.73\pm0.08$ respectively, somewhat
closer to 5/3 than 3/2. Similar results were obtained by Marsch and
Tu \cite{MarsTu90} for $E^{\pm}(k)$ and $E^{u,b}(k)$ at various
heliocentric distances. Fig. \ref{Fig:Ek-SW} illustrates the energy
spectra $E^{\pm,u,b}(k)$ of a typical solar wind stream. This is
surprising because $B_{0}\gg\sqrt{kE^{\pm}(k)}$ for inertial range
wavenumbers in the solar wind, and according to KID's phenomenology,
the exponent should be 3/2 (see Section \ref{sub:MHD-Turbulence-Models}).
The phenomenological model of Matthaeus and Zhou, and Zhou and Matthaeus
\cite{ZhouMatt,MattZhou} predicts that KID's phenomenology should
hold when $\sqrt{kE^{\pm}(k)}\ll B_{0}$(high $k$), and Kolmogorov-like
phenomenology should be to be applicable when $\sqrt{kE^{\pm}(k)}\gg B_{0}$(low
$k$). We do not find any such break from 5/3 to 3/2 spectrum in the
observed spectrum, thus ruling out phenomenological model of Matthaeus
and Zhou, and Zhou and Matthaeus \cite{ZhouMatt,MattZhou}. 

The observational studies of Tu and Marsch \cite{TuMars:book} show
that the spectral index for large cross helicity is also close to
$5/3$. This is in contradiction to Grappin et al.'s predictions $\Pi^{+}/\Pi^{-}\approx m^{+}/m^{-}=(3-m^{-})/(3-m^{+})$
\cite{Grap82,Grap83}. Hence the solar wind observations invalidate
the phenomenology of Grappin et al. as well. On the whole, the solar
wind data appears to indicate that Kolmogorov-like model (5/3) is
applicable in MHD turbulence, even when the mean magnetic field is
large as compared to the fluctuations. As we will discuss in later
sections, numerical simulations and analytic arguments also support
this observation. 

As discussed above, the Alfvén ratio ($r_{A}=$$E^{u}/E^{b})$ is
high ($\approx5$) in the inner heliosphere, and it decreases to near
0.5 at 1AU. Similar evolution is seen in numerical simulations as
well. In Section \ref{sub:Field-theoretic-Calculation-of-shell} we
will discuss a plausible argument why Alfvén ratio evolves toward
1 or lower in decaying turbulence. 

Temperature of the solar wind decreases slower that adiabatic cooling,
implying that solar wind is heated as it evolves. Some of the studied
mechanism for heating are turbulence, shocks, neutral ions etc. Tu
\cite{Tu:Dissipation}, Verma et al. \cite{MKV:SWheat}, Matthaeus
et al. \cite{Matt:Dissipation}, Verma \cite{MKV:MHD_Flux}, and others
have estimated the turbulent dissipation rate in the solar wind from
the observational data and modelling. They argued that turbulent heating
can contribute significantly to heating of the solar wind.

There are interesting studies on coherent structures, compressibility,
density spectrum etc. in the solar wind. Due to lack of space, we
will not discuss them here and refer the readers to excellent reviews
on solar wind \cite{Gold:RevAnnual,TuMars:book}.

\section{Numerical Investigation of MHD Turbulence \label{sec:Numerical-Investigation-MHD}}

Like experiments, numerical simulations help us test existing models
and theories, and inspire new one. In addition, numerical simulations
can be performed for conditions which may be impossible in real experiments,
and all the field components can be probed everywhere, and at all
times. Recent exponential growth in computing power has fueled major
growth in this area of research. Of course, numerical simulations
have limitations as well. Even the best computers of today cannot
resolve all the scales in a turbulent flow. We will investigate these
issues in this section.

There are many numerical methods to simulate turbulence on a computer.
Engineers have devised many clever schemes to simulate flows in complex
geometries; however, their attention is typically at large scales.
Physicists normally focus on intermediate and small scales in a simple
geometry because these scales obey universal laws. Since nonlinear
equations are generally quite sensitive, one needs to compute both
the spatial and temporal derivatives as accurately as possible. It
has been shown that spatial derivative could be computed {}``exactly''
using Fourier transforms given enough resolutions \cite{Canu:book}.
Therefore, physicists typically choose spectral method to simulate
turbulence. Note however that several researchers (for example, Brandenburg
\cite{Bran:rev_computation}) have used higher order finite-difference
scheme and have obtained comparable results.

\subsection{Numerical Solution of MHD Equations using Pseudo-Spectral Method}

In this subsection we will briefly sketch the spectral method for
3D flows. For details refer to Canuto et al. \cite{Canu:book}. The
MHD equations in Fourier space is written as

\begin{eqnarray*}
\frac{\partial\mathbf{z^{\pm}}\left(\mathbf{k},t\right)}{\partial t} & = & \pm i\left(\mathbf{B}_{0}\cdot\mathbf{k}\right)\mathbf{z^{\pm}}\left(\mathbf{k},t\right)-i\mathbf{k}p\left(\mathbf{k},t\right)-FT\left[\mathbf{z^{\mp}}\left(\mathbf{k},t\right)\cdot\nabla\mathbf{\mathbf{z^{\pm}}\left(\mathbf{k},t\right)}\right]\\
 &  & -\nu_{\pm}k^{2}\mathbf{z^{\pm}}\left(\mathbf{k},t\right)-\nu_{\mp}k^{2}\mathbf{z^{\mp}}\left(\mathbf{k},t\right)+\mathbf{f}^{\pm}(\mathbf{k},t),\end{eqnarray*}
where $FT$ stands for Fourier transform, and $\mathbf{f}^{\pm}(\mathbf{k},t)$
are forcing functions. The flow is assumed to be incompressible, i.
e., $\mathbf{k}\cdot\mathbf{z^{\pm}}\left(\mathbf{k},t\right)=0$.
We assume periodic boundary condition with real-space box size as
$(2\pi)\times(2\pi)\times(2\pi)$, and Fourier-space box size as $(nx,ny,nz)$.
The allowed wavenumbers are ${\textbf{k}}=(k_{x},k_{y},k_{z})$ with
$k_{x}=(-n_{x}/2:n_{x}/2),k_{y}=(-n_{y}/2:n_{y}/2),k_{z}=(-n_{z}/2:n_{z}/2)$.
The reality condition implies that $\mathbf{z^{\pm}}\left(\mathbf{-k}\right)=\mathbf{z^{\pm*}}\left(\mathbf{k}\right)$,
therefore, we need to consider only half of the modes \cite{Canu:book}.
Typically we take $(-n_{x}/2:n_{x}/2,-n_{y}/2:n_{y}/2,0:n_{z}/2)$,
hence, we have $N=n_{x}*n_{y}*(n_{z}/2+1)$ coupled ordinary differential
equations. The objective is to solve for the field variables at a
later time given initial conditions. The following important issues
are involved in this method:

\begin{enumerate}
\item The MHD equations are converted to nondimensionalized form, and then
solved numerically. The parameter $\nu$ is inverse Reynold's number.
Hence, for turbulent flows, $\nu$ is chosen to be quite small (typically
$10^{-3}$ or $10^{-4}$). In Section \ref{sub:Kolmogorov's-1941-Theory}
we deduced using Kolmogorov's phenomenology that the number of active
modes are\[
N\sim\nu^{-9/4}.\]
 If we choose a moderate Reynolds number $\nu^{-1}=10^{4}$, $N$
will be $10^{9}$, which is a very large number even for the most
powerful supercomputers. To overcome this difficulty, researchers
apply some tricks; the most popular among them are introduction of
hyperviscosity and hyperresistivity, and large-eddy simulations. Hyperviscous
(hyperresistive) terms are of the form $(\nu_{j},\eta_{j})k^{2j}\mathbf{\mathbf{z^{\pm}}\left(\mathbf{k}\right)}$
with $j\ge2$; these terms become active only at large wavenumbers,
and are expected not to affect the inertial range physics, which is
of interest to us. Because of this property, the usage of hyperviscosity
and hyperresistivity has become very popular in turbulence simulations.
Large-eddy simulations will be discussed in Section \ref{sec:Miscellaneous-Topics}
of this paper. Just to note, one of the highest resolution fluid turbulence
simulation is by Gotoh \cite{Goto:DNS} on a $4096^{3}$ grid; this
simulation was done on Fujitsu VPP5000/56 with 32 processors with
8 Gigabytes of RAM on each processor, and it took 500 hours of computer
time.
\item The computation of the nonlinear terms is the most expensive part
of turbulence simulation. A naive calculation involving convolution
will take $O(N^{2})$ floating point operations. It is instead efficiently
computed using Fast Fourier Transform (FFT) as follows: \\
(a) Compute $\mathbf{\mathbf{z^{\pm}}\left(x\right)}$ from $\mathbf{\mathbf{z^{\pm}}\left(\mathbf{k}\right)}$
using Inverse FFT. \\
(b) Compute $z_{i}^{\mp}({\textbf{x}})z_{j}^{\pm}({\textbf{x}})$
in real space by multiplying the fields at each space points.\\
(c) Compute $FFT[z_{i}^{\mp}({\textbf{x}})z_{j}^{\pm}({\textbf{x}})]$
using FFT.\\
(d) Compute $ik_{j}FFT[z_{i}^{\mp}({\textbf{x}})z_{j}^{\pm}({\textbf{x}})]$
by multiplying by $k_{j}$ and summing over all $j$. This vector
is $-FFT\left[\mathbf{z^{\mp}}\left(\mathbf{k},t\right)\cdot\nabla\mathbf{\mathbf{z^{\pm}}\left(\mathbf{k},t\right)}\right]$.\\
Since FFT takes $O(N\log N)$, the above method is quite efficient.
The multiplication is done in real space, therefore this method is
called pseudo-spectral method instead of just spectral method.
\item Products $z_{i}^{\mp}({\textbf{x}})z_{j}^{\pm}({\textbf{x}})$ produce
modes with wavenumbers larger than $k_{max}$. On FFT, these modes
get aliased with $k<k_{max}$ and will provide incorrect value for
the convolution. To overcome this difficulty, last 1/3 modes of fields
$z_{i}^{\pm}(\mathbf{k})$ are set to zero (zero padding), and then
FFTs are performed. This scheme is called $2/3$ rule. For details
refer to Canuto et al. \cite{Canu:book}.
\item Pressure is computed by taking the dot product of MHD equation with
$\mathbf{k}$. Using incompressibility condition one obtains\[
p\left(\mathbf{k},t\right)=\frac{i\mathbf{k}}{k^{2}}\cdot FT\left[\mathbf{z^{\mp}}\left(\mathbf{k},t\right)\cdot\nabla\mathbf{\mathbf{z^{\pm}}\left(\mathbf{k},t\right)}\right].\]
To compute $p({\textbf{k}})$ we use already computed nonlinear term. 
\item Once the right-hand side of the MHD equation could be computed, we
could time advance the equation using one of the standard techniques.
The viscous terms are advanced using an implicit method called Crank-Nicholson's
scheme. However, the nonlinear terms are advanced using Adam-Bashforth
or Runge-Kutta scheme. One uses either second or third order scheme.
Choice of $dt$ is determined by CFL criteria $(dt<(\bigtriangleup x)/U_{rms})$.
By repeated application of time-advancing, we can reach the desired
final time.
\item The MHD turbulence equations can be solved either using $\mathbf{z^{\pm}}$
or $\mathbf{\left(\mathbf{u},b\right)}$. The usage of $\mathbf{z^{\pm}}$
turns out to be more efficient because they involve less number of
FFT operations.
\item When forcing $\mathbf{f}^{\pm}=0$, the total energy gets dissipated
due to viscosity and resistivity. This is called decaying simulation.
On the contrary, forced simulation have nonzero forcing $(\mathbf{f}\ne0)$,
which feed energy into the system, and the system typically reaches
a steady-state in several eddy turnover time. Forcing in astrophysical
and terrestrial systems are typically at large-scale eddies (shocks,
explosions etc.). Therefore, in forced MHD equations $\mathbf{f}^{u}$
is typically applied at small wavenumbers, which could feed both kinetic
energy and kinetic helicity. For details refer to Brandenburg \cite{Bran:Alpha}.
\end{enumerate}
Spectral method has several disadvantages as well. This method can
not be easily applied to nonperiodic flows. That is the reason why
engineers hardly use spectral method. Note however that even in aperiodic
flows with complex boundaries, the flows at small length-scale can
be quite homogeneous, and can be simulated using spectral method.
Spectral simulations are very popular among physicists who try to
probe universal features of small-scale turbulent flows. Since the
MHD equations are solved directly (without any modeling), this method
is called Direct Numerical Simulation (DNS).

Many researchers have done spectral simulation of MHD turbulence.
In this section we will mention some of the main results concerning
energy spectra and cascade rates. Numerical results on dynamo and
intermittency will be discussed later in this paper. Some numerical
results on the evolution of global quantities (e. g., dynamic alignment
by Matthaeus et al. \cite{Matt:DynaicAlign}) were discussed were
discussed in Section \ref{sec:MHD-Turbulence-Models}, and they will
not be repeated here.

\subsection{Numerical Results on Energy Spectra (3/2 or 5/3) \label{sub:Numerical-Results-on-spectra}}

In Sec. \ref{sec:MHD-Turbulence-Models} we discussed various MHD
turbulence phenomenologies, which predict the exponents to be 3/2,
or 5/3, or mix of both. Grappin et al. \cite{Grap83} predicted the
exponents to be cross helicity dependent; for small $\sigma_{c}$,
$m^{+}\approx m^{-}\approx3/2$, but for large $\sigma_{c}$, $m^{+}\rightarrow3$,
and $m^{-}\rightarrow0$. Many researchers tried to test these predictions
numerically. 

One-dimensional energy spectrum $E(k)$ is computed by summing over
all the modes in the shell $(k-1/2:k+1/2)$, i.e.,\[
E^{X}(k)=\frac{1}{2}\sum_{k-1/2<|\mathbf{s}|<k+1/2}\left|\hat{\mathbf{X}}(\mathbf{s})\right|^{2},\]
where $\mathbf{X}=\mathbf{u},\mathbf{b},\mathbf{z}^{\pm}$. The Energy
spectrum is computed for both decaying or forced simulations. In the
final state (after 2-10 eddy turnover time), Alfvén ratio is typically
found to be close to 1/2. Most of the MHD turbulence simulations have
been done for zero cross helicity; for these cases, normalized cross
helicity typically fluctuates in the range of $-0.1$ to 0.1.

Most of the high resolution simulations till early 1990s were done
in 2D due to lack of computing resources. Biskamp and Welter \cite{BiskWelt}
performed numerical studies of 2D MHD turbulence on grid up to $1024^{2}$
under small cross helicity limit. They reported the spectral index
to be close to $3/2$ in agreement with the models of Kraichnan, Iroshnikov,
and Dobrowolny et al. (KID), with a caveat that the exponents may
be close to 5/3 in transition states, in which turbulence is concentrated
in regions of weak magnetic field. In summary, the numerical simulations
till early 1990s supported 3/2 spectral index. Note that according
to absolute equilibrium theory, 2D and 3D are expected to have the
same energy spectra. So we can test the turbulence models in 2D as
well.

\begin{figure}
\includegraphics[%
  scale=1.0,bb=0 0  550 400 ]{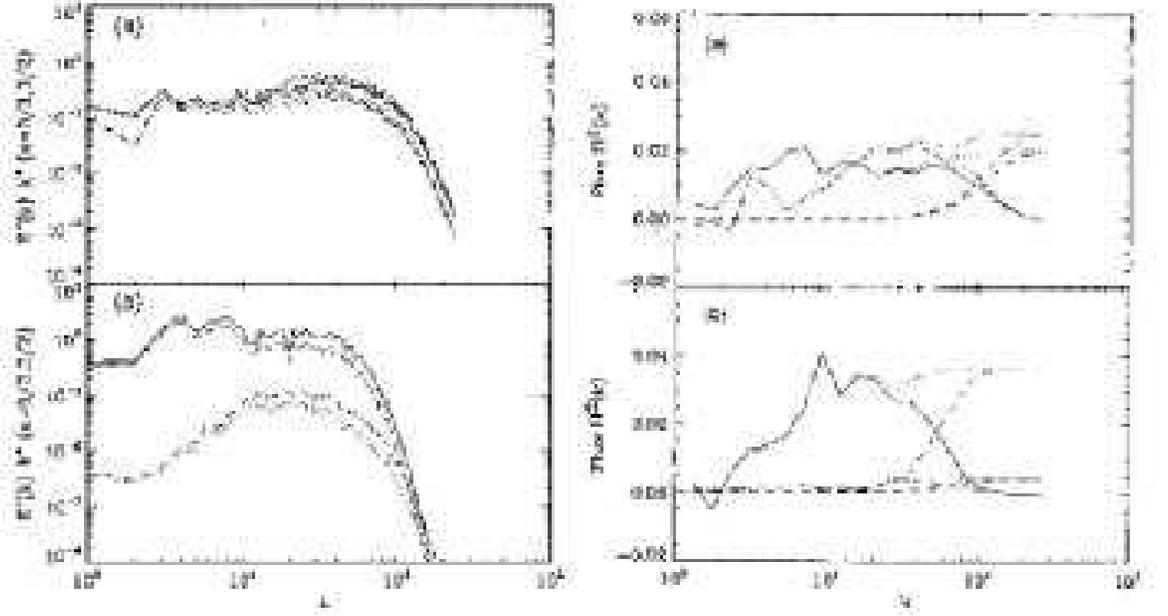}

\caption{\label{Fig:MKV-sim} $E^{\pm}(k)k^{a}$ vs. $k$ (left panel) and
$\Pi^{\pm}(k)$ vs. $k$ (right panel) for 2D runs with (a) $B_{0}=0$
and initial $\sigma_{c}=0.0$, and (b) $B_{0}=1.0$ and initial $\sigma_{c}=0.9$.
In the left panel the solid $(E^{+})$ and dashed line $(E^{-})$
correspond to $a=5/3$, and chained $(E^{+})$ and dotted $(E^{-})$
correspond to $a=3/2$. In the right panel solid and chained lines
represent $\Pi^{\pm}$ respectively, while the dashed and dotted lines
represent dissipation rate and the energy loss in the sphere of radius
$k$. The numerical results favour Kolmogorov's phenomenology over
KID's phenomenology. Adopted from Verma et al. \cite{MKV:MHD_Simulation}.}
\end{figure}

Since 5/3 and 3/2 are very close, there is a practical difficulty
in resolving the spectral index. They can be resolved with certainty
only in a high-resolution simulations. Verma \cite{MKV:Thesis}, and
Verma et al. \cite{MKV:MHD_Simulation} approached this problem indirectly.
They tested the energy cascade rates $\Pi^{\pm}$ for nonzero cross
helicity in $512^{2}$ DNS. Recall that KID's model (3/2) predicts
$\Pi^{+}=\Pi^{-}$(Eq. {[}\ref{eq:Dobro}{]}) independent of $E^{+}/E^{-}$
ratio, while Kolmogorov-like theories (5/3) predict (Eq. {[}\ref{eq:MHD_Kolm_zpm}{]})\[
\frac{E^{-}(k)}{E^{+}(k)}=\frac{K^{-}}{K^{+}}\left(\frac{\Pi^{-}}{\Pi^{+}}\right)^{2}.\]
Verma \cite{MKV:Thesis} and Verma et al. \cite{MKV:MHD_Simulation}
computed both energy spectra and cascade rates; their plots of the
energy spectra and fluxes are reproduced in Fig. \ref{Fig:MKV-sim}.

Regarding spectral indices, no particular claim could be made because
the numerically computed indices were within the error bars of both
3/2 and 5/3. However, the study of energy fluxes showed that underlying
turbulent dynamics is closer to Kolmogorov-like. The energy flux of
majority species (larger of $E^{+}$and $E^{-}$, here taken to be
$E^{+}$) was always greater that that of minority species, even in
situations where $z_{rms}^{\pm}\ll B_{0}$. When we look at the values
of cascade rates more closely (see Table 1 of Verma et al. \cite{MKV:MHD_Simulation}),
we find that ,\begin{equation}
\frac{E^{-}(k)}{E^{+}(k)}\approx\left(\frac{\Pi^{-}}{\Pi^{+}}\right)^{2}\label{eq:MKV_Kolm_flux}\end{equation}
for initial $\sigma_{c}=0.25$. However, for initial $\sigma_{c}=0.9$,
they were off by a factor of 10. The Eq. (\ref{eq:MKV_Kolm_flux})
assumes $K^{+}=K^{-}$, which is not a valid assumption for large
$\sigma_{c}$. Verma \cite{MKV:MHD_Flux} has shown that $K^{\pm}$
depend on $\sigma_{c}$, and the factor of $K^{-}/K^{+}$ is of the
order of 4. With this input, the numerical results come closer to
the analytical results, but the agreement is still poor. This indicates
that physics at large cross helicity (Alfvénic turbulence) is still
unresolved.

The above numerical results imply that the relevant time scale for
MHD turbulence is nonlinear time-scale $(kz_{k}^{\pm})^{-1}$, not
the Alfvén time-scale $(kB_{0})^{-1}$. Verma's \cite{MKV:Thesis}
and Verma et al.'s \cite{MKV:MHD_Simulation} provided one of the
first numerical evidence that Kolmogorov's scaling is preferred over
KID's 3/2 scaling in MHD turbulence.

\begin{figure}
\includegraphics[%
  scale=0.6,bb=14 14 450 300 ]{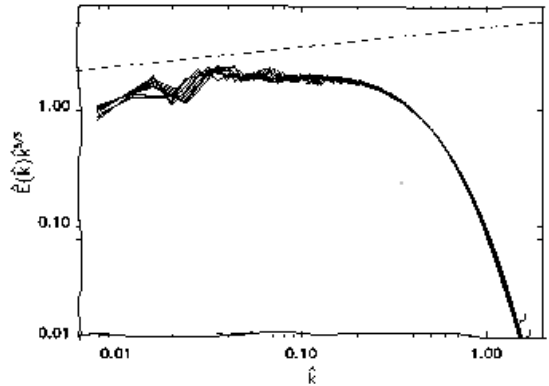}

\caption{\label{Fig:Bisk-Ek} Plot of normalized energy spectrum compensated
by $k^{5/3}$for a 3D MHD simulation on $512^{3}$grid. Flatness of
the plot indicates that Kolmogorov's 5/3 index fits better than 3/2
(dashed line). Adopted from Biskamp and Müller \cite{Bisk:Kolm2}.}
\end{figure}

High resolution three-dimensional simulations soon became possible
due to availability of powerful computers. Müller and Biskamp \cite{Bisk:Kolm1},
and Biskamp and Müller \cite{Bisk:Kolm2} performed $512^{3}$ DNS
with both normal and hyperdiffusive terms. They showed that the energy
spectrum follows a $k^{-5/3}$ law, steeper than $k^{-3/2}$ as previously
thought (see Fig. \ref{Fig:Bisk-Ek}). %
The runs with hyperdiffusivity had a bump at large wavenumbers. The
Kolmogorov's constant was found to be 2.3. The range of 5/3 powerlaw
is close to 1 decade. Numerical results of Cho \cite{Cho:Simulation},
Cho et al. \cite{ChoLaz:Viscosity-damped}, and others are consistent
with Kolmogorov's scaling.

Biskamp and Müller also computed the intermittency exponents and showed
that they are consistent with Kolmogorov scaling and sheet-like dissipative
structure. We will discuss these issues later in this paper. 

Biskamp and Schwarz \cite{Bisk:2D} performed DNS on two-dimensional
MHD turbulence on grids of $2048^{2}$ to $8192^{2}$. They claimed
that the energy spectrum agrees with KID's law (3/2), contrary to
3D case. They have also computed the structure functions, and reported
a strong anomalous bottleneck effect. Note that Biskamp and Schwarz's
results contradicts Verma et al.'s \cite{MKV:MHD_Simulation} results,
where the energy fluxes follow Kolmogorov's predictions. This issue
needs a closer look. It is possible that the dynamics is Kolmogorov-like,
but they are strongly modified by intermittency effects. Refer to
Verma et al. \cite{MKV:Bisk_Comment}, Biskamp \cite{Bisk:MKV_Comment},
and Section \ref{sec:Intermittency-in-MHD} for further details.

\subsection{Numerical Results on Anisotropic Energy Spectra}

Shebalin et al. \cite{Sheb} performed DNS in 2D and studied the anisotropy
resulting from the application of a mean magnetic field. They quantified
anisotropy using the angle $\theta_{Q}$ defined by\[
\tan^{2}\theta_{Q}=\frac{\sum k_{\perp}^{2}\left|Q\left(\mathbf{k},t\right)\right|^{2}}{\sum k_{||}^{2}\left|Q\left(\mathbf{k},t\right)\right|^{2}},\]
where $Q$ represents any one of the vector fields like $\mathbf{u},\mathbf{b},\nabla\times\mathbf{u}$
etc. They found turbulence to be anisotropic. Later Oughton et al.
\cite{Ough} carried out the anisotropic studies in 3D. They found
that with the increase of $B_{0}$, anisotropy increases up to $B_{0}\sim3,$
then it saturated. They also found that anisotropy increases with
increasing mechanical and magnetic Reynolds numbers, and also with
increasing wavenumbers. $B_{0}$ also tended to suppress energy cascade.
Matthaeus et al. \cite{Matt:AnisotropyPRL} numerically show that
the anisotropy scales linearly with the ratio of fluctuating to total
magnetic field strength.

\begin{figure}
\includegraphics[%
  scale=0.5,bb=14 14 600 700 ]{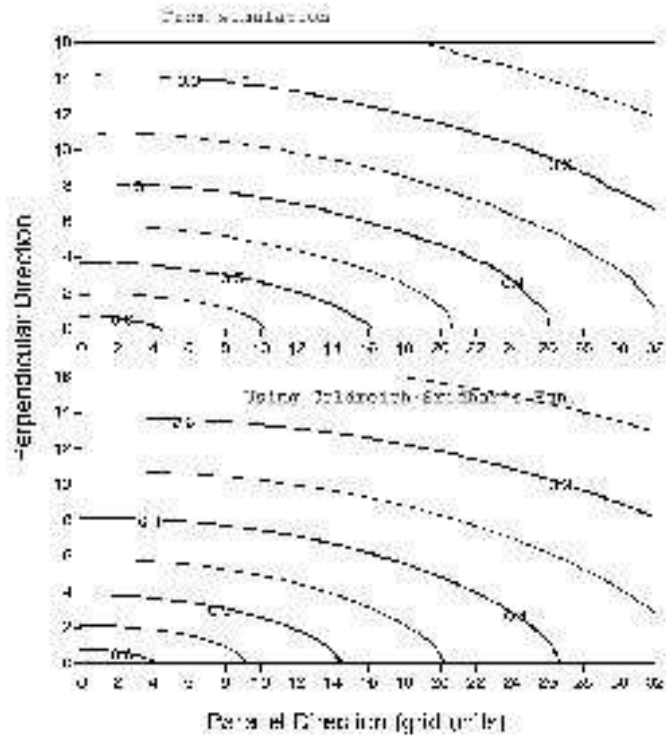}

\caption{\label{Fig:Cho-Laz} Cho et al. \cite{ChoVish:localB} compared the
velocity correlation function from simulation (top panel) with the
predictions of Goldreich and Sridhar's theory (bottom panel). The
results are in very good agreement. Adopted from Cho et al. \cite{ChoVish:localB}.}
\end{figure}

Cho et al. \cite{ChoVish:anis,ChoVish:localB} performed 3D DNS and
studied anisotropic spectrum. They found that anisotropy of eddies
depended on their size: along {}``local'' magnetic field lines,
the smaller eddies are more elongated than the larger ones. See Fig.
\ref{Fig:Cho-Laz} for an illustration of numerically computed velocity
correlation function. %
The numerical value matched quite closely with the predictions of
Goldreich and Sridhar \cite{Srid2}. Their result was also consistent
with the scaling law $k_{||}\sim k_{\perp}^{2/3}$ proposed by Goldreich
and Sridhar \cite{Srid2}. Here $k_{||}$ and $k_{\perp}$ are the
wavenumbers measured relative to the local magnetic field direction.
The {}``local'' magnetic field is in the same spirit as the {}``effective''
mean magnetic field of Verma \cite{MKV:B0_RG}.

Maron and Goldreich \cite{Maro:Simulation} performed a detailed DNS
of MHD turbulence. Their grid size ranged from $(64^{2}\times256)$
to $(256^{2}\times512)$. They numerical results are in general agreement
with the Goldreich and Sridhar prediction $k_{||}\sim k_{\perp}^{2/3}$.
However, their 1D spectral index was closer to 3/2 than 5/3, contrary
to Cho et al. 's \cite{ChoVish:anis,ChoVish:localB} numerical results
that $E(k_{\perp})\propto k_{\perp}^{-5/3}$.

After a review of energy spectra in MHD turbulence, we now turn to
studies on energy fluxes in MHD turbulence.

\subsection{Numerical Results on Energy Fluxes}

Computation of energy fluxes using DNS has done by Verma et al. \cite{MKV:MHD_Simulation},
Dar et al. \cite{Dar:flux}, and Ishizawa and Hattori \cite{Ishi:EDQNM,Ishi:flux}
(using wavelet basis). Verma et al. numerically computed $\Pi^{\pm}$,
while Dar et al., and Ishizawa and Hattori computed various fluxes
$\Pi_{Y>}^{X<}$ ($X,Y=u,b)$ in 2D MHD turbulence.

Dar et al. performed numerical DNS on a $512^{2}$ grid with random
kinetic forcing over a wavenumber annulus $4\le k\le5$. Theoretically,
the magnetic energy in two-dimensional MHD decays in the long run
even with steady kinetic energy forcing \cite{Zeld:book}. However,
we find that the magnetic energy remains steady for sufficiently long
time before it starts to decay. For $128^{2}$ simulation, the decay
of magnetic energy starts only after $t=25-30$ time units, and for
$512^{2}$, the decay starts much later. We compute the energy fluxes
in this quasi steady-state. In the quasi steady-state, the Alfvén
ratio fluctuates between 0.4 and 0.56, and the normalized Cross helicity
$\sigma_{c}$ approximately equal to 0.1. They found however that
a variation of $\sigma_{c}$ does not change the behaviour of flux
appreciably.

We illustrate the flux computation using an example. The flux $\Pi_{b>}^{u<}(k_{0})$,
which is energy flux from inside of $u$-sphere of radius $k_{0}$,
to outside of $b$-sphere of radius $k_{0}$, is\begin{equation}
\Pi_{b>}^{u<}(k_{0})=\sum_{\mathbf{\left|k'\right|<k_{0}}}\sum_{\left|\mathbf{p}\right|<k_{0}}\Im\left(\left[\mathbf{k}'\cdot\mathbf{b}(\mathbf{q})\right]\left[\mathbf{b}(\mathbf{k'})\cdot\mathbf{u}(\mathbf{p})\right]\right).\label{eq:Piulbg_num}\end{equation}
Now we define two {}``truncated'' variables ${\textbf{u}}^{<}$
and ${\textbf{b}}^{>}$ as \begin{equation}
{\textbf{u}}^{<}({\textbf{k}})=\left\{ \begin{array}{ll}
{\textbf{u}}({\textbf{k}}) & \mbox{if $|{\textbf{k}}|<k_{0}$}\\
0 & \mbox{if $|{\textbf{k}}|>k_{0}$}\end{array}\right.\end{equation}
 and \begin{equation}
{\textbf{b}}^{>}({\textbf{p}})=\left\{ \begin{array}{ll}
0 & \mbox{if $|{\textbf{p}}|<K$}\\
{\textbf{b}}({\textbf{p}}) & \mbox{if $|{\textbf{p}}|>K$}\end{array}\right.\end{equation}
 Eq.~(\ref{eq:Piulbg_num}) written in terms of ${\textbf{u}}^{<}$
and ${\textbf{b}}^{>}$ reads as \begin{equation}
\Pi_{b>}^{u<}(K)=\Im\left[\sum_{\textbf{k}}k_{j}b_{i}^{>}({\textbf{k}})\sum_{\textbf{p}}b_{j}({\textbf{k-p}})u_{i}^{<}({\textbf{p}})\right]\label{eq:Piulbg_num_FFT}\end{equation}
 The ${\textbf{p}}$ summation in the above equation is a convolution,
which is computed using FFT. After FFT, $\mathbf{k}$ sum is performed.
We compute energy flux this way. Ishizawa and Hattori \cite{Ishi:EDQNM,Ishi:flux}
used Meyor wavelets as basis vectors and have computed the energy
fluxes. Their approach is very similar to that of Dar et al.

\begin{figure}
\includegraphics[bb= 14 14 312 226]{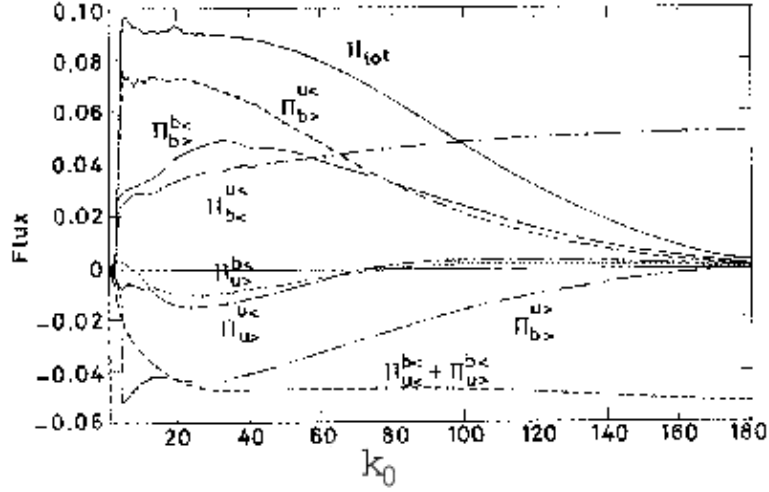}

\caption{\label{Fig:Dar-flux-vs-k} The plots of various energy fluxes in
2D-MHD versus wavenumber for a $512^{2}$ simulation. Adopted from
Dar et al. \cite{Dar:flux}.}
\end{figure}

\begin{figure}
\includegraphics[%
  scale=0.6,bb= 91 156 521 635 ]{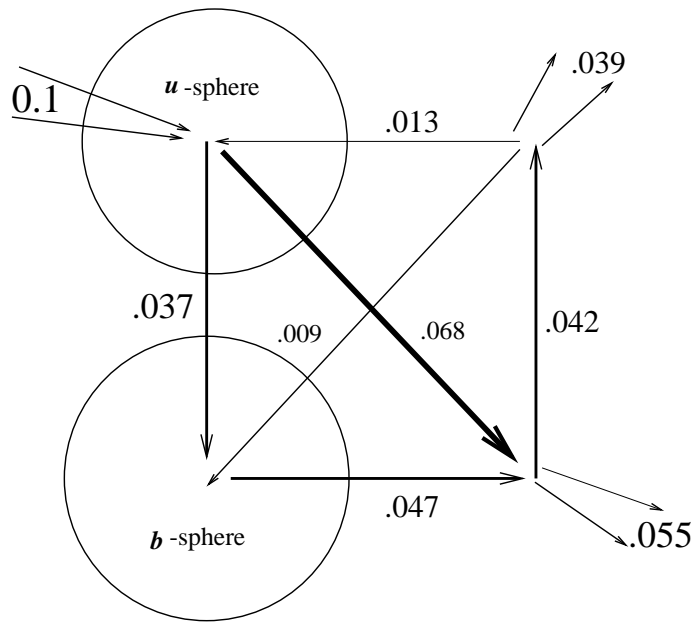}

\caption{\label{Fig:Dar-schematic} A schematic illustration of the numerically
evaluated values of the fluxes averaged over 15 time units. The values
shown here are for $k=20,$ a wavenumber within the inertial range.
Taken from Dar et al. \cite{Dar:flux}}
\end{figure}

The energy fluxes as a function of wavenumber spheres are plotted
in Fig. \ref{Fig:Dar-flux-vs-k}.%
 The inertial range $20<k<50$ can be deduced from the approximate
constancy of the energy fluxes. In Fig. \ref{Fig:Dar-schematic} we
schematically illustrate the numerical values of fluxes for $k=20$,
a wavenumber within the inertial range. %
The main conclusions of Dar et al.'s calculation are as follows:

\begin{enumerate}
\item The sources for the large-scale magnetic field are large-scale velocity
field ($\Pi_{b<}^{u<}$) and the small-scale velocity field ($\Pi_{u>}^{b<}$).
The former transfer is of a greater magnitude than the latter, hence
the magnetic field enhancement is primarily caused by a transfer from
the \textit{u}-sphere to the \textit{b}-sphere. Indeed the first few
\textit{b}-modes get most of this energy. 
\item Energy from the large-scale velocity field gets transferred to the
small-scale magnetic field ($\Pi_{b>}^{u<}$). The small-scale magnetic
field transfers energy to the small-scale velocity field ($\Pi_{u>}^{b>}$).
\item Small-scale velocity field transfers energy to large-scale velocity
as well as magnetic field. Hence, there is an \textit{inverse} cascade
of kinetic energy. This result is consistent with the EDQNM closure
calculations \cite{Pouq:EDQNM2D,Ishi:EDQNM}. 
\item There is a \textit{forward} cascade of magnetic energy toward the
small scales. EDQNM closure calculations also yield a magnetic energy
transfer to the small-scales \cite{Ishi:EDQNM}.
\end{enumerate}
Ishizawa and Hattori \cite{Ishi:EDQNM,Ishi:flux} obtained very similar
results in their DNS using wavelet basis. 

\begin{figure}
\includegraphics[%
  scale=0.5, bb=24 185 548 593 ]{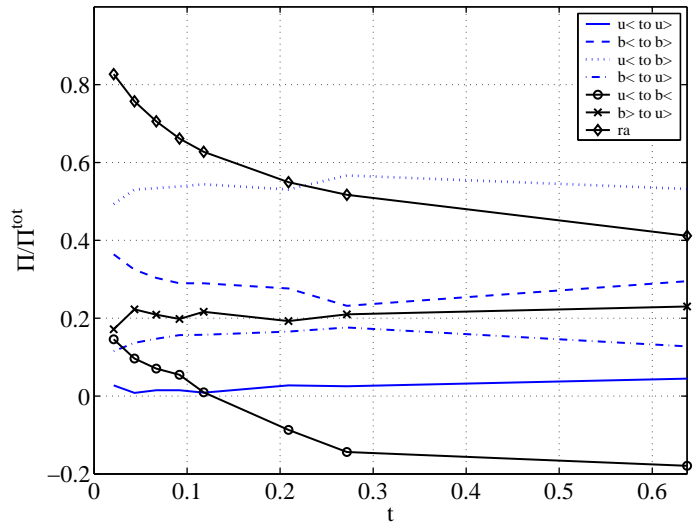}

\caption{\label{Fig:Olivier} Values of inertial-range energy fluxes at different
time in a $512^{3}$ decaying 3D-MHD turbulence simulation. Note the
change in sign for some of the fluxes near $r_{A}=0.6$. }
\end{figure}

We have performed a preliminary 3D simulation on $512^{3}$ grid size
and calculated the energy fluxes \cite{Oliv:Simulation}. Ours was
a decaying simulation in which Alfvén ratio saturated near 0.4. The
energy fluxes at $k=11$, one of the inertial range wavenumbers, is
shown in Fig. \ref{Fig:Olivier}. %
 All the outgoing fluxes are positive, but $\Pi_{b<}^{u<}$ is negative
for $r_{A}<0.6$. These are in qualitative agreement with Verma's
analytic results \cite{MKV:MHD_Flux,Ayye:MHD}, which will be discussed
in Section \ref{sec:analytic-energy}.

Haugen et al. \cite{Bran:Flux} have computed the dissipation rates
of kinetic and magnetic energies. They find that for $E^{b}/E^{u}\approx0.5$,
$\epsilon^{u}/\Pi=(\Pi_{u>}^{u<}+\Pi_{u>}^{b<})/\Pi\approx0.3$, and
$\epsilon^{b}/\Pi=(\Pi_{b>}^{u<}+\Pi_{b>}^{b<})/\Pi\approx0.7$ (see
Figures 11 and 12 of Haugen et al. \cite{Bran:Flux}). Note however
that Haugen et al. 's schematics of energy fluxes are missing $\Pi_{u>}^{b<}$
and $\Pi_{b>}^{u<}$. We will compare Haugen et al.'s numerical values
with theoretical predictions in Section \ref{sub:Nonhelical-nonAlfvenic-MHD-flux}.

There are major differences between the energy fluxes in 2D and 3D
MHD turbulence. Primarily, for $r_{A}<0.6$, $\Pi_{b<}^{u<}<0$ in
3D MHD but is positive in 2D MHD. Also, $\Pi_{u>}^{u<}$ and $\Pi_{u>}^{b<}$
are positive in 3D, but are negative in 2D MHD. In Section \ref{sec:analytic-energy}
we will compare 3D numerical results with their analytical counterparts.
Unfortunately, we do not have analytical results for 2D MHD turbulence.

The energy fluxes give us information about the overall energy transfer
from inside/outside $u/b$-sphere to inside/outside $u/b$-sphere.
To obtain a more detailed account of the energy transfer, Dar et al.
\cite{Dar:flux} also studied energy exchange between the wavenumber
shells; Ishizawa and Hattori \cite{Ishi:EDQNM,Ishi:flux} performed
the same studies using wavelet basis.

\subsection{Shell-to-Shell energy Transfer-rates in MHD Turbulence}

\begin{figure}
\includegraphics[scale=0.8,bb=14 14 600 500]{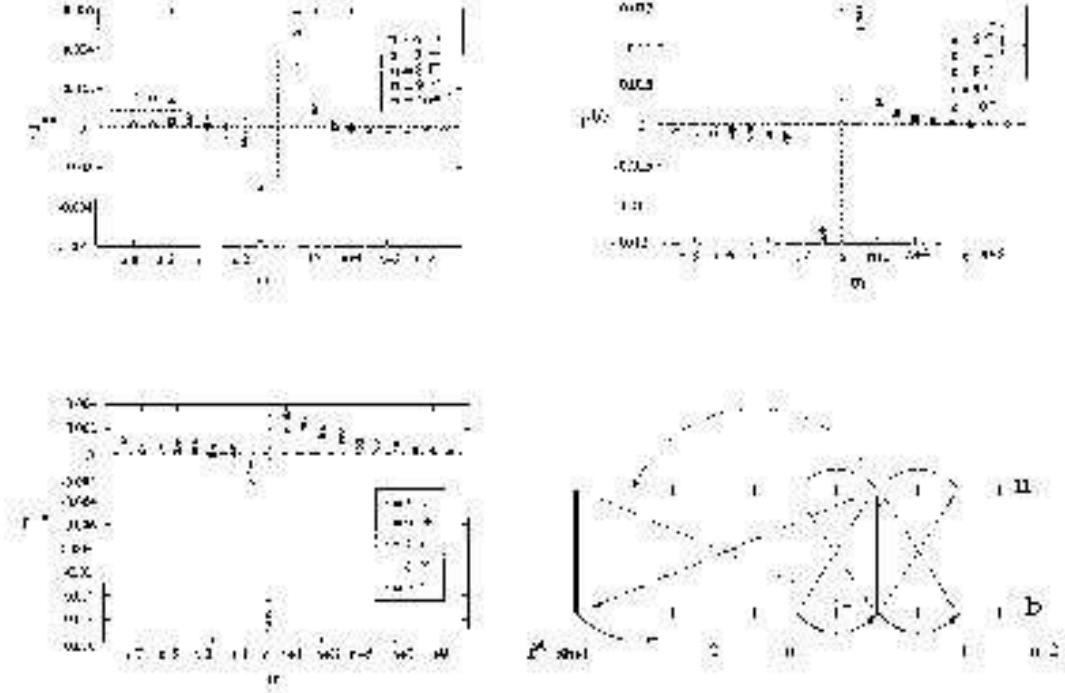}
\caption{\label{Fig:Dar-T} Shell-to-shell energy transfer $T_{nm}^{uu},
T_{nm}^{bb},T_{nm}^{bu}$ (in order)
for Dar et al.'s $512^{2}$ run. 
 They are schematically illustrated
in the last diagram. Adopted from Dar et al. \cite{Dar:flux}.}
\end{figure}

Dar et al. \cite{Dar:flux} partitioned the k-space into shells with
boundaries at wavenumbers $k_{n}(n=1,2,3,...)=1,16,19.02,22.62,...,2^{(n+14)/4}$,
and computed the shell-to-shell energy transfer rates $T_{nm}^{YX}$
defined by Eq. (\ref{eq:T_MHD}). The results are shown in Fig. \ref{Fig:Dar-T}.

The main conclusions of Dar et al. are as follows:
\begin{enumerate}
\item The energy fluxes $T_{nm}^{YX}$ are virtually independent of the
individual values of $m$ and $n$, but only dependent on their differences.
Hence, the transfer rates in the inertial range are \emph{self-similar.}
\item $T_{nm}^{bb}$ is positive for $n>m$, and negative for $n<m$, and
maximum for $n=m+1.$ Hence magnetic to magnetic energy transfer is
forward and local. 
\item The most dominant $T_{nm}^{uu}$ is from $(m-1)$-th shell to $m$-th
shell and $m$-th shell to $(n=m+1)$-th shell, thus $u$-$u$ energy
transfer to the nearest shell is forward. However, $T_{nm}^{uu}$
is negative for $n>m+3$, which may contribute to the inverse cascade
of kinetic energy. This result is consistent with analytic results
on 2D shell-to-shell energy transfers in fluid turbulence (see Appendix
\ref{sec:Digression-to-Fluid}). There are significant energy transfers
from inertial wavenumbers to small wavenumbers (nonlocal), which will
also contribute to the inverse cascade of energy.
\item $T_{nm}^{bu}$ is positive for all $n$, except for $n=m-1$ and $m$,
implying that $n$-th $u$ shell loses energy to all but $n$th and
$(n-1)$th $b$-shell. However, energy gained from these two shells
is larger than the total loss. Consequently there is a net gain of
energy by $u$-shells in the inertial range.
\item $T_{1m}^{bu}$ is positive implying that the first $b$-shell gains
energy from the inertial range $u$-shells through a nonlocal transfer. 
\item $T_{n1}^{bu}$ is positive implying that the first $u$-shell loses
energy to the inertial range $b-$shells through a nonlocal transfer. 
\end{enumerate}
All the above results are schematically illustrated in Fig. \ref{Fig:Dar-T}(d).

The above results on shell-to-shell energy transfers provide us with
important insights into the dynamics of MHD turbulence. Unfortunately,
we do not have enough results on shell-to-shell energy transfer in
3D.

In this section we described the methodology of spectral method and
some important results on energy spectra, fluxes, and shell-to-shell
transfers. In Section \ref{sec:Miscellaneous-Topics} we will also
present large-eddy simulation (LES), which enables us to perform turbulence
simulations on smaller grids. In the next three sections we will describe
the field-theoretic calculation of renormalized viscosity and resistivity,
energy fluxes, and shell-to-shell energy transfer rates.

\section{Renormalization Group Analysis of MHD Turbulence \label{sec:Renormalization-Group-Analysis-MHD}}

In Section \ref{sec:MHD-Turbulence-Models} we discussed various existing
MHD turbulence models. Till early 1990s, KID's model (3/2 spectral
index) used be the accepted model of MHD turbulence. However, solar
wind observations and numerical results in the last decade are in
better agreement with the predictions of Kolmogorov-like models (5/3
spectral index). In this and the next two sections we will present
computation of energy spectrum and energy cascade rates starting from
the MHD equations using field-theoretic techniques. In this section
we will present some of the important renormalization group calculations
applied to MHD turbulence. Most recent RG calculations favour 5/3
spectral index for energy spectrum.

Field theory is well developed, and has been applied to many areas
of physics, e.g., Quantum Electrodynamics, Condensed Matter Physics
etc. In this theory, the equations are expanded perturbatively in
terms of nonlinear term, which are considered small. In fluid turbulence
the nonlinear term is not small; the ratio of nonlinear to linear
(viscous) term is Reynolds numbers, which is large in turbulence regime.
However in MHD turbulence, when $B_{0}\gg z^{\pm}$, the nonlinear
term is small compared to the linear (Alfvén propagation term $\mathbf{B}_{0}\cdot\nabla\mathbf{z}^{\pm}$)
term. This is the weak turbulence limit, and the perturbative expansion
makes sense here. On the other hand when $z^{\pm}\gg B_{0}$ (the
strong turbulence limit), the nonlinear term is not small, and the
perturbative expansion is questionable. This problem appears in many
areas of physics including Quantum Chromodynamics (QCD), Strongly
Correlated Systems, Quantum Gravity etc., and is largely unsolved.
Several interesting methods, Direct Interaction Approximation, Renormalization
Groups (RG), Eddy-damped quasi-normal Markovian approximations, have
been attempted in turbulence. We discuss some of them below. 

A simple-minded calculation of Green's function shows divergence at
small wavenumbers (infrared divergence). One way to solve problem
is by introducing an infrared cutoff for the integral. The reader
is referred to Leslie \cite{Lesl:book} for details. RG technique,
to be described below, is a systematic procedure to cure this problem.

\subsection{Renormalization Groups in Turbulence \label{sub:Renormalization-Groups-in-Turb}}

Renormalization Group Theory (RG) is a technique which is applied
to complex problems involving many length scales. Many researchers
have applied RG to fluid and MHD turbulence. Over the years, several
different RG applications for turbulence has been discovered. Broadly
speaking, they fall in three different categories:

\subsubsection*{Yakhot-Orszag (YO) Perturbative approach }

Yakhot and Orszag's \cite{YakhOrsz} work, motivated by Forster \emph{et
al.} \cite{FNS} and Fournier and Frisch \cite{FourFrisRG}, is the
first comprehensive application of RG to turbulence. It is based on
Wilson's shell-elimination procedure. Also refer to  Smith and Woodruff
\cite{Smit} for details. Here the renormalized parameter is function
of forcing noise spectrum $D(k)=D_{0}k^{-y}$. It is shown that the
local Reynolds number $\bar{\lambda}$ is\[
\bar{\lambda}=\frac{\lambda_{0}^{2}D_{0}}{\nu^{3}(\Lambda)\Lambda^{\epsilon}},\]
where $\lambda_{0}$ is the expansion parameter, $\Lambda$ is the
cutoff wavenumber, and $\epsilon=4+y-d$ \cite{YakhOrsz}. It is found
that $\nu(\Lambda)$ increases as $\Lambda$ decreases, therefore,
$\bar{\lambda}$ remains small (may not be less that one though) compared
to $Re$ as the wavenumber shells are eliminated. Hence, the {}``effective''
expansion parameter is small even when the Reynolds number may be
large.

The RG analysis of Yakhot and Orszag \cite{YakhOrsz} yielded Kolmogorov's
constant $K_{Ko}=1.617$, turbulent Prandtl number for high-Reynolds-number
heat transfer $P_{t}=0.7179$, Batchelor constant $Ba=1.161$ etc.
These numbers are quite close to the experimental results. Hence,
Yakhot and Orszag's method appears to be highly successful. However
there are several criticisms to the YO scheme. Kolmogorov's spectrum
results in the YO scheme for $\epsilon=4$, far away from $\epsilon=0$,
hence epsilon-expansion is questionable. YO proposed that higher order
nonlinearities are {}``irrelevant'' in the RG sense for $\epsilon=0$,
and are marginal when $\epsilon=4$. Eyink \cite{Eyin:RG} objected
to this claim and demonstrated that the higher order nonlinearities
are marginal regardless of $\epsilon$. Kraichnan \cite{Krai:YO}
compared YO's procedure with Kraichnan's Direct Interaction Approximation
\cite{Krai:59} and raised certain objections regarding distant-interaction
in YO scheme. For details refer to Zhou et al. \cite{ZhouMcCo:RGrev}
and Smith and Woodruff \cite{Smit}.

There are several RG calculations applied to MHD turbulence based
on YO procedure. These calculations will be described in Section \ref{sub:MHD-RG-YO}.

\subsubsection*{Self-consistent approach of McComb and Zhou}

This is one of the nonperturbative method, which is often used in
Quantum Field theory. In this method, a self-consistent equation of
the full propagator is written in terms of itself and the proper vertex
part. The equation may contain many (possibly infinite) terms, but
it is truncated at some order. Then the equation is solved iteratively.
McComb \cite{McCo:book}, Zhou and coworkers \cite{ZhouVaha88} have
applied this scheme to fluid turbulence, and have calculated renormalized
viscosity and Kolmogorov's constant successfully. Direct Interaction
Approximation of Kraichnan is quite similar to self-consistent theory
(Smith and Woodruff \cite{Smit}).

The difficulty with this method is that it is not rigorous. In McComb
and Zhou's procedures, the vertex correction is not taken into account.
Verma \cite{MKV:B0_RG,MKV:MHD_PRE,MKV:MHD_RG} has applied the self-consistent
theory to MHD turbulence.

\subsubsection*{Callan-Symanzik Equation for Turbulence}

DeDominicis and Martin \cite{DeDo} and Teodorovich \cite{Teod} obtained
the RG equation using functional integral. Teodorovich obtained $K_{Ko}=2.447$,
which is in worse agreement with the experimental data, though it
is not too far away. 

It has been shown that Wilson's shell renormalization and RG through
Callan-Symanzik equation are equivalent procedure. However, careful
comparison of RG schemes in turbulence is not completely worked out. 

The renormalization of viscosity, resistivity, and {}``mean magnetic
field'' will be discussed below. The self-consistent approach will
be discussed at somewhat greater length because it is one of the most
recent and exhaustive work. After renormalization, in Section \ref{sec:analytic-energy}
we will discuss the computation of energy fluxes in MHD turbulence.
These calculations are done using self-consistent field theory, a
scheme very similar to DIA. At the end we will describe Eddy-damped
quasi-normal Markovian approximation, which is very similar to the
energy flux calculation.

\subsection{Physical Meaning of Renormalization in Turbulence}

The field theorists have been using renormalization techniques since
1940s. However, the physical meaning of renormalization became clear
after path-breaking work of Wilson \cite{WilsKogu}. Here renormalization
is a variation of parameters as we go from one length scale to the
next. Following Wilson, renormalized viscosity and resistivity can
also be interpreted as scale-dependent parameters. We coarse-grain
the physical space and look for an effective theory at a larger scale.
In this method, we sum up all the interactions at smaller scales,
and as a outcome we obtain terms that can be treated as a correction
to viscosity and resistivity. The corrected viscosity and resistivity
are called {}``effective'' or renormalized dissipative parameters.
This procedure of coarse graining is also called shell elimination
in wavenumber space. We carry on with this averaging process till
we reach inertial range. In the inertial range the {}``effective''
or renormalized parameters follow a universal powerlaw, e. g., renormalized
viscosity $\nu(l)\propto l^{4/3}$. This is the renormalization procedure
in turbulence. Note that the renormalized parameters are independent
of microscopic viscosity or resistivity. 

\begin{figure}
\includegraphics[%
  scale=0.5,bb=0 150 575 630 ]{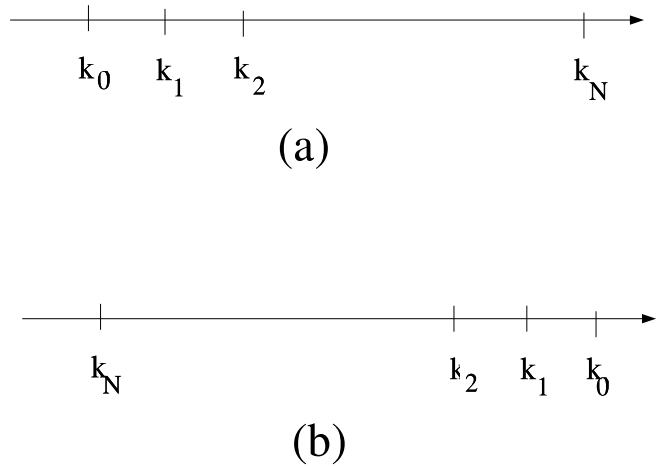}

\caption{\label{Fig:k0kN} The wavenumber shells to be averaged during renormalization
procedure. (a) In {}``mean magnetic field'' renormalization, averaging
starts from small wavenumber, first shell being $(k_{0},k_{1})$.
(b) In viscosity and resistivity, averaging starts from large wavenumbers.}
\end{figure}

In viscosity and resistivity renormalization the large wavenumber
shells are eliminated, and the interaction involving these shells
are summed. Hence, we move from larger wavenumbers to smaller wavenumbers.
However, it is also possible to go from smaller wavenumbers to larger
wavenumber by summing the smaller wavenumber shells. This process
is not coarse-graining, but it is a perfectly valid RG procedure,
and is useful when the small wavenumber modes (large length scales)
are linear. This scheme is followed in Quantum Electrodynamics (QED),
where the electromagnetic field is negligible at a large distance
(small wavenumbers) from a charge particle, while the field becomes
nonzero at short distances (large wavenumber). In QED, the charge
of a particle gets renormalized when we come closer to the charge
particle, i. e., from smaller wavenumbers to larger wavenumbers. In
MHD too, the large-scale Alfvén modes are linear, hence we can apply
RG procedure from smaller wavenumbers to larger wavenumbers. Verma
\cite{MKV:B0_RG} has precisely done this to compute the {}``effective
or renormalized mean magnetic field'' in MHD turbulence. See Fig.
\ref{Fig:k0kN} for an illustration of wavenumber shells to be averaged.%

\subsection{{}``Mean Magnetic Field'' Renormalization in MHD Turbulence \label{sub:Mean-Magnetic-Field-RG}}

In this subsection we describe the self-consistent RG procedure of
Verma \cite{MKV:B0_RG}, which is similar to that used by McComb \cite{McCo},
McComb and Shanmugsundaram \cite{McCoShan}, McComb and Watt \cite{McCoWatt},
and Zhou \textit{et al.} \cite{ZhouVaha88} for fluid turbulence.
However, one major difference between the two is that Verma \cite{MKV:B0_RG}
integrates the small wavenumber modes instead of integrating the large
wavenumber modes, as done by earlier authors. At small wavenumbers
the MHD equations are linear in $B_{0}$, the mean magnetic field.
Verma applied RG procedure to compute the renormalized mean magnetic
field.

The basic idea of the calculation is that the effective mean magnetic
field is the magnetic field of the next-largest eddy (local field),
contrary to the KID's phenomenology where the effective mean magnetic
field at any scale is a constant. This argument is based on a physical
intuition that the scattering of the Alfv\'{e}n waves at a wavenumber
$k$ is effected by the magnetic field of the next-largest eddy, rather
than the external magnetic field. The mean magnetic field at the largest
scale will simply convect the waves, whereas the local inhomogeneities
contribute to the scattering of waves which leads to turbulence (note
that in WKB method, the local inhomogeneity of the medium determines
the amplitude and the phase evolution). The calculation shows that
$E(k)\propto k^{-5/3}$, and the mean magnetic field $B_{0}(k)\propto k^{-1/3}$
are the self-consistent solutions of the RG equations. Thus $B_{0}$
appearing in KID's phenomenology should be $k$-dependent.

Verma made one drastic assumption that the mean magnetic field at
large-scales are randomly oriented. This assumption simplifies the
calculation tremendously because the problem remains isotropic. Physically,
the above assumption may be approximately valid several scales below
the largest length scale. Now, Verma's procedure follows.

The MHD equations in the Fourier space is (see Eq. {[}\ref{eq:MHDzkw}{]})\begin{equation}
\left(-i\omega\mp i\left({\textbf{B}}_{0}\cdot{\textbf{k}}\right)\right)z_{i}^{\pm}(\hat{k})=-iM_{ijm}({\textbf{k}})\int d\hat{{p}}z_{j}^{\mp}(\hat{{p}})z_{m}^{\pm}(\hat{{k}}-\hat{{p}})\end{equation}
 where \begin{equation}
M_{ijm}({\textbf{k}})=k_{j}P_{im}({\textbf{k}});\,\, P_{im}({\textbf{k}})=\delta_{im}-\frac{k_{i}k_{m}}{k^{2}},\end{equation}
 and $\hat{{k}}=(\mathbf{k},\omega)$. We ignore the viscous terms
because they are effective at large wavenumbers. We take the mean
magnetic field $\mathbf{B}_{0}$ to be random. Hence,\[
\left(-i\omega+\hat{{\Sigma}}_{(0)}\right)\left[\begin{array}{c}
z_{i}^{+}(\hat{k})\\
z_{i}^{-}(\hat{k})\end{array}\right]=-iM_{ijm}({\textbf{k}})\int d\hat{{k}}\left[\begin{array}{c}
z_{j}^{-}(\hat{{p}})z_{m}^{+}(\hat{{k}}-\hat{{p}})\\
z_{j}^{+}(\hat{{p}})z_{m}^{-}(\hat{{k}}-\hat{{p}})\end{array}\right]\]
with the self-energy matrix $\hat{{\Sigma}}_{(0)}$ given by\[
\hat{\Sigma}_{(0)}=\left[\begin{array}{cc}
-ikB_{0} & 0\\
0 & ikB_{0}\end{array}\right]\]

We logarithmically divide the wavenumber range $(k_{0},k_{N})$ into
$N$ shells. The $n$th shell is $(k_{n-1},k_{n})$ where $k_{n}=s^{n}k_{o}(s>1)$.
The modes in the first few shells will be the energy containing eddies
that will force the turbulence. For keeping our calculation procedure
simple, we assume that the external forcing maintains the energy of
the first few shells to the initial values. The modes in the first
few shells are assumed to be random with a gaussian distribution with
zero mean (see Items 3 and 4 below).

First we eliminate the first shell $(k_{0},k_{1})$, and then obtain
the modified the MHD equations. Subsequently higher wavenumber shells
are eliminated, and a general expression for the modified MHD equations
after elimination of $n$th shell is obtained. The details of each
step are as follows:

\begin{enumerate}
\item We decompose the modes into the modes to be eliminated $(k^{<})$
and the modes to be retained $(k^{>}).$ In the first iteration $(k_{0},k_{1})=k^{<}$
and $(k_{1},k_{N})=k^{>}$. Note that $B_{0}(k)$ is the mean magnetic
field before the elimination of the first shell. 
\item We rewrite the Eq. (\ref{eq:MHDzkw}) for $k^{<}$ and $k^{>}$. The
equations for $z_{i}^{\pm>}(\hat{{k}})$ modes are\begin{eqnarray}
\left(-i\omega\mp i\left(B_{0}k\right)\right)z_{i}^{\pm>}(\hat{k}) & = & -iM_{ijm}({\textbf{k}})\int d\hat{{k}}\left[z_{j}^{\mp>}(\hat{{p}})z_{m}^{\pm>}(\hat{{k}}-\hat{{p}})\right]\nonumber \\
 &  & +\left[z_{j}^{\mp>}(\hat{{p}})z_{m}^{\pm<}(\hat{{k}}-\hat{{p}})+z_{j}^{\mp<}(\hat{{p}})z_{m}^{\pm>}(\hat{{k}}-\hat{{p}})\right]\nonumber \\
 &  & +\left[z_{j}^{\mp<}(\hat{{p}})z_{m}^{\pm<}(\hat{{k}}-\hat{{p}})\right]\label{eq:klarge}\end{eqnarray}
 while the equation for $z_{i}^{\pm<}({\textbf{k}},t)$ modes can
be obtained by interchanging $<$ and $>$ in the above equation.
\item The terms given in the second and third brackets in the RHS of Eq.
(\ref{eq:klarge}) are calculated perturbatively. The details of the
perturbation expansion is given in Appendix B. We perform ensemble
average over the first shell, which is to be eliminated. We assume
that $z_{i}^{\pm<}(\hat{{k}})$ has a gaussian distribution with zero
mean. Hence, \begin{equation}
\begin{array}{c}
\left\langle z_{i}^{\pm<}(\hat{{k}})\right\rangle =0\\
\left\langle z_{i}^{\pm>}(\hat{{k}})\right\rangle =z_{i}^{\pm>}(\hat{{k}})\end{array}\end{equation}
 and \begin{equation}
\left\langle z_{s}^{a<}(\hat{{p}})z_{m}^{b<}(\hat{{q}})\right\rangle =P_{sm}({\textbf{p)}}C^{ab}(\hat{{p}})\delta\left(\hat{{p}}+\hat{{q}}\right)\end{equation}
 where $a,b=\pm$ or $\mp$. Also, the triple order correlations $\left\langle z_{s}^{a,b<}(\hat{{k}})z_{s}^{a,b<}(\hat{{p}})z_{s}^{a,b<}(\hat{{q}})\right\rangle $
are zero due to the Gaussian nature of the fluctuations. The experiments
show that gaussian approximation for $z_{i}^{\pm<}(\hat{{k}})$ is
not quite correct, however it is a good approximation (refer to Sections
\ref{sec:Intermittency-in-MHD}).  A popular method called EDQNM calculation
also makes this assumption (see Sections \ref{sub:EDQNM-Calculation}). 
\item As shown in Appendix B, to first order in perturbation, the second
bracketed term of Eq. (\ref{eq:klarge}) vanishes except the terms
of the type $\left\langle z_{s}^{a,b>}(\hat{{k}})z_{s}^{a,b>}(\hat{{p}})z_{s}^{a,b>}(\hat{{k}})\right\rangle $
(called triple nonlinearity). Verma ignored this term. The effects
of triple nonlinearity can be included using the procedure of Zhou
and Vahala \cite{ZhouVaha88}, but they are expected to be of higher
order. For averaging, we also hypothesize that\[
\left\langle z^{>}z^{<}z^{<}\right\rangle =z^{>}\left\langle z^{<}z^{<}\right\rangle ,\]
which cannot be strictly correct. This is one of the major assumption
of RG procedure \cite{ZhouMcCo:RGrev}. After performing the perturbation
we find that the third bracketed term of Eq. (\ref{eq:klarge}) is
nonzero, and yields corrections $\delta\hat{\Sigma}_{(0)}$to the
self energy $\hat{\Sigma}_{(0)}$:\[
\left(-i\omega+\hat{{\Sigma}}_{(0)}+\hat{{\delta\Sigma}_{(0)}}\right)\left[\begin{array}{c}
z_{i}^{+>}(\hat{k})\\
z_{i}^{->}(\hat{k})\end{array}\right]=-iM_{ijm}({\textbf{k}})\int d\hat{{k}}\left[\begin{array}{c}
z_{j}^{->}(\hat{{p}})z_{m}^{+>}(\hat{{k}}-\hat{{p}})\\
z_{j}^{+>}(\hat{{p}})z_{m}^{->}(\hat{{k}}-\hat{{p}})\end{array}\right]\]
with \begin{eqnarray}
\delta\Sigma_{(0)}^{++} & = & \int_{\hat{p}+\hat{{q}}=\hat{{k}}}^{\Delta}d\hat{{p}}[S_{1}(k,p,q)G_{(0)}^{++}(\hat{{p})}C_{(0)}^{--}(\hat{{q}})+S_{2}(k,p,q)G_{(0)}^{+-}(\hat{{p})}C_{(0)}^{--}(\hat{{q}})\nonumber \\
 &  & +S_{3}(k,p,q)G_{(0)}^{-+}(\hat{{p})}C_{(0)}^{+-}(\hat{{q}})+S_{4}(k,p,q)G_{(0)}^{--}(\hat{{p})}C_{(0)}^{+-}(\hat{{q}})],\label{eq:Sig++}\\
\delta\Sigma_{(0)}^{+-} & = & \int_{\hat{p}+\hat{{q}}=\hat{{k}}}^{\Delta}d\hat{{p}}[S_{1}(k,p,q)G_{(0)}^{+-}(\hat{{p})}C_{(0)}^{-+}(\hat{{q}})+S_{2}(k,p,q)G_{(0)}^{++}(\hat{{p})}C_{(0)}^{-+}(\hat{{q}})\nonumber \\
 &  & +S_{3}(k,p,q)G_{(0)}^{--}(\hat{{p})}C_{(0)}^{++}(\hat{{q}})+S_{4}(k,p,q)G_{(0)}^{-+}(\hat{{p})}C_{(0)}^{++}(\hat{{q}})].\label{eq:Sig+-}\end{eqnarray}
where the integral is to performed over the first shell $(k_{0},k_{1})$,
denoted by region $\Delta$, and $S_{i}(k,p,q)$ are given in Appendix
B. The equations for the other two terms $\Sigma^{--}$and $\Sigma^{-+}$
can be obtained by interchanging $+$ and $-$ signs. Note that \[
\left[\begin{array}{cc}
\Sigma_{(0)}^{++} & \Sigma_{(0)}^{+-}\\
\Sigma_{(0)}^{-+} & \Sigma_{(0)}^{--}\end{array}\right]=\left[\begin{array}{cc}
-ikB_{(0)}^{++} & -ikB_{(0)}^{+-}\\
ikB_{(0)}^{-+} & ikB_{(0)}^{--}\end{array}\right]\]
with $B_{(0)}^{\pm\mp}=0$.
\item The full-fledged calculation of $\Sigma$'s are quite involved. Therefore,
to simplify the calculation by solving the equations in the limit
$C^{\pm\mp}=C^{R}=C^{uu}(k)-C^{bb}(k)=0$ and $E^{+}(k)=E^{-}(k)$.
Under this approximation we have $+-$ symmetry in our problem, hence
$B_{(0)}^{++}=B_{(0)}^{--}$ and $B_{(0)}^{+-}=B_{(0)}^{-+}$. In
the first iteration, $B_{(0)}^{+-}=B_{(0)}^{-+}=0$, but they become
nonzero after the first iteration, hence we will keep the expressions
$B_{(0)}^{+-}$ intact. 
\item The expressions for $\delta\Sigma'$s involve Green's functions and
correlation functions, which are themselves functions of $\Sigma'$s.
We need to solve for $\Sigma$'s and $G$'s self-consistently. Green's
function after first iteration is \begin{equation}
\hat{G}_{(0)}^{-1}(k,\omega)=\left[\begin{array}{cc}
-i\omega-ikB_{(0)}^{++} & -ikB_{(0)}^{+-}\\
ikB_{(0)}^{-+} & -i\omega+ikB_{(0)}^{--}\end{array}\right],\end{equation}
 which implies that \begin{eqnarray*}
G_{(0)}^{\pm\pm}(k,t-t') & = & \frac{X_{(0)}(k)+B_{(0)}^{++}(k)}{2X_{(0)}(k)}\exp{\pm ikX_{(0)}(k)(t-t')}\\
G_{(0)}^{\pm\mp}(k,t-t') & = & \frac{B_{(0)}^{+-}}{2X_{(0)}(k)}\exp{\pm ikX_{(0)}(k)(t-t')},\end{eqnarray*}
where $X_{(0)}(k)=\sqrt{{B_{(0)}^{++2}-B_{(0)}^{+-2}}}$. The frequency
dependence of correlation function are taken as $\hat{C}(k,\omega)=2\Re[\hat{{G}}(k,\omega)]\hat{C}(k)$,
which is one of the generalizations of fluctuation-dissipation theorem
to nonequilibrium systems. In terms of time difference, $\hat{C}(k,t-t')=\hat{{G}}(k,t-t')C(k,t,t)$,
which yields\[
\hat{C}(k,t-t')=\Re\left[\begin{array}{cc}
\frac{X_{(0)}(k)+B_{(0)}^{++}(k)}{2X_{(0)}(k)}C(k)\exp{i\Phi} & \frac{B_{(0)}^{+-}}{2X_{(0)}(k)}C(k)\exp{i\Phi}\\
\frac{B_{(0)}^{+-}}{2X_{(0)}(k)}C(k)\exp{-i\Phi} & \frac{X_{(0)}(k)+B_{(0)}^{++}(k)}{2X_{(0)}(k)}C(k)\exp{-i\Phi}\end{array}\right].\]
where $\Phi=kX_{(0)}(k)(t-t')$. To derive the above, we use the fact
that $C^{R}=C^{uu}(k)-C^{bb}(k)=0$, and $C^{++}(k)=C^{--}(k)=C(k)$.
While doing the integral, the choice of the pole is dictated by the
direction of the waves.
\item Above Green's functions and correlation functions are substituted
in Eqs. (\ref{eq:Sig++},\ref{eq:Sig+-}), and the frequency integral
is performed. These operations yield\begin{eqnarray}
k\delta B_{(0)}^{++} & = & \frac{1}{d-1}\int\frac{d\mathbf{p}}{\left(2\pi\right)^{d}}C(q)[-S_{1}(k,p,q)\frac{X_{(0)}(p)+B_{(0)}^{++}(p)}{2X_{(0)}(p)}-S_{2}(k,p,q)\frac{B_{(0)}^{+-}(p)}{2X_{(0)}(p)}\nonumber \\
 &  & +S_{3}(k,p,q)\frac{B^{+-}(p)}{2X_{(0)}(p)}-S_{4}(k,p,q)\frac{X_{(0)}(p)-B_{(0)}^{++}(p)}{2X_{(0)}(p)}]/denr,\label{eq:deltaB++}\\
k\delta B_{0}^{+-} & = & \frac{1}{d-1}\int\frac{d\mathbf{p}}{\left(2\pi\right)^{d}}C(q)[-S_{1}(k,p,q)\frac{B_{(0)}^{++}(p)}{2X_{(0)}(p)}-S_{2}(k,p,q)\frac{X_{(0)}(p)-B_{(0)}^{++}(p)}{2X_{(0)}(p)}\nonumber \\
 &  & +S_{3}(k,p,q)\frac{X_{(0)}(p)+B_{(0)}^{++}(p)}{2X_{(0)}(p)}+S_{4}(k,p,q)\frac{B_{(0)}^{+-}(p)}{2X_{(0)}(p)}]/denr,\label{eq:deltaB+-}\end{eqnarray}
with \[
denr=[-kX_{(0)}(k)+pX_{(0)}(p)-qX_{(0)}(q)]\]
The frequency integral in the above are done using contour integral.
It is also possible to obtain the above using $t'$ integral \cite{Lesl:book}.
Also note that $\omega_{k}^{\pm}=\mp kB_{0}^{\pm\pm}(k)$, which is
equivalent to using $\omega=k^{z}$.
\item Let us denote $\hat{B}_{(1)}(k)$ as the effective mean magnetic field
after the elimination of the first shell. Therefore, \begin{equation}
B_{(1)}^{ab}(k)=B_{(0)}^{ab}(k)+\delta B_{(0)}^{ab}(k),\label{eq:B1}\end{equation}
Recall that $a,b=\pm1$. We keep eliminating the shells one after
the other by the above procedure, and obtain the following recurrence
relation after $n+1$ iterations:\begin{equation}
B_{(n+1)}^{ab}(k)=B_{(n)}^{ab}(k)+\delta B_{(n)}^{ab}(k),\label{eq:Bn}\end{equation}
 where the equations for $\delta B_{(n)}^{\pm\pm}(k)$ and $\delta B_{(n)}^{\pm\mp}(k)$
are the same as the equations (\ref{eq:deltaB++},\ref{eq:deltaB+-})
except that the terms $B_{(0)}^{ab}(k)$ and $X_{(0)}^{ab}(k)$ are
to be replaced by $B_{(n)}^{ab}(k)$ and $X_{(n)}^{ab}(k)$ respectively$.$
Clearly $B_{(n+1)}(k)$ is the effective mean magnetic field after
the elimination of the $(n+1)$th shell. The set of RG equations to
be solved are Eqs. (\ref{eq:deltaB++},\ref{eq:deltaB+-}) with $B_{(0)}$
replaced by $B_{(n)}$s, and Eq. (\ref{eq:Bn}).
\item In YO's perturbative RG calculation, the correlation function depends
of the noise (forcing) spectrum. In the self-consistent procedure,
we assume that we are in the inertial range, and the energy spectrum
is proportional to Kolmogorov's $5/3$ power law, i.e., \[
C(k)=\frac{2\left(2\pi\right)^{d}}{S_{d}(d-1)}k^{_{(d-1)}}E(k)\]
where \begin{equation}
E(k)=K\Pi^{2/3}k^{-5/3}\end{equation}
Here, $S_{d}$ is the surface area of $d$-dimensional unit sphere,
$\Pi$ is the total energy cascade rate, and $K$ is Kolmogorov's
constant. Note that $\Pi^{+}=\Pi^{-}=\Pi$ due to symmetry. We substitute
the following form of $B_{(n)}(k)$ in the modified equations (\ref{eq:deltaB++},\ref{eq:deltaB+-})
\begin{equation}
B_{(n)}^{ab}(k_{n}k^{\prime})=K^{1/2}\Pi^{1/3}k_{n}^{-1/3}B_{(n)}^{*ab}(k^{\prime})\label{eq:Bn2}\end{equation}
 with $k=k_{(n+1)}k^{\prime}$ ($k^{\prime}>1$). We expect $B_{(n)}^{*ab}(k^{\prime})$
to be a universal function for large $n$$.$ After the substitution
we obtain the equations for $B_{(n)}^{*ab}(k^{\prime})$ that are
\begin{eqnarray}
\delta B_{n}^{++*}(k^{\prime}) & = & \frac{1}{d-1}\int d\mathbf{p}'\frac{2}{S_{d}(d-1)}\frac{E(q')}{q'^{d-1}}[-S_{1}(k',p',q')\frac{X_{(0)}(sp')+B_{(0)}^{++}(sp')}{2X_{(0)}(sp')}\nonumber \\
 &  & -S_{2}(k',p',q')\frac{B_{(0)}^{+-}(sp')}{2X_{(0)}(sp')}+S_{3}(k',p',q')\frac{B^{+-}(sp')}{2X_{(0)}(sp')}\nonumber \\
 &  & -S_{4}(k',p',q')\frac{X_{(0)}(sp')-B_{(0)}^{++}(sp')}{2X_{(0)}(sp')}]/denr',\label{eq:rg1}\\
\delta B_{n}^{+-*}(k^{\prime}) & = & \frac{1}{d-1}\int d\mathbf{p}'\frac{2}{S_{d}(d-1)}\frac{E(q')}{q'^{d-1}}[-S_{1}(k',p',q')\frac{B_{(0)}^{++}(sp')}{2X_{(0)}(sp')}\nonumber \\
 &  & -S_{2}(k',p',q')\frac{X_{(0)}(sp')-B_{(0)}^{++}(sp')}{2X_{(0)}(sp')}+S_{4}(k',p',q')\frac{B_{(0)}^{+-}(sp')}{2X_{(0)}(sp')}\nonumber \\
 &  & +S_{3}(k',p',q')\frac{X_{(0)}(sp')+B_{(0)}^{++}(sp')}{2X_{(0)}(sp')}]/denr',\label{eq:rg2}\end{eqnarray}
 where\[
denr'=[-k'X_{(0)}(sk')+p'X_{(0)}(sp')-q'X_{(0)}(sq')],\]
 The integrals in the Eqs. (\ref{eq:rg1},\ref{eq:rg2}) are performed
over a region $1/s\leq p',q'\leq1$ with the constraint that $\mathbf{p'}+\mathbf{q}'=\mathbf{k}'$.
The recurrence relation for $B_{(n)}$ is \begin{equation}
B_{(n+1)}^{*ab}(k')=s^{1/3}B_{(n)}^{*ab}(sk')+s^{-1/3}\delta B_{(n)}^{*ab}(k')\label{eq:rg3}\end{equation}

\item Now we need to solve the above three equations iteratively. Here we
take the space dimensionality $d=3$. We use Monte Carlo technique
to solve the integrals. Since the integrals are identically zero for
$k^{\prime}>2$, the initial $B_{(0)}^{*}(k_{i}^{\prime})=B_{(0)}^{*initial}$
for $k_{i}^{\prime}<2$ and $B_{(0)}^{*}(k_{i}^{\prime})=B_{(0)}^{*initial}*(k_{i}^{\prime}/2)^{-1/3}$
for $k_{i}^{\prime}>2$. We take $B_{(0)}^{+-}=0$. The Eqs. (\ref{eq:rg1},\ref{eq:rg2},\ref{eq:rg3})
are solved iteratively. We continue iterating the equations till $B_{(n+1)}^{*}(k^{\prime})\approx B_{(n)}^{*}(k^{\prime})$,
that is, till the solution converges. The $B_{(n)}^{*}$s for various
$n$ ranging from $0..3$ is shown in Fig. \ref{Fig:B0RG}. %
\begin{figure}
\includegraphics[bb=140 328 412 463  ]{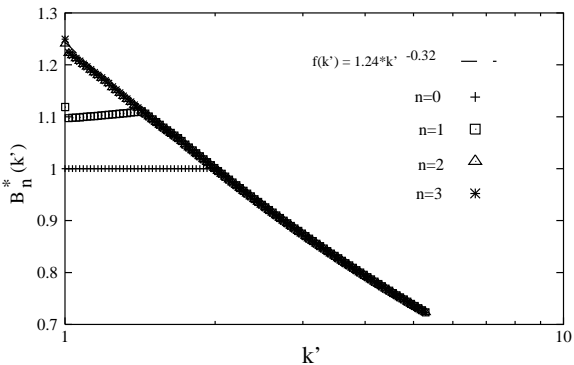}

\caption{\label{Fig:B0RG} $B_{n}^{*}(k')$ for $n=0..3$. The line of best
is $k'^{-1/3}$.}
\end{figure}
 Here the convergence is very fast, and after $n=3-4$ iterations
$B_{(n)}^{*}(k)$ converges to an universal function \[
f(k')=1.24*B_{(0)}^{*initial}k^{\prime-0.32}\approx B_{(0)}^{*initial}(k'/2)^{-1/3}.\]
 The other parameter $B_{(n)}^{*+-}(k^{\prime})$ remains close to
zero. Since $B_{(n)}^{*}(k')$ converges, the universal function is
an stable solution in the RG sense. The substitution of the function
$B_{(n)}^{*}(k')$ in Eq. (\ref{eq:Bn2}) yields and \[
B_{(n+1)}(k)=K^{1/2}\Pi^{1/2}B_{(0)}^{*initial}(k/2)^{-1/3}=B_{0}\left(\frac{k}{2k_{0}}\right)^{-1/3},\]
for $k>k_{n+1}$ when $n$ is large (stable RG solution). Hence we
see that $B_{n}(k)\propto k^{-1/3}$ in our self-consistent scheme.
\end{enumerate}
To summarize, we have shown that the mean magnetic field $B_{0}$
gets renormalized due to the nonlinear term. As a consequence, the
energy spectrum is Kolmogorov-like, not $k^{-3/2}$ as predicted by
KID's phenomenology, i. e., 

\[
E(k)=K\Pi^{2/3}k^{-5/3}.\]
Since, $B_{0}$is corrected by renormalization, we can claim that
KID's phenomenology is not valid for MHD turbulence.

The physical idea behind our argument is that scattering of the Alfv\'{e}n
waves at a wavenumber $k$ is caused by the {}``effective or renormalized
magnetic field'', rather than the mean magnetic field effective at
the largest scale. The effective field turns out to be $k$-dependent
or local field, and can be interpreted as the field due to the next
largest eddy. The above theoretical result can be put in perspective
with the numerical results of Cho et al. \cite{ChoVish:localB} where
they show that turbulent dynamics is determined by the {}``local''
mean magnetic field. Note that KID take $\tau_{A}\approx(kB_{0})^{-1}$
to be the effective time scale for the nonlinear interactions that
gives $E(k)\propto k^{-3/2}$. However the timescale $\tau_{NL}^{B_{(n)}}$,
which is of the same order as the nonlinear time-scales of $z_{k}^{\pm}$,
$\tau_{NL}^{\pm}\approx(kz_{k}^{\pm})^{-1}$, yields $E(k)\propto k^{-5/3}$.
The quantity $\tau_{NL}^{B_{(n)}}$ can possibly be obtained numerically
from the time evolution of the Fourier components; this test will
validate the theoretical assumptions made in the above calculation.

The above calculation shows that Kolmogorov-like energy spectrum is
one of the solution of RG equation. However, we cannot claim this
to be the unique solution. Further investigation in this direction
is required. Also, the above RG calculation was done for $E^{+}=E^{-}$
and $r_{A}=1$ for simplicity of the calculation. The generalization
to arbitrary field configuration is not yet done. The mean magnetic
field is assumed to be isotropic, which is unrealistic. In addition,
self-consistent RG scheme has other fundamental problems, as described
in Section \ref{sub:Renormalization-Groups-in-Turb}.

In the above RG scheme, averaging of wavenumber has been performed
for small wavenumbers in contrast to the earlier RG analysis of turbulence
in which higher wavenumbers were averaged out. Here a self-consistent
power-law energy spectrum was obtained for smaller length scales,
and the spectrum was shown to be independent of the small wavenumber
forcing states. This is in agreement with the Kolmogorov's hypothesis
which states that the energy spectrum of the intermediate scale is
independent of the large-scale forcing. Any extension of this scheme
to fluid turbulence in the presence of large-scale shear etc. will
yield interesting insights into the connection of energy spectrum
with large-scale forcing.

After the discussion on the renormalization of mean magnetic field,
we move to renormalization of dissipative parameters.

\subsection{Renormalization of viscosity and resistivity using self-consistent
procedure \label{sub:Renormalization-of-viscosity}}

In this subsection we compute renormalized viscosity and resistivity
using self-consistent procedure. This work was done by Verma \cite{MKV:MHD_RG,MKV:MHD_Helical},
and Chang and Lin \cite{Chan:RG}. Here the mean magnetic field is
assumed to be zero, and renormalization of viscosity and resistivity
is performed \emph{from large wavenumber to smaller wavenumbers.}
This is the major difference between the calculation of Subsection
\ref{sub:Mean-Magnetic-Field-RG} and the present calculation. The
RG calculation for arbitrary cross helicity, Alfvén ratio, magnetic
helicity, and kinetic helicities is very complex, therefore Verma
performed the calculation in the following three limiting cases: (1)
Nonhelical nonAlfvénic MHD ($H_{M}=H_{K}=H_{c}=0$), (2) Nonhelical
Alfvénic MHD ($H_{M}=H_{K}=0,\sigma_{c}\rightarrow1$), and (3) Helical
nonAlfvénic MHD ($H_{M}\ne0,H_{K}\ne0,H_{c}=0$). These generic cases
provide us with many useful insights into the dynamics of MHD turbulence.

\subsubsection{Nonhelical nonAlfvénic MHD ($H_{M}=H_{K}=H_{c}=0$):\label{sub:Nonhelical-nonAlfvenic-MHD}}

In this case, the RG calculations are done in terms of $\mathbf{u}$
and $\mathbf{b}$ variables because the matrix of Green's function
becomes diagonal in these variables. We take the following form of
Kolmogorov's spectrum for kinetic energy {[}$E^{u}(k)${]} and magnetic
energy {[}$E^{b}(k)${]}\begin{eqnarray}
E^{u}(k) & = & K^{u}\Pi^{2/3}k^{-5/3},\label{eq:Euk}\\
E^{b}(k) & = & E^{u}(k)/r_{A},\label{eq:Ebk}\end{eqnarray}
where $K^{u}$ is Kolmogorov's constant for MHD turbulence, and $\Pi$
is the total energy flux. In the limit $\sigma_{c}=0$, we have $E^{+}=E^{-}$
and $\Pi^{+}=\Pi^{-}=\Pi$ {[}cf. Eq. (\ref{eq:MHD_Kolm_zpm}){]}.
Therefore, $E_{total}(k)=E^{+}(K)=E^{u}(k)+E^{b}(k)$ and \begin{equation}
K^{+}=K^{u}(1+r_{A}^{-1})\label{eq:Kplus}\end{equation}

With these preliminaries we start our RG calculation. The incompressible
MHD equations in the Fourier space are \begin{eqnarray}
\left(-i\omega+\nu k^{2}\right)u_{i}\left(\hat{{k}}\right) & = & -\frac{i}{2}P_{ijm}^{+}({\textbf{k}})\int_{\hat{{p}}+\hat{{q}}=\hat{{k}}}d\hat{{p}}\left[u_{j}(\hat{{p}})u_{m}(\hat{{q}})-b_{j}(\hat{{p}})b_{m}(\hat{{q}})\right],\label{eq:udot}\\
\left(-i\omega+\eta k^{2}\right)b_{i}\left(\hat{{k}}\right) & = & -iP_{ijm}^{-}({\textbf{k}})\int_{\hat{{p}}+\hat{{q}}=\hat{{k}}}d\hat{{p}}\left[u_{j}(\hat{{p}})b_{m}(\hat{{q}})\right],\label{eq:bdot}\end{eqnarray}
where\begin{eqnarray}
P_{ijm}^{+}({\textbf{k}}) & = & k_{j}P_{im}({\textbf{k}})+k_{m}P_{ij}({\textbf{k}}),\label{eq:Pp}\\
P_{ijm}^{-}({\textbf{k}}) & = & k_{j}\delta_{im}-k_{m}\delta_{ij}.\label{eq:Pm}\end{eqnarray}
Here $\nu$ and $\eta$ are the viscosity and the resistivity respectively,
and $d$ is the space dimensionality.

In our RG procedure the wavenumber range $(k_{N},k_{0})$ is divided
logarithmically into $N$ shells. The $n$th shell is $(k_{n},k_{n-1})$
where $k_{n}=h^{n}k_{0}\,\,(h<1)$. In the following discussion, we
carry out the elimination of the first shell $(k_{1},k_{0})$ and
obtain the modified MHD equations. We then proceed iteratively to
eliminate higher shells and get a general expression for the modified
MHD equations. The renormalization group procedure is as follows:

\begin{enumerate}
\item We divide the spectral space into two parts: 1. the shell $(k_{1},k_{0})=k^{>}$,
which is to be eliminated; 2. $(k_{N},k_{1})=k^{<}$, set of modes
to be retained. Note that $\nu_{(0)}$ and $\eta_{(0)}$ denote the
viscosity and resistivity before the elimination of the first shell.
\item We rewrite Eqs.~(\ref{eq:udot}, \ref{eq:bdot}) for $k^{<}$ and
$k^{>}$. The equations for $u_{i}^{<}(\hat{k})$ and $b_{i}^{<}(\hat{k})$
modes are \begin{eqnarray}
\left(-i\omega+\Sigma_{(0)}^{uu}(k)\right)u_{i}^{<}(\hat{k})+\Sigma_{(0)}^{ub}(k)b_{i}^{<}(\hat{k}) & = & -\frac{i}{2}P_{ijm}^{+}({\textbf{k}})\int d\hat{p}([u_{j}^{<}(\hat{p})u_{m}^{<}(\hat{k}-\hat{p})]\nonumber \\
 &  & +2[u_{j}^{<}(\hat{p})u_{m}^{>}(\hat{k}-\hat{p})]+[u_{j}^{>}(\hat{p})u_{m}^{>}(\hat{k}-\hat{p})]\nonumber \\
 &  & -\mbox{Similar terms for $b$})\label{eqn:ukless}\\
\left(-i\omega+\Sigma_{(0)}^{bb}(k)\right)b_{i}^{<}(\hat{k})+\Sigma_{(0)}^{bu}(k)u_{i}^{<}(\hat{k}) & = & -iP_{ijm}^{-}({\textbf{k}})\int d\hat{p}([u_{j}^{<}(\hat{p})b_{m}^{<}(\hat{k}-\hat{p})]\nonumber \\
 &  & +[u_{j}^{<}(\hat{p})b_{m}^{>}(\hat{k}-\hat{p})+u_{j}^{>}(\hat{p})b_{m}^{<}(\hat{k}-\hat{p})]\nonumber \\
 &  & +[u_{j}^{>}(\hat{p})b_{m}^{>}(\hat{k}-\hat{p})])\label{eqn:bkless}\end{eqnarray}
 The $\Sigma$s appearing in the equations are usually called the
{}``self-energy'' in Quantum field theory language. In the first
iteration, $\Sigma_{(0)}^{uu}=\nu_{(0)}k^{2}$ and $\Sigma_{(0)}^{bb}=\eta_{(0)}k^{2}$,
while the other two $\Sigma$s are zero. The equation for $u_{i}^{>}(\hat{k})$
modes can be obtained by interchanging $<$ and $>$ in the above
equations.
\item The terms given in the second and third brackets in the Right-hand
side of Eqs.~(\ref{eqn:ukless}, \ref{eqn:bkless}) are calculated
perturbatively. Since we are interested in the statistical properties
of ${\textbf{u}}$ and ${\textbf{b}}$ fluctuations, we perform the
usual ensemble average of the system \cite{YakhOrsz}. We assume that
${\textbf{u}}^{>}(\hat{k})$ and ${\textbf{b}}^{>}(\hat{k})$ have
gaussian distributions with zero mean, while ${\textbf{u}}^{<}(\hat{k})$
and ${\textbf{b}}^{<}(\hat{k})$ are unaffected by the averaging process.
Hence,\begin{eqnarray}
\left\langle u_{i}^{>}(\hat{k})\right\rangle  & = & 0\label{eqn:avgbegin}\\
\left\langle b_{i}^{>}(\hat{k})\right\rangle  & = & 0\\
\left\langle u_{i}^{<}(\hat{k})\right\rangle  & = & u_{i}^{<}(\hat{k})\\
\left\langle b_{i}^{<}(\hat{k})\right\rangle  & = & b_{i}^{<}(\hat{k})\end{eqnarray}
and\begin{eqnarray}
\left\langle u_{i}^{>}(\hat{p})u_{j}^{>}(\hat{q})\right\rangle  & = & P_{ij}({\textbf{p)}}C^{uu}(\hat{p})\delta(\hat{p}+\hat{q})\label{eq:nonhelical-uu}\\
\left\langle b_{i}^{>}(\hat{p})b_{j}^{>}(\hat{q})\right\rangle  & = & P_{ij}({\textbf{p)}}C^{bb}(\hat{p})\delta(\hat{p}+\hat{q})\label{eq:nonhelical-bb}\\
\left\langle u_{i}^{>}(\hat{p})b_{j}^{>}(\hat{q})\right\rangle  & = & P_{ij}({\textbf{p)}}C^{ub}(\hat{p})\delta(\hat{p}+\hat{q})\label{eqn:avgend}\end{eqnarray}
The triple order correlations $\left\langle X_{i}^{>}(\hat{k})X_{j}^{>}(\hat{p})X_{m}^{>}(\hat{q})\right\rangle $
are zero due to Gaussian nature of the fluctuations. Here, $X$ stands
for $u$ or $b$. In addition, we also neglect the contribution from
the triple nonlinearity $\left\langle X^{<}(\hat{k})X_{j}^{<}(\hat{p})X_{m}^{<}(\hat{q})\right\rangle $,
as done in many of the turbulence RG calculations \cite{YakhOrsz,McCo:book}.
The effects of triple nonlinearity can be included following the scheme
of Zhou and Vahala \cite{ZhouVaha88}.
\item To the first order, the second bracketed terms of Eqs.~(\ref{eqn:ukless},
\ref{eqn:bkless}) vanish, but the nonvanishing third bracketed terms
yield corrections to $\Sigma$s. Refer to Appendix C for details.
Eqs. (\ref{eqn:ukless}, \ref{eqn:bkless}) can now be approximated
by \begin{eqnarray}
\left(-i\omega+\Sigma_{(0)}^{uu}+\delta\Sigma_{(0)}^{uu}\right)u_{i}^{<}(\hat{k})+\left(\Sigma_{(0)}^{ub}+\delta\Sigma_{(0)}^{ub}\right)b_{i}^{<}(\hat{k}) & = & -\frac{i}{2}P_{ijm}^{+}({\textbf{k}})\int d\hat{p}[u_{j}^{<}(\hat{p})u_{m}^{<}(\hat{k}-\hat{p})\nonumber \\
 &  & -b_{j}^{<}(\hat{p})b_{m}^{<}(\hat{k}-\hat{p})]\\
\left(-i\omega+\Sigma_{(0)}^{bb}+\delta\Sigma_{(0)}^{bb}\right)b_{i}^{<}(\hat{k})+\left(\Sigma_{(0)}^{bu}+\delta\Sigma_{(0)}^{bu}\right)u_{i}^{<}(\hat{k}) & = & -iP_{ijm}^{-}({\textbf{k}})\int d\hat{p}[u_{j}^{<}(\hat{p})b_{m}^{<}(\hat{k}-\hat{p})]\end{eqnarray}
 with \begin{eqnarray}
\delta\Sigma_{(0)}^{uu}(k) & = & \frac{1}{(d-1)}\int_{\hat{p}+\hat{q}=\hat{k}}^{\Delta}d\hat{p}[S(k,p,q)G^{uu}(\hat{p})C^{uu}(\hat{q})-S_{6}(k,p,q)G^{bb}(\hat{p})C^{bb}(\hat{q})\nonumber \\
 &  & +S_{6}(k,p,q)G^{ub}(\hat{p})C^{ub}(\hat{q})-S(k,p,q)G^{bu}(\hat{p})C^{ub}(\hat{q})]\label{eqn:sigmauu}\\
\delta\Sigma_{(0)}^{ub}(k) & = & \frac{1}{(d-1)}\int_{\hat{p}+\hat{q}=\hat{k}}^{\Delta}d\hat{p}[-S(k,p,q)G^{uu}(\hat{p})C^{ub}(\hat{q})+S_{5}(k,p,q)G^{ub}(\hat{p})C^{uu}(\hat{q})\nonumber \\
 &  & +S(k,p,q)G^{bu}(\hat{p})C^{bb}(\hat{q})-S_{5}(k,p,q)G^{bb}(\hat{p})C^{ub}(\hat{q})]\label{eqn:sigmaub}\\
\delta\Sigma_{(0)}^{bu}(k) & = & \frac{1}{(d-1)}\int_{\hat{p}+\hat{q}=\hat{k}}^{\Delta}d\hat{p}[S_{8}(k,p,q)G^{uu}(\hat{p})C^{ub}(\hat{q})+S_{10}(k,p,q)G^{bb}(\hat{p})C^{ub}(\hat{q})\nonumber \\
 &  & +S_{12}(k,p,q)G^{ub}(\hat{p})C^{bb}(\hat{q})-S_{7}(k,p,q)G^{bu}(\hat{p})C^{uu}(\hat{q})]\label{eqn:sigmabu}\\
\delta\Sigma_{(0)}^{bb}(k) & = & \frac{1}{(d-1)}\int_{\hat{p}+\hat{q}=\hat{k}}^{\Delta}d\hat{p}[-S_{8}(k,p,q)G^{uu}(\hat{p})C^{bb}(\hat{q})+S_{9}(k,p,q)G^{bb}(\hat{p})C^{uu}(\hat{q})\nonumber \\
 &  & +S_{11}(k,p,q)G^{ub}(\hat{p})C^{ub}(\hat{q})-S_{9}(k,p,q)G^{bu}(\hat{p})C^{ub}(\hat{q})]\label{eqn:sigmabb}\end{eqnarray}
 The quantities $S_{i}(k,p,q)$ are given in the Appendix C. The integral
$\Delta$ is to be done over the first shell.
\item The full-fledge calculation of $\Sigma$'s is quite involved. Therefore,
we take two special cases: (1) NonAlfvénic: $C^{ub}=0$ or $\sigma_{c}=0$;
and (2) Alfvénic: $C^{ub}\approx C^{uu}\approx C^{bb}$ or $\sigma_{c}\rightarrow1$.
In this subsubsection we will discuss only the case $\sigma_{c}=0$.
The other case will be taken up in the next subsubsection. A word
of caution is in order here. In our calculation the parameters used
$\sigma_{c}(k)=2C^{ub}(k)/(C^{uu}(k)+C^{bb}(k))$ and $r_{A}(k)=E^{u}(k)/E^{b}(k)$
are taken to be constants, even though they could be a function of
$k$. Also note that these parameters could differ from the global
$\sigma_{c}$ and $r_{A}$, yet we assume that they are probably closer
to the global value. \\
When $\sigma_{c}=0$, an inspection of the self-energy diagrams shows
that $\Sigma^{ub}=\Sigma^{bu}=0$, and $G^{ub}=G^{bu}=0$. Clearly,
the equations become much simpler because of the diagonal nature of
matrices $G$ and $\Sigma$, and the two quantities of interest $\delta\Sigma_{(0)}^{uu}$
and $\delta\Sigma_{(0)}^{bb}$ are given by\begin{eqnarray}
\delta\Sigma_{(0)}^{uu}(\hat{k}) & = & \frac{1}{d-1}\int_{\hat{p}+\hat{q}=\hat{k}}^{\Delta}d\hat{p}\left(S(k,p,q)G^{uu}(p)C^{uu}(q)-S_{6}(k,p,q)G^{bb}(p)C^{bb}(q)\right)\label{eq:delta-Sigmauu}\\
\delta\Sigma_{(0)}^{bb}(\hat{k}) & = & \frac{1}{d-1}\int_{\hat{p}+\hat{q}=\hat{k}}^{\Delta}d\hat{p}\left(-S_{8}(k,p,q)G^{uu}(p)C^{bb}(q)+S_{9}(k,p,q)G^{bb}(p)C^{uu}(q)\right)\label{eq:delta-Sigmabb}\end{eqnarray}

\item The frequency dependence of the correlation function are taken as:
$C^{uu}(k,\omega)=2C^{uu}(k)\Re(G^{uu}(k,\omega))$ and $C^{bb}(k,\omega)=2C^{bb}(k)\Re(G^{bb}(k,\omega))$.
In other words, the relaxation time-scale of correlation function
is assumed to be the same as that of corresponding Green's function.
Since we are interested in the large time-scale behaviour of turbulence,
we take the limit $\omega$ going to zero. Under these assumptions,
the frequency integration of the above equations yield \begin{eqnarray}
\delta\nu_{(0)}(k) & = & \frac{1}{(d-1)k^{2}}\int_{\textbf{p+q=k}}^{\Delta}\frac{d{\textbf{p}}}{(2\pi)^{d}}\nonumber \\
 &  & \left[\frac{S(k,p,q)C^{uu}(q)}{\nu_{(0)}(p)p^{2}+\nu_{(0)}(q)q^{2}}-\frac{S_{6}(k,p,q)C^{bb}(q)}{\eta_{(0)}(p)p^{2}+\eta_{(0)}(q)q^{2}}]\right]\label{eq:delta-nu}\\
\delta\eta_{(0)}(k) & = & \frac{1}{(d-1)k^{2}}\int_{\textbf{p+q=k}}^{\Delta}\frac{d{\textbf{p}}}{(2\pi)^{d}}\nonumber \\
 &  & \left[-\frac{S_{8}(k,p,q)C^{bb}(q)}{\nu_{(0)}(p)p^{2}+\eta_{(0)}(q)q^{2}}+\frac{S_{9}(k,p,q)C^{uu}(q)}{\eta_{(0)}(p)p^{2}+\nu_{(0)}(q)q^{2}}]\right]\label{eq:delta-eta}\end{eqnarray}
Note that $\nu(k)=\Sigma^{uu}(k)/k^{2}$ and $\eta(k)=\Sigma^{bb}(k)/k^{2}$.
There are some important points to remember in the above step. The
frequency integral in the above is done using contour integral. It
can be shown that the integrals are nonzero only when both the components
appearing the denominator are of the same sign. For example, first
term of Eq. (\ref{eq:delta-eta}) is nonzero only when both $\nu_{(0)}(p)$
and $\eta_{(0)}(q)$ are of the same sign.
\item Let us denote $\nu_{(1)}(k)$ and $\eta_{(1)}(k)$ as the renormalized
viscosity and resistivity respectively after the first step of wavenumber
elimination. Hence, \begin{eqnarray}
\nu_{(1)}(k) & = & \nu_{(0)}(k)+\delta\nu_{(0)}(k);\\
\eta_{(1)}(k) & = & \eta_{(0)}(k)+\delta\eta_{(0)}(k)\label{eqn:eta_1}\end{eqnarray}
We keep eliminating the shells one after the other by the above procedure.
After $n+1$ iterations we obtain \begin{eqnarray}
\nu_{(n+1)}(k)=\nu_{(n)}(k)+\delta\nu_{(n)}(k)\label{nu_n}\\
\eta_{(n+1)}(k)=\eta_{(n)}(k)+\delta\eta_{(n)}(k)\end{eqnarray}
where the equations for $\delta\nu_{(n)}(k)$ and $\delta\eta_{(n)}(k)$
are the same as the Eqs.~(\ref{eq:delta-nu}, \ref{eq:delta-eta})
except that $\nu_{(0)}(k)$ and $\eta_{(0)}(k)$ appearing in the
equations are to be replaced by $\nu_{(n)}(k)$ and $\eta_{(n)}(k)$
respectively. Clearly $\nu_{(n+1)}(k)$ and $\eta_{(n+1)}$(k) are
the renormalized viscosity and resistivity after the elimination of
the $(n+1)$th shell.
\item We need to compute $\delta\nu_{(n)}$ and $\delta\eta_{(n)}$ for
various $n$. These computations, however, require $\nu_{(n)}$ and
$\eta_{(n)}$. In our scheme we solve these equations iteratively.
In Eqs.~(\ref{eq:delta-nu}, \ref{eq:delta-eta}) we substitute $C(k)$
by one dimensional energy spectrum $E(k)$\[
C^{(uu,bb)}(k)=\frac{2(2\pi)^{d}}{S_{d}(d-1)}k^{-(d-1)}E^{(u,b)}(k)\]
 where $S_{d}$ is the surface area of $d$-dimensional spheres. We
assume that $E^{u}(k)$ and $E^{b}(k)$ follow Eqs.~(\ref{eq:Euk},
\ref{eq:Ebk}) respectively. Regarding $\nu_{(n)}$ and $\eta_{(n)}$,
we attempt the following form of solution \[
(\nu,\eta)_{(n)}(k_{n}k')=(K^{u})^{1/2}\Pi^{1/3}k_{n}^{-4/3}(\nu^{*},\eta^{*})_{(n)}(k')\]
 with $k=k_{n+1}k'\,(k'<1)$. We expect $\nu_{(n)}^{*}(k')$ and $\eta_{(n)}^{*}(k')$
to be a universal functions for large $n$. The substitution of $C^{uu}(k),C^{bb}(k),\nu_{(n)}(k)$,
and $\eta_{(n)}(k)$ yields the following equations: \begin{eqnarray}
\delta\nu_{(n)}^{*}(k') & = & \frac{1}{(d-1)}\int_{\textbf{p'+q'=k'}}d{\textbf{q'}}\frac{2}{(d-1)S_{d}}\frac{E^{u}(q')}{q'^{d-1}}[S(k',p',q')\frac{1}{\nu_{(n)}^{*}(hp')p'^{2}+\nu_{(n)}^{*}(hq')q'^{2}}\nonumber \\
 &  & -S_{6}(k',p',q')\frac{r_{A}^{-1}}{\eta_{(n)}^{*}(hp')p'^{2}+\eta_{(n)}^{*}(hq')q'^{2}}]\label{eq:delta_nu*}\\
\delta\eta_{(n)}^{*}(k') & = & \frac{1}{(d-1)}\int_{\textbf{p'+q'=k'}}d{\textbf{q'}}\frac{2}{(d-1)S_{d}}\frac{E^{u}(q')}{q'^{d-1}}[-S_{8}(k',p',q')\frac{1}{\nu_{(n)}^{*}(hp')p'^{2}+\eta_{(n)}^{*}(hq')q'^{2}}\nonumber \\
 &  & +S_{9}(k',p',q')\frac{r_{A}^{-1}}{\eta_{(n)}^{*}(hp')p'^{2}+\nu_{(n)}^{*}(hq')q'^{2}}]\label{eq:delta_eta*}\\
\nu_{(n+1)}^{*}(k') & = & h^{4/3}\nu_{(n)}^{*}(hk')+h^{-4/3}\delta\nu_{(n)}^{*}(k')\\
\eta_{(n+1)}^{*}(k') & = & h^{4/3}\eta_{(n)}^{*}(hk')+h^{-4/3}\delta\eta_{(n)}^{*}(hk')\end{eqnarray}
 where the integrals in the above equations are performed iteratively
over a region $1\leq p',q'\leq1/h$ with the constraint that $\mathbf{p}'+\mathbf{q}'=\mathbf{k}'$.
Fournier and Frisch \cite{FourFris} showed the above volume integral
in $d$ dimension to be \begin{equation}
\int_{\mathbf{p}'+\mathbf{q}'=\mathbf{k}'}d\mathbf{p'}=S_{d-1}\int dp'dq'\left(\frac{p'q'}{k'}\right)^{d-2}\left(\sin\alpha\right)^{d-3},\label{eq:volume-integral}\end{equation}
where $\alpha$ is the angle between vectors $\mathbf{p}'$ and $\mathbf{q}'$.
\item Now we solve the above four equations self consistently for various
$r_{A}$s. We have taken $h=0.7$. This value is about middle of the
range (0.55-0.75) estimated to be the reasonable values of $h$ by
Zhou \emph{et al.} \cite{ZhouMcCo:RGrev}. We start with constant
values of $\nu_{(0)}^{*}$ and $\eta_{(0)}^{*}$, and compute the
integrals using Gauss quadrature technique. Once $\delta\nu_{(0)}^{*}$
and $\delta\eta_{(0)}^{*}$ have been computed, we can calculate $\nu_{(1)}^{*}$
and $\eta_{(1)}^{*}$. We iterate this process till $\nu_{(m+1)}^{*}(k')\approx\nu_{(m)}^{*}(k')$
and $\eta_{(m+1)}^{*}(k')\approx\eta_{(m)}^{*}(k')$, that is, till
they converge. We have reported the limiting $\nu^{*}$ and $\eta^{*}$
whenever the solution converges. The criterion for convergence is
that the error must be less than 1\%. This criterion is usually achieved
by $n=10$ or so. The result of our RG analysis is given below.
\end{enumerate}
Verma carried out the RG analysis for various space dimensions and
found that the solution converged for all $d>d_{c}\approx2.2$. Hence,
the RG fixed-point for MHD turbulence is stable for $d\geq d_{c}$.
For illustration of convergent solution, see the plot of $\nu_{(n)}^{*}(k')$
and $\eta_{(n)}^{*}(k')$ for $d=3,r_{A}=1$ in Fig.~\ref{Fig:d3ra1}.
\begin{figure}
\includegraphics[%
  scale=0.6,bb= 0 250 525 560 ]{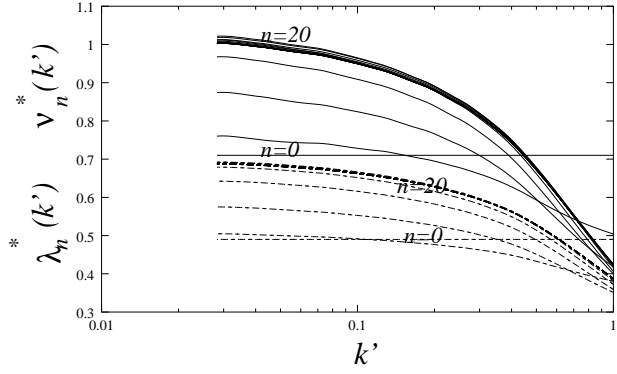}

\caption{\label{Fig:d3ra1} The plots of $\nu^{*}(k')$ (solid) and $\eta^{*}(k')$
(dashed) vs. $k'$ for $d=3$ and $\sigma_{c}=0,$ $r_{A}=1$. The
values converge. Adopted from Verma \cite{MKV:MHD_RG}.}
\end{figure}
The RG fixed point for $d<d_{c}$ is unstable. Refer to Fig. \ref{Fig:d2}
for $d=2,r_{A}=1$ as an example of an unstable solution. %
\begin{figure}
\includegraphics[%
  scale=0.6, bb=0 240 612 560 ]{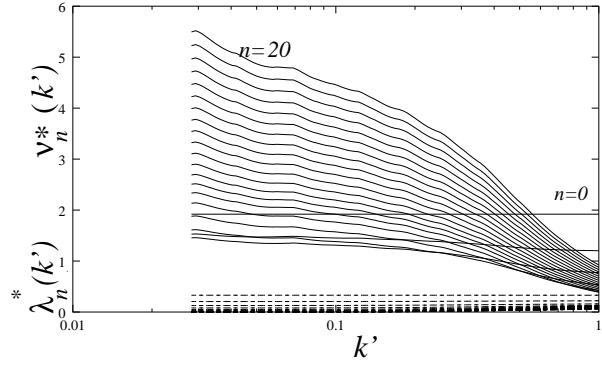}

\caption{\label{Fig:d2} The plots of $\nu^{*}(k')$ (solid) and $\eta^{*}(k')$
(dashed) vs. $k'$ for $d=2$ and $\sigma_{c}=0,$ $r_{A}=1$. There
is no convergence. Adopted from Verma \cite{MKV:MHD_RG}.}
\end{figure}
From this observation we can claim that Kolmogorov's powerlaw is a
consistent solution of MHD RG equations at least for $d\geq d_{c}$.
Verma also computed the contribution to renormalized viscosity and
resistivity from each of the four terms $\mathbf{u}\cdot\nabla\mathbf{u},-\mathbf{b}\cdot\nabla\mathbf{b},-\mathbf{u}\cdot\nabla\mathbf{b},\mathbf{b}\cdot\nabla\mathbf{u}$.
These quantities are denoted by $\nu^{uu}$, $\nu^{ub}$, $\eta^{bu}$,
and $\eta^{bb}$ respectively. The values of asymptotic ($k'\rightarrow0$
limit) $\nu^{*}$,$\eta^{*}$, $\nu^{uu}$, $\nu^{ub}$, $\eta^{bu}$,
and $\eta^{bb}$ for various $d$ and $r_{A}=1$ are displayed in
Table \ref{table:alldra1}.%
\begin{table}

\caption{\label{table:alldra1} The values of $\nu^{*}$, $\eta^{*}$, $\nu^{uu*}$,
$\nu^{ub*}$, $\eta^{bu*}$, $\eta^{bb*}$ for various space dimensions
$d$ with $r_{A}=1$ and $\sigma_{c}=0$.}

\begin{tabular}{|c|c|c|c|c|c|c|c|}
\hline 
$d$&
$\nu^{*}$&
$\eta^{*}$&
$Pr$&
$\nu^{uu*}$&
$\nu^{ub*}$&
$\eta^{bu*}$&
$\eta^{bb*}$\tabularnewline
\hline
\hline 
2.1&
...&
...&
...&
...&
...&
...&
...\tabularnewline
\hline 
2.2&
1.9&
0.32&
6.0&
-0.041
&
1.96&
-0.44
&
0.76\tabularnewline
\hline 
2.5&
1.2&
0.57&
2.1&
0.089&
1.15&
$-0.15$&
0.72\tabularnewline
\hline 
3.0&
1.00&
0.69&
1.4&
0.20&
0.80&
0.078&
0.61\tabularnewline
\hline 
4.0&
0.83&
0.70&
1.2&
0.27&
0.56&
0.21&
0.49\tabularnewline
\hline 
10.0&
0.51&
0.50&
1.0&
0.23&
0.28&
0.22&
0.28\tabularnewline
\hline 
50.0&
0.23&
0.23&
1.0&
0.11&
0.12&
0.11&
0.12\tabularnewline
\hline 
100.0&
0.14&
0.14&
1.0&
0.065&
0.069&
0.066&
0.069\tabularnewline
\hline
\end{tabular}
\end{table}
The MHD equations can be written in terms of these renormalized parameters
as\begin{eqnarray*}
\left(\frac{\partial}{\partial t}+\nu^{uu}k^{2}+\nu^{ub}k^{2}\right)u_{i}^{<}(\mathbf{k},t) & = & -\frac{i}{2}P_{ijm}^{+}(\mathbf{k})\int\frac{d\mathbf{p}}{(2\pi)^{d}}[u_{j}^{<}(\mathbf{p},t)u_{m}^{<}(\mathbf{k-p},t)\\
 &  & -b_{j}^{<}(\mathbf{p},t)b_{m}^{<}(\mathbf{k-p},t)]\\
\left(\frac{\partial}{\partial t}+\eta^{bu}k^{2}+\eta^{bb}k^{2}\right)b_{i}^{<}(\mathbf{k},t) & = & -P_{ijm}^{-}(\mathbf{k})\int\frac{d\mathbf{p}}{(2\pi)^{d}}[u_{j}^{<}(\mathbf{p},t)b_{m}^{<}(\mathbf{k-p},t)]\end{eqnarray*}

We multiply the above equations by $u_{i}^{<*}(\mathbf{k},t)$ and
$b_{i}^{<*}(\mathbf{k},t)$ respectively and obtain the energy equation.
When we integrate the terms up to the last wavenumbers $k_{N}$, the
terms in the RHS vanish because of {}``detailed conservation of energy
in a triad interaction'' (see Section \ref{sub:Mode-to-Mode-Energy-Transfer-MHD}).
Therefore from the definition, we deduce that the energy cascade rate
from inside of the $X$ sphere ($X<$) to outside of the $Y$ sphere
($Y>$) is \begin{equation}
\Pi_{Y>}^{X<}=\int_{0}^{k_{N}}2\nu^{XY}(k)k^{2}E^{X}(k)dk\label{eq:fluxRG}\end{equation}
 where $X,Y$ denote $u$ or $b$. From Table 1, we see that the sign
of $\nu^{uu}$ changes from positive to negative at $d=2.2$; this
result is consistent with the conclusions of Fournier and Frisch \cite{FourFris},
where they predict the reversal of the sign of eddy viscosity at $d=2.208$.
Even though Verma's RG calculation could not be extended to $d=2$
(because of instability of the fixed point), it is reasonable to expect
that for $d=2$, $\nu^{uu}$will be negative, and $\Pi_{u>}^{u<}$
will be negative consistent with the EDQNM results of Pouquet et al.
\cite{Pouq:EDQNM2D}, Ishizawa and Hattori \cite{Ishi:EDQNM}, and
numerical results of Dar et al. \cite{Dar:flux}.

\begin{figure}
\includegraphics[%
  scale=0.6,bb= 0 240 612 580 ]{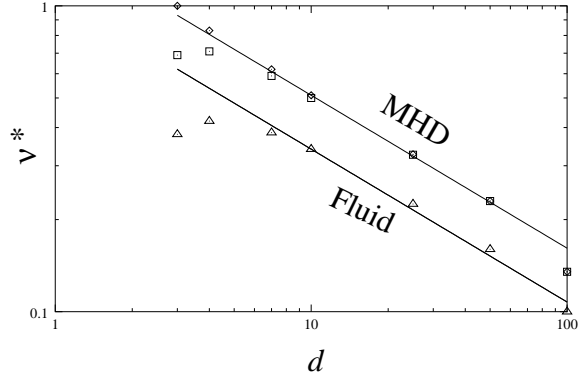}

\caption{\label{Fig:nuvsd} The plot of asymptotic $\nu^{*}$(square) and
$\eta^{*}$(diamond) vs. $d$ for $\sigma_{c}=0$ and $r_{A}=1$.
The fluid $\nu^{*}$ (triangle) is also plotted for reference. For
large $d$, these values fit quite well with predicted $d^{-1/2}$
curve. Adopted from Verma \cite{MKV:MHD_RG}.}
\end{figure}

For large $d$ , $\nu^{*}=\eta^{*}$, and it decreases as $d^{-1/2}$
(see Fig. \ref{Fig:nuvsd});%
 $\nu^{*}$ for pure fluid turbulence also decreases as $d^{-1/2}$,
as shown in the same figure. This is evident from Eqs. (\ref{eq:delta_nu*},\ref{eq:delta_eta*})
using the following arguments of Fournier et al. \cite{Four:Inf}.
For large $d$ \begin{eqnarray}
\int dp'dq'\left(\frac{p'q'}{k'}\right)^{d-2}\left(\sin\alpha\right)^{d-3}... & \sim & d^{-1/2},\\
\frac{S_{d-1}}{\left(d-1\right)^{2}S_{d}} & \sim & \frac{1}{d^{2}}\left(\frac{d}{2\pi}\right)^{1/2},\nonumber \\
S,-S_{6},-S_{8},S_{9}(k',p',q') & = & kpd(z+xy),\label{eq:SSSS}\end{eqnarray}
which leads to\[
\nu^{*}\delta\nu^{*}\propto\frac{1}{d^{2}}\left(\frac{d}{2\pi}\right)^{1/2}d^{-1/2}d\]
hence $\nu^{*}\propto d^{-1/2}$. Also, from Eq. (\ref{eq:SSSS})
it can be deduced that $\nu^{uu*}=\nu^{ub*}=\eta^{bu*}=\eta^{bb*}$
for large $d$, as is seen from Table \ref{table:alldra1}.

Verma \cite{MKV:MHD_RG} also observed that the stability of RG fixed
point in a given space dimension depends on Alfv\'{e}n ratio and
normalized cross helicity. For example, for $d=2.2$ the RG fixed
point is stable for $r_{A}\geq1$, but unstable for $r_{A}<1$. A
detailed study of stability of the RG fixed point is required to ascertain
the boundary of stability.

The values of renormalized parameters for $d=3$ and various $r_{A}$
are shown in Table \ref{table:d3}. %
\begin{table}

\caption{\label{table:d3}The values of $\nu^{*}$, $\eta^{*}$, $\nu^{uu*}$,
$\nu^{ub*}$, $\eta^{bu*}$, $\eta^{bb*}$ for various $r_{A}$ with
$d=3$ and $\sigma_{c}=0$.}

\begin{tabular}{|c|c|c|c|c|c|c|c|}
\hline 
$r_{A}$&
$\nu^{*}$&
$\eta^{*}$&
$Pr$&
$\nu^{uu*}$&
$\nu^{ub*}$&
$\eta^{bu*}$&
$\eta^{bb*}$\tabularnewline
\hline
\hline 
$\infty$&
0.38&
...&
...&
$0.38$&
...&
...&
...\tabularnewline
\hline 
5000&
0.36&
0.85&
0.42&
0.36&
$1.4\times10^{-4}$&
$-0.023$&
0.87\tabularnewline
\hline 
100&
0.36&
0.85&
0.42&
0.36&
$7.3\times10^{-3}$&
$-0.022$&
0.87\tabularnewline
\hline 
5&
0.47&
0.82&
0.57&
0.32&
0.15&
$4.7\times10^{-4}$&
0.82\tabularnewline
\hline 
2&
0.65&
0.78&
0.83&
0.27&
0.38&
0.031&
0.75\tabularnewline
\hline 
1&
1.00&
0.69&
1.40&
0.20&
0.80&
0.078&
0.61\tabularnewline
\hline 
0.5&
2.1&
0.50&
4.2&
0.11&
2.00&
0.15&
0.35\tabularnewline
\hline 
0.3&
11.0&
0.14&
78&
0.022&
11.0&
0.082&
0.053\tabularnewline
\hline 
0.2&
...&
...&
...&
...&
...&
...&
...\tabularnewline
\hline
\end{tabular}
\end{table}
 For large $r_{A}$ (fluid dominated regime), $\nu^{*}$ is close
to renormalized viscosity of fluid turbulence ($r_{A}=\infty$), but
$\eta^{*}$ is also finite. As $r_{A}$ is decreased, $\eta^{*}$
decreases but $\nu^{*}$ increases, or the Prandtl number $Pr=\nu/\eta$
increases. This trend is seen till $r_{A}\approx0.25$ when the RG
fixed point with nonzero $\nu^{*}$ and $\eta^{*}$ becomes unstable,
and the trivial RG fixed point with $\nu^{*}=\eta^{*}=0$ becomes
stable. This result suggests an absence of turbulence for $r_{A}$
below 0.25 (approximately). Note that in the $r_{A}\rightarrow0$
(fully magnetic) limit, the MHD equations become linear, hence there
is no turbulence. Surprisingly, our RG calculation suggests that turbulence
disappear near $r_{A}=0.25$ itself.

Using the flux interpretation of renormalized parameters (Eq. {[}\ref{eq:fluxRG}{]}),
and from the values of renormalized parameters in Table \ref{table:d3},
we can deduce that energy fluxes from kinetic to kinetic, magnetic
to magnetic, and kinetic to magnetic energies are always positive.
The energy fluxes from magnetic energy to kinetic energy is positive
for $0.3<r_{A}<2$, but changes sign on further increase of $r_{A}$.
The negative value of $\eta^{bu*}$ indicates that the kinetic energy
at large wavenumbers are transferred to the magnetic energy at smaller
wavenumbers (inverse transfer). 

Verma found that the final $\nu^{*}(k')$ and $\eta^{*}(k')$ are
constant for small $k'$ but shifts toward zero for larger $k'$ (see
Fig. \ref{Fig:d3ra1}). Similar behaviour has been seen by McComb
and coworkers \cite{McCoWatt} for fluid turbulence, and is attributed
to the neglect of triple nonlinearity. Triple nonlinearity for fluid
turbulence was first included in the RG calculation by Zhou and Vahala
\cite{ZhouVaha93}; similar calculation for MHD turbulence is yet
to done. 

Pouquet \cite{Pouq:EDQNM2D} and Ishizawa and Hattori \cite{Ishi:EDQNM}
calculated $\nu^{uu}$, $\nu^{ub}$, $\eta^{bu}$, $\eta^{bb}$ for
$d=2$ using EDQNM (Eddy-damped quasi-normal Markovian) approximation.
Pouquet argued that $\eta^{bb}$ is negative, while Ishizawa found
it to be positive. Unfortunately Verma's procedure cannot be extended
to $d=2$. However, Verma claimed that the magnetic energy cascade
rate ($\Pi_{b>}^{b<}$) is positive for all $d>d_{c}$ because $\eta^{bb}>0$. 

In the following subsection we present Verma's calculation of renormalized
viscosity and resistivity for $\sigma_{c}\rightarrow1$ limit \cite{MKV:MHD_RG}.

\subsubsection{Nonhelical Alfvénic MHD ($H_{M}=H_{K};\sigma_{c}\rightarrow1$):
\label{sub:Nonhelical-Alfvenic-MHD}}

Alfvénic MHD has high $\mathbf{u}$-\textbf{b} correlation or $\left\langle |z^{+}|^{2}\right\rangle \gg\left\langle |z^{-}|^{2}\right\rangle $.
For this case it is best to work with Els\"{a}sser variables $\mathbf{z}^{\pm}=\mathbf{u}\pm\mathbf{b}$.
These types of fluctuations have been observed in the solar wind near
the Sun. However, by the time the solar wind approaches the Earth,
the normalized cross helicity is normally close to zero. In this section
we will briefly discuss the RG treatment for the above case. For the
following discussion we will denote $\left\langle |z^{-}|^{2}\right\rangle /\left\langle |z^{+}|^{2}\right\rangle =r=(1-\sigma_{c})/(1+\sigma_{c})$.
Clearly $r\ll1$.

MHD equations in terms of Els\"{a}sser variables are \[
\left(-i\omega+\nu_{(0)\pm\pm}k^{2}\right)z_{i}^{\pm}(\hat{{k}})+\nu_{(0)\pm\mp}k^{2}z_{i}^{\mp}(\hat{{k}})=-iM_{ijm}({\textbf{k}})\int d\hat{{k}}z_{j}^{\mp}(\hat{{p}})z_{m}^{\pm}(\hat{{k}}-\hat{{p}}).\]
Note that the above equations contain four dissipative coefficients
$\nu_{\pm\pm}$ and $\nu_{\pm\mp}$ instead of usual two constants
$\nu_{\pm}=(\nu\pm\eta)/2$. The $+-$ symmetry is broken when $r\ne1$.
RG generates the other two constants. We carry out the same procedure
as outlined in the previous RG calculation. After $n+1$ steps of
the RG calculation, the above equations become \begin{eqnarray}
\left[-i\omega+\left(\nu_{(n)\pm\pm}(k)+\delta\nu_{(n)\pm\pm}(k)\right)k^{2}\right]z_{i}^{\pm<}(\hat{k})\nonumber \\
+\left(\nu_{(n)\pm\mp}(k)+\delta\nu_{(n)\pm\mp}(k)\right)k^{2}z_{i}^{\mp<}(\hat{k}) & = & -iM_{ijm}({\textbf{k}})\int d\hat{p}z_{j}^{\mp<}(\hat{p})z_{m}^{\pm<}(\hat{k}-\hat{p})\end{eqnarray}
 with \begin{eqnarray}
\delta\nu_{(n)++}(k) & = & \frac{1}{(d-1)k^{2}}\int_{\hat{p}+\hat{q}=\hat{k}}^{\Delta}d\hat{p}[S_{1}(k,p,q)G_{(n)}^{++}(\hat{p})C^{--}(\hat{q})+S_{2}(k,p,q)G_{(n)}^{+-}(\hat{p})C^{--}(\hat{q})\nonumber \\
 &  & +S_{3}(k,p,q)G_{(n)}^{-+}(\hat{p})C^{+-}(\hat{q})+S_{4}(k,p,q)G_{(n)}^{--}(\hat{p})C^{+-}(\hat{q})]\label{eqn:nupp}\\
\delta\nu_{(n)+-}(k) & = & \frac{1}{(d-1)k^{2}}\int_{\hat{p}+\hat{q}=\hat{k}}^{\Delta}d\hat{p}[S_{1}(k,p,q)G_{(n)}^{+-}(\hat{p})C^{-+}(\hat{q})+S_{2}(k,p,q)G_{(n)}^{++}(\hat{p})C^{-+}(\hat{q})\nonumber \\
 &  & +S_{3}(k,p,q)G_{(n)}^{--}(\hat{p})C^{++}(\hat{q})+S_{4}(k,p,q)G_{(n)}^{-+}(\hat{p})C^{++}(\hat{q})]\label{eqn:nupm}\end{eqnarray}
 where the integral is performed over the $(n+1)$th shell $(k_{n+1},k_{n})$.
The equations for the other two $\delta\nu$s can be obtained by interchanging
$+$ and $-$ signs. Now we assume that the Alfv\'{e}n ratio is one,
i.e., $C^{+-}=E^{u}-E^{b}=0$. Under this condition, the above equations
reduce to \begin{eqnarray}
\delta\nu_{(n)++}(k) & = & \frac{1}{(d-1)k^{2}}\int_{\hat{p}+\hat{q}=\hat{k}}^{\Delta}d\hat{p}[S_{1}(k,p,q)G_{(n)}^{++}(\hat{p})+S_{2}(k,p,q)G_{(n)}^{+-}(\hat{p})]C^{--}(\hat{q})\label{eqn:nuppsig0}\\
\delta\nu_{(n)+-}(k) & = & \frac{1}{(d-1)k^{2}}\int_{\hat{p}+\hat{q}=\hat{k}}^{\Delta}d\hat{p}[S_{3}(k,p,q)G_{(n)}^{--}(\hat{p})+S_{4}(k,p,q)G_{(n)}^{-+}(\hat{p})]C^{++}(\hat{q})\\
\delta\nu_{(n)-+}(k) & = & \frac{1}{(d-1)k^{2}}\int_{\hat{p}+\hat{q}=\hat{k}}^{\Delta}d\hat{p}[S_{3}(k,p,q)G_{(n)}^{++}(\hat{p})+S_{4}(k,p,q)G_{(n)}^{+-}(\hat{p})]C^{--}(\hat{q})\\
\delta\nu_{(n)--}(k) & = & \frac{1}{(d-1)k^{2}}\int_{\hat{p}+\hat{q}=\hat{k}}^{\Delta}d\hat{p}[S_{1}(k,p,q)G_{(n)}^{--}(\hat{p})+S_{2}(k,p,q)G_{(n)}^{-+}(\hat{p})]C^{++}(\hat{q})\label{eqn:nummsig0}\end{eqnarray}
 The inspection of Eqs. (\ref{eqn:nuppsig0}--\ref{eqn:nummsig0})
reveal that $\nu_{++}$ and $\nu_{-+}$ are of the order of $r$.
Hence, we take the $\hat{\nu}$ matrix to be of the form \begin{equation}
\hat{\nu}(k,\omega)=\left(\begin{array}{cc}
r\zeta & \alpha\\
r\psi & \beta\end{array}\right)\end{equation}

It is convenient to transform the frequency integrals in Eqs. (\ref{eqn:nuppsig0}-\ref{eqn:nummsig0})
into temporal integrals, which yields\begin{eqnarray}
\delta\nu_{(n)++}(k) & = & \frac{1}{(d-1)k^{2}}\int_{\textbf{p+q=k}}^{\Delta}\frac{d{\textbf{p}}}{(2\pi)^{d}}\int_{-\infty}^{t}dt'[S_{1}(k,p,q)G_{(n)}^{++}({\textbf{p}},t-t')\nonumber \\
 &  & +S_{2}(k,p,q)G_{(n)}^{+-}({\textbf{p}},t-t')]C^{--}({\textbf{q}},t-t')]\end{eqnarray}
 and similar forms for equations for other $\nu$s. Green's function
$\hat{G}(k,t-t')=\exp{-[\hat{\nu}k^{2}(t-t')]}$ can be easily evaluated
by diagonalizing the matrix $\hat{\nu}$. The final form of $\hat{G}(k,t-t')$
to leading order in $r$ is \[
\hat{G}(k,t-t')=\left(\begin{array}{cc}
1-\frac{r\alpha\psi}{\beta^{2}}\left(1-\exp{\left(-\beta(t-t')\right)}\right) & -\left\{ \frac{\alpha}{\beta}+\frac{r\alpha}{\beta}\left(\frac{\zeta}{\beta}-\frac{2\alpha\psi}{\beta^{2}}\right)\right\} \left(1-\exp{\left(-\beta(t-t')\right)}\right)\\
-\frac{r\psi}{\beta}\left(1-\exp{\left(-\beta(t-t')\right)}\right) & \exp{\left(-\beta(t-t')\right)}+\frac{r\alpha\psi}{\beta^{2}}\left(1-\exp{\left(-\beta(t-t')\right)}\right)\end{array}\right)\]

The correlation matrix $\hat{C}(k,t-t')$ is given by \begin{eqnarray}
\left(\begin{array}{cc}
C^{++}(k,t-t') & C^{+-}(k,t-t')\\
C^{-+}(k,t-t') & C^{--}(k,t-t')\end{array}\right) & = & \hat{G}(k,t-t')\left(\begin{array}{cc}
C^{++}(k) & C^{+-}(k)\\
C^{-+}(k) & C^{--}(k)\end{array}\right)\end{eqnarray}
 The substitution of correlation functions and Green's functions yield
the following expressions for the elements of $\delta\hat{\nu}$\begin{eqnarray}
\delta\zeta_{(n)}(k) & = & \frac{1}{(d-1)k^{2}}\int^{\Delta}\frac{d{\textbf{p}}}{(2\pi)^{d}}C^{+}(q)\{ S_{1}(k,p,q)\frac{1}{\beta_{(n)}(q)q^{2}}\nonumber \\
 &  & \hspace{1in}+S_{2}(k,p,q)\frac{\alpha_{(n)}(p)}{\beta_{(n)}(p)}\left(\frac{1}{\beta_{(n)}(p)p^{2}+\beta_{(n)}(q)q^{2}}-\frac{1}{\beta_{(n)}(q)q^{2}}\right)\nonumber \\
 &  & \hspace{1in}-S_{3}(k,p,q)\frac{\alpha_{(n)}(q)}{\beta_{(n)}(q)}\left(\frac{1}{\beta_{(n)}(p)p^{2}+\beta_{(n)}(q)q^{2}}-\frac{1}{\beta_{(n)}(p)p^{2}}\right)\}\\
\delta\alpha_{(n)}(k) & = & \frac{1}{(d-1)k^{2}}\int^{\Delta}\frac{d{\textbf{p}}}{(2\pi)^{d}}S_{3}(k,p,q)\frac{C^{+}(q)}{\beta_{(n)}(p)p^{2}}\\
\delta\psi_{(n)}(k) & = & \frac{1}{(d-1)k^{2}}\int^{\Delta}\frac{d{\textbf{p}}}{(2\pi)^{d}}C^{+}(q)\{ S_{3}(k,p,q)\frac{1}{\beta_{(n)}(q)q^{2}}\nonumber \\
 &  & +S_{2}(k,p,q)\frac{\alpha_{(n)}(q)}{\beta_{(n)}(q)}\left(\frac{1}{\beta_{(n)}(p)p^{2}+\beta_{(n)}(q)q^{2}}-\frac{1}{\beta_{(n)}(p)p^{2}}\right)\nonumber \\
 &  & +S_{4}(k,p,q)\frac{\alpha_{(n)}(p)}{\beta_{(n)}(p)}\frac{1}{\beta_{(n)}(q)q^{2}}\}\\
\delta\beta_{(n)}(k) & = & \frac{1}{(d-1)k^{2}}\int^{\Delta}\frac{d{\textbf{p}}}{(2\pi)^{d}}S_{1}(k,p,q)\frac{C^{+}(q)}{\beta_{(n)}(p)p^{2}}\end{eqnarray}

Note that $\delta\alpha$, $\delta\beta$, $\delta\zeta$, $\delta\psi$,
and hence $\alpha,\beta,\zeta,\psi$, are all independent of $r$.
To solve the above equations we substitute the following one-dimensional
energy spectra in the above equations: \begin{eqnarray}
E^{+}(k) & = & K^{+}\frac{\left(\Pi^{+}\right)^{4/3}}{\left(\Pi^{-}\right)^{2/3}}k^{-5/3}\label{eq:Kolm-E+-}\\
E^{-}(k) & = & rE^{+}(k),\end{eqnarray}
 For the elements of $\hat{\nu}$ we substitute \begin{equation}
Z_{(n)}(k)=Z_{(n)}^{*}\sqrt{K^{+}}\frac{\left(\Pi^{+}\right)^{2/3}}{\left(\Pi^{-}\right)^{1/3}}k^{-4/3}\end{equation}
 where $Z$ stands for $\zeta,\alpha,\psi,\beta$. The renormalized
$Z^{*}$s are calculated using the procedure outlined in the previous
section. For large $n$ their values for $d=3$ are \begin{equation}
\hat{Z^{*}}=\left(\begin{array}{cc}
0.86r & 0.14\\
0.16r & 0.84\end{array}\right),\end{equation}
 and for $d=2$ they are \begin{equation}
\hat{Z^{*}}=\left(\begin{array}{cc}
0.95r & 0.54\\
1.10r & 0.54\end{array}\right)\end{equation}
 Note that the solution converges for both $d=2$ and $d=3$.

As discussed in the earlier section, the cascade rates $\Pi^{\pm}$
can be calculated from the renormalized parameters discussed above.
Using the energy equations we can easily derive the equations for
the cascade rates, which are \begin{eqnarray}
\Pi^{+} & = & \int_{0}^{k_{N}}2r\zeta k^{2}E^{+}(k)+\int_{0}^{k_{N}}2\alpha k^{2}(E^{u}(k)-E^{b}(k))\\
\Pi^{-} & = & \int_{0}^{k_{N}}2\beta k^{2}E^{-}(k)+\int_{0}^{k_{N}}2r\psi k^{2}(E^{u}(k)-E^{b}(k))\end{eqnarray}
 Under the assumption that $r_{A}=1$, the parts of $\Pi^{\pm}$ proportional
to $(E^{u}(k)-E^{b}(k))$ vanish. Hence, the total cascade rate will
be \begin{eqnarray}
\Pi & = & \frac{1}{2}(\Pi^{+}+\Pi^{-})\\
 & = & r\int_{0}^{k_{N}}(\zeta+\beta)k^{2}E^{+}(k)\end{eqnarray}
 Since $\zeta$ and $\beta$ are independent of $r$, the total cascade
rate is proportional to $r$ (for $r$ small). Clearly the cascade
rate $\Pi$ vanishes when $r=0$ or $\sigma_{c}=1$. This result is
consistent with the fact that the nonlinear interactions vanishes
for pure Alfv\'{e}n waves ($z^{+}$ or $z^{-}$). The detailed calculation
of the cascade rates $\Pi^{\pm}$ and the constants $K^{\pm}$ is
presented in Section \ref{sub:Nohelical-Alfvenic-MHD-flux}.

Now we will present the renormalization group analysis for helical
MHD.

\subsubsection{Helical nonAlfvénic MHD ($H_{M}\ne0;H_{K}\ne0;\sigma_{c}=0$):}

Helical MHD is defined for space dimension $d=3$. Verma \cite{MKV:MHD_Helical}
performed the RG analysis for helical MHD. His method moves along
the same lines as that applied for nonhelical MHD (Section \ref{sub:Nonhelical-nonAlfvenic-MHD}).
All the steps are the same except Eqs. (\ref{eq:nonhelical-uu},\ref{eq:nonhelical-bb})
are replaced by\begin{eqnarray}
\left\langle u_{i}^{>}(\hat{p})u_{j}^{>}(\hat{q})\right\rangle  & = & \left[P_{ij}({\textbf{p)}}C^{uu}(\hat{p})-i\epsilon_{ijl}\frac{p_{l}}{p^{2}}H_{K}(\hat{p})\right](2\pi)^{4}\delta(\hat{p}+\hat{q})\\
\left\langle b_{i}^{>}(\hat{p})b_{j}^{>}(\hat{q})\right\rangle  & = & \left[P_{ij}({\textbf{p)}}C^{bb}(\hat{p})-i\epsilon_{ijl}p_{l}H_{M}(\hat{p})\right](2\pi)^{4}\delta(\hat{p}+\hat{q})\end{eqnarray}
 Note that $u$-$b$ correlation has been taken to be zero in our
calculation. Because of helicities, the equations for change in renormalized
self-energy (\ref{eq:delta-nu}, \ref{eq:delta-eta}) get altered
to\begin{eqnarray*}
\delta\nu_{(0)}(k) & = & \frac{1}{(d-1)k^{2}}\int_{\textbf{p+q=k}}^{\Delta}\frac{d{\textbf{p}}}{(2\pi)^{d}}[\frac{S(k,p,q)C^{uu}(q)+S'(k,p,q)H_{K}(q)}{\nu_{(0)}(p)p^{2}+\nu_{(0)}(q)q^{2}}\\
 &  & \hspace{1.5in}-\frac{S_{6}(k,p,q)C^{bb}(q)+S'_{6}(k,p,q)H_{M}(q)}{\eta_{(0)}(p)p^{2}+\eta_{(0)}(q)q^{2}}],\\
\delta\eta_{(0)}(k) & = & \frac{1}{(d-1)k^{2}}\int_{\textbf{p+q=k}}^{\Delta}\frac{d{\textbf{p}}}{(2\pi)^{d}}[-\frac{S_{8}(k,p,q)C^{bb}(q)+S'_{8}(k,p,q)H_{M}(q)}{\nu_{(0)}(p)p^{2}+\eta_{(0)}(q)q^{2}}\\
 &  & \hspace{1.5in}+\frac{S_{9}(k,p,q)C^{uu}(q)+S'_{9}(k,p,q)H_{K}(q)}{\eta_{(0)}(p)p^{2}+\nu_{(0)}(q)q^{2}}],\end{eqnarray*}
where $S'_{i}$ defined below can be shown to be zero.\begin{eqnarray*}
S'(k,p,q) & = & P_{bjm}^{+}(k)P_{mab}^{+}(p)\epsilon_{jal}q_{l}=0,\\
S'_{6}(k,p,q) & = & P_{ajm}^{+}(k)P_{mba}^{-}(p)\epsilon_{jal}q_{l}=0,\\
S'_{8}(k,p,q) & = & P_{ijm}^{-}(k)P_{jab}^{+}(p)\epsilon_{mal}q_{l}P_{ib}(k)=0,\\
S'_{9}(k,p,q) & = & P_{ijm}^{-}(k)P_{mab}^{-}(p)\epsilon_{jal}q_{l}P_{ib}(k)=0.\end{eqnarray*}
The argument for vanishing of $S'$ is follows. Since $\delta\nu$
and $\delta\eta$ are proper scalars and $H_{M,K}$ are pseudo scalars,
$S'_{i}(k,p,q)$ will be pseudo scalars. In addition, $S'_{i}(k,p,q)$
are also linear in $k,p$ and $q$. This implies that $S'_{i}(k,p,q)$
must be proportional to $\mathbf{q}\cdot(\mathbf{k}\times\mathbf{p})$,
which will be zero because $\mathbf{k}=\mathbf{p}+\mathbf{q}$. Hence
all $S'_{i}(k,p,q)$ turn out to be zero. \emph{Hence, helicities
do not alter the already calculated $\delta(\nu,\eta)_{(n)}(k)$ in
Section \ref{sub:Nonhelical-nonAlfvenic-MHD}}. Zhou \cite{Zhou:HK}
arrived at a similar conclusion while calculating the renormalized
viscosity for helical fluid turbulence.

\subsection{RG Calculations of MHD Turbulence using YO's Perturbative Scheme
\label{sub:MHD-RG-YO}}

In YO's perturbative scheme for fluid turbulence, corrections to viscosity,
vertex, and noise are computed on shell elimination. After that recurrence
relations are written for these quantities, and fixed points are sought.
The nontrivial fixed point provides us with spectral exponents etc.

In MHD turbulence there are more variables than fluid turbulence.
If cross helicity is zero, we can manage with corrections to (1) viscosity
and resistivity, (2) three vertices corresponding to $(\mathbf{u}\cdot\nabla)\mathbf{u},$
$(\mathbf{b}\cdot\nabla)\mathbf{b}$, and $\mathbf{(u}\cdot\nabla)\mathbf{b}-(\mathbf{b}\cdot\nabla)\mathbf{u}$,
and (3) two noise parameters corresponding to the velocity and magnetic
fields respectively \cite{FourSule}. In terms of Elsässer variables,
we get similar terms. These calculations have been performed by Lee
\cite{Lee}, Fournier et al. \cite{FourSule}, Camargo and Tasso \cite{Cama},
Liang and Diamond \cite{Lian}, Berera and Hochberg \cite{Bere},
Longcope and Sudan \cite{Long}, and Basu \cite{Basu:Fieldth}. Note
that in 1965 itself Lee \cite{Lee} had written all the Feynman diagrams
for dressed Green's functions, noise, and vertex, but could not compute
the dressed Green's function or correlation function.

A brief comments on all the above work are as follows. In almost all
the following work, cross helicity is taken to be zero.

\subsubsection{Fournier, Sulem, and Pouquet}

Fournier et al. \cite{FourSule} were the first to perform RG calculation
for MHD turbulence in 1982. Different regimes were obtained depending
on space dimension, external driving (noise), and fluid characteristics
like Prandtl number. The trivial and kinetic regimes exist in any
space dimension. Here, the dissipative coefficients, viscosity and
resistivity, are renormalized, and they have the same scaling. Turbulent
magnetic Prandtl number depends on space dimension only and tends
to 1 when $d\rightarrow\infty$.

The magnetic regime is found only for $d>d_{c}\approx2.8$. The effect
of the small-scales kinetic energy on the large scales is negligible,
and the renormalization of the coupling is only due to the small scales
magnetic energy. The turbulent magnetic Prandtl number is infinite
for $d_{c}<d<d_{c}'\approx4.7$, while for $d>d_{c}'$, it has a finite
value which tends to 1 as $d\rightarrow\infty$. 

No magnetic regime can be computed by the RG for $d<d_{c}$. Also,
in $d<3$, the contribution of the magnetic small scales to the turbulent
diffusivity is negative and tends to destabilize the magnetic large
scales. In $d=2$ or close to 2, the electromagnetic force produces
unbounded nonlinear effects on large scales, making RG inapplicable.

\subsubsection{Camargo and Tasso}

Camargo and Tasso \cite{Cama} performed RG analysis using $\mathbf{z}^{\pm}$
variables. They derived flow equations for the Prandtl number. They
showed that effective resistivity could be negative, but effective
viscosity is always positive.

\subsubsection{Liang and Diamond}

Liang and Diamond \cite{Lian} applied RG for 2D fluid and MHD turbulence.
They found that no RG fixed point exists for both these systems. They
attributed this phenomena to dual-cascade.

\subsubsection{Berera and Hochberg \cite{Bere}}

Berera and Hochberg's \cite{Bere} RG analysis showed that Kolmogorov-like
5/3, KID's 3/2, or any other energy spectra can be obtained by a suitable
choice of the spectrum of the injected noise. They also report forward
cascade for both energy and magnetic helicity.

\subsubsection{Longcope and Sudan}

Longcopee and Sudan \cite{Long} applied RG analysis to Reduced Magnetohydrodynamics
(RMHD) and obtained effective values of the viscosity and resistivity
at large-scales.

\subsection{Callan-Symanzik Equation for MHD Turbulence}

This scheme is equivalent to Wilson's RG scheme of shell elimination.
For details of this scheme, refer to the book by Adzhemyan \cite{Adzh:book}.
Hnatich, Honkonen, and Jurcisin \cite{Hnat} performed RG analysis
based on two parameters, space dimension and noise spectral index.
They showed that the kinetic fixed point is stable for $d\ge2$, but
the magnetic fixed point is stable only for $d>d_{c}\approx2.46$.
Adzhemyan et al. \cite{Adzh:MHD} applied quantum-field approach to
MHD turbulence and performed a detailed RG analysis.

\subsection{Other Analytic Techniques in MHD Turbulence}

Direct Interaction Approximation \cite{Krai:59} is very popular in
fluid turbulence. In fact, some researchers (e.g., Kraichnan \cite{Krai:RG})
argue in favour of DIA over RG. One problem of DIA is that the integral
for Green's function diverges (infrared divergence), and one needs
to introduce a infrared cutoff \cite{Lesl:book}. In any case, there
are only a few DIA calculations for MHD turbulence. Verma and Bhattacharjee
\cite{MKVJKB} applied DIA to compute Kolmogorov's constant in MHD
turbulence assuming 5/3 energy spectrum. Note however that their self-energy
matrix is not quite correct, and should be replaced by that given
in Section \ref{sub:Nonhelical-Alfvenic-MHD}. Nakayama \cite{Naka}
performed one such calculation based on KID's scaling for Green's
function and correlation functions. 

There are some interesting work by Montgomery and Hatori \cite{MontHato},
and Montgomery and Chen \cite{MontChen1,MontChen2} using scale separation.
They computed the effects of small scales on the large-scale magnetic
field, and found that helicity could enhance the magnetic field. They
have also computed the corrections to the transport parameters due
to small-scale fields. Note that RG schemes are superior to these
schemes because they include all the interaction terms. For details,
the reader is referred to the original papers.

Now let us compare the various results discussed above. One common
conclusion is that the magnetic (dominated) fixed point near $d=2$
is unstable, however, authors report different critical dimension
$d_{c}$. Both Fournier et al. \cite{FourSule} and Verma find that
magnetic $Pr=1$ as $d\rightarrow\infty$. For 2D fluid turbulence,
Liang and Diamond's \cite{Lian} argued that RG fixed point is unstable.
This result is in disagreement with our self-consistent RG (see Appendix
\ref{sec:Digression-to-Fluid}). To sum up, RG calculations for MHD
turbulence appears to be quite involved, and there are many unresolved
issues.

In fluid turbulence, there are some other interesting variations of
field-theoretic calculations by DeDominicis and Martin \cite{DeDo},
Bhattacharjee \cite{JKB:mode}, Carati \cite{Cara:RG} and others.
In MHD turbulence, however, these types of calculations are less.

In the next section we will compute energy fluxes for MHD turbulence
using field-theoretic techniques.

\section{Field-theoretic Calculation of Energy Fluxes and Shell-to-shell Energy
Transfer \label{sec:analytic-energy}}

In this section we present calculation of energy fluxes in MHD turbulence.
The computation was carried out by Verma \cite{MKV:MHD_PRE,MKV:MHD_Flux,MKV:MHD_Helical}
for the \emph{inertial-range wavenumbers} using perturbative self-consistent
field-theoretic technique. He assumed the turbulence to be homogeneous
and isotropic. Even though the real-world turbulence do not satisfy
these properties, many conclusions drawn using these assumption provide
us with important insights into the energy transfer mechanisms at
small scales. Verma assumed that the mean magnetic field is absent;
this assumption was made to ensure that the turbulence is isotropic.
The field-theoretic procedure requires Fourier space integrations
of functions involving products of energy spectrum and the Greens
functions. Since there is a general agreement on Kolmogorov-like spectrum
for MHD turbulence, Verma took $E(k)\propto k^{-5/3}$ for all the
energy spectra. For the Greens function, he substituted the {}``renormalized''
or {}``dressed'' Greens function computed by Verma \cite{MKV:MHD_RG}
(see Section \ref{sub:Renormalization-of-viscosity}).

\subsection{Field-theoretic Calculation of Energy Fluxes}

The field-theoretic calculation for arbitrary cross helicity, Alfvén
ratio, magnetic helicity, and kinetic helicities is quite intractable,
therefore Verma performed the calculation in the following three limiting
cases: (1) Nonhelical nonAlfvénic MHD ($H_{M}=H_{K}=H_{c}=0$), (2)
Nonhelical Alfvénic MHD ($H_{M}=H_{K}=0,\sigma_{c}\rightarrow1$),
and (3) Helical nonAlfvénic MHD ($H_{M}\ne0,H_{K}\ne0,H_{c}=0$).
Energy flux calculation for each of these cases is described below.

\subsubsection{Nonhelical nonAlfvénic MHD ($H_{M}=H_{K}=H_{c}=0$) \label{sub:Nonhelical-nonAlfvenic-MHD-flux}}

As described in Section \ref{sub:Energy-Cascade-Rates-formula-MHD}
the energy flux from inside of the $X$-sphere of radius $k_{0}$
to outside of the $Y$-sphere of the same radius is \begin{eqnarray}
\Pi_{Y>}^{X<}(k_{0}) & = & \frac{1}{(2\pi)^{d}\delta(\mathbf{k'+p+q})}\int_{k'>k_{0}}\frac{d\mathbf{k'}}{(2\pi)^{d}}\int_{p<k_{0}}\frac{d\mathbf{p}}{(2\pi)^{d}}\left\langle S(\mathbf{k'|p|q})\right\rangle \label{eq:flux-MHD}\end{eqnarray}
 where $X$ and $Y$ stand for $u$ or $b$. Verma assumed that the
kinetic energy is forced at small wavenumbers.

Verma \cite{MKV:MHD_Flux} analytically calculated the above energy
fluxes in the inertial range to leading order in perturbation series.
It was assumed that $\mathbf{u}(\mathbf{k})$ is quasi-gaussian as
in EDQNM approximation. Under this approximation, the triple correlation
$\left\langle XXX\right\rangle $ is zero to zeroth order, but nonzero
to first oder. To first order $\left\langle XXX\right\rangle $ is
written in terms of $\left\langle XXXX\right\rangle $, which is replaced
by its Gaussian value, a sum of products of second-order moment. Consequently,
the ensemble average of $S^{YX}$, $\left\langle S^{YX}\right\rangle $,
is zero to the zeroth order, but is nonzero to the first order. The
first order terms for $\left\langle S^{YX}(k|p|q)\right\rangle $
in terms of Feynman diagrams are given in Appendix C. They are given
below in terms of Green's functions and correlation functions:

\begin{eqnarray}
\left\langle S^{uu}(k|p|q)\right\rangle  & = & \int_{-\infty}^{t}dt'(2\pi)^{d}[T_{1}(k,p,q)G^{uu}(k,t-t')C^{uu}(p,t,t')C^{uu}(q,t,t')\nonumber \\
 &  & \hspace{1cm}+T_{5}(k,p,q)G^{uu}(p,t-t')C^{uu}(k,t,t')C^{uu}(q,t,t')\nonumber \\
 &  & \hspace{1cm}+T_{9}(k,p,q)G^{uu}(q,t-t')C^{uu}(k,t,t')C^{uu}(p,t,t')]\delta(\mathbf{k'}+\mathbf{p}+\mathbf{q})\label{eqn:Suu-nonhelical}\\
\left\langle S^{ub}(k|p|q)\right\rangle  & = & -\int_{-\infty}^{t}dt'(2\pi)^{d}[T_{2}(k,p,q)G^{uu}(k,t-t')C^{bb}(p,t,t')C^{bb}(q,t,t')\nonumber \\
 &  & \hspace{1cm}+T_{7}(k,p,q)G^{bb}(p,t-t')C^{uu}(k,t,t')C^{bb}(q,t,t')\nonumber \\
 &  & \hspace{1cm}+T_{11}(k,p,q)G^{uu}(q,t-t')C^{uu}(k,t,t')C^{bb}(p,t,t')]\delta(\mathbf{k'}+\mathbf{p}+\mathbf{q})\\
\left\langle S^{bu}(k|p|q)\right\rangle  & = & -\int_{-\infty}^{t}dt'(2\pi)^{d}[T_{3}(k,p,q)G^{bb}(k,t-t')C^{uu}(p,t,t')C^{bb}(q,t,t')\nonumber \\
 &  & \hspace{1cm}+T_{6}(k,p,q)G^{uu}(p,t-t')C^{bb}(k,t,t')C^{bb}(q,t,t')\nonumber \\
 &  & \hspace{1cm}+T_{12}(k,p,q)G^{bb}(q,t-t')C^{bb}(k,t,t')C^{uu}(p,t,t')]\delta(\mathbf{k'}+\mathbf{p}+\mathbf{q})\\
\left\langle S^{bb}(k|p|q)\right\rangle  & = & \int_{-\infty}^{t}dt'(2\pi)^{d}[T_{4}(k,p,q)G^{bb}(k,t-t')C^{bb}(p,t,t')C^{uu}(q,t,t')\nonumber \\
 &  & \hspace{1cm}+T_{8}(k,p,q)G^{bb}(p,t-t')C^{bb}(k,t,t')C^{uu}(q,t,t')\nonumber \\
 &  & \hspace{1cm}+T_{10}(k,p,q)G^{uu}(q,t-t')C^{bb}(k,t,t')C^{bb}(p,t,t')]\delta(\mathbf{k'}+\mathbf{p}+\mathbf{q})\label{eq:Sbb-nonhelical}\end{eqnarray}
 where $T_{i}(k,p,q)$ are functions of wavevectors $k,p$, and $q$
given in Appendix C.

The Greens functions can be written in terms of {}``effective''
or {}``renormalized'' viscosity $\nu(k)$ and resistivity $\eta(k)$
computed in Section \ref{sub:Nonhelical-nonAlfvenic-MHD} \[
G^{uu,bb}(k,t-t')=\theta{(t-t')}\exp\left(-[\nu(k),\eta(k)]k^{2}(t-t')\right)\]
 The relaxation time for $C^{uu,}(k,t,t')$ $[C^{bb}(k,t,t')]$ is
assumed to be the same as that of $G^{uu}(k,t,t')$ {[}$G^{bb}(k,t,t')${]}.
Therefore the time dependence of the unequal-time correlation functions
will be \[
C^{uu,bb}(k,t,t')=\theta{(t-t')}\exp\left(-[\nu(k),\eta(k)]k^{2}(t-t')\right)C^{uu,bb}(k,t,t)\]
 The above forms of Green's and correlation functions are substituted
in the expression of $\left\langle S^{YX}\right\rangle $, and the
$t'$ integral is performed. Now Eq. (\ref{eq:flux-MHD}) yields the
following flux formula for $\Pi_{u>}^{u<}(k_{0})$: \begin{eqnarray}
\Pi_{u>}^{u<}(k_{0}) & = & \int_{k>k_{0}}\frac{d{\textbf{k}}}{(2\pi)^{d}}\int_{p<k_{0}}\frac{d{\textbf{p}}}{(2\pi)^{d}}\frac{1}{\nu(k)k^{2}+\nu(p)p^{2}+\nu(q)q^{2}}\times[T_{1}(k,p,q)C^{uu}(p)C^{uu}(q)\nonumber \\
 &  & \hspace{3cm}+T_{5}(k,p,q)C^{uu}(k)C^{uu}(q)+T_{9}(k,p,q)C^{uu}(k)C^{uu}(p)].\label{eqn:Pi_mhd}\end{eqnarray}
 The expressions for the other fluxes can be obtained similarly.

The equal-time correlation functions $C^{uu}(k,t,t)$ and $C^{bb}(k,t,t)$
at the steady-state can be written in terms of one dimensional energy
spectrum as \[
C^{uu,bb}(k,t,t)=\frac{2(2\pi)^{d}}{S_{d}(d-1)}k^{-(d-1)}E^{u,b}(k),\]
 where $S_{d}$ is the surface area of $d$-dimensional unit spheres.
We are interested in the fluxes in the inertial range. Therefore,
Verma substituted Kolmogorov's spectrum {[}Eqs.(\ref{eq:Euk},\ref{eq:Ebk}){]}
for the energy spectrum. The effective viscosity and resistivity are
proportional to $k^{-4/3}$, i.e., \[
[\nu,\eta](k)=(K^{u})^{1/2}\Pi^{1/3}k^{-4/3}[\nu^{*},\eta^{*}],\]
 and the parameters $\nu^{*}$ and $\eta^{*}$ were calculated in
Section \ref{sub:Nonhelical-nonAlfvenic-MHD}.

Verma nondimensionalized Eq.~(\ref{eqn:Pi_mhd}) by substituting
\cite{Lesl:book}\begin{equation}
k=\frac{k_{0}}{u};\,\,\,\,\, p=\frac{k_{0}}{u}v;\,\,\,\,\, q=\frac{k_{0}}{u}w.\end{equation}
 Application of Eq. (\ref{eq:volume-integral}) yields\begin{equation}
\Pi_{Y>}^{X<}=(K^{u})^{3/2}\Pi\left[\frac{4S_{d-1}}{(d-1)^{2}S_{d}}\int_{0}^{1}dv\ln{(1/v)}\int_{1-v}^{1+v}dw(vw)^{d-2}(\sin\alpha)^{d-3}F_{Y>}^{X<}(v,w)\right],\end{equation}
 where the integrals $F_{Y>}^{X<}(v,w)$ are \begin{eqnarray}
F_{u>}^{u<} & = & \frac{1}{\nu^{*}(1+v^{2/3}+w^{2/3})}[t_{1}(v,w)(vw)^{-d-\frac{2}{3}}+t_{5}(v,w)w^{-d-\frac{2}{3}}+t_{9}(v,w)v^{-d-\frac{2}{3}}],\\
F_{u>}^{b<} & = & -\frac{1}{\nu^{*}+\eta^{*}(v^{2/3}+w^{2/3})}[t_{2}(v,w)(vw)^{-d-\frac{2}{3}}r_{A}^{-2},\nonumber \\
 &  & \hspace{4cm}+t_{7}(v,w)w^{-d-\frac{2}{3}}r_{A}^{-1}+t_{11}(v,w)v^{-d-\frac{2}{3}}r_{A}^{-1}],\\
F_{b>}^{u<} & =- & \frac{1}{\nu^{*}v^{2/3}+\eta^{*}(1+w^{2/3})}[t_{3}(v,w)(vw)^{-d-\frac{2}{3}}r_{A}^{-1},\nonumber \\
 &  & \hspace{4cm}+t_{6}(v,w)w^{-d-\frac{2}{3}}r_{A}^{-2}+t_{12}(v,w)v^{-d-\frac{2}{3}}r_{A}^{-1}]\\
F_{b>}^{b<} & = & \frac{1}{\nu^{*}w^{2/3}+\eta^{*}(1+v^{2/3})}[t_{4}(v,w)(vw)^{-d-\frac{2}{3}}r_{A}^{-1}\nonumber \\
 &  & \hspace{4cm}+t_{8}(v,w)w^{-d-\frac{2}{3}}r_{A}^{-1}+t_{10}(v,w)v^{-d-\frac{2}{3}}r_{A}^{-2}].\end{eqnarray}
 Here $t_{i}(v,w)=T_{i}(k,kv,kw)/k^{2}$. Note that the energy fluxes
are constant, consistent with the Kolmogorov's picture. Verma computed
the bracketed terms (denoted by $I_{Y>}^{X<}$) numerically using
Gaussian-quadrature method, and found all of them to be convergent.
Let us denote $I=I_{u>}^{u<}+I_{u>}^{b<}+I_{b>}^{u<}+I_{b>}^{b<}$.
Using the fact that the total flux $\Pi$ is \begin{equation}
\Pi=\Pi_{u>}^{u<}+\Pi_{u>}^{b<}+\Pi_{b>}^{u<}+\Pi_{b>}^{b<},\end{equation}
 constant $K^{u}$ can be calculated as \begin{equation}
K^{u}=(I)^{-2/3}.\end{equation}
 The energy flux ratios can be computed using $\Pi_{Y>}^{X<}/\Pi=I_{Y>}^{X<}/I$.
The value of constant $K$ can be computed using Eq. (\ref{eq:Kplus}).
The flux ratios and Kolmogorov's constant for $d=3$ and various $r_{A}$
are listed in Table \ref{table:flux_ra}. %
\begin{table}

\caption{\label{table:flux_ra}The computed values of energy fluxes in MHD
turbulence for various $r_{A}$ when $d=3$ and $\sigma_{c}=0$.}

\begin{tabular}{|c|c|c|c|c|c|c|c|}
\hline 
$\Pi/r_{A}$&
5000&
100&
5&
1&
0.5&
0.3&
Trend\tabularnewline
\hline
\hline 
$\Pi_{u>}^{u<}/\Pi$&
1&
0.97&
0.60&
0.12&
0.037&
0.011&
Decreases\tabularnewline
\hline 
$\Pi_{b>}^{u<}/\Pi$&
$3.4\times10^{-4}$&
$1.7\times10^{-2}$&
0.25&
0.40&
0.33&
0.36&
Increases then saturates\tabularnewline
\hline 
$\Pi_{u>}^{b<}/\Pi$&
$-1.1\times10^{-4}$&
$-5.1\times10^{-3}$&
$-0.05$&
0.12&
0.33&
0.42&
Increases then saturates\tabularnewline
\hline 
$\Pi_{b>}^{b<}/\Pi$&
$2.7\times10^{-4}$&
$1.3\times10^{-2}$&
0.20&
0.35&
0.30&
0.21&
Increases then dips\tabularnewline
\hline 
$K^{+}$&
1.53&
1.51&
1.55&
1.50&
1.65&
3.26&
Approx. const till $r_{A}\approx0.5$\tabularnewline
\hline 
$K^{u}$&
1.53&
1.50&
1.29&
0.75&
0.55&
0.75&
Decrease\tabularnewline
\hline
\end{tabular}
\end{table}

The following trends can be inferred by studying Table \ref{table:flux_ra}.
We find that for $d=3$, $\Pi_{u>}^{u<}/\Pi$ starts from 1 for large
$r_{A}$ and decreases nearly to zero near $r_{A}=0.3$. All other
fluxes start from zero or small negative values, and increase up to
some saturated values. Near $r_{A}\approx1$, all the energy fluxes
become significant. All the fluxes are positive except for $\Pi_{u>}^{b<}$,
which is negative for $r_{A}$ greater than 3 (kinetic energy dominated
state). Hence, \emph{}both kinetic and magnetic energies flow from
large length scale to small length scale. However, in kinetic energy
dominated situations, there is an energy transfer from small-scale
kinetic energy to large-scale magnetic field (Inverse energy cascade)\emph{.}
Negative $\Pi_{u>}^{b<}$ for $r_{A}>3$ is consistent with negative
value of $\eta^{bu*}$ computed in Sec. \ref{sub:Nonhelical-nonAlfvenic-MHD}. 

The quantity of special interest is $\Pi_{b>}^{b<}$, which is positive
indicating that magnetic energy cascades from large length-scale to
small length-scale. The Kolmogorov constant $K$ for $d=3$ is listed
in Table \ref{table:flux_ra}. For all $r_{A}$ greater than 0.5,
$K$ is approximately constant and is close to 1.6, same as that for
fluid turbulence ($r_{A}=\infty$). 

Verma's method mentioned above cannot be used to compute energy transfer
$\Pi_{b<}^{u<}$ from the large-scale kinetic energy to the large-scale
magnetic energy because the forcing wavenumbers (large scales) do
not obey Kolmogorov's powerlaw. Verma \cite{MKV:MHD_Flux} attempted
to compute these using steady-state assumption \[
\Pi_{b<}^{u<}=\Pi_{b>}^{b<}+\Pi_{u>}^{b<}.\]
Unfortunately, shell-to-shell energy transfer studies (Section \ref{sub:Field-theoretic-Calculation-of-shell})
reveal that in kinetic regime ($r_{A}>1$), $u$-shells lose energy
to $b$-shells; while in magnetically dominated situations ($r_{A}<1$),
$b$-shells lose energy to $u$-shells. Hence, steady-state MHD is
not possible except near $r_{A}=1$. Therefore, Verma's prediction
of $\Pi_{b<}^{u<}$ based on steady-state assumption is incorrect.
However, $\Pi_{b>}^{b<}+\Pi_{u>}^{b<}$ can still be used as an estimate
for $\Pi_{b<}^{u<}$.

We compute total kinetic energy flux $\Pi^{u}=\Pi_{u>}^{u<}+\Pi_{b>}^{u<}$,
and total magnetic energy flux $\Pi^{b}=\Pi_{b>}^{b<}+\Pi_{u>}^{b<}$.
We find that $E^{u}\propto(\Pi^{u})^{2/3}$ and $E^{b}\propto(\Pi^{b})^{2/3}$
for all $r_{A}$. Hence,\begin{equation}
E^{u,b}=K^{1,2}\left(\Pi^{u,b}\right)^{2/3}k^{-5/3},\label{eq:EubPiub}\end{equation}
where the constants $K^{1,2}$ are somewhat independent of $r_{A}$
unlike $K^{u}$. 

Now let us compare the theoretical values with their numerical counterparts
reported in Section \ref{sec:Numerical-Investigation-MHD}. Haugen
et al. \cite{Bran:Flux} reported that for $E^{b}/E^{u}\approx0.5$,
$\Pi^{u}\approx0.3$ and $\Pi^{b}\approx0.7$. These numbers match
with our theoretical values very well. However, the preliminary numerical
results of Olivier et al. \cite{Oliv:Simulation} agree with the theoretical
values only qualitatively. Further numerical simulations are required
to test the above theoretical predictions.

The values of energy flux-ratios and Kolmogorov's constant for various
space dimensions (when $r_{A}=1$) are listed in Table \ref{table:flux_d}.
\begin{table}

\caption{\label{table:flux_d}The computed values of energy fluxes in MHD
turbulence for various space dimensions $d$ when $r_{A}=1$ and $\sigma_{c}=0$.}

\begin{tabular}{|c|c|c|c|c|c|c|c|}
\hline 
$\Pi/d$&
2.1&
2.2&
2.5&
3&
4&
10&
100\tabularnewline
\hline
\hline 
$\Pi_{u>}^{u<}/\Pi$&
-&
0.02&
0.068&
0.12&
0.17&
0.23&
0.25\tabularnewline
\hline 
$\Pi_{b>}^{u<}/\Pi$&
-&
0.61&
0.49&
0.40&
0.34&
0.27&
0.25\tabularnewline
\hline 
$\Pi_{u>}^{b<}/\Pi$&
-&
-0.027
&
0.048&
0.12&
0.18&
0.23&
0.25\tabularnewline
\hline 
$\Pi_{b>}^{b<}/\Pi$&
-&
0.40&
0.39&
0.35&
0.31&
0.27&
0.25\tabularnewline
\hline 
$K^{+}$&
-&
1.4&
1.4&
1.5&
1.57&
1.90&
3.46\tabularnewline
\hline 
$K^{u}$&
-&
0.69&
0.72&
0.75&
0.79&
0.95&
1.73\tabularnewline
\hline
\end{tabular}
\end{table}
In Section \ref{sub:Nonhelical-nonAlfvenic-MHD} it has been shown
that for $d<2.2$, the RG fixed point is unstable, and the renormalized
parameters could not be determined. Due to that reason we have calculated
fluxes and Kolmogorov's constant for $d\geq2.2$ only. For large $d$
it is observed that all the flux ratios are equal, and Kolmogorov's
constant is proportional to $d^{1/3}$. It can be explained by observing
that for large $d$\begin{eqnarray}
\int dp'dq'\left(\frac{p'q'}{k'}\right)^{d-2}\left(\sin\alpha\right)^{d-3}... & \sim & d^{-1/2},\\
\frac{S_{d-1}}{\left(d-1\right)^{2}S_{d}} & \sim & \frac{1}{d^{2}}\left(\frac{d}{2\pi}\right)^{1/2},\nonumber \\
\nu^{*}=\eta^{*} & \sim & d^{-1/2}\\
t_{1}=-t_{3}=-t_{5}=t_{7} & = & kpd(z+xy),\label{eq:t-large-d}\end{eqnarray}
and $t_{9}=t_{11}=0$. Using Eq. (\ref{eq:t-large-d}) it can be shown
that all $F_{Y>}^{X<}$ are equal for large $d$, which implies that
all the flux ratios will be equal. By matching the dimensions, it
can be shown that $K\propto d^{-1/3}$. This result is a generalization
of theoretical analysis of Fournier \emph{et al.} \cite{FourFris}
for fluid turbulence.

In this subsection we calculated the cascade rates for $\sigma_{c}=0$.
In the next subsection we take the other limit $\sigma_{c}\rightarrow1$.

\subsubsection{Nonhelical Alfvénic MHD ($H_{M}=H_{K}=0,\sigma_{c}\rightarrow1$)\label{sub:Nohelical-Alfvenic-MHD-flux}}

In $z^{\pm}$ variables, there are only two types of fluxes $\Pi^{\pm}$,
one for the $z^{+}$ cascade and the other for $z^{-}$ cascade. These
energy fluxes $\Pi^{\pm}$ can be computed using \begin{eqnarray}
\Pi^{\pm}(k_{0}) & = & \frac{1}{(2\pi)^{d}\delta(\mathbf{k'+p+q})}\int_{k'>k_{0}}\frac{d\mathbf{k'}}{(2\pi)^{d}}\int_{p<k_{0}}\frac{d\mathbf{p}}{(2\pi)^{d}}\left\langle S^{\pm}(\mathbf{k'|p|q})\right\rangle .\label{eqn:flux_zpm}\end{eqnarray}
 Verma \cite{MKV:MHD_Flux} calculated the above fluxes to the leading
order in perturbation series. The Feynman diagrams are given in Appendix
B. To first order, $\left\langle S^{\pm}(\mathbf{k'|p|q})\right\rangle $
are \begin{eqnarray}
\left\langle S^{\pm}(\mathbf{k'|p|q})\right\rangle  & = & \int_{-\infty}^{t}dt'(2\pi)^{d}[T_{13}(k,p,q)G^{\pm\pm}(k,t-t')C^{\pm\mp}(p,t,t')C^{\mp\pm}(q,t,t')\nonumber \\
 &  & +T_{14}(k,p,q)G^{\pm\pm}(k,t-t')C^{\pm\pm}(p,t,t')C^{\mp\mp}(q,t,t')\nonumber \\
 &  & +T_{15}(k,p,q)G^{\pm\mp}(k,t-t')C^{\pm\pm}(p,t,t')C^{\mp\mp}(q,t,t')\nonumber \\
 &  & +T_{16}(k,p,q)G^{\pm\mp}(k,t-t')C^{\pm\mp}(p,t,t')C^{\mp\pm}(q,t,t')\nonumber \\
 &  & +T_{17}(k,p,q)G^{\pm\pm}(p,t-t')C^{\pm\mp}(k,t,t')C^{\mp\pm}(q,t,t')\nonumber \\
 &  & +T_{18}(k,p,q)G^{\pm\pm}(p,t-t')C^{\pm\pm}(k,t,t')C^{\mp\mp}(q,t,t')\nonumber \\
 &  & +T_{19}(k,p,q)G^{\pm\mp}(p,t-t')C^{\pm\pm}(k,t,t')C^{\mp\mp}(q,t,t')\nonumber \\
 &  & +T_{20}(k,p,q)G^{\pm\mp}(p,t-t')C^{\pm\pm}(k,t,t')C^{\mp\mp}(q,t,t')\nonumber \\
 &  & +T_{21}(k,p,q)G^{\mp\pm}(q,t-t')C^{\pm\mp}(k,t,t')C^{\pm\pm}(p,t,t')\nonumber \\
 &  & +T_{22}(k,p,q)G^{\mp\pm}(q,t-t')C^{\pm\pm}(k,t,t')C^{\pm\mp}(p,t,t')\nonumber \\
 &  & +T_{23}(k,p,q)G^{\mp\mp}(q,t-t')C^{\pm\pm}(k,t,t')C^{\pm\mp}(p,t,t')\nonumber \\
 &  & +T_{24}(k,p,q)G^{\mp\mp}(q,t-t')C^{\pm\mp}(k,t,t')C^{\pm\pm}(p,t,t')]\delta(\mathbf{k'}+\mathbf{p}+\mathbf{q}),\label{eq:S-Alfvenic}\end{eqnarray}
 where $T_{i}(k,p,q)$ are given in Appendix A.

Verma considered the case when $r=E^{-}(k)/E^{+}(k)$ is small. In
terms of renormalized $\hat{\nu}$ matrix, Green's function and correlation
functions calculated in Section \ref{sub:Nonhelical-Alfvenic-MHD},
we obtain the following expression for $\Pi^{\pm}$ in leading order
in $r$: 

\vspace{0.3cm}\begin{equation}
\Pi^{\pm}=r\frac{(\Pi^{+})^{2}}{\Pi^{-}}(K^{+})^{3/2}\left[\frac{4S_{d-1}}{(d-1)^{2}S_{d}}\int_{0}^{1}dv\ln{(1/v)}\int_{1-v}^{1+v}dw(vw)^{d-2}(\sin\alpha)^{d-3}F^{\pm}(v,w)\right],\label{eqn:Pipm}\end{equation}
 \vspace{0.3cm} where the integrand $F^{\pm}$ are \begin{eqnarray}
F^{+} & = & t_{13}(v,w)(vw)^{-d-2/3}\frac{1}{\beta^{*}w^{2/3}}+t_{14}(v,w)(vw)^{-d-2/3}\frac{\alpha^{*}}{\beta^{*}}\left\{ \frac{1}{\beta^{*}(1+w^{2/3})}-\frac{1}{\beta^{*}w^{2/3}}\right\} \nonumber \\
 &  & +t_{15}(v,w)w^{-d-2/3}\frac{1}{\beta^{*}w^{2/3}}+t_{16}(v,w)w^{-d-2/3}\frac{\alpha^{*}}{\beta^{*}}\left\{ \frac{1}{\beta^{*}(v^{2/3}+w^{2/3})}-\frac{1}{\beta^{*}w^{2/3}}\right\} \nonumber \\
 &  & +t_{17}(v,w)v^{-d-2/3}\frac{\alpha^{*}}{\beta^{*}}\left\{ \frac{1}{\beta^{*}(v^{2/3}+w^{2/3})}-\frac{1}{\beta^{*}w^{2/3}}\right\} \nonumber \\
 &  & +t_{18}(v,w)v^{-d-2/3}\frac{\alpha^{*}}{\beta^{*}}\left\{ \frac{1}{\beta^{*}(1+w^{2/3})}-\frac{1}{\beta^{*}w^{2/3}}\right\} ,\\
F^{-} & = & t_{13}(v,w)(vw)^{-d-2/3}\frac{1}{\beta^{*}(1+v^{2/3})}+t_{15}(v,w)w^{-d-2/3}\frac{1}{\beta^{*}(1+v^{2/3})},\end{eqnarray}
 where $t_{i}(v,w)=T_{i}(k,kv,kw)/k^{2}$. Here we assumed that $r_{A}=1$.
We find that some of the terms of Eq.~(\ref{eq:S-Alfvenic}) are
of higher order, and they have been neglected.

The bracketed term of Eq.~(\ref{eqn:Pipm}), denoted by $I^{\pm}$,
are computed numerically. The integrals are finite for $d=2$ and
3. Also note that $I^{\pm}$ are independent of $r$. Now the constant
$K^{\pm}$ of Eq.~(\ref{eqn:Pipm}) is computed in terms of $I^{\pm}$;
they are listed in Table \ref{table:Kpm}. The constants $K^{\pm}$
depend very sensitively on $r$.%
\begin{table}

\caption{\label{table:Kpm}The computed values of Kolmogorov's constants for
$\sigma_{c}\rightarrow1$ and $r_{A}=1$ limit for various $r=E^{-}/E^{+}$
$(d=2,3)$.}

\begin{tabular}{|c|c|c|c|}
\hline 
$d$&
$r$&
$K^{+}$&
$K^{-}$\tabularnewline
\hline
\hline 
&
0.17&
1.4&
1.4\tabularnewline
\hline 
&
0.10&
2.1&
1.2\tabularnewline
\hline 
3&
0.07&
2.7&
1.07\tabularnewline
\hline 
&
$10^{-3}$&
45&
0.26\tabularnewline
\hline 
&
0.2&
0.72&
3.1\tabularnewline
\hline
\end{tabular} ~~~~~~~~~~\begin{tabular}{|c|c|c|c|}
\hline 
$d$&
$r$&
$K^{+}$&
$K^{-}$\tabularnewline
\hline
\hline 
&
0.1&
1.2&
2.4\tabularnewline
\hline 
&
0.07&
1.5&
2.2\tabularnewline
\hline 
2&
0.047&
1.9&
1.9\tabularnewline
\hline 
&
$10^{-3}$&
25&
0.52\tabularnewline
\hline 
&
$10^{-6}$&
2480&
0.052\tabularnewline
\hline
\end{tabular}
\end{table}
 Also, there is a change of behaviour near $r=(I^{-}/I^{+})^{2}=r_{c}$;
$K^{-}<K^{+}$ for $r<r_{c}$, whereas inequality reverses for $r$
beyond $r_{c}$.

Many important relationships can be deduced from the equations derived
above. For example, \begin{equation}
\frac{\Pi^{-}}{\Pi^{+}}=\frac{I^{-}}{I^{+}}\end{equation}
 Since $I^{\pm}$ are independent of $r$, we can immediately conclude
that the ratio $\Pi^{-}/\Pi^{+}$ is also \emph{independent of $r$}.
This is an important conclusion derivable from the above calculation.
From the above equations one can also derive \begin{eqnarray}
K^{+} & = & \frac{1}{r^{2/3}}\frac{(I^{-})^{2/3}}{(I^{+})^{4/3}}\\
K^{-} & = & r^{1/3}\frac{(I^{+})^{2/3}}{(I^{-})^{4/3}}\\
\frac{K^{-}}{K^{+}} & = & r\left(\frac{I^{+}}{I^{-}}\right)^{2}\end{eqnarray}

The total energy cascade rate can be written in terms of $E^{+}(k)$
as \begin{equation}
\Pi=\frac{1}{2}(\Pi^{+}+\Pi^{-})=\frac{r}{2}(I^{+}+I^{-})(E^{+}(k))^{3/2}k^{5/2}\label{eqn:flux_sigc1}\end{equation}
 Since $I^{\pm}$ are independent of $r$, $\Pi$ is a linear function
of $r$. As expected the energy flux vanishes when $r=0$.

In this section, we have dealt with strong turbulence. For weak turbulence,
Lithwick and Goldreich \cite{Gold:Imbalanced} have solved Alfvénic
MHD equations using kinetic theory.

\subsubsection{Helical nonAlfvénic MHD ($H_{M}\ne0,H_{K}\ne0,H_{c}=0$): \label{sub:Helical-nonAlfvenic-MHD-flux}}

Now we present computation of cascade rates of energy and magnetic
helicity for helical MHD $(H_{M}\ne0,H_{K}\ne0)$ \cite{MKV:MHD_Helical}.
Here $d=3$. To simplify the equation, we consider only nonAlfvénic
fluctuations ($\sigma_{c}=0$). We start with the flux formulas of
energy (Eq. {[}\ref{eq:flux-MHD}{]}) and magnetic helicity\begin{eqnarray}
\Pi_{H_{M}}(k_{0}) & = & \frac{1}{(2\pi)^{d}\delta(\mathbf{k'+p+q})}\int_{k'>k_{0}}\frac{d\mathbf{k'}}{(2\pi)^{3}}\int_{p<k_{0}}\frac{d\mathbf{p}}{(2\pi)^{3}}\left\langle S^{H_{M}}(\mathbf{k'|p|q})\right\rangle .\label{eq:flux-HM}\end{eqnarray}
The calculation procedure is identical to that of nonhelical nonAlfvénic
MHD. The only difference is that additional terms appear in $\left\langle S^{YX}(\mathbf{k'}|\mathbf{p}|\mathbf{q})\right\rangle $
(Eq. {[}\ref{eqn:Suu-nonhelical}-\ref{eq:Sbb-nonhelical}{]}) because
terms $\left\langle u_{i}(\mathbf{k},t)u_{j}(\mathbf{k},t')\right\rangle $
and $\left\langle b_{i}(\mathbf{k},t)b_{j}(\mathbf{k},t')\right\rangle $
contain helicities in addition to correlation functions:\begin{eqnarray*}
\left\langle u_{i}(\mathbf{p},t)u_{j}(\mathbf{q},t')\right\rangle  & = & \left[P_{ij}\left(\mathbf{p}\right)C^{uu}(\mathbf{p},t,t')-i\epsilon_{ijl}k_{l}H_{M}(k,t,t')\right]\delta(\mathbf{p}+\mathbf{q})\left(2\pi\right)^{3},\\
\left\langle b_{i}(\mathbf{p},t)b_{j}(\mathbf{q},t')\right\rangle  & = & \left[P_{ij}\left(\mathbf{p}\right)C^{bb}(\mathbf{p},t,t')-i\epsilon_{ijl}k_{l}\frac{H_{M}(k,t,t')}{k^{2}}\right]\delta(\mathbf{p}+\mathbf{q})\left(2\pi\right)^{3}.\end{eqnarray*}
Substitutions of these functions in perturbative series yield\begin{eqnarray}
\left\langle S^{uu}(k'|p|q)\right\rangle  & = & \int_{-\infty}^{t}dt'(2\pi)^{3}\left[T_{1}(k,p,q)G^{uu}(k,t-t')C^{u}(p,t,t')C^{u}(q,t,t')\right.\nonumber \\
 &  & \hspace{1cm}+T'_{1}(k,p,q)G^{uu}(k,t-t')\frac{H_{K}(p,t,t')}{p^{2}}\frac{H_{K}(q,t,t')}{q^{2}}\nonumber \\
 &  & \hspace{1cm}+T_{5}(k,p,q)G^{uu}(p,t-t')C^{u}(k,t,t')C^{u}(q,t,t')\nonumber \\
 &  & \hspace{1cm}+T'_{5}(k,p,q)G^{uu}(p,t-t')\frac{H_{K}(k,t,t')}{k^{2}}\frac{H_{K}(q,t,t')}{q^{2}}\nonumber \\
 &  & \hspace{1cm}+T_{9}(k,p,q)G^{uu}(q,t-t')C^{u}(k,t,t')C^{u}(p,t,t')\nonumber \\
 &  & \hspace{1cm}+T'_{9}(k,p,q)G^{uu}(q,t-t')\frac{H_{K}(k,t,t')}{k^{2}}\frac{H_{K}(p,t,t')}{p^{2}}\delta(\mathbf{k'}+\mathbf{p}+\mathbf{q})\\
\left\langle S^{ub}(k'|p|q)\right\rangle  & = & -\int_{-\infty}^{t}dt'(2\pi)^{3}\left[T_{2}(k,p,q)G^{uu}(k,t-t')C^{b}(p,t,t')C^{b}(q,t,t')\right.\nonumber \\
 &  & \hspace{1cm}+T'_{2}(k,p,q)G^{uu}(k,t-t')H_{M}(p,t,t')H_{M}(q,t,t')\nonumber \\
 &  & \hspace{1cm}+T_{7}(k,p,q)G^{bb}(p,t-t')C^{u}(k,t,t')C^{b}(q,t,t')\nonumber \\
 &  & \hspace{1cm}+T'_{7}(k,p,q)G^{bb}(p,t-t')\frac{H_{K}(k,t,t')}{k^{2}}H_{M}(q,t,t')\nonumber \\
 &  & \hspace{1cm}+T_{11}(k,p,q)G^{uu}(q,t-t')C^{u}(k,t,t')C^{b}(p,t,t')\nonumber \\
 &  & \hspace{1cm}\left.+T'_{11}(k,p,q)G^{uu}(q,t-t')\frac{H_{K}(k,t,t')}{k^{2}}H_{M}(p,t,t')\right]\delta(\mathbf{k'}+\mathbf{p}+\mathbf{q})\\
\left\langle S^{bu}(k'|p|q)\right\rangle  & = & -\int_{-\infty}^{t}dt'(2\pi)^{3}\left[T_{3}(k,p,q)G^{bb}(k,t-t')C^{u}(p,t,t')C^{b}(q,t,t')\right.\nonumber \\
 &  & \hspace{1cm}+T'_{3}(k,p,q)G^{bb}(k,t-t')\frac{H_{K}(p,t,t')}{p^{2}}H_{M}(q,t,t')\nonumber \\
 &  & \hspace{1cm}+T_{6}(k,p,q)G^{uu}(p,t-t')C^{b}(k,t,t')C^{b}(q,t,t')\nonumber \\
 &  & \hspace{1cm}+T'_{6}(k,p,q)G^{uu}(p,t-t')H_{M}(k,t,t')H_{M}(q,t,t')\nonumber \\
 &  & \hspace{1cm}+T_{12}(k,p,q)G^{bb}(q,t-t')C^{b}(k,t,t')C^{u}(p,t,t')\nonumber \\
 &  & \hspace{1cm}\left.+T'_{12}(k,p,q)G^{bb}(q,t-t')H_{M}(k,t,t')\frac{H_{K}(p,t,t')}{p^{2}}\right]\delta(\mathbf{k'}+\mathbf{p}+\mathbf{q})\label{eq:Sbu-helical}\\
\left\langle S^{bb}(k'|p|q)\right\rangle  & = & \int_{-\infty}^{t}dt'(2\pi)^{3}\left[T_{4}(k,p,q)G^{bb}(k,t-t')C^{b}(p,t,t')C^{u}(q,t,t')\right.\nonumber \\
 &  & \hspace{1cm}+T'_{4}(k,p,q)G^{bb}(k,t-t')H_{M}(p,t,t')\frac{H_{K}(q,t,t')}{q^{2}}\nonumber \\
 &  & \hspace{1cm}+T_{8}(k,p,q)G^{bb}(p,t-t')C^{b}(k,t,t')C^{u}(q,t,t')\nonumber \\
 &  & \hspace{1cm}+T'_{8}(k,p,q)G^{bb}(p,t-t')H_{M}(k,t,t')\frac{H_{K}(q,t,t')}{q^{2}}\nonumber \\
 &  & \hspace{1cm}+T_{10}(k,p,q)G^{uu}(q,t-t')C^{b}(k,t,t')C^{b}(p,t,t')\nonumber \\
 &  & \hspace{1cm}\left.+T'_{10}(k,p,q)G^{uu}(q,t-t')H_{M}(k,t,t')H_{M}(p,t,t')\right]\delta(\mathbf{k'}+\mathbf{p}+\mathbf{q})\label{eq:Sbb-helical}\end{eqnarray}
The functions $T_{i}(k,p,q)$ and $T'_{i}(k,p,q)$ are given in Appendix
C. Note that $T'_{i}(k,p,q)$ are the additional terms as compared
to nonhelical flux (see Eqs. {[}\ref{eqn:Suu-nonhelical}-\ref{eq:Sbb-nonhelical}{]}). 

The quantity $\left\langle S^{H_{M}}(\mathbf{k'|p|q})\right\rangle $
of Eq. (\ref{eq:Skpq_HM}) simplifies to \begin{eqnarray}
\left\langle S^{H_{M}}(\mathbf{k'|p|q})\right\rangle  & = & \frac{1}{2}\Re\left[\epsilon_{ijm}\left\langle b_{i}(k')u_{j}(p)b_{m}(q)\right\rangle \right.\nonumber \\
 &  & -\epsilon_{jlm}\frac{k_{i}k_{l}}{k^{2}}\left\langle u_{i}(q)b_{m}(k')b_{j}(p)\right\rangle \nonumber \\
 &  & \left.+\epsilon_{jlm}\frac{k_{i}k_{l}}{k^{2}}\left\langle b_{i}(q)b_{m}(k')u_{j}(p)\right\rangle \right],\end{eqnarray}
 which is computed perturbatively to the first order. The corresponding
Feynman diagrams are given in Appendix C. The resulting expression
for $\left\langle S^{H_{M}}({\textbf{k'|p|q}})\right\rangle $ is
\begin{eqnarray}
\left\langle S^{H_{M}}({\textbf{k'|p|q}})\right\rangle  & = & \int_{-\infty}^{t}dt'(2\pi)^{3}\left[T_{31}(k,p,q)G^{bb}(k,t-t')\frac{H_{K}(p,t-t')}{p^{2}}C^{b}(q,t-t')\right.\nonumber \\
 &  & +T_{32}(k,p,q)G^{bb}(k,t-t')C^{uu}(p,t-t')H_{M}(q,t-t')\nonumber \\
 &  & +T_{33}(k,p,q)G^{uu}(p,t-t')H_{M}(k,t-t')C^{bb}(q,t-t')\nonumber \\
 &  & +T_{34}(k,p,q)G^{uu}(p,t-t')C^{bb}(k,t-t')H_{M}(q,t-t')\nonumber \\
 &  & +T_{35}(k,p,q)G^{bb}(q,t-t')H_{M}(k,t-t')C^{uu}(p,t-t')\nonumber \\
 &  & +T_{36}(k,p,q)G^{bb}(q,t-t')C^{bb}(k,t-t')\frac{H_{K}(p,t-t')}{p^{2}}\nonumber \\
 &  & +T_{37}(k,p,q)G^{bb}(k,t-t')H_{M}(p,t-t')C^{uu}(q,t-t')\nonumber \\
 &  & +T_{38}(k,p,q)G^{bb}(k,t-t')C^{bb}(p,t-t')\frac{H_{K}(q,t-t')}{q^{2}}\nonumber \\
 &  & +T_{39}(k,p,q)G^{bb}(p,t-t')H_{M}(k,t-t')C^{uu}(q,t-t')\nonumber \\
 &  & +T_{40}(k,p,q)G^{bb}(p,t-t')C^{bb}(k,t-t')\frac{H_{K}(q,t-t')}{q^{2}}\nonumber \\
 &  & +T_{41}(k,p,q)G^{uu}(q,t-t')H_{M}(k,t-t')C^{bb}(p,t-t')\nonumber \\
 &  & +T_{42}(k,p,q)G^{uu}(q,t-t')C^{bb}(k,t-t')H_{M}(p,t-t')\}\nonumber \\
 &  & +T_{43}(k,p,q)G^{bb}(k,t-t')\frac{H_{K}(p,t-t')}{p^{2}}C^{bb}(q,t-t')\nonumber \\
 &  & +T_{44}(k,p,q)G^{bb}(k,t-t')C^{uu}(p,t-t')H_{M}(q,t-t')\}\nonumber \\
 &  & +T_{45}(k,p,q)G^{uu}(p,t-t')H_{M}(k,t-t')C^{bb}(q,t-t')\nonumber \\
 &  & +T_{46}(k,p,q)G^{uu}(p,t-t')C^{bb}(k,t-t')H_{M}(q,t-t')\nonumber \\
 &  & +T_{47}(k,p,q)G^{bb}(q,t-t')H_{M}(k,t-t')C^{uu}(p,t-t')\nonumber \\
 &  & +T_{48}(k,p,q)G^{bb}(q,t-t')C^{bb}(k,t-t')\frac{H_{K}(p,t-t')}{p^{2}}\delta(\mathbf{k'}+\mathbf{p}+\mathbf{q})\end{eqnarray}
For Greens' functions and correlation functions the same substitutions
were made as in nonhelical case. For helicities, the following assumptions
were made: the relaxation time-scales for $H_{K}(k,t,t')$ and $H_{M}(k,t,t')$
are $(\nu(k)k^{2})^{-1}$and $(\eta(k)k^{2})^{-1}$ respectively,
i.e.,\[
H_{K,M}(k,t,t')=H_{K,M}(k,t,t)\theta(t-t')\exp{\left\{ -[\nu,\eta]k^{2}(t-t')\right\} }.\]
The spectra of helicities are tricky. In the presence of magnetic
helicity, the calculations based on absolute equilibrium theories
suggest that the energy cascades forward, and the magnetic helicity
cascades backward \cite{Fris:HM}. Verma did not consider the inverse
cascade region of magnetic helicity, and computed energy fluxes for
the forward energy cascade region (5/3 powerlaw). 

The helicities were written in terms of energy spectra as \begin{eqnarray}
H_{K}(k) & = & r_{K}kE^{u}(k)\label{eqn:HK(k)}\\
H_{M}(k) & = & r_{M}\frac{E^{b}(k)}{k}\label{eqn:HM(k)}\end{eqnarray}
The ratios $r_{A},r_{M}$, and $r_{K}$ were treated as constants.
In pure fluid turbulence, kinetic helicity spectrum is proportional
to $k^{-5/3}$, contrary to the assumption made here. The cascade
picture of magnetic helicity is also not quite clear. Therefore, Verma
\cite{MKV:MHD_Helical} performed the calculations for the simplest
spectra assumed above.

The above form of correlation and Green's functions were substituted
in the expressions for $\left\langle S^{YX}(k'|p|q)\right\rangle $
and $\left\langle S^{H_{M}}(k'|p|q)\right\rangle $. These $S$'s
were then substituted in the flux formulas (Eqs.~{[}\ref{eq:flux-MHD},
\ref{eq:flux-HM}{]}). After performing the following change of variable:
\begin{equation}
k=\frac{k_{0}}{u};p=\frac{k_{0}}{u}v;q=\frac{k_{0}}{u}w\end{equation}
 one obtains the following nondimensional form of the equation in
the $5/3$ region \begin{eqnarray}
\frac{\Pi_{Y>}^{X<}(k_{0})}{\left|\Pi(k_{0})\right|} & = & (K^{u})^{3/2}\left[\frac{1}{2}\int_{0}^{1}dv\ln{(1/v)}\int_{1-v}^{1+v}dw(vw)\sin\alpha F_{Y>}^{X<}\right]\label{eqn:piE_ratio1}\\
\nonumber \\\frac{\Pi_{H_{M}}(k_{0})}{\left|\Pi(k_{0})\right|} & = & \frac{1}{k_{0}}(K^{u})^{3/2}\left[\frac{1}{2}\int_{0}^{1}dv(1-v)\int_{1-v}^{1+v}dw(vw)\sin\alpha F_{H_{M}}\right]\label{eqn:piHm_ratio1}\end{eqnarray}
 where the integrands $(F_{Y>}^{X<},F_{H_{M}})$ are function of $v$,
$w$, $\nu^{*}$, $\eta^{*}$, $r_{A},r_{K}$ and $r_{M}$ \cite{MKV:MHD_Flux}.

Verma \cite{MKV:MHD_Helical} computed the terms in the square brackets,
$I_{Y>}^{X<}$, using Gaussian-quadrature method. The constant $K^{u}$
was calculated using the fact that the total energy flux $\Pi$ is
sum of all $\Pi_{Y>}^{X<}$. For parameters ($r_{A}=5000,r_{K}=0.1,r_{M}=-0.1$),
$K^{u}=1.53$, while for ($r_{A}=1,r_{K}=0.1,r_{M}=-0.1$), $K^{u}=0.78$.
After that the energy flux ratios $\Pi_{Y>}^{X<}/\Pi$ could be calculated.
Table \ref{table:hm_rA1_5000} contains these values for $r_{A}=1$
and $r_{A}=5000$. %
\begin{table}

\caption{\label{table:hm_rA1_5000}The values of $I_{Y}^{X}=(\Pi_{Y}^{X}/\Pi)/(K^{u})^{1.5}$
calculated using Eqs. (\ref{eqn:piE_ratio1}, \ref{eqn:piHm_ratio1})
for $r_{A}=1$ and 5000.}

\begin{tabular}{|c|c|c|}
\hline 
&
$r_{A}=1$&
$r_{A}=5000$\tabularnewline
\hline
\hline 
$I_{u>}^{u<}$&
$0.19-0.10r_{K}^{2}$&
$0.53-0.28r_{K}^{2}$\tabularnewline
\hline 
$I_{b>}^{u<}$&
$0.62+0.3r_{M}^{2}+0.095r_{K}r_{M}$&
$1.9\times10^{-4}+1.4\times10^{-9}r_{M}^{2}+2.1\times10^{-5}r_{K}r_{M}$\tabularnewline
\hline 
$I_{u>}^{b<}$&
$0.18-2.04r_{M}^{2}+1.93r_{K}r_{M}$&
$-5.6\times10^{-5}-1.1\times10^{-7}r_{M}^{2}+5.4\times10^{-4}r_{K}r_{M}$\tabularnewline
\hline 
$I_{b>}^{b<}$&
$0.54-1.9r_{M}^{2}+2.02r_{K}r_{M}$&
$1.4\times10^{-4}-1.02\times10^{-7}r_{M}^{2}+5.4\times10^{-4}r_{K}r_{M}$\tabularnewline
\hline 
$I_{H_{M}}$&
-$25r_{M}+0.35r_{K}$&
$-4.1\times10^{-3}r_{M}+8.1\times10^{-5}r_{K}$\tabularnewline
\hline 
$K^{u}$&
0.78&
1.53\tabularnewline
\hline
\end{tabular}
\end{table}
These ratios for some of the specific values of $r_{A}$, $r_{K}$
and $r_{M}$ are listed in Table \ref{table:hm_rkrm}. The energy
flux has been split into two parts: nonhelical (independent of helicity,
the first term of the bracket) and helical (dependent on $r_{K}$
and/or $r_{M}$, the second term \emph{}of the bracket).%
\begin{table}

\caption{\label{table:hm_rkrm}The values of energy ratios $\Pi_{Y}^{X}/\Pi$
for various values of $r_{A},r_{K}$ and $r_{M}$ for $k^{-5/3}$
region. The first and second entries are the nonhelical and helical
contributions respectively.}

\begin{tabular}{|c|c|c|c|c|}
\hline 
$(r_{A},r_{K},r_{M})$&
$\Pi_{u>}^{u<}/\Pi$&
$\Pi_{b>}^{u<}/\Pi$&
$\Pi_{u>}^{b<}/\Pi$&
$\Pi_{b>}^{b<}/\Pi$\tabularnewline
\hline
\hline 
(1,0.1,-0.1)&
$(0.13,-6.9\times10^{-4})$&
$(0.43,-4.4\times10^{-4})$&
$(0.13,-0.027)$&
$(0.37,-0.027)$\tabularnewline
\hline 
$(1,0.1,0.1)$&
$(0.12,-6.5\times10^{-4})$&
$(0.40,8.1\times10^{-4})$&
$(0.12,-7.7\times10^{-4})$&
$(0.35,8.3\times10^{-4})$\tabularnewline
\hline 
$(1,1,-1)$&
$(0.029,-0.015)$&
$(0.095,-9.9\times10^{-3})$&
$(0.028,-0.61)$&
$(0.083,-0.60)$\tabularnewline
\hline 
$(1,1,1)$&
$(0.12,-0.064)$&
$(0.39,0.079)$&
$(0.12,-0.075)$&
$(0.34,0.081)$\tabularnewline
\hline 
$(1,0,1)$&
$(0.081,0)$&
$(0.26,0.013)$&
$(0.078,-0.86)$&
$(0.23,-0.8)$\tabularnewline
\hline
\hline 
$(5000,0.1,-0.1)$&
$(1.0,$&
$(3.2\times10^{-4},$&
$(-9.7\times10^{-5},$&
$(2.5\times10^{-4},$\tabularnewline
&
$-5.3\times10^{-3})$&
$-3.7\times10^{-7})$&
$-9.0\times10^{-6})$&
$-9.4\times10^{-4}$\tabularnewline
\hline 
$(5000,0.1,0.1)$&
$(1.0,$&
$(3.2\times10^{-4},$&
$(-9.7\times10^{-5},$&
$(2.5\times10^{-4},$\tabularnewline
&
$-5.3\times10^{-3})$&
$3.7\times10^{-7})$&
$9.0\times10^{-6})$&
$9.4\times10^{-6})$\tabularnewline
\hline
\end{tabular}
\end{table}

An observation of the results shows some interesting patterns. The
values of nonhelical part of all the flux-ratios are quite similar
to those discussed in Section \ref{sub:Nonhelical-nonAlfvenic-MHD-flux}.
All the fluxes except $\Pi_{u>nonhelical}^{b<}$ ($\Pi_{u>nonhelical}^{b<}<0$
for $r_{A}>3$ ) are always positive. As a consequence, in nonhelical
channel, magnetic energy cascades from large scales to small scales
for $r_{A}<3$. 

The sign of $\Pi_{u>helical}^{u<}$ is always negative, i.e., kinetic
helicity reduces the Kinetic energy flux. But the sign of helical
component of other energy fluxes depends quite crucially on the sign
of helicities. From the entries of Table \ref{table:hm_rA1_5000},
we see that \begin{equation}
\Pi_{(b>,u>)helical}^{b<}=-ar_{M}^{2}+br_{M}r_{K},\label{eqn:rM_dependence}\end{equation}
 where $a$ and $b$ are positive constants. If $r_{M}r_{K}<0$, the
large-scale magnetic field will get positive contribution from both
the terms in the right-hand-side of the above equation. The EDQNM
work of Pouquet et al. \cite{Pouq:EDQNM} and numerical simulations
of Brandenburg \cite{Bran:Alpha} with forcing (kinetic energy and
kinetic helicity) typically have $r_{K}r_{M}<0$. Hence, we can claim
that helicity typically induces an inverse energy cascade via $\Pi_{b>}^{b<}$
and $\Pi_{u>}^{b<}$. These fluxes will enhance the large-scale magnetic
field.

The helical and nonhelical contributions to the fluxes for $r_{A}=5000,r_{K}=0.1,r_{M}=-0.1$
is shown in Table \ref{table:hm_rkrm}. The flux ratios shown in the
table do not change appreciably as long as $r_{A}>100$ or so. The
three fluxes responsible for the growth of large-scale magnetic energy
are nonhelical $(\Pi_{b<}^{u<}\approx\Pi_{b>}^{b<}+\Pi_{u>}^{b<})/\Pi\approx2.6\times10^{-4}$,
and helical $\Pi_{b>helical}^{b<}/\Pi\approx-10^{-5}$ and $\Pi_{u>helical}^{b<}/\Pi\approx-10^{-5}$.
The ratio of nonhelical to helical contribution is of the order of
10. Hence, for the large-scale magnetic energy growth, the nonhelical
contribution is comparable to the helical contribution. Note that
in the earlier papers on dynamo, the helical part is strongly emphasized,
and nonhelical component is typically ignored.

From the entries of Table \ref{table:hm_rkrm} we can infer that the
for small and moderate $r_{K}$ and $r_{M}$, the inverse energy cascade
into the large-scale magnetic field is less than the the forward nonhelical
energy flux $\Pi_{b>}^{b<}$. While for helical MHD ($r_{K},r_{M}\rightarrow1$),
the inverse helical cascade dominates the nonhelical magnetic-to-magnetic
energy cascade.

The flux ratio $\Pi_{H_{M}}/\Pi$ of Eqs. (\ref{eqn:piHm_ratio1})
can be computed using the same procedure. The numerical values of
the integrals are shown in Tables \ref{table:hm_rA1_5000} and \ref{table:hm_rkrm}.
Clearly, \begin{equation}
\Pi_{H_{M}}(k_{0})=\frac{1}{k_{0}}\left(-dr_{M}+er_{K}\right)\label{eq:flux-HM-value}\end{equation}
 where $d$ and $e$ are positive constants. In 5/3 regime, the magnetic
helicity is not constant, and is inversely proportional to $k_{0}$.
The contribution from$H_{M}$ dominates that from $H_{K}$ and is
of opposite sign. For positive $H_{M}$, the magnetic helicity cascade
is backward. This result is in agreement with Frisch et al.'s \cite{Fris:HM}
prediction of an inverse cascade of magnetic helicity using Absolute
equilibrium theory. Verma's theoretical result on inverse cascade
of $H_{M}$ is in agreement with the results derived using EDQNM calculation
\cite{Pouq:EDQNM} and numerical simulations \cite{Pouq:num}. Reader
is also referred to Oughton and Prandi \cite{Ough:HK} for the effects
of kinetic helicity on the decay of magnetic energy.

When the system is forced with positive kinetic helicity ($r_{K}>0$),
Eq. (\ref{eq:flux-HM-value}) indicates a forward cascade of magnetic
helicity. This effect could be the reason for the observed production
of positive magnetic helicity at small scales by Brandenburg \cite{Bran:Alpha}.
Because of magnetic helicity conservation, he also finds negative
magnetic helicity at large-scales. Now, positive kinetic helicity
and negative magnetic helicity at large-scales may yield an inverse
cascade of magnetic energy (see Eq. {[}\ref{eqn:rM_dependence}{]}).
This could be one of the main reason for the growth of magnetic energy
in the simulations of Brandenburg \cite{Bran:Alpha}.

After completing the discussion on energy fluxes for MHD turbulence,
we now move on to theoretical computation of shell-to-shell energy
transfer.

\subsection{Field-theoretic Calculation of Shell-to-shell Energy Transfer \label{sub:Field-theoretic-Calculation-of-shell}}

Energy transfers between wavenumber shells provide us with important
insights into the dynamics of MHD turbulence. Kolmogorov's fluid turbulence
model is based on local energy transfer between wavenumber shells.
There are several quantitative theories in fluid turbulence about
the amount of energy transfer between neighbouring wavenumber shells.
For examples, Kraichnan \cite{Krai:71} showed that $35$\% of the
energy transfer comes from wavenumber triads where the smallest wave-number
is greater than one-half of the middle wavenumber. In MHD turbulence,
Pouquet et al. \cite{Pouq:EDQNM} estimated the contributions of local
and nonlocal interactions using EDQNM calculation. They argued that
large-scale magnetic energy brings about equipartition of kinetic
and magnetic excitations by the Alfvén effect. The small-scale {}``residual
helicity'' ($H_{K}-H_{M}$) induces growth of large-scale magnetic
energy. These results will be compared with our field-theoretic results
described below.

In this subsection we will compute the shell-to-shell energy transfer
in MHD turbulence using field-theoretic method \cite{Ayye:MHD}. The
procedure is identical to the one described for MHD fluxes. We will
limit ourselves to nonAlfvénic MHD (both nonhelical and helical).
Recall that the energy transfer rates from $m$-th shell of field
$X$ to $n$-th shell of field $Y$ is\[
T_{nm}^{YX}=\sum_{\mathbf{k'}\in n}\sum_{\mathbf{p}\in m}S^{YX}(\mathbf{k'|p|q}).\]
The $\mathbf{p}$-sum is over $m$-th shell, and the $\mathbf{k'}$-sum
is over $n$-th shell (Section \ref{sec:Mode-to-mode-Energy-Transfer}).
The terms of $S^{YX}$'s remain the same as in flux calculation, however,
the limits of the integrals are different. The shells are binned logarithmically
with $n$-th shell being $(k_{0}s^{n-1},k_{0}s^{n})$. We nondimensionalize
the equations using the transformation \cite{Lesl:book}\begin{equation}
k=\frac{a}{u};\hspace{1cm}p=\frac{a}{u}v;\hspace{1cm}q=\frac{a}{u}w,\end{equation}
 where $a=k_{0}s^{n-1}$. The resulting equation is\begin{equation}
\frac{T_{nm}^{YX}}{\Pi}=K_{u}^{3/2}\frac{4S_{d-1}}{(d-1)^{2}S_{d}}\int_{s^{-1}}^{1}\frac{du}{u}\int_{us^{m-n}}^{us^{m-n+1}}dv\int_{|1-v|}^{1+v}dw\left(vw\right)^{d-2}\left(\sin{\alpha}\right)^{d-3}F^{YX}(v,w),\label{eqn:shell_final}\end{equation}
where$F^{YX}(v,w)$ was computed for helical nonAlfvénic MHD flows
(see Eq. {[}\ref{eqn:piE_ratio1}{]}). It includes both nonhelical
$F_{nonhelical}^{YX}(v,w)$ and helical $F_{helical}^{YX}(v,w)$ components.
The renormalized parameters $\nu^{*}$, $\lambda^{*}$ and Kolmogorov's
constant $K^{u}$ required to compute $T_{nm}^{YX}/\Pi$ are taken
from the previous calculations. From Eq. (\ref{eqn:shell_final})
we can draw the following inferences:

\begin{enumerate}
\item The shell-to-shell energy transfer rate is a function of $n-m$, that
is, $\Phi_{nm}=\Phi_{(n-i)(m-i)}$. Hence, the turbulent energy transfer
rates in the inertial range are all self-similar. Of course, this
is true only in the inertial range.
\item $T_{nm}^{ub}/\Pi=-T_{mn}^{bu}/\Pi$. Hence $T_{nm}^{bu}/\Pi$ can
be obtained from $T_{mn}^{ub}/\Pi$ by inversion at the origin. 
\item $\Pi_{Y>}^{X<}=\sum_{n=m+1}^{\infty}(n-m)T_{nm}^{YX}$.
\item Net energy gained by a $u$-shell from $u$-to-$u$ transfer is zero
because of self similarity. However, a $u$-shell can gain or lose
a net energy due to imbalance between $u$-to-$b$ and $b$-to-$u$
energy transfers. By definition, we can show that net energy gained
by an inertial $u$-shell is\begin{equation}
\sum_{n}\left(T_{nm}^{ub}-T_{nm}^{bu}\right)+T_{nn}^{ub}.\label{eq:Sum-Tub}\end{equation}
Similarly, net energy gained by a $b$-shell from $b$-to-$b$ transfer
is zero. However, net energy gained by an inertial $b$-shell due
to $u$-to-\textbf{$b$} and $b$-to-$u$ transfers is \begin{equation}
\sum_{n}\left(T_{nm}^{bu}-T_{nm}^{ub}\right)+T_{nn}^{bu}.\label{eq:Sum-Tbu}\end{equation}

\end{enumerate}
We compute the integral of Eq. (\ref{eqn:shell_final}). we describe
the results in two separate parts: (1) nonhelical contributions, (2)
helical contributions.

\subsubsection{Nonhelical Contributions: }

We compute nonhelical contributions by turning off kinetic and magnetic
helicities in $F^{YX}$. We have chosen $s=2^{1/4}$. This study has
been done for various values of Alfvén ratios. %
\begin{figure}
\includegraphics[bb= 100 260 520 540]{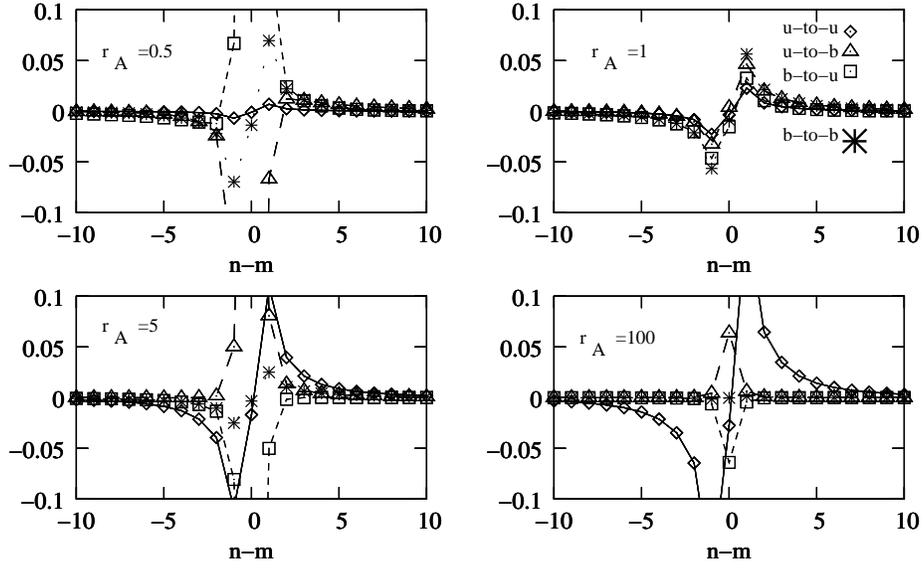}

\caption{\label{Fig:T-MHD-theory} Theoretically calculated values of $T_{nm}^{YX}/\Pi$
vs. $n-m$ for zero helicities $(\sigma_{c}=r_{K}=r_{M}=0)$ and Alfvén
ratios $r_{A}=0.5,1,4,100$. }
\end{figure}
 Fig. \ref{Fig:T-MHD-theory} contains plots of $T_{nm}^{YX}/\Pi$
vs $n-m$ for four typical values of $r_{A}=0.5,1,5,100$. The numbers
represent energy transfer rates from shell $m$ to shells $m+1,m+2,...$in
the right, and to shells $m-1,m-2,...$ in the left. All the plots
are to same scale. For $r_{A}=0.5$, maxima of $T_{nm}^{ub}/\Pi$
and $T_{nm}^{bu}/\Pi$ occurs at $m=n$, and its values are $\pm1.40$
respectively. The corresponding values for $r_{A}=5$ are $\mp0.78$.
By observing the plots we find the following interesting patterns:

\begin{enumerate}
\item $T_{nm}^{uu}/\Pi$ is positive for $n>m$, and negative for $n<m$.
Hence, a $u$-shell gains energy from smaller wavenumber $u$-shells,
and loses energy to higher wavenumber $u$-shells, implying that energy
cascade is forward. Also, the absolute maximum occurs for $n=m\pm1$,
hence the energy transfer is local. For kinetic dominated regime,
$s=2^{1/2}$ yields $T_{nm}^{uu}/\Pi\approx34\%$, similar to Kraichnan's
Test Mean Field model (TFM) predictions \cite{Krai:71}. 
\item $T_{nm}^{bb}/\Pi$ is positive for $n>m$, and negative for $n<m$,
and maximum for $n=m\pm1.$ Hence magnetic to magnetic energy transfer
is forward and local. This result is consistent with the forward magnetic-to-magnetic
cascade $(\Pi_{b>}^{b<}>0)$ discussed in Section \ref{sub:Nonhelical-nonAlfvenic-MHD-flux}.
\item For $r_{A}>1$ (kinetic energy dominated), kinetic to magnetic energy
transfer rate $T_{nm}^{bu}/\Pi$ is positive for $n\ge m-1$, and
negative otherwise\emph{.} These transfers have been illustrated in
Fig. \ref{Fig:ub-shell}(a).%
\begin{figure}
\includegraphics[%
  scale=0.4,bb= 0 0 575 600]{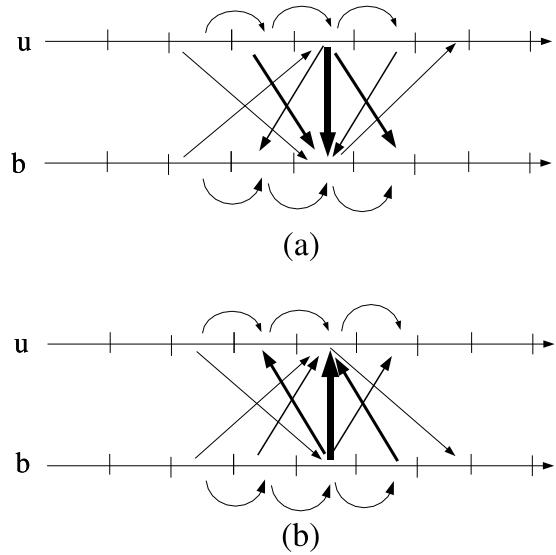}

\caption{\label{Fig:ub-shell} Schematic illustration of nonhelical $T_{nm}^{YX}/\Pi$
in the inertial range for (a) kinetic-energy dominated regime, and
(b) magnetic-energy dominated regime. In (a) $T_{nm}^{bu}/\Pi$ is
positive for $n\ge m-1$, and negative otherwise, while in (b) $T_{nm}^{ub}/\Pi$
is positive for $n\ge m-1$, and negative otherwise. $T_{nm}^{uu}$
and $T_{nm}^{bb}$ is forward and local. }
\end{figure}
 \emph{Using Eq. (\ref{eq:Sum-Tbu}) we find there is a net energy
transfer from kinetic to magnetic, and the net energy transfer rate
decreases as we go toward} $r_{A}=1$. Here, each $u$-shell loses
net energy to $b$-shells, hence the turbulence is not steady. This
phenomena is seen for all $r_{A}>1$.
\item For $r_{A}=0.5$ (magnetically dominated), magnetic to kinetic energy
transfer rate $T_{nm}^{ub}/\Pi$ is positive for $n\ge m-1$, and
negative otherwise (see Fig. \ref{Fig:T-MHD-theory}). \emph{There
is a net energy transfer from magnetic to kinetic energy; its magnitude
decreases as $r_{A}\rightarrow1$.} In addition, using Eq. (\ref{eq:Sum-Tub})
we find that each $b$-shell is losing net energy to $u$-shells,
hence the turbulence cannot be steady. This phenomena is seen for
all $r_{A}<1$.
\item The observations of (3) and (4) indicate that kinetic to magnetic
or the reverse energy transfer rate almost vanishes near $r_{A}=1$.
\emph{We believe that MHD turbulence evolves toward $r_{A}\approx1$
due to above reasons.} For $r_{A}\ne1$, MHD turbulence is not steady.
This result is same as the prediction of equipartition of kinetic
and magnetic energy due to Pouquet et al.'s using EDQNM calculation
\cite{Pouq:EDQNM}. Note however that the steady-state value of $r_{A}$
in numerical simulations and solar wind is around 0.5-0.6. The difference
is probably because realistic flows have more interactions than discussed
above, e. g., nonlocal coupling with forcing wavenumbers. Careful
numerical simulations are required to clarify this issue.
\item When $r_{A}$ is not close to 1 ($r_{A}\le0.5$ or $r_{A}>5$), $u$-to-$b$
shell-to-shell transfer involves many neighbouring shells (see Fig.
\ref{Fig:T-MHD-theory}). This observation implies that $u-$$b$
energy transfer is somewhat nonlocal as predicted by Pouquet et al.
\cite{Pouq:EDQNM}.
\item We compute energy fluxes using $T_{nm}^{YX}$, and find them to be
the same as that computed in Section \ref{sub:Nonhelical-nonAlfvenic-MHD-flux}.
Hence both the results are consistent with each other.
\end{enumerate}
After the above discussion on nonhelical MHD, we move to helical MHD.

\subsubsection{Helical Contributions:}

Now we present computation of shell-to-shell energy transfer for helical
MHD $(H_{M}\ne0,H_{K}\ne0)$ \cite{MKV:MHD_Helical}. To simplify
the equation, we consider only nonAlfvénic fluctuations ($\sigma_{c}=0$).
In order to compare the helical contributions with nonlocal ones,
we have chosen $r_{A}=1,r_{K}=0.1,r_{M}=-0.1$. These are also the
typical parameters chosen in numerical simulations. For these parameters,
Kolmogorov's constant $K^{u}=0.78$ (Section \ref{sub:Helical-nonAlfvenic-MHD-flux}).
In Figure we have plotted $T_{nm}^{uu}/\Pi$ vs $n-m$. %
\begin{figure}
\includegraphics[bb= 100 250 530 560]{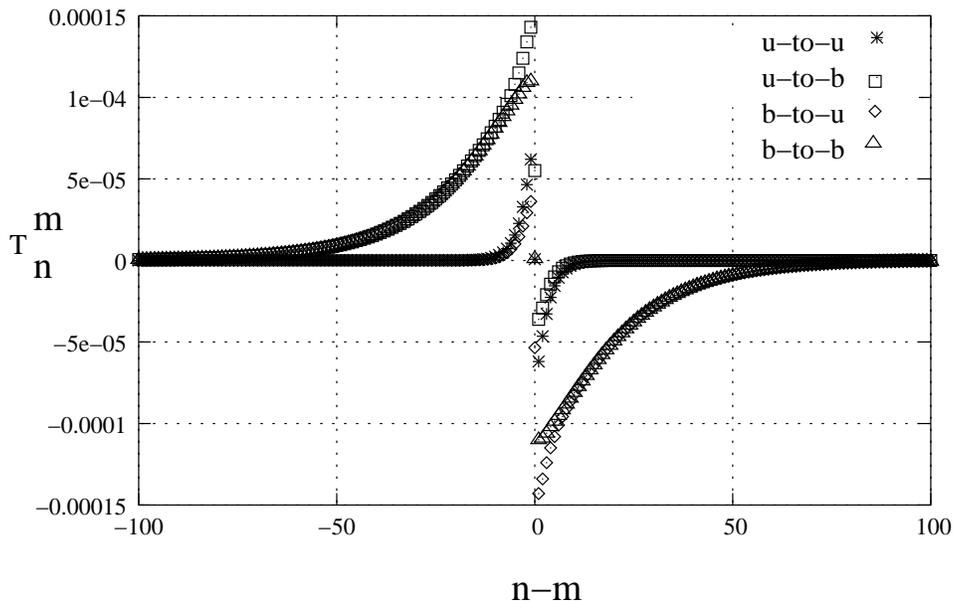}

\caption{\label{Fig:T-MHD-helical-theory} Theoretically calculated values
of \emph{helical contributions to $T_{nm}^{YX}/\Pi$} vs. $n-m$ in
helical MHD with $r_{A}=1,r_{K}=0.1,r_{M}=-0.1$ and $\sigma_{c}=0$.}
\end{figure}
Our results of helical shell-to-shell transfers are given below:

\begin{enumerate}
\item Comparison of Fig. \ref{Fig:T-MHD-helical-theory} with Fig. \ref{Fig:T-MHD-theory}
($r_{A}=1$) shows that helical transfers are order-of-magnitude lower
that nonhelical ones for the parameters chosen here ($r_{A}=1,r_{K}=0.1,r_{M}=-0.1$).
When the helicities are large $(r_{K}\rightarrow1,r_{M}\rightarrow-1)$,
then the helical and nonhelical values become comparable.
\item All the helical contributions are negative for $n>m$, and positive
for $n<m$. Hence, helical transfers are from larger wavenumbers to
smaller wavenumbers. This is consistent with the inverse cascade of
energy due to helical contributions, as discussed in \ref{sub:Helical-nonAlfvenic-MHD-flux}.
\item We find that $T_{nm-helical}^{ub}$ and $T_{nm-helical}^{bb}$ is
significant positive values for $-50<n-m\le0$. This signals a nonlocal
$b$-to-$b$ and $u$-to-$b$ inverse energy transfers. Hence, helicity
induces nonlocal energy transfer between wavenumber shells. This is
in agreement with Pouquet et al.'s result \cite{Pouq:EDQNM} that
{}``residual helicity'' induces growth of large-scale magnetic field
by nonlocal interactions. 
\end{enumerate}
With this we conclude our discussion on shell-to-shell energy transfer
rates in MHD turbulence.

\subsection{EDQNM Calculation of MHD Turbulence \label{sub:EDQNM-Calculation}}

Eddy-damped quasi-normal Markovian (EDQNM) calculation of turbulence
is very similar to the field-theoretic calculation of energy evolution.
This scheme was first invented by Orszag \cite{Orsz:Rev} for Fluid
turbulence. Pouquet et al. \cite{Pouq:EDQNM} were the first to apply
EDQNM scheme to MHD turbulence. Grappin et al. \cite{Grap82,Grap83},
Pouquet \cite{Pouq:EDQNM2D}, Ishizawa and Hattori \cite{Ishi:EDQNM}
and others have performed further analysis in this area. In the following
discussion we will describe some of the key results.

In 1976 Pouquet et al. \cite{Pouq:EDQNM} constructed a scheme to
compute evolution of MHD turbulence. See Pouquet et al. \cite{Pouq:EDQNM}
for details. Here Navier-Stokes or the MHD equations are symbolically
written as \[
\left(\frac{d}{dt}+\zeta k^{2}\right)X(\mathbf{k},t)=\sum_{\mathbf{p+q=k}}X(\mathbf{p},t)X(\mathbf{q},t),\]
where $X$ stands for the fields \textbf{u} or \textbf{b}, $X(\mathbf{p},t)X(\mathbf{q},t)$
represents all the nonlinear terms, and $\zeta$ is the dissipation
coefficient $(\nu$ or $\eta$). The evolution of second and third
moment would be

\begin{eqnarray}
\left(\frac{d}{dt}+2\zeta k^{2}\right)\left\langle X(\mathbf{k},t)X(\mathbf{-k},t)\right\rangle  & = & \sum_{\mathbf{p+q=k}}\left\langle X(\mathbf{-k},t)X(\mathbf{p},t)X(\mathbf{q},t)\right\rangle \label{eq:EDQNM-C}\\
\left(\frac{d}{dt}+\zeta(k^{2}+p^{2}+q^{2})\right)\left\langle X(\mathbf{-k},t)X(\mathbf{p},t)X(\mathbf{q},t)\right\rangle  & = & \sum_{\mathbf{p+q+r+s=0}}\left\langle X(\mathbf{q},t)X(\mathbf{p},t)X(\mathbf{r},t)X(\mathbf{s},t)\right\rangle \nonumber \end{eqnarray}
If $X$ were Gaussian, third-order moment would vanish. However, quasi-normal
approximation gives nonzero triple correlation; here we replace $\left\langle XXXX\right\rangle $
by its Gaussian value, which is a sum of products of second-order
moments. Hence, \begin{eqnarray*}
\left\langle X(\mathbf{-k},t)X(\mathbf{p},t)X(\mathbf{q},t)\right\rangle  & = & \int_{0}^{t}d\tau\exp{-\zeta(k^{2}+p^{2}+q^{2})(t-\tau)}\sum_{\mathbf{p+q=k}}\\
 &  & [\left\langle X(\mathbf{q},\tau)X(-\mathbf{q},\tau)\right\rangle \left\langle X(\mathbf{p},\tau)X(\mathbf{p},\tau)\right\rangle +...],\end{eqnarray*}
where $...$ refers to other products of second-order moments. The
substitution of the above in Eq. (\ref{eq:EDQNM-C}) yields a closed
form equation for second-order correlation functions. Orszag \cite{Orsz:Rev}
discovered that the solution of the above equation was plagued by
problems like negative energy. To cure this problem, a suitable linear
relaxation operator of the triple correlation (denoted by $\mu$)
was introduced (Eddy-damped approximation). In addition, it was assumed
that the characteristic evolution time of $\left\langle XX\right\rangle \left\langle XX\right\rangle $
is larger than $\left(\mu_{kpq}+\nu(k^{2}+p^{2}+q^{2})\right)^{-1}$(Markovian
approximation). As a result Pouquet et al. obtained the following
kind of energy evolution equation\begin{equation}
\left(\frac{d}{dt}+2\zeta k^{2}\right)\left\langle X(\mathbf{k},t)X(\mathbf{-k},t)\right\rangle =\int d\mathbf{p}\theta_{kpq}(t)\sum_{\mathbf{p+q=k}}[\left\langle X(\mathbf{q},t)X(-\mathbf{q},t\right\rangle \left\langle X(\mathbf{p},t)X(-\mathbf{p},t\right\rangle +...],\label{eq:EDQNM-final-eqn}\end{equation}
where \[
\theta_{kpq}(t)=\left(1-\exp{-(\mu_{k}+\mu_{p}+\mu_{q})t}\right)/\left(\mu_{k}+\mu_{p}+\mu_{q}\right)\]
with \begin{equation}
\mu_{k}=\left(\nu+\eta\right)k^{2}+C_{s}\left(\int_{0}^{k}dq\left(E^{u}(q)+E^{b}(q)\right)q^{2}\right)^{1/2}+\frac{1}{\sqrt{3}}k\left(2\int_{0}^{k}dqE^{b}(q)\right)^{1/2}.\label{eq:mu-EDQNM}\end{equation}
The first, second, and third terms represent viscous and resistive
dissipation rate, nonlinear eddy-distortion rate, and Alfvén effect
respectively. Pouquet et al. \cite{Pouq:EDQNM} also wrote the equations
for kinetic and magnetic helicities, then they evolved the equations
for appropriate initial spectra and forcing. Note that homogeneity
and isotropy are assumed in EDQNM analysis too.

The right-hand side of Eq. (\ref{eq:EDQNM-final-eqn}) is very similar
to the perturbative expansion of $S(k|p|q)$ (under $t\rightarrow\infty$).
The term $\mu_{k}$ of Eq. (\ref{eq:mu-EDQNM}) is nothing but the
renormalized dissipative parameters. Thus, field-theoretic techniques
for turbulence is quite similar to EDQNM calculation. There is a bit
of difference however. In field-theory, we typically compute asymptotic
energy fluxes in the inertial range. On the contrary, energy is numerically
evolved in EDQNM calculations. 

To obtain insights into the dynamics of turbulence, Pouquet et al.
\cite{Pouq:EDQNM} computed the contributions of local and nonlocal
mode interactions. In their convention local meant triads whose largest
wavenumber is less than double of the smallest wavenumber. In a triad
$(k,p,q)$ with $k\leq p\leq q$, a locality parameter $a$ is defined
using inequalities $k\geq p/a$ and $q\leq ap$. Also note that Pouquet
et al.'s flow is nonAlfvénic ($\sigma_{c}=0$).

The main results of Pouquet et al.'s \cite{Pouq:EDQNM} are as follows:

\subsubsection{Pouquet et al. on Nonhelical Flows ($H^{M}=H^{K}=0$)}

In these flows an inertial-range develops with a cascade of energy
to small scales. To the lowest order, the energy spectra was -3/2
 powerlaw with an equipartition of kinetic and magnetic energy. There
was a slight excess of magnetic energy with spectral index equals
to $-2$. 

Pouquet et al. studied the local and nonlocal interactions carefully.
Local interactions cause the energy cascade, but the nonlocal ones
lead to an equipartition of kinetic and magnetic energies. They obtained
the following evolution equations for energies:

\[
\partial_{t}E_{k}^{K}|_{nl}=-k\Gamma_{k}\left(E_{k}^{K}-E_{k}^{M}\right),\,\,\,\,\,\partial_{t}E_{k}^{M}|_{nl}=k\Gamma_{k}\left(E_{k}^{K}-E_{k}^{M}\right),\]
where $\Gamma_{k}\sim(E^{M})^{1/2}\sim C_{A}$. Here $nl$ stands
for nonlocal effect. The above equations clearly indicate that kinetic
and magnetic energy get equipartitioned. 

Note that equipartition of kinetic and magnetic energy is also borne
out in our field-theoretic calculation (based on shell-to-shell energy
transfer). However, field-theoretic calculation shows that nonhelical
MHD has predominantly local energy transfer.

\subsubsection{Pouquet et al. on Helical Flows}

When kinetic helicity is injected, an inverse cascade of magnetic
helicity is obtained leading to the appearance of magnetic energy
and helicity at larger scales. At smaller wavenumbers magnetic helicity
converges to a quasi-stationary spectrum with spectral index of -2
. Pouquet et al. derived the following evolution equations:\[
\partial_{t}E_{k}^{M}|_{nl}=\alpha_{k}^{R}k^{2}H_{k}^{M},\,\,\,\,\,\,\,\partial_{t}H_{k}^{M}|_{nl}=\alpha_{k}^{R}E_{k}^{M},\]
with\[
\alpha_{k}^{R}=-\frac{4}{3}\int_{k/a}^{\infty}\theta_{kpq}\left(H_{q}^{K}-q^{2}H_{q}^{M}\right)dq.\]
$\alpha^{R}$ is called residual helicity. Pouquet et al. provide
the following argument for magnetic energy growth at smaller wavenumbers.
They argue that the inverse-cascade process results from the competition
between helicity and Alfvén effect. The residual helicity in the energy
range (say $k\sim k_{E}$) induces a growth of magnetic energy and
helicity at smaller wavenumber, say at $k\sim k_{E}/2,$ due to helicity
effect. However, growth of helicity near $k_{E}/2$ tends to reduce
residual helicity due to Alfvén effect. As a combined effect, the
inverse cascade advances further to smaller wavenumbers.

Our field-theoretic calculation predicts inverse magnetic-energy cascade
due to helicity. The dependence of growth rate of magnetic energy
on $H_{K,M}$ are qualitatively similar, however, there are quantitative
differences (see Eq. {[}\ref{eqn:rM_dependence}{]}). Our field-theory
calculation show \emph{nonlocal} energy transfer for helical MHD similar
to those obtained in EDQNM calculations. However, the present field-theoretic
calculation cannot take into account helicities with both signs; this
feature needs further examination. 

Brandenburg \cite{Bran:Alpha} studied the above process using direct
numerical simulation. His results are in qualitative agreement with
the EDQNM calculations.

\subsubsection{Grappin et al. on Alfvénic MHD flows}

Grappin et al. \cite{Grap82,Grap83} applied EDQNM scheme to Alfvénic
MHD ($\sigma_{c}\ne0$). They claimed that the spectral exponents
deviate strongly from KID's 3/2 exponent for strongly correlated flows
($m^{+}\rightarrow3$ and $m^{-}\rightarrow0$). Also refer to Section
\ref{sub:Grappin-et-al.-Phenomenology} for some of the phenomenological
arguments of Grappin et al.

Let us compare Grappin et al.'s energy evolution equation (Table 2
of \cite{Grap82}) with our field-theoretic analysis of Alfvénic MHD
(see Eq. {[}\ref{eq:S-Alfvenic}{]}). Clearly, Grappin et al.'s relaxation
time-scale is much more simplified, and all terms of expansion are
not included. Also, choice of KID's 3/2 powerlaw for energy spectrum
is erroneous. These assumptions lead to inconsistent conclusions,
which do not appear to agree with the numerical results and the solar
wind observations.

\subsubsection{EDQNM for 2D MHD Flows}

Pouquet \cite{Pouq:EDQNM2D} applied EDQNM scheme to 2D MHD turbulence.
She found a forward energy cascade to small scales, but an inverse
cascade of squared magnetic potential. She also claimed that small-scale
magnetic field acts like a negative eddy viscosity on large-scale
magnetic field.

Ishizawa and Hattori \cite{Ishi:EDQNM} also performed EDQNM calculation
on 2D MHD and deduced that the eddy diffusive parameters $\nu^{uu}<0,$
$\nu^{ub}>0$, and $\nu^{bu}<0$ (see Section \ref{sub:Nonhelical-nonAlfvenic-MHD}
for definition). However, $\nu^{bb}$ is positive if $E^{b}(p)$ decays
faster than $p^{-1}$ for large $p$, which would be the case for
Kolmogorov-like flows. The above results are consistent with Dar et
al.'s \cite{Dar:flux} numerical findings for 2D MHD. Thus Ishizawa
and Hattori's \cite{Ishi:EDQNM} and Dar et al. \cite{Dar:flux} results
that $\nu^{bb}>0$ are inconsistent with the Pouquet's above conclusions.

Here we close our discussion on EDQNM and energy fluxes. In the next
section we will discuss spectral properties of anisotropic MHD turbulence.

\section{Field Theory of Anisotropic MHD Turbulence \label{sec:Field-Theor-of-anisotropic}}

In Section \ref{sec:MHD-Turbulence-Models} we had a preliminary discussion
on anisotropy in MHD turbulence. In this subsection, we will apply
field-theoretic techniques to anisotropic turbulence. The main results
in this area are (1) Galtier et al.'s weak turbulence analysis where
$E(k)\propto k^{-2}$, and (2) Goldreich and Sridhar's theory of strong
turbulence where \[
E(k_{\perp},k_{||})\sim\Pi^{2/3}k_{\perp}^{-10/3}g\left(k_{||}/k_{\perp}^{2/3}\right).\]
Here we will describe their work. For consistency and saving space,
we have reworked their calculation in our notation. In fluid turbulence,
Carati and Brenig \cite{Cara:AnisotropyRG} applied renormalization-group
method for anisotropic flows.

\subsection{Galtier et al.'s weak turbulence analysis: \label{sub:Galtiers-et-al.'s analytic}}

We start with Eq. (\ref{eqn:flux_zpm}). To first order $\left\langle S^{\pm}(\mathbf{k'}|\mathbf{p}|\mathbf{q})\right\rangle $
has many terms (see Eq. {[}\ref{eq:S-Alfvenic}{]}). However, we take
$\left\langle |z^{+}|^{2}\right\rangle =\left\langle |z^{-}|^{2}\right\rangle $
and $\left\langle \mathbf{z}^{+}(\mathbf{k})\cdot\mathbf{z}^{-}(\mathbf{k})\right\rangle =0$,
which yields the following two nonzero terms:\begin{eqnarray}
\left\langle S^{+}(\mathbf{k'|p|q})\right\rangle  & = & -\Im\int_{\infty}^{t}dt'\int d\mathbf{p'}d\mathbf{q'}[\nonumber \\
 &  & k'_{i}(-iM_{jab}(\mathbf{k'}))G^{++}(k',t-t')\left\langle z_{i}^{-}(\mathbf{q},t)z_{a}^{-}(\mathbf{q}',t')\right\rangle \left\langle z_{j}^{+}(\mathbf{p},t)z_{b}^{+}(\mathbf{p}',t')\right\rangle \nonumber \\
 &  & +k'_{i}(-iM_{jab}(\mathbf{p}))G^{++}(p,t-t')\left\langle z_{i}^{-}(\mathbf{q},t)z_{a}^{-}(\mathbf{q}',t')\right\rangle \left\langle z_{j}^{+}(\mathbf{k'},t)z_{b}^{+}(\mathbf{p}',t')\right\rangle ]\label{eq:S-anisotropic}\end{eqnarray}
Note that $\mathbf{k'}=-\mathbf{k}$, and $\left\langle S^{+}(\mathbf{k'}|\mathbf{p}|\mathbf{q})\right\rangle =\left\langle S^{-}(\mathbf{k'}|\mathbf{p}|\mathbf{q})\right\rangle $.

Because of Alfvénic nature of fluctuations, the time dependence of
Green's function and correlation function will be\begin{eqnarray*}
G^{\pm\pm}(\mathbf{k},t-t') & = & \theta{(t-t')}\exp{[\pm i\mathbf{k}\cdot\mathbf{B}_{0}(t-t')]},\\
\left\langle z_{i}^{\pm}(\mathbf{k},t)z_{j}^{\pm}(\mathbf{k}',t')\right\rangle  & = & \theta{(t-t')}C_{ij}^{\pm\pm}(\mathbf{k},t,t)\exp{[\pm i\mathbf{k}\cdot\mathbf{B}_{0}(t-t')]}.\end{eqnarray*}
The anisotropic correlation correlations $C_{ij}^{\pm\pm}(\mathbf{k},t,t)$
is written as\begin{equation}
C_{ij}^{\pm\pm}(\mathbf{k},t,t)=\left(2\pi\right)^{d}\delta(\mathbf{k+k'})\left[P_{ij}(\mathbf{k})C_{1}(k)+P_{ij}^{'}(\mathbf{k'},\mathbf{n})C_{2}(k)\right]\label{eq:C-anisotropic}\end{equation}
with

\begin{equation}
P'_{ij}(\mathbf{k},\mathbf{n})=\left(n_{i}-\frac{\mathbf{n}\cdot\mathbf{k}}{k^{2}}k_{i}\right)\left(n_{j}-\frac{\mathbf{n}\cdot\mathbf{k}}{k^{2}}k_{j}\right)\label{eq:P'ij}\end{equation}
Here $\mathbf{n}$ is the unit vector along the mean magnetic field.
Along $\mathbf{t}_{1}$ and $\mathbf{t}_{2}$ of Fig. \ref{Fig:Alfven},
the correlation functions are $C_{11}=C_{1}(k)+C_{2}(k)\sin^{2}{\theta}$
and $C_{22}=C_{1}(k)$ respectively. These functions are also called
poloidal and toroidal correlations respectively, and they correspond
to Galtier et al.'s functions $\Phi$ and $\Psi$. The substitutions
of the above expressions in Eq. (\ref{eq:S-anisotropic}) yields $\left\langle S^{+}(\mathbf{k'}|\mathbf{p}|\mathbf{q})\right\rangle $
in terms of $C_{1,2}(k)$. 

The $dt'$ integral of Eq. (\ref{eq:S-anisotropic}) is\begin{eqnarray}
\int_{-\infty}^{t}dt'\theta{(t-t')}\exp{[i(\mathbb{-\mathbf{k}+\mathbf{p}-\mathbf{q})}\cdot\mathbf{B}_{0}(t-t')]} & = & \frac{1-\exp{i(-q_{||}B_{0}+i\epsilon)t}}{i(-2q_{||}B_{0}+i\epsilon)}\nonumber \\
 &  & =i\Pr\frac{1}{2q_{||}B_{0}}+\pi\delta(2q_{||}B_{0}),\label{eq:int-dt'}\end{eqnarray}
where $\Pr$ stand for Principal value, and $\epsilon>0$. In the
above calculation we have taken $t\rightarrow\infty$ limit. Note
that the above integral makes sense only when $\epsilon$ is nonzero.
When $dt'$ integral is substituted in Eq. (\ref{eq:S-anisotropic}),
$\left\langle S^{+}(\mathbf{k'}|\mathbf{p}|\mathbf{q})\right\rangle $
will be nonzero through $\pi\delta(2q_{||}B_{0})$ of Eq. (\ref{eq:int-dt'}).
The term $\delta(q_{||})$ in $\left\langle S^{+}(\mathbf{k'}|\mathbf{p}|\mathbf{q})\right\rangle $
indicates that the energy transfer in weak MHD takes place in a plane
formed by $\mathbf{p}_{\perp}$ and $\mathbf{k}_{\perp}$, as seen
in Fig. \ref{Fig:anisotropy}. Energy transfer across the planes are
not allowed in weak MHD turbulence.

Galtier et al. \cite{Galt:Weak} correct KID phenomenological model,
and Sridhar and Goldreich's argument discussed in Section \ref{sec:MHD-Turbulence-Models}.
The $dt'$ integral provides inverse of the effective timescale for
the nonlinear interaction. KID's model assumes it to be $(kB_{0})^{-1},$
differing from the corrected expression $\delta(q_{||}B_{0})$ of
Galtier et al. If we wrongly set $\epsilon$ to zero in Eq. (\ref{eq:int-dt'}),
the $dt'$ integral will be zero, and from Eq. (\ref{eq:S-anisotropic})
$\left\langle S^{+}(\mathbf{k'}|\mathbf{p}|\mathbf{q})\right\rangle $
will become zero; this was the basic argument of Sridhar and Goldreich's
\cite{Srid1} claim that the triad interaction is absent in weak MHD
turbulence. Galtier et al. \cite{Galt:Weak} modified Sridhar and
Goldreich's argument by correctly performing the $dt'$ integral. 

Galtier et al. \cite{Galt:Weak} also observed that since the energy
transfer is in a plane perpendicular to the mean magnetic field, the
perpendicular components of interacting wavenumbers are much larger
than their corresponding parallel component. Geometrically, the wavenumber
space is pancake-like with a spread along $k_{\perp}$ ($k_{||}/k_{\perp}\rightarrow0$)
. This simplifies Eq. (\ref{eq:P'ij}) to\[
P'_{ij}(\mathbf{n,k})=n_{i}n_{j},\]
and yields\begin{equation}
\left\langle S^{+}(\mathbf{k'|p|q})\right\rangle =\frac{\pi\delta(q_{||})}{2B_{0}}k_{\perp}^{2}(1-y^{2})\left[1+z^{2}+C_{2}(p)/C_{1}(q)\right]C_{1}(q)\left[C_{1}(p_{\perp})-C_{1}(k_{\perp})\right].\label{eq:S-weak-final}\end{equation}
 Now we substitute the above expression in Eq. (\ref{eqn:flux_zpm})
and obtain the following expression for the flux: \begin{eqnarray*}
\Pi & \sim & \int d\mathbf{k}\int d\mathbf{q}\frac{\pi\delta(q_{||})}{2B_{0}}k_{\perp}^{2}(1-y^{2})\left[1+z^{2}+C_{2}(p)/C_{1}(q)\right]C_{1}(\mathbf{q})\left[C_{1}(\mathbf{p})-C_{1}(\mathbf{k})\right]\\
 & = & k_{||}\left\{ \int d\mathbf{k}_{\perp}\int d\mathbf{q}_{\perp}dq_{||}\frac{\pi\delta(q_{||})}{2B_{0}}k_{\perp}^{2}(1-y^{2})\left[1+z^{2}+C_{2}(p)/C_{1}(q)\right]C_{1}(\mathbf{q})\left[C_{1}(\mathbf{p})-C_{1}(\mathbf{k})\right]\right\} ,\end{eqnarray*}
The above energy transfer process has cylindrical symmetry, and the
term within the curly bracket represents the energy flux passing through
circles in the perpendicular planes (see Fig. \ref{Fig:anisotropy}).
Under steady state, the energy flux passing through any circle should
be independent of its radius. This immediately implies that\[
\Pi\sim k_{||}k_{\perp}^{6}C_{1}^{2}(\mathbf{k})/B_{0}.\]
The correlation functions $C_{1,2}(\mathbf{k})$ is essentially cylindrical,
hence $C_{1,2}(k_{\perp},k_{||})=E_{1,2}(k_{\perp})/(2\pi k_{\perp}k_{||})$.
Therefore,\begin{equation}
E_{1,2}(k_{\perp})\sim\left(\Pi B_{0}\right)^{1/2}k_{||}^{1/2}k_{\perp}^{-2}.\label{eq:Galtier-E(k)}\end{equation}
This was how Galtier et al. \cite{Galt:Weak} obtained the $k_{\perp}^{-2}$
energy spectrum for weak turbulence. The above derivation differs
from Galtier et al. on one count. Here we have used constancy of flux
rather than applying Zakharov transform. Both these conditions ensure
steady-state turbulence.

Now let us look at the dynamical equation once more. In one Alfvén
timescale, the fractional change in $z_{k_{\perp}}$ induced by collision
is \cite{Srid2}\begin{equation}
\chi\sim\frac{\delta z_{k_{\perp}}}{z_{k_{\perp}}}\sim\frac{k_{\perp}z_{k_{\perp}}}{k_{||}B_{0}}.\label{eq:chi-GS}\end{equation}
When $\chi$ is small (or $z_{k_{\perp}}$ is small), we have weak
turbulence theory. However, when $\delta z_{k_{\perp}}\ge z_{k_{\perp}}(\chi\ge1)$,
the fluctuations become important; this is called strong turbulence
limit, which will be discussed in the next subsection.

\subsection{Goldreich and Sridhar's Theory for Strong Anisotropic MHD Turbulence:
\label{sub:Goldreich-and-Sridhar's analytic}}

Goldreich and Sridhar (GS) \cite{Srid2} have studied MHD turbulence
under strong turbulence limit. From Eq. (\ref{eq:Galtier-E(k)}) we
can derive that $z_{k_{\perp}}\sim k_{\perp}^{-1/2}$. Therefore,
according to GS theory, $\chi$ will become order 1 for large enough
$k_{\perp}$. However, when the energy cascades to higher $k_{\perp}$
($\chi\gg1)$, $k_{||}$ also tends to increase from its initial small
value of $L^{-1}$ because of a {}``nonlinear renormalization of
frequencies'' \cite{Srid2}. Hence, the parameter $\chi$ approaches
unity from both sides with $k_{\perp}z_{k_{\perp}}\sim k_{||}B_{0}$;
this was termed as {}``critical balanced state''. 

For strong turbulence, Goldreich and Sridhar \cite{Srid2} included
a damping term with the following eddy damping rate\[
\eta(\mathbf{k})=\eta_{0}k_{\perp}^{2}\left[k_{||}E(k,t)\right]^{1/2},\]
where $\eta_{0}$ is a dimensionless constant of order unity. Then
they attempted the following anisotropic energy spectrum for the kinetic
equation \[
C(\mathbf{k})=Ak_{\perp}^{-(\mu+\xi)}f(k_{||}/\Lambda k_{\perp}^{\xi}).\]
Here we state Goldreich and Sridhar's result \cite{Srid2} in terms
of energy flux,\begin{eqnarray}
\Pi(k_{0}) & \sim & \int d\mathbf{k'}\int d\mathbf{p}\Im\left[(-i)k_{\perp}^{2}t_{i}(v,w)C(\mathbf{q})(C(\mathbf{p})-C(\mathbf{k}))\frac{1}{-i\omega(\mathbf{k})+\eta(\mathbf{k})}\right]\nonumber \\
 & \sim & \int\int dk_{\perp}dk_{||}dp_{\perp}dp_{||}k_{\perp}^{3}p_{\perp}C(\mathbf{q})(C(\mathbf{p})-C(\mathbf{k}))\frac{1}{\omega}\frac{(\eta(\mathbf{k})/\omega(\mathbf{k}))}{1+(\eta(\mathbf{k})/\omega(\mathbf{k}))^{2}}.\label{eq:Pi-GS}\end{eqnarray}

Since $\omega\sim k_{||}B_{0}$, the constraint that $\eta(\mathbf{k})/\omega(\mathbf{k})$
is dimensionless yields\begin{equation}
\xi=2-\mu/2.\label{eq:xi}\end{equation}
Now constraint that $\Pi(k_{0})$ is independent of $k_{0}$ provides
\begin{equation}
6-3\xi-2\mu=0.\label{eq:mu}\end{equation}
The solution of Eqs. (\ref{eq:xi}, \ref{eq:mu}) is \[
\mu=8/3,\,\,\,\xi=2/3.\]
Therefore, \[
C(\mathbf{k})\sim\Pi^{2/3}k_{\perp}^{-10/3}L^{-1/3}f\left(\frac{k_{||}L^{1/3}}{k_{\perp}^{2/3}}\right).\]
Here $L$ is the large length scale. The factors involving $\Pi$
and $\  L$ have been deduced dimensionally. Note that the exponent
10/3 appears because of $k_{\perp}^{-5/3}/(k_{\perp}k_{||})$.

The damping term $\eta(\mathbf{k})$ has been postulated in GS model.
Verma \cite{MKV:B0_RG} attempted to deduce a similar term using RG
procedure in {}``random mean magnetic field'' (see Section \ref{sub:Mean-Magnetic-Field-RG}).
Extension of Verma's model to anisotropic situation will shed important
insights into the dynamics. The {}``critical balanced state'' in
the inertial range is based on phenomenological arguments; it will
be useful to have analytic understanding of this argument. 

Let us contrast the above conclusions with the earlier results of
Kraichnan \cite{Krai:65} and Iroshnikov \cite{Iros} where effective
time-scale is determined by the mean magnetic field $B_{0}$, and
the energy spectrum is $k^{-3/2}$. Kraichnan's and Iroshnikov's phenomenology
is weak turbulence theory under isotropic situations. This is contradictory
because strong mean magnetic field will create anisotropy. This is
why 3/2 theory is inapplicable to MHD turbulence.

There are many numerical simulations connected to anisotropy in MHD
turbulence. Matthaeus et al. \cite{Matt:AnisotropyPRL} showed that
anisotropy scales linearly with the ratio of fluctuating to total
magnetic field strength $(b/B_{0})$, and reaches the maximum value
for $b/B_{0}\approx1$. Hence, the turbulence will appear almost isotropic
when the fluctuations become of the order of the mean magnetic field.
In another development, Cho et al. \cite{ChoVish:anis,ChoVish:localB}
found that the anisotropy of eddies depended on their size: Along
the {}``local'' magnetic field lines, the smaller eddies are more
elongated than the larger ones, a result consistent with the theoretical
predictions of Goldreich and Sridhar \cite{Srid1,Srid2}.

After studying anisotropic turbulence, we move on to the problem of
generation of magnetic field in MHD turbulence.

\section{Magnetic Field Growth in MHD Turbulence \label{sec:Magnetic-Field-Growth}}

Scientists have been puzzled by the existence of magnetic field in
the Sun, Earth, galaxies, and other astrophysical objects. It is believed
that any cosmic body that is fluid and rotating possess a magnetic
field. It is also known that the cosmic magnetic field is neither
due to some permanent magnet, not due to any remnants of the past.
Scientists believe that the generation of magnetic field is due to
the motion of the electrically conducting fluid within these bodies
such that the flow generated by the inductive action generates just
those current required to provide the same magnetic field. This is
positive feedback or {}``bootstrap'' effect (until some sort of
saturation occurs), technically known as {}``dynamo'' mechanism.
Larmer \cite{Larm} was the first scientist to suggest this mechanism
in 1919.

A quantitative implementation of the above idea is very hard and still
unsolved because of the nonlinear and dynamic interactions between
many scales involved. There are many important results in this challenging
area, but all of them cannot be discussed here due to lack of space.
In this paper we will focus on some of the recent results on dynamic
dynamo theory. For detailed discussion, refer to books by Moffatt
\cite{Moff:book}, Krause and Rädler \cite{Krau:book}, and recent
review article by Gilbert \cite{Gilb:inbook}, and Brandenburg and
Subramanian \cite{Bran:PR}. The statements of some of the main results
in this area are listed below.

\begin{enumerate}
\item Larmer (1919) \cite{Larm}: He was first to suggest that the self-generation
of magnetic field in cosmic bodies may be possible by bootstrap mechanism:
magnetic field driving currents, and then currents driving the magnetic
field.
\item Cowling (1934) \cite{Cowl:book}: The above idea of Larmer was shaken
by Cowling who showed that steady axisymmetric magnetic field could
not be maintained by axisymmetric motions. The above statement is
called {}``anti-dynamo'' theorem. It has been shown that dynamo
action is absent in two-dimension flows and other geometries as well.
Therefore, for dynamo action, the field and flow have to be sufficiently
complicated, breaking certain symmetries.
\item Elsässer (1946) \cite{Elsa:Dynamo}: He considered conducting fluid
within a rigid spherical boundary with a non-axisymmetric velocity
field. He emphasized the importance of differential rotation and non-axisymmetric
motion for dynamo action.
\item Parker (1955) \cite{Park:Dynamo}: He showed that in the Sun, buoyantly
rising or descending fluid will generate a helical flow under the
influence of Coriolis force. Helicity and differential rotation in
a star can produce both toroidal and poloidal magnetic field.
\item Steenbeck, Krause and Rädler (1966) \cite{Stee}, Braginskii (1964)
\cite{Brag:Dynamo1,Brag:Dynamo2}: They separated the fields into
two part, the mean and and the turbulent, using scale separation.
The evolution of the mean magnetic field was expressed in terms of
mean EMF obtained by averaging the turbulent fields. In this model,
known as \emph{kinematic dynamo,} the \emph{}random velocity field
generates magnetic field, but the {}``back-reaction'' of magnetic
field on velocity field was ignored. Here the growth rate of magnetic
field is characterized by a parameter called {}``alpha'' parameter,
which is found to be proportional to kinetic helicity. See Section
\ref{sub:Steenbeck-et-al.'s alpha}.
\item Pouquet et al. (1976) \cite{Pouq:EDQNM}: They solved full MHD equations
using EDQNM approximation, hence keeping the effect of back-reaction
of magnetic field on velocity field. Thus, their model is dynamic.
Pouquet et al. found that the growth of the magnetic field is induced
by {}``residual helicity'', which is the difference of kinetic helicity
and magnetic helicity.
\item Kulsrud and Anderson (1992) \cite{Kuls1}: They solved the equation
for energy spectrum when kinetic energy dominates the magnetic energy
(kinematic dynamo). They claimed that the small-scale magnetic energy
grows very fast, and get dissipated by Joule heating. This process
prevents the growth of large-scale magnetic field.
\item Chou \cite{Chou:num} and Recent Numerical Simulations ($\sim2000$)
: Chow and others have performed direct numerical simulations of dynamo
like situations, and studied the growth of magnetic field. For small-scale
seed magnetic field, the numerical results are in agreement with those
of Kulsrud and Anderson in early phase, but differ widely at later
times. For large-scale seed magnetic field, the magnetic energy grows
with the time-scale of the largest eddy.
\item Brandenburg (2001) \cite{Bran:Alpha}: Brandenburg investigated the
role of magnetic and kinetic helicity in dynamo mechanism. He found
a buildup of negative magnetic helicity and magnetic energy at large-scales.
He has also studied the fluxes of these quantities.
\item Recent theoretical Development ($\sim2000$): Field et al. \cite{Fiel:Dynamo},
Chou \cite{Chou:theo}, Schekochihin et al. \cite{Maro:DynamoNlin}
and Blackman \cite{Blac:Rev_Dynamo} have constructed theoretical
models of dynamics dynamo, and studied their nonlinear evolution and
saturation mechanisms. Verma \cite{MKV:MHD_Helical} used energy fluxes
in nonhelical and helical MHD to construct a dynamic model.
\end{enumerate}
The items (6,8,9,10) are based on dynamic models. 

In dynamo research, there are calculations of magnetic field growth
in specific geometry of interest, e. g., solving MHD equations in
a spherical shell to mimic solar dynamo. In addition there are papers
addressing fundamental issues (e.g., role of helicity), which are
applicable to all geometries. Most of the calculations of the later
type assume turbulence to be homogeneous and isotropic, and use turbulence
models for predictions. This line of thinking is valid at intermediate
scales of the system, and expected to provide insights into the dynamics
of dynamo. In this paper we will focus on calculations of the later
type.

We divide our discussion in this section on two major parts: Kinematic
dynamo, and Dynamic dynamo.

\subsection{Kinematic Dynamo:}

In MHD velocity and magnetic field affect each other. The early models
of dynamo simplified the dynamics by assuming that a fully-developed
turbulent velocity field amplifies a weak magnetic field, and the
weak magnetic field does not back-react to modify the velocity field.
This assumption is called \emph{Kinematic approximation}, and the
dynamo is called \emph{kinematic dynamo}. In this subsection, we discuss
kinematic dynamo models of Steenbeck et al. \cite{Stee} and Kulsrud
and Anderson's \cite{Kuls1}. Note that kinematic approximation breaks
down when the magnetic field has grown to sufficiently large value.

\subsubsection{Steenbeck et al.'s model for $\alpha$-effect \label{sub:Steenbeck-et-al.'s alpha}}

Steenbeck et al. \cite{Stee} separated the magnetic field into two
parts: $\bar{\mathbf{B}}$ on large scale $L$, and $\mathbf{b}$
at small scale $l$ ($\mathbf{B}=\bar{\mathbf{B}}+\mathbf{b})$, and
assumed that $l\ll L$. They provided a formula for the growth rate
of $\bar{\mathbf{B}}$ under the influence of homogeneous and isotropic
random velocity field.

Steenbeck et al. averaged the fields over scales intermediate between
$L$ and $l$; the averages are denoted by $\left\langle .\right\rangle $.
Now the induction equation can be separated into a mean and a fluctuating
part,\begin{eqnarray}
\frac{\partial\bar{\mathbf{B}}}{\partial t} & = & \bigtriangledown\times\bar{\varepsilon}+\eta\bigtriangledown^{2}\bar{\mathbf{B}},\label{eq:B-bar}\\
\frac{\partial\mathbf{b}}{\partial t} & = & \nabla\times\left(\mathbf{u}\times\bar{\mathbf{B}}\right)+\nabla\times\left(\mathbf{u}\times\mathbf{b}-\left\langle \mathbf{u}\times\mathbf{b}\right\rangle \right)+\eta\nabla^{2}\mathbf{b},\label{eq:b}\end{eqnarray}
where the mean electromotive force (EMF) $\bar{\varepsilon}$ is given
by\[
\bar{\varepsilon}=\left\langle \mathbf{u}\times\mathbf{b}\right\rangle .\]
Steenbeck et al. assumed $\mathbf{b}$ to be small, hence neglected
the second term of Eq. (\ref{eq:b}). Eq. (\ref{eq:b}) is linear,
with a source term proportional to $\bar{\mathbf{B}}$. For a given
random velocity field, $\mathbf{b}$ is linear in $\bar{\mathbf{B}}$.
Therefore, the mean EMF will also be linear in $\bar{\mathbf{B}}$,
and is written in the form \[
\bar{\varepsilon}_{i}=\alpha_{ij}\bar{B}_{j}+\beta_{ijk}\partial_{k}\bar{B}_{j}.\]
Here $\alpha_{ij}$ and $\beta_{ijk}$ are pseudo-tensors. For homogeneous,
isotropic, and random $\mathbf{u}(\mathbf{x},t)$ field varying with
time scale $\tau$, it can be shown that \cite{Moff:book}\[
\alpha=-\frac{1}{3}\tau\left\langle \mathbf{u}\cdot\nabla\times\mathbf{u}\right\rangle ,\,\,\,\,\,\beta=\frac{1}{3}\tau\left\langle \left|\mathbf{u}\right|^{2}\right\rangle .\]
See Moffatt \cite{Moff:book}, Krause and Rädler \cite{Krau:book},
and Gilbert \cite{Gilb:inbook} for the growth rate as a function
of $\alpha$ and $\beta$. This model has been used to study the evolution
of large-scale magnetic field in the Sun and other cosmic bodies (see
Gilbert \cite{Gilb:inbook} for details). 

In this kinematic dynamo theory, the magnetic field does not react
back to affect the velocity field. In reality, however, when magnetic
field has grown to some level, it affects the velocity field by Lorentz
force. Therefore, alpha is modified to \[
\alpha=\alpha_{0}\frac{1}{1+c\left|\bar{\mathbf{B}}\right|^{2}/B_{eq}^{2}},\]
where $B_{eq}$ is the saturation value of the magnetic field, and
$c$ is a constant.

Kulsrud and Anderson \cite{Kuls1} studied the evolution of energy
spectrum of $\mathbf{b}$ under the influence of random velocity field
using analytical technique. We will describe their results in the
next subsection.

\subsubsection{Kulsrud and Anderson's Model for the Evolution of Magnetic Energy
Spectrum}

The equations for magnetic energy spectrum were derived in Section
\ref{sub:Digression-to-Infinite-box:flux} as\[
\left(\frac{\partial}{\partial t}+2\eta k^{2}\right)C^{bb}\left(\mathbf{k},t\right)=\frac{2}{\left(d-1\right)\delta\left(\mathbf{k+k'}\right)}\int_{\mathbf{k'+p+q=0}}\frac{d\mathbf{p}}{(2\pi)^{2d}}\left[S^{bu}(\mathbf{k'}|\mathbf{p}|\mathbf{q})+S^{bb}(\mathbf{k'}|\mathbf{p}|\mathbf{q})\right]\]
In Section \ref{sec:analytic-energy} we computed $\left\langle S^{YX}(k,p,q)\right\rangle $
using field-theory technique. Substitution of $S$'s in the above
yields an equation of the following form. \begin{eqnarray}
\left(\frac{\partial}{\partial t}+2\eta k^{2}\right)C^{bb}\left(\mathbf{k},t\right) & = & Const\int dt'\int d\mathbf{p}[T(k,p,q)G^{bb}(k,t-t')C^{bb}(p,t,t')C^{uu}(q,t,t')\nonumber \\
 &  & \,\,\,\,\,\,\,+T(k,p,q)G^{uu}(k,t-t')C^{bb}(p,t,t')C^{bb}(q,t,t')]\label{eq:Cbb(k)-KA}\end{eqnarray}
Kulsrud and Anderson (KA) \cite{Kuls1} made the following assumptions
to simplify the above equation:

\begin{enumerate}
\item the second term of Eq. (\ref{eq:Cbb(k)-KA}) was dropped because $C^{bb}(q)\ll C^{uu}(q)$. 
\item The velocity field was assumed to uncorrelated in time, i.e.,\[
\left\langle u_{i}(\mathbf{k},t)u_{j}(\mathbf{k}',t')\right\rangle =\left[P_{ij}(\mathbf{k})C^{uu}(k)-i\epsilon_{ijl}k_{l}\frac{H_{K}(k)}{k^{2}}\right]\delta(\mathbf{k}+\mathbf{k}')\delta(t-t').\]

\item $q\ll k$, so that the integral of Eq. (\ref{eq:Cbb(k)-KA}) could
be performed analytically. Note that this is an assumption of nonlocality
and scale separation.
\end{enumerate}
Under the above assumptions, KA could reduce the Eq. (\ref{eq:Cbb(k)-KA})
to\begin{equation}
\frac{\partial E^{b}(k,t)}{\partial t}=\int K_{m}(k,p)E^{b}(p,t)dp-2k^{2}\frac{\eta_{T}}{4\pi}E^{b}(k,t)-2k^{2}\frac{\eta}{4\pi}E^{b}(k,t)\label{eq:Eb(k)-KA}\end{equation}
 where\begin{eqnarray*}
K_{m}(k,p) & \sim & k^{4}\int d\theta\sin^{3}{\theta}\left(k^{2}+p^{2}-kp\cos{\theta}\right)\frac{C^{uu}(q)}{q^{2}}\\
\frac{\eta_{T}}{4\pi} & \sim & -\int d\mathbf{q}H_{K}(q)\end{eqnarray*}
with $q=(k^{2}+p^{2}-2kp\cos{\theta})^{1/2}$.

Using the definition that the total magnetic energy $E^{b}=\int E^{b}(k)dk$,
KA deduced that\[
\frac{\partial E^{b}(t)}{\partial t}=2\gamma E^{b}(t),\]
where \[
\gamma\sim-\int d\mathbf{q}H_{K}(q).\]
By assuming $q\ll k$, KA expanded $p$ near $k$, and obtained (by
integrating by parts)\[
\frac{\partial E^{b}(k,t)}{\partial t}=\frac{\gamma}{5}\left(k^{2}\frac{\partial E^{b}(k)}{\partial k^{2}}-2\frac{\partial E^{b}(k)}{\partial k}+6E^{b}(k)\right)-2k^{2}\frac{\eta}{4\pi}E^{b}(k,t).\]
From the above equation KA deduced that\[
E^{b}(k)\sim k^{3/2}f(k/k_{R})\exp{[(3/4)\gamma t]}\]
for $k$ much less than the resistive wavenumber $k_{R}\approx(4\pi\gamma/\eta)^{1/2}$.
The fluctuating magnetic energy will flow to small scales, and then
to $k_{R}$, and get dissipated by Joule heating. Thus, according
to KA, magnetic energy at large length scale does not build up. 

Chou \cite{Chou:num} performed numerical simulation to test KA's
predictions. He finds that in early phase, $E(k)\propto k^{3/2}$,
and that energy grows exponentially in time, thus verifying KA's model
prediction. However, at later phase of evolution, the magnetic field
back-reacts on the velocity field. Consequently, the energy growth
saturates, and the energy spectrum also evolves differently from $k^{3/2}$.
Clearly these discrepancies are due to the kinematic approximation
made by KA.

\subsection{Dynamic Dynamo:}

The kinematic approximation described above breaks down when the magnetic
field becomes comparable to the velocity field. In \emph{dynamic dynamos}
the back-reaction of the magnetic field on the velocity field is accounted
for. There are several analytic theories in this area, but the final
word is still awaited. Researchers are trying to understand these
types of dynamo using direct numerical simulations. Here we will present
some of the main results.

\subsubsection{Pouquet et al.'s EDQNM Calculation}

Pouquet et al. \cite{Pouq:EDQNM} solved MHD equations with large-scale
forcing under EDQNM approximation. For details, refer to Section \ref{sub:EDQNM-Calculation}.
Pouquet et al. observed that \emph{for nonhelical flows}, the magnetic
energy first grows at the highest wavenumbers, where equipartition
is obtained. After that the magnetic energy at smaller wavenumbers
start to grow. 

Pouquet et al. analyzed the helical flows by forcing kinetic energy
and helicity at forcing wavenumber. They find that the magnetic helicity
has an inverse cascade, and negative magnetic helicity and magnetic
energy grow at wavenumbers smaller than the forcing wavenumber. 

Pouquet et al. estimated the contributions of helicities to the growth
of magnetic energy, and concluded that\begin{equation}
\alpha\approx\alpha_{u}+\alpha_{b}=\frac{1}{3}\tau\left[-\mathbf{u}\cdot\mathbf{\omega}+\mathbf{b}\cdot\left(\nabla\times\mathbf{b}\right)\right]\label{eq:alpha-Pouq}\end{equation}
where $\tau$ is a typical coherence time of the small-scale magnetic
energy. The second term of the above equation is due to the back-reaction
of magnetic field.

\subsubsection{Direct Numerical Simulation}

\begin{figure}
\includegraphics[scale=0.5,bb=14 14 600 530]{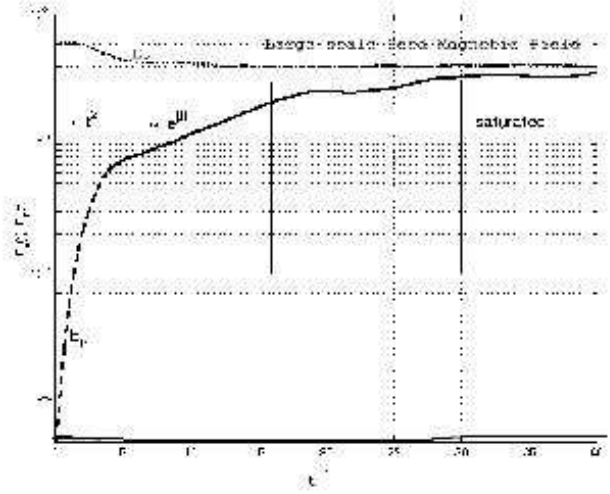}

\caption{\label{Fig:Chou-4} Evolution of kinetic energy ($E^{u}$) and magnetic
energy $(E^{b})$ for initial weak, large-scale seed magnetic field.
Initially $E^{b}$ grows as $t^{2}$, then exponentially, after which
it saturates. Adopted from Chou \cite{Chou:num}.}
\end{figure}

\begin{figure}
\includegraphics[%
  scale=0.5, bb=0 0 600 550]{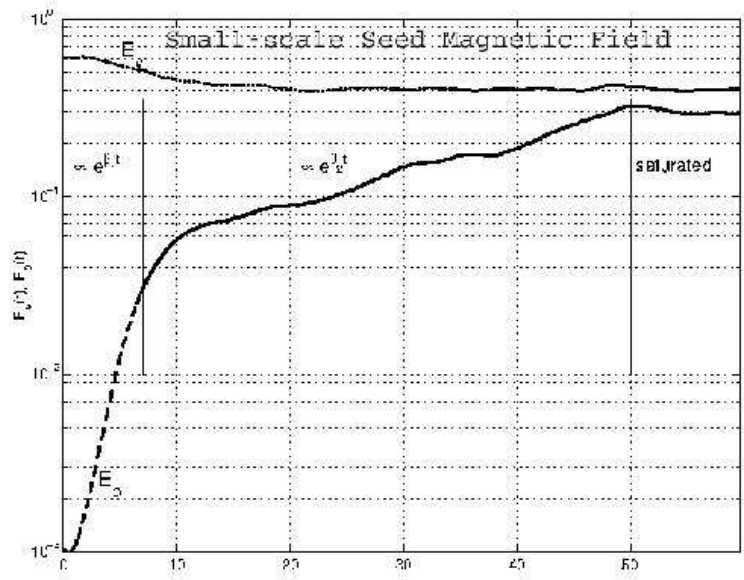}

\caption{\label{Fig:Chou-9}Evolution of kinetic energy ($E^{u}$) and magnetic
energy $(E^{b})$ for initial weak, \textbf{small-scale} seed magnetic
field. $E^{b}$exponentially, then it saturates. Adopted from Chou
\cite{Chou:num}.}
\end{figure}

Chou \cite{Chou:num} performed direct numerical simulation of 3D
incompressible MHD turbulence using pseudo-spectral method (see Section
\ref{sec:Numerical-Investigation-MHD}), and analyzed the growth of
(a) initial weak, large-scale seed magnetic field, and (b) initial
weak, small-scale seed magnetic field. In both the cases the magnetic
energy grows at all scales. For the initial condition of type (a),
the magnetic energy grows as $t^{2}$ for the first few turbulence
eddy turnover times, followed by an exponential growth, in which the
growth time-scale is approximately the large-scale eddy turnover time.
After sometime the magnetic field saturates (see Fig. \ref{Fig:Chou-4}).
 For the initial condition of type (b), initial growth of magnetic
energy is determined by the eddy turnover time of the smallest scale
of turbulence, as predicted by KA, and then by the eddy turnover time
of inertial range modes (See Fig. \ref{Fig:Chou-9}); finally the
growth saturates. %

\begin{figure}
\includegraphics[%
  scale=0.5,bb=0 0 600 550]{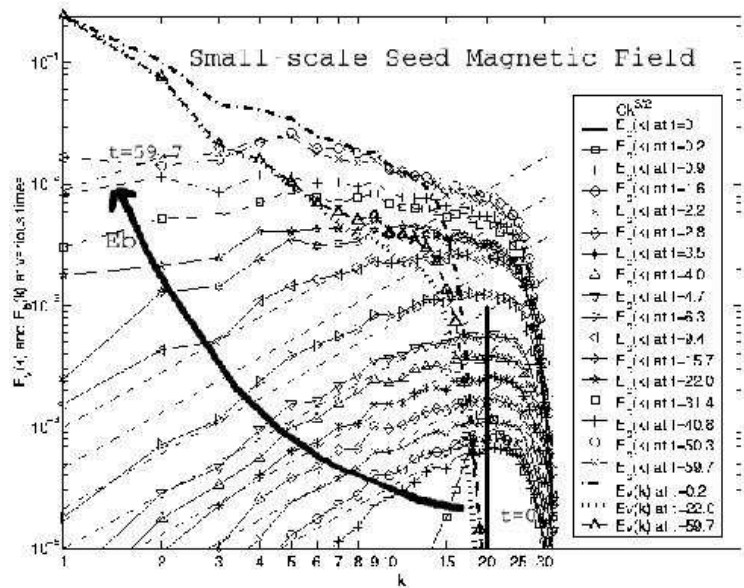}

\caption{\label{Fig:Chou-7} Magnetic and kinetic energy spectrum at various
times. Initial seed magnetic field is concentrated at $k=20$. $E^{b}(k)\propto k^{3/2}$
for $2<t<10$, reminiscent of Kulsrud and Anderson's predictions.
It shifts to flatter and then to Kolmogorov-like spectrum at later
time. Adopted from Chou \cite{Chou:num}.}
\end{figure}

When the initial seed magnetic energy is at narrow bandwidth of large
wavenumbers, the magnetic energy quickly gets spread out, extending
to both larger and smaller wavenumbers. The evolution of energy spectrum
is shown in Fig. \ref{Fig:Chou-7}. In the early phase, the magnetic
energy spectrum is proportional to $k^{3/2}$, confirming KA's predictions.
However, at a later time, the energy spectrum is very different, which
is due to the dynamic aspect of dynamo. %

Recently Cho and Vishniac (CV) \cite{ChoVish:gene} performed numerical
simulation of nonhelical MHD turbulence and arrived at the following
conclusion based on their numerical results. In our language, their
results for large $r_{A}$ can be rephrased as (1) $\Pi_{u>}^{u<}\approx U^{3}$;
(2) $\Pi_{(b<+b>)}^{u<}\approx UB^{2}$; (3) $\Pi_{b<}^{u<}\approx(U-cB)B^{2}$,
where $U$ and $B$ are the large-scale velocity and magnetic field
respectively, and $c$ is a constant. These results are somewhat consistent
with the field-theoretic flux calculations of Verma \cite{MKV:MHD_Flux}.

\subsubsection{Brandenburg's Calculations \label{sub:Brandenburg's-Calculations}}

Brandenburg \cite{Bran:Alpha} performed direct numerical simulation
of compressible MHD (Mach number around 0.1-0.3) on maximum grid of
$128^{3}$. He applied kinetic energy and kinetic helicity forcing
in the wavenumber band $(4.5,5.5)$. He obtained many interesting
results, some of which are given below.

\begin{enumerate}
\item Magnetic helicity grows at small wavenumbers, but it has a negative
sign. Brandenburg explains this phenomena by invoking conservation
of magnetic helicity. For a closed or periodic system the net magnetic
helicity is conserved, except for dissipation at small scales. Thus,
for magnetic field to be helical, it must have equal amount of positive
and negative helicity. The helicity at small scales will get destroyed
by dissipation, while magnetic helicity at large scales will survive
with negative sign.
\item Brandenburg computed the magnetic-helicity flux and found that to
be positive. Note that injection of kinetic helicity induces a flux
of magnetic-helicity (see Eq. {[}\ref{eq:flux-HM-value}{]}).
\item Brandenburg argued that most of the energy input to the large scale
field is from scales near the forcing. He claimed the above process
to be alpha-effect, not an inverse cascade (local). Now the built-up
energy at large scales cascades to neighbouring scales by forward
cascade ($k^{3/2}$ region). Once the large scale fields have grown,
Kolmogorov's direct cascade will take place. Above observations are
illustrated in Fig \ref{Fig:Brand-26}. %
\begin{figure}
\includegraphics[%
  scale=0.5,bb=14 14 554 386]{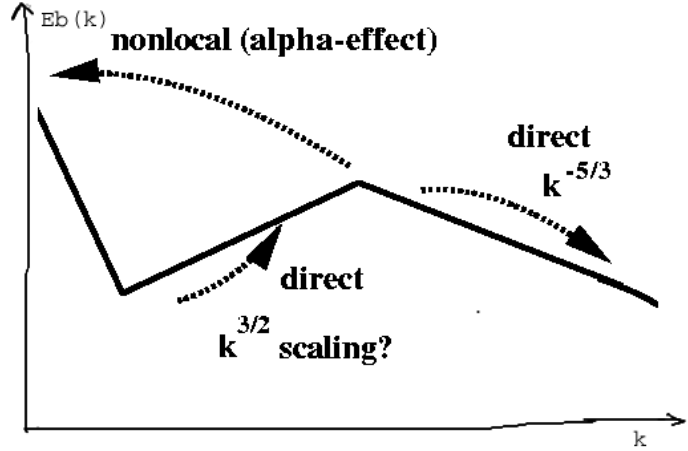}

\caption{\label{Fig:Brand-26} Cascade processes of magnetic energy in Helical
MHD turbulence. Adopted from Brandenburg \cite{Bran:Alpha}.}
\end{figure}

\end{enumerate}
Brandenburg has done further work on open boundaries, and applied
these ideas to solar dynamo. For details, refer to review paper by
Brandenburg \cite{Bran:rev_helicity}. Verma \cite{MKV:MHD_Flux,MKV:MHD_Helical}
has computed energy fluxes and shell-to-shell energy transfers in
MHD turbulence using field-theoretic calculations. Below we show how
Verma's results are consistent with Brandenburg's results.

\subsubsection{Dynamo using Energy Fluxes}

In Section \ref{sec:analytic-energy} we discussed various energy
fluxes and shell-to-shell energy transfer in MHD turbulence. Here
we will apply those ideas to estimate the growth rate of magnetic
energy. These calculations are based on homogeneous and isotropic
turbulence, so the predictions made here are probably valid for galactic
dynamo, or at small scales in solar and planetary dynamo.

Verma has calculated energy flux for both nonhelical and helical MHD.
It has been shown in Section \ref{sub:Steenbeck-et-al.'s alpha} that
when $E^{u}(k)>E^{b}(k)$, kinetic energy gets transferred to magnetic
energy, hence turbulence is not steady. Now we compute the energy
transfer rate to the large-scale magnetic field. In absence of helicity,
the source of energy for the large-scale magnetic field is $\Pi_{b<}^{u<}$.
When helicity is present, the fluxes $\Pi_{b<helical}^{b>}+\Pi_{b<helical}^{u>}$
provide additional energy to the large-scale magnetic field (see Sec.
\ref{sec:analytic-energy}). Hence, the growth rate of magnetic energy
is\begin{equation}
\frac{dE^{b}(t)}{dt}=\Pi_{b<}^{u<}+\Pi_{b<helical}^{b>}+\Pi_{b<helical}^{u>}\label{eq:dynamo_Ebdot}\end{equation}
Since there is no external forcing for large-scale magnetic field,
we assume\[
\Pi_{b<}^{u<}\approx\Pi_{b>}^{b<}+\Pi_{u>}^{b<}.\]
In typical astrophysical situations, magnetic and kinetic helicities
are typically small, with negative magnetic helicity and positive
kinetic helicity. For this combination of helicities, both the helical
fluxes are negative, thus become a source of energy for the growth
for large-scale magnetic field.

Since the magnetic energy starts with a small value (large $r_{A}$
limit), all the fluxes appearing in Eq. (\ref{eq:dynamo_Ebdot}) are
proportional to $r_{A}^{-1}$ {[}cf. Eqs. (\ref{eq:Sbu-helical},
\ref{eq:Sbb-helical}){]}, i.e., \begin{equation}
\Pi_{b<}^{u<}+\Pi_{b<helical}^{b>}+\Pi_{b<helical}^{u>}=c\Pi\frac{E^{b}}{E^{u}}\end{equation}
 where $E^{u}$ is the large-scale kinetic energy, $\Pi$ is the kinetic
energy flux, and $c$ is the constant of order 1. Hence, \begin{eqnarray}
\frac{1}{\Pi}\frac{dE^{b}}{dt} & \approx & \frac{E^{b}}{E^{u}}\end{eqnarray}
 Using $E^{u}=K^{u}\Pi^{2/3}L^{2/3}$, where $L$ is the large length-scale
of the system, we obtain \begin{equation}
\frac{1}{\sqrt{E^{u}}E^{b}}\frac{dE^{b}}{dt}\approx\frac{1}{L(K^{u})^{3/2}}\end{equation}
 We assume that $E^{u}$ does not change appreciably in the early
phase. Therefore, \begin{equation}
E^{b}(t)\approx E^{b}(0)\exp{\left(\frac{\sqrt{E^{u}}}{L(K^{u})^{3/2}}t\right)}\end{equation}
 Hence, the magnetic energy grows exponentially in the early periods,
and the time-scale of growth is of the order of $L(K^{u})^{3/2}/\sqrt{E^{u}}$,
which is the large-scale eddy turnover time. 

In Section \ref{sub:Helical-nonAlfvenic-MHD-flux} we derived the
following expression (Eq. {[}\ref{eq:flux-HM-value}{]}) for the flux
of magnetic helicity:\begin{equation}
\Pi_{H_{M}}(k_{0})=\frac{1}{k_{0}}\left(-dr_{M}+er_{K}\right),\end{equation}
where $r_{K}=H_{K}(k)/(kE^{u}(k))$, $r_{M}=kH_{M}(k)/E^{b}(k)$,
and $d$ and $e$ are positive constants. When kinetic helicity is
forced $(r_{K}>0)$ at forcing wavenumber, magnetic helicity flux
will be positive. But the total magnetic helicity is conserved, so
positive $H_{M}$ will flow to larger wavenumber, and negative $H_{M}$
will flow to smaller wavenumber. The negative $H_{M}$ ($r_{K}<0$)
will further enhance the positive magnetic helicity flux, further
increasing negative $H_{M}$ at lower wavenumbers. The above observation
explains the numerical findings of Brandenburg \cite{Bran:Alpha}
discussed above. 

The negative magnetic helicity described above contributes to the
growth of magnetic energy. Note that for small wavenumber $H_{M}$
and $H_{K}$ have opposite sign, and according to formula (\ref{eqn:rM_dependence})
derived in Section \ref{sub:Helical-nonAlfvenic-MHD-flux}\begin{equation}
\frac{dE^{b}}{dt}=ar_{M}^{2}-br_{M}r_{K},\label{eq:Ebdot}\end{equation}
 ($a$ and $b$ are positive constants) magnetic energy will grow.
This result is consistent with the  numerical simulation of Brandenburg
\cite{Bran:Alpha} and EDQNM calculation of Pouquet et al. \cite{Pouq:EDQNM}.
It is important to contrast the above equation with the growth equation
of Pouquet et al. \cite{Pouq:EDQNM}) (cf. Eq. {[}\ref{eq:alpha-Pouq}{]})
, and test which of the two better describes the dynamo. The direct
numerical simulation of Pouquet and Patterson \cite{Pouq:num} indicate
that $H_{M}$ helps the growth of magnetic energy considerably, but
that is not the case with $H_{K}$ alone. This numerical result is
somewhat inconsistent with results of Pouquet et al.~and others \cite{Pouq:EDQNM}
(Eq. {[}\ref{eq:alpha-Pouq}{]}), but it fits better our formula (\ref{eq:Ebdot})
($dE^{b}/dt=0$ if $r_{M}=0$). Hence, the formula (\ref{eq:Ebdot})
probably is a better model for the dynamically consistent dynamo.
We need more careful numerical tests and analytic investigations to
settle these issues.

In Section \ref{sec:analytic-energy} we studied the shell-to-shell
energy transfer in MHD turbulence assuming powerlaw energy spectrum
for all of wavenumber space. Since magnetic helicity changes sign,
and its spectrum does not follow a powerlaw, the above assumption
is not realistic. However, some of the shell-to-shell energy transfer
results are in tune with Brandenburg's numerical results. For example,
we found that helicity induces energy transfers across distant wavenumber
shells, in the same lines as $\alpha$-effect. More detailed analytic
calculation of shell-to-shell energy transfer is required to better
understand dynamo mechanism.

\subsubsection{Theoretical Dynamic Models}

Field et al. \cite{Fiel:Dynamo} and Chou \cite{Chou:theo} constructed
a theoretical dynamical model of dynamo. They use scale-separation
and perturbative techniques to compute the effects of back-reaction
of magnetic field on $\alpha$. Schekochihin et al. \cite{Maro:DynamoNlin}
and Blackman \cite{Blac:Rev_Dynamo} discussed various models of nonlinear
evolution and saturation for both small-scale and large-scale dynamo.
Basu \cite{Basu:FieldthDynamo} has applied field-theoretic methods
to compute $\alpha$. For details refer to the original papers and
review by Brandenburg and Subramanian \cite{Bran:PR}.

In summary, dynamo theory has come a long way. Early calculations
assumed kinematic approximations. For last fifteen years, there have
been attempts to construct dynamic dynamo models, both numerically
and theoretically. Role of magnetic and kinetic helicity is becoming
clearer. Yet, we are far away from fully-consistent dynamo theory.

\section{Intermittency in MHD Turbulence \label{sec:Intermittency-in-MHD}}

The famous Kolmogorov's turbulence model assumes a constant energy
flux or dissipation rate at all scales, i. e., $\Pi(k)\sim\int dk\nu(k)k^{2}E(k)$
is independent of $k$. The renormalized viscosity $\nu(k)\sim k^{-4/3}$
and $E(k)\sim k^{-5/3}$ are consistent with the above assumption.
Landau \cite{LandFlui:book} pointed out that the dissipation rate,
which is proportional to the square of vorticity, is singular and
quite inhomogeneous. Thus Kolmogorov's theory of turbulence needs
modification. The above phenomena in which strong dissipation is localized
both in time and space is called intermittency.

\subsection{Quantitative Measures of Intermittency}

There are several quantitative measures of intermittency. Consider
the increment of the velocity, or some other field, between two points
separated by $\mathbf{l}$,\[
\delta\mathbf{u}(\mathbf{x,l})=\mathbf{u}(\mathbf{x}+\mathbf{l})-\mathbf{u}(\mathbf{x}).\]
 The longitudinal component of $\delta\mathbf{u}(\mathbf{x,l})$ will
be given by\[
\delta u_{||}(l)=\delta\mathbf{u}(\mathbf{x,l}).\mathbf{l}/l,\]
and the transverse component is $\delta u_{\perp}(l)=\delta\mathbf{u}(\mathbf{x,l})-\delta u_{||}(l)\mathbf{l}/l$.
Here we have assumed homogeneity and isotropy for turbulence, so that
the increment in velocities depend only on $l$, not on $\mathbf{x}$.
Now we define longitudinal and transverse structure functions using\[
S^{(n)}(l)=\left\langle \left[\delta u_{||}(l)\right]^{n}\right\rangle ,\,\,\,\, U^{(n)}(l)=\left\langle \left[\delta u_{\perp}(l)\right]^{n}\right\rangle \]
respectively. The structure function $S^{(n)}(l)$ is expected to
have a power law behaviour for $l$ in the inertial range, \begin{equation}
S^{(n)}(l)=a_{n}l^{\zeta_{n}},\label{eq:Sn-defn}\end{equation}
where $a_{n}$ and $\zeta_{n}$ are universal numbers. The exponents
$\zeta_{n}$ are called the intermittency exponent. 

Moments and probability density function (pdf) are equivalent description
of random variables. Note that if $P(\delta u_{||}(l))$ were gaussian,
i. e., \[
P(\delta u_{||}(l))=\frac{1}{\sigma_{r}\sqrt{{\pi}}}\exp{-\frac{(\delta u_{||}(l))^{2}}{\sigma_{r}^{2}}}\]
then, it is easy to verify that\[
\left\langle (\delta u_{||}(l))^{n}\right\rangle \propto\sigma_{r}^{n}.\]
Kolmogorov's model of turbulence predicts that \[
\sigma_{r}\sim\epsilon^{1/3}l^{1/3},\]
For constant $\epsilon$, we obtain\[
S^{(n)}(l)\propto\epsilon^{n/3}l^{n/3}.\]
Systems with gaussian probability distribution or equivalently $S^{(n)}(l)\propto l^{cn}$
($c=$ constant) are called non-intermittent system. For intermittent
systems, the tails of pdf decays slower that gaussian, and could follow
a powerlaw.

Structure function can be written in terms of local dissipation rate
\cite{Obuk} \[
\delta u_{l}\sim\epsilon_{l}^{1/3}l^{1/3}.\]
Kolmogorov \cite{Kolm:Intermittency} introduced the \emph{refined
similarity hypothesis} relating structure function to $\epsilon_{l}$
as\[
S_{||}^{(n)}(l)=d_{n}\left\langle \epsilon_{l}^{n/3}\right\rangle l^{n/3}.\]
If \[
\left\langle \epsilon_{l}^{n}\right\rangle \sim l^{\mu_{n}},\]
then \[
\zeta_{n}=\frac{n}{3}+\mu_{n/3}.\]
Many researchers have attempted to model $\epsilon_{l}$. 

In any numerical simulation or experiment, the powerlaw range is quite
limited. However, when we plot $S^{(n)}(l)$ vs. $S^{(3)}(l)$, we
obtain a much larger scaling range. This phenomena is called Extended
self-similarity (ESS). Since, $S^{(3)}(l)\propto l$ \cite{K41b},
$\zeta_{n}$ measured using Eq. (\ref{eq:Sn-defn}) or ESS are expected
to be the same.

There have been some ingenious attempts to theoretically compute the
intermittency exponents (e.g., see series of papers by L'vov and Procaccia
\cite{LvovProc:Intermittency}). Yet, this problem is unsolved. There
are several phenomenological models. Even here, phenomenological models
have been better developed for fluid turbulence than MHD turbulence.
We will describe some of them in the following discussion, first for
fluids and then for MHD turbulence.

\subsection{Results on Intermittency in Fluid Turbulence}

In fluid turbulence, the pdf of velocity increment deviates from gaussian
\cite{Fris:book}. In experiments and simulations one finds that $\zeta_{n}$
vs. $n$ is a nonlinear function of $n$. Hence, fluid turbulence
shows intermittency. Note that $\zeta_{2}\approx0.71,$ which yields
a correction of approximately 0.04 to Kolmogorov's spectral index
of 5/3. However, the correction for large $n$ is much more. See Frisch
\cite{Fris:book} for further details. 

Remarkably, starting from Navier-Stokes equation, Kolmogorov \cite{K41b}
obtained an exact relation\[
S_{||}^{(3)}(l)=-\frac{4}{5}\epsilon l\]
 under $\nu\rightarrow0$ limit (also see \cite{Fris:book,LandFlui:book}).
Note that $\epsilon$ is the mean dissipation rate. Unfortunately,
similar relationship could not be derived for other structure functions.
In the following discussion we will discuss some of the prominent
intermittency models for fluid turbulence.

\subsubsection{Kolmogorov's log-normal model}

Obukhov \cite{Obuk} and Kolmogorov \cite{Kolm:Intermittency} claimed
that the dissipation rate in turbulent fluid is log-normal. As a consequence,\[
\zeta_{n}=\frac{n}{3}-\mu\frac{n(n-3)}{18},\]
where \[
\left\langle \epsilon(\mathbf{x})\epsilon(\mathbf{x+l})\right\rangle \sim l^{-\mu}.\]
Numerical simulations and experiments give $\mu\approx0.2.$ 

The predictions of this model agree well with the experimental results
up to $n\approx10$, but fails for higher values of $n$. In Fig.
\ref{Fig:Intermiitency} we have plotted the above $\zeta_{n}$ along
with other model predictions given below. %
\begin{figure}
\includegraphics[%
  scale=0.8,bb= 00 250 550 550]{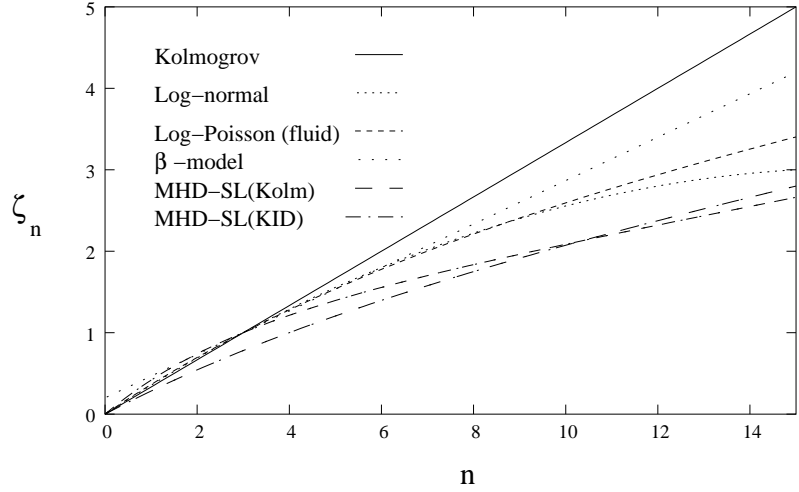}

\caption{\label{Fig:Intermiitency} Plots of $\zeta_{n}$vs. $n$ for various
intermittency models in fluids and MHD. She-Leveque's log-Poission
model fits best with the experimental data in both fluid and MHD.
For MHD turbulence, Kolmogorov-like models are in better agreement
than KID's like model.}
\end{figure}

\subsubsection{The $\beta-$model}

Novikov and Stewart \cite{Novi} and Frisch et al. \cite{Fris:Beta}
proposed that smaller scales in turbulent fluid is less space filling.
In each step of the cascade an eddy $\delta u_{n}$ of scale $l_{n}$
splits into $2^{D}\beta$ eddies of scale $l_{n+1}=l_{n}/2$, where
$D$ is the space dimensionality, and $\beta$ is a fixed parameter
with $0<\beta\le1$. In this model \[
\zeta_{n}=\frac{n}{3}-\frac{\delta}{3}(n-3),\]
where $\beta=2^{-\delta}$. 

Note that $\zeta_{n}$ is linear in $n$, and it does not match with
experimental and numerical data for large $n$ (see Fig. \ref{Fig:Intermiitency}).

\subsubsection{The Multifractal Models}

Parisi and Frisch \cite{FrisPari} developed a multifractal model
of turbulence. Maneveau and Sreenivasan \cite{Sree1} constructed
an intuitive model. Here the energy cascade rate $\epsilon_{l}$ is
not distributed equally into smaller eddies, say, in each cascade
it gets divided into $p\epsilon_{l}$ and $(1-p)\epsilon_{l}$. After
several cascades, one finds that energy distribution is very skewed
or intermittent. The intermittency exponent in this model is \[
\zeta_{n}=\left(\frac{n}{3}-1\right)D_{n}+1,\]
with\[
D_{n}=\log_{2}\left(p^{n}+(1-p)^{n}\right)^{1/(1-n)}.\]
For $p$ near 0.7, $\zeta_{n}$ fits quite well with the experimental
data. The deficiency of this model is that it requires an adjustable
parameter $p$. For more detailed discussion, refer to Stolovitzky
and Sreenivasan\cite{Sree:RMP}.

\subsubsection{The Log-Poisson Model}

She and Leveque \cite{SheLeve} proposed a model involving a hierarchy
of fluctuating structures associated with the vortex filament. In
their model\begin{equation}
\zeta_{n}=\frac{n}{3}(1-x)+C_{0}\left(1-\beta^{n/3}\right)\label{eq:SheLeve}\end{equation}
where $C_{0}$ is co-dimension of the dissipative eddies, and $x$
and $\beta$ are parameters connected by\begin{equation}
C_{0}=\frac{x}{1-\beta}\label{eq:ShelLeve-C0}\end{equation}
(see Biskamp \cite{BiskTurb:book} for details; also see Politano
and Pouquet \cite{Poli:Intermittency}). For Kolmogorov scaling, $x=\beta=2/3$.
In hydrodynamic turbulence, the dissipative eddies are vortex filaments,
i.e., one-dimensional structures. Therefore, the co-dimension is $C_{0}=2$.
Hence, for fluid turbulence\begin{equation}
\zeta_{n}^{SL}=\frac{n}{9}+2\left[1-\left(\frac{2}{3}\right)^{n/3}\right].\label{eq:SL-fluid}\end{equation}
The above prediction fits remarkably well with experimental results.
All the above functions have been plotted in Fig. \ref{Fig:Intermiitency}
for comparison. 

After the above introductory discussion on intermittency in fluid
turbulence, we move on to intermittency in MHD turbulence.

\subsection{Results on Intermittency in MHD Turbulence}

In MHD turbulence, the pdf of increment of velocity, magnetic, and
Elsässer variables are all nongaussian. The $\zeta_{n}$ vs. $n$
is a nonlinear function of $n$, hence MHD turbulence also exhibits
intermittency. The theoretical and phenomenological understanding
of intermittency in MHD turbulence is more uncertain than that in
fluid turbulence because the nature of energy dissipation rates in
MHD turbulence is still quite obscure. 

Following the similar lines as Kolmogorov \cite{K41b}, Politano and
Pouquet \cite{Poli:S3exact} derived an exact relationship:\[
\left\langle \delta z_{||}^{\mp}\delta z_{i}^{\pm}\delta z_{i}^{\pm}\right\rangle =-\frac{4}{3}\epsilon^{\pm}l.\]
This result is consistent with Kolmogorov-like model (Eq. {[}\ref{eq:MHD_Kolm_zpm}{]})
that\[
\delta z_{l}^{\pm}\sim\left(\epsilon^{\pm}\right)^{2/3}\left(\epsilon^{\mp}\right)^{-1/3}l^{1/3}.\]

There are more than one set of exponents in MHD because of presence
of more number of variables. For $\mathbf{z}^{\pm}$ variables we
have\[
S_{n}^{\pm}(l)=\left\langle \left|\delta z_{l}^{\pm}\right|^{n}\right\rangle \sim l^{\zeta_{n}^{\pm}}.\]
In the following, we will discuss Log-Poission model and numerical
results on intermittency in MHD turbulence.

\subsubsection{The log-Poisson model}

Politano and Pouquet \cite{Poli:Intermittency} extended She and Leveque's
formula (\ref{eq:SheLeve}) to MHD turbulence. They argued that smallest
eddies in fully developed MHD turbulence are micro-current sheets,
hence the codimension will be $C_{0}=1$. Kolmogorov's scaling yields
$x=2/3$ and $\beta=1/3$. Therefore \[
\zeta_{n}^{MHD}=\frac{n}{9}+1-\left(\frac{1}{3}\right)^{n/3}.\]
If KID's scaling were to hold for MHD, then $x=1/2$ and $\beta=1/2$.
Consequently,\[
\zeta_{n}^{KID}=-\frac{n}{2}+1-\left(\frac{1}{2}\right)^{n/3}.\]
For details refer to Biskamp \cite{BiskTurb:book}. Now we compare
these predictions with the numerical results.

\subsubsection{Numerical Results}

\begin{figure}
\includegraphics[bb=14 14 402 350]{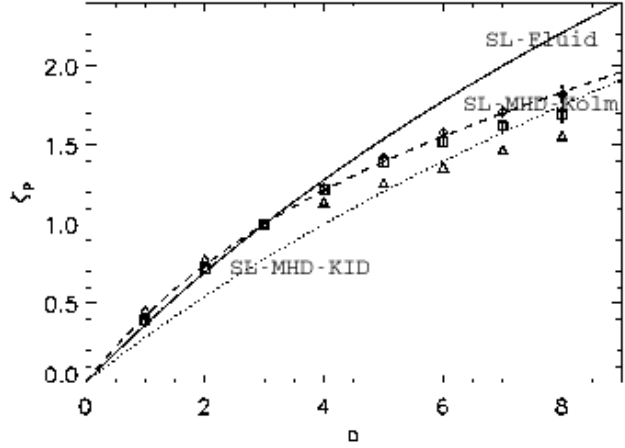}
\caption{\label{Fig:Bisk-Intermittency} 
Numerically computed intermittency
exponents $\zeta_{n}^{+}$(diamond) and $\zeta_{n}^{-}$ (square)
for 3D MHD turbulence, and $\zeta_{n}^{+}/\zeta_{3}^{+}$ (triangle)
for 2D MHD turbulence. The numerical values matches quite well with
the She-Leveque model based on Kolmogorov-like spectrum (dashed line).
Adopted from Biskamp and Müller \cite{Bisk:Kolm2}.}
\end{figure}

Biskamp and Müller \cite{Bisk:Kolm2} have computed the exponents
$\zeta_{n}^{\pm}$ for 3D MHD, and they are shown in Fig. \ref{Fig:Bisk-Intermittency}.
In the same plot, $\zeta_{n}^{MHD}$and $\zeta_{n}^{KID}$ have also
been plotted. Clearly, $\zeta_{n}^{MHD}$ agrees very well with 3D
MHD numerical data. This again shows that Kolmogorov-like phenomenology
models the dynamics of MHD turbulence better that KID's phenomenology.
In Fig. \ref{Fig:Bisk-Intermittency} She-Leveque's predictions for
fluid (solid line) and KID's model (dotted line) are also shown for
reference. 2D MHD appears to be more intermittent than 3D MHD. A point
to note that the plots of Figs. \ref{Fig:Intermiitency}, \ref{Fig:Bisk-Intermittency}
are for small cross helicity ($\sigma_{c}\rightarrow0$); the equality
of $\zeta_{n}^{+}$ and $\zeta_{n}^{-}$ may not hold for higher cross
helicity.

Müller et al. \cite{Mull:IntermittencyB0} numerically computed the
intermittency exponents in the presence of mean magnetic field. They
found that a mean magnetic field reduces the parallel-field dynamics,
while in the perpendicular direction a gradual transition toward 2D
MHD turbulence is observed.

Biskamp and Schwarz \cite{Bisk:2D} computed the intermittency exponents
for 2D MHD turbulence (see Table III of Biskamp and Schwarz \cite{Bisk:2D}).
The exponents are much lower than $\zeta_{n}^{MHD}$. The exponent
$\zeta_{2}$ is close to 0.5, which prompted Biskamp and Schwarz to
infer that 2D MHD follows KID's phenomenology with $E(k)\sim k^{-\zeta_{2}-1}\sim k^{-3/2}$
power spectrum. However, $\zeta_{4}$ is much below 1, which makes
the claim less certain. Earlier, using flux analysis Verma et al.
\cite{MKV:MHD_Simulation} had shown that Kolmogorov-like phenomenology
is a better model for 2D MHD than KID's phenomenology (see Section
\ref{sec:Numerical-Investigation-MHD}). Hence, Biskamp and Schwarz
\cite{Bisk:2D} and Verma et al.'s \cite{MKV:MHD_Simulation} conclusions
appear contradictory. It may be possible that 2D MHD turbulence is
highly intermittent, with 5/3 exponent still applicable. In any case,
further work is required to clarify these issues. Refer to Verma et
al. \cite{MKV:Bisk_Comment} and Biskamp \cite{Bisk:MKV_Comment}
for further details.

Basu et al. \cite{Basu:Intermittency} numerically computed the intermittency
exponents for velocity and magnetic fields. They showed that $\zeta^{b}>\zeta^{\pm}>\zeta^{u}$,
i.e., magnetic field is more intermittent than the velocity field.
They also find that $\zeta^{b}\approx\zeta^{SL}$. For theoretical
arguments regarding $\zeta^{u}$ and $\zeta^{b}$ we refer to Eq.
(\ref{eq:EubPiub}), which implies that\begin{eqnarray*}
\delta u_{l} & \sim & \left(\Pi^{u}\right)^{1/3}l^{1/3}\\
\delta b_{l} & \sim & \left(\Pi^{b}\right)^{1/3}l^{1/3},\end{eqnarray*}
where $\Pi^{u}=\Pi_{u>}^{u<}+\Pi_{b>}^{u<}$ is the total kinetic
energy flux, and $\Pi^{b}=\Pi_{b>}^{b<}+\Pi_{u>}^{b<}$ is the total
magnetic energy flux. Clearly,\begin{eqnarray*}
S^{u(n)}(l) & = & \left\langle \left(\delta u_{l}\right)^{n}\right\rangle =\left\langle \left(\Pi^{u}\right)^{n/3}\right\rangle l^{n/3}\sim l^{\zeta_{n}^{u}}\\
S^{b(n)}(l) & = & \left\langle \left(\delta b_{l}\right)^{n}\right\rangle =\left\langle \left(\Pi^{b}\right)^{n/3}\right\rangle l^{n/3}\sim l^{\zeta_{n}^{b}}.\end{eqnarray*}
Hence, $\zeta^{u}$ and $\zeta^{b}$ depend on the small-scale properties
of $\Pi^{u}$ and $\Pi^{b}$. From the numerical results of Basu et
al. \cite{Basu:Intermittency} it appears that $\Pi^{b}$ is more
intermittent that $\Pi^{u}$. Note that Basu et al.'s result was derived
from magnetically dominated run, So we need to test the above hypothesis
for various ratios of kinetic and magnetic energies.

Some of the earlier work on intermittency in MHD turbulence has been
done by Carbone \cite{Carb:IntermittencyFirst}. His work is based
on KID's model, and he alludes that the spectral index of solar wind
is close to 1.7 because of intermittency correction of approximately
0.2 over 3/2. There is also an extensive investigation of intermittency
in solar wind data. Refer to Burlaga \cite{Burl}, Marsch and Tu \cite{MarsTu:Nongaussian},
and Tu et al. \cite{MarsTu:StructureFn}.

It is evident from the above discussion that physical understanding
of intermittency is quite weak. We need to better understand dissipation
rates in MHD turbulence. With these remarks, we close our discussion
on intermittency.

\section{Miscellaneous Topics \label{sec:Miscellaneous-Topics}}

In this section we will briefly discuss the following topics connected
to the spectral theory of MHD turbulence: (a) Large-Eddy simulations
of MHD Turbulence, (b) Energy decay of MHD turbulence, (c) Shell model
of MHD turbulence, and (d) Compressible Turbulence.

\subsection{Large-Eddy Simulations (LES) of MHD Turbulence}

Basic idea of LES is to resolve only the large scales of turbulent
flow. The effect of smaller scale interactions are modeled appropriately
using the existing theories. Let $u_{K}^{<}$ and $b_{K}^{<}$ represent
the filtered fields at filter width of $l$. The filtered MHD equations
are\begin{eqnarray*}
\frac{\partial\mathbf{u}^{<}}{\partial t} & = & -\nabla\cdot\left(\mathbf{u}^{<}\mathbf{u}^{<}-\mathbf{b}^{<}\mathbf{b}^{<}+\mathbf{\tau}^{u}\right)-\nabla p^{<}+\nu\nabla^{2}\mathbf{u}^{<}\\
\frac{\partial\mathbf{b}^{<}}{\partial t} & = & -\nabla\cdot\left(\mathbf{u}^{<}\mathbf{b}^{<}-\mathbf{b}^{<}\mathbf{u}^{<}+\mathbf{\tau}^{b}\right)+\nu\nabla^{2}\mathbf{u}^{<}\\
\nabla\cdot\mathbf{u}^{<} & = & \nabla\cdot\mathbf{b}^{<}=0,\end{eqnarray*}
where $\mathbf{\tau}^{u}=(\mathbf{uu})^{<}-\mathbf{u}^{<}\mathbf{u}^{<}-(\mathbf{bb})^{<}+\mathbf{b}^{<}\mathbf{b}^{<}$,
and $\mathbf{\tau}^{b}=(\mathbf{ub})^{<}-\mathbf{u}^{<}\mathbf{b}^{<}-(\mathbf{bu})^{<}+\mathbf{b}^{<}\mathbf{u}^{<}$
are the filtered-scale stress tensors. Main task in LES is to model
these tensors. A class of models assume that \cite{MullCara_pop1,MullCara_pop2}\[
\mathbf{\tau}^{u}=-2\nu_{t}\mathbf{S}^{<},\,\,\,\,\,\mathbf{S}^{<}=\left(\nabla\mathbf{u}^{<}+[\nabla\mathbf{u}^{<}]^{T}\right)/2,\]
\[
\mathbf{\tau}^{b}=-2\eta_{t}\mathbf{J}^{<},\,\,\,\,\,\mathbf{J}^{<}=\left(\nabla\mathbf{b}^{<}+[\nabla\mathbf{b}^{<}]^{T}\right)/2,\]
where {}``$T$'' denotes the transposed matrix, and $\nu_{t}$ and
$\eta_{t}$ are the eddy-viscosity and eddy-resistivity respectively.
Agullo et al. \cite{MullCara_pop1} and Müller and Carati \cite{MullCara_pop2}
prescribed $\nu_{t}$ and $\eta_{t}$ using two different models $M_{1}$
and $M_{2}$:\begin{eqnarray*}
M_{1}: & \nu_{t}=C_{1}(t)l^{4/3}, & \eta_{t}=D_{1}(t)l^{4/3},\\
M_{2}: & \nu_{t}=C_{2}(t)l^{2}(2\mathbf{S^{<}}:\mathbf{S^{<}})^{1/2}, & \eta_{t}=D_{2}(t)l^{2}|\mathbf{j}^{<}|.\end{eqnarray*}
Both models contain two unknown parameters $C_{i}$ and $D_{i}$.
Agullo et al. \cite{MullCara_pop1} and Müller and Carati \cite{MullCara_pop2}
determined these parameters using dynamic LES, in which a test filter
is used \cite{Germ}. After determining $\nu_{t}$ and $\eta_{t}$,
the velocity and magnetic fields were updated using DNS. Their evolution
of kinetic and magnetic energy using Models $M_{1}$ and $M_{2}$
agree quite well with DNS. The decay of the magnetic energy in DNS
and $M_{1,2}$ are quite close, but there is a slight discrepancy.
Note that $M_{0}:\mathbf{\tau}^{u,b}=0$ fares quite badly.

We \cite{Shis} employ DNS to MHD equations, but viscosity and resistivity
are replaced by renormalized viscosity and renormalized resistivity
given below:\begin{eqnarray}
\nu_{r}(k_{C}) & = & (K^{u})^{1/2}\Pi^{1/3}k_{C}^{-4/3}\nu^{*}\label{eqn:nuk}\\
\eta_{r}(k_{C}) & = & (K^{u})^{1/2}\Pi^{1/3}k_{C}^{-4/3}\eta^{*}.\label{eqn:etak}\end{eqnarray}
 Here $K^{u}$ is Kolmogorov's constant for MHD, $\Pi$ is the total
energy flux, and $\nu^{*},\eta^{*}$ are the renormalized parameters.
The parameters $\nu^{*},\eta^{*}$, and $K^{u}$ depend on the Alfv\'{e}n
ratio $r_{A}$. In our decaying MHD turbulence simulation, we start
with unit total energy and $r_{A}=100.0$. The ratio of magnetic to
kinetic energy grows as a function of time, as expected. Therefore,
we need to compute the renormalized parameters for various values
of $r_{A}$. The energy cascade rates are computed following the method
described in Section \ref{sec:Numerical-Investigation-MHD}. We take
$\nu_{r}(k_{C})$ and $\eta_{r}(k_{C})$ from Eqs. (\ref{eqn:nuk},
\ref{eqn:etak}). The energy flux $\Pi$ changes with time; we compute
$\Pi$ dynamically every $0.01$ time-unit. We carried out LES for
MHD up to 25 nondimensional time units. McComb et al. \cite{McCo:LES}
had done a similar LES calculation.

The evolution of kinetic energy using LES is quite close to that using
DNS. However, the evolution of magnetic energy does not match very
well. Comparatively, LES of Agullo et al. \cite{MullCara_pop1} and
M\"{u}ller and Carati \cite{MullCara_pop2} yield a better fit to
the temporal evolution of magnetic energy. Hence, refinements are
required in our modelling.

In summary, LES of MHD turbulence is in its infancy, and More work
is required in modelling of eddy-viscosity and resistivity.

\subsection{Energy Decay of MHD Turbulence}

The models of energy decay in MHD turbulence are motivated by the
decay laws of fluid turbulence. In these models, the energy loss is
due to Kolmogorov's energy flux. In addition, conservation laws are
used to close the equation. Biskamp and Müller \cite{Bisk:Kolm2}
first proposed that\begin{equation}
E^{b}l_{0}=H_{M},\label{eq:Decay-E-HM}\end{equation}
where $l_{0}$ is the integral scale, $E^{b}$ is the total magnetic
energy, and $H_{M}$ is magnetic helicity. The corresponding equation
for fluid turbulence is $EL^{s+1}=const$, with $s=4$. Assuming advection
term to be the dominant nonlinearity for energy flux, Biskamp and
Mülller suggested that the dissipation rate $\epsilon$ is\[
\epsilon=-\frac{dE}{dt}\sim\mathbf{u}\cdot\nabla E\sim\left(E^{u}\right)^{1/2}\frac{E}{l_{0}}.\]
A substitution of $l_{0}$ of Eq. (\ref{eq:Decay-E-HM}) into the
above equation yields\[
\frac{E^{5/2}}{\epsilon H_{M}}\frac{r_{A}}{(1+r_{A})^{3/2}}=const.\]
This phenomenological formula was found to be in very good agreement
with numerical result. Alfvén ratio $r_{A}$ itself is varies with
time; Biskamp and Müller numerically found its variation to be $r_{A}\approx1.5(E/H_{M})$.
Using this result and taking the limit $r_{A}\ll1$, they obtained
\[
-\frac{dE}{dt}\approx0.5\frac{E^{3}}{H^{3/2}}\]
with the similarity solution $E\sim t^{-1/2}$. The relationship $r_{A}\approx1.5(E/H_{M})$
yields $E^{u}\sim(E^{b})^{2}\sim t^{-1}$. For finite $r_{A}$, the
evolution is expected to be somewhat steeper. 

For nonhelical MHD ($H_{M}=0$), Biskamp and Müller found a different
decay law\[
\frac{dE}{dt}E^{-2}=const\]
yielding $E\sim t^{-1}$; this result was verified in numerical simulations.
Note that all the above arguments are valid for zero or vanishing
cross helicity. When cross helicity is finite, it decays with a finite
rate. Galtier et al. \cite{Galt:Decay} reached to the similar conclusions
as Biskamp. Note that similar arguments for fluid turbulence shows
that kinetic energy decays as $t^{-10/7}$. 

In the light of current results on evolution of kinetic and magnetic
energy discussed in Section \ref{sub:Field-theoretic-Calculation-of-shell},
some new deductions can made regarding the energy evolution in MHD
turbulence. Since the energy fluxes $\Pi^{\pm}$ are not coupled (see
Fig. \ref{Fig:MHD-flux-z}), we expect $E^{\pm}$ to decay in the
same way as fluid turbulence. However, the evolution of kinetic and
magnetic energy is more complex because of cross transfers of energy
between velocity and magnetic fields (see Fig. \ref{Fig:MHD-flux-ub}).
Their evolution in nonhelical MHD could be modelled as\begin{eqnarray*}
\frac{dE^{b}}{dt} & = & \Pi_{b<}^{u<}-\Pi^{b},\\
\frac{dE^{u}}{dt} & = & \Pi_{u<}^{b<}-\Pi^{u}.\end{eqnarray*}
In kinematic regime ($E^{u}\gg E^{b}$), $E^{b}$ grows exponentially
(dynamo) due to $\Pi_{b<}^{u<}$, and $E^{u}$ should decay as in
fluid turbulence. However, in magnetic regime ($E^{b}\gg E^{u}$),
magnetic energy should decay faster than kinetic energy due to positive
$\Pi_{u<}^{b<}$. Since $E^{u}$ and $E^{b}$ are equipartitioned
in the asymptotic state, it may be possible that $\Pi_{b<}^{u<}=|E^{u}-E^{b}|^{h}$.
Using these ideas, we may get further insights into the physics of
decay in MHD turbulence.

\subsection{Shell Models of MHD Turbulence}

Shell models of turbulence were introduced as an attempt to solve
hydrodynamic equations using much fewer degrees of freedom. In these
models, one variable is used to represent all the modes in wavenumber
shell $(k_{n},k_{n+1})$. The shell radius is given by $k_{n}=k_{0}s^{n}$
with $s$ as a parameter, typically taken to be 2. The coupling between
shells is local with constraints of preserving conserved quantities.
One type of shell-model for MHD turbulence is given below \cite{Fric}:\begin{eqnarray}
\left(\frac{d}{dt}+\nu k^{2}\right)U_{n} & = & ik_{n}[(U_{n+1}^{*}U_{n+2}^{*}-B_{n+1}^{*}B_{n+2}^{*})-\frac{\epsilon}{2}(U_{n-1}^{*}U_{n+1}^{*}-B_{n-1}^{*}B_{n+1}^{*})\nonumber \\
 &  & -\frac{(1-\epsilon)}{4}(U_{n-2}^{*}U_{n-1}^{*}-B_{n-2}^{*}B_{n-1}^{*})]+f_{n}\label{eq:Shell-u}\\
\left(\frac{d}{dt}+\eta k^{2}\right)B_{n} & = & ik_{n}[(1-\epsilon-\epsilon_{m})(U_{n+1}^{*}B_{n+2}^{*}-B_{n+1}^{*}U_{n+2}^{*})+\frac{\epsilon_{m}}{2}(U_{n-1}^{*}B_{n+1}^{*}-B_{n-1}^{*}U_{n+1}^{*})\nonumber \\
 &  & \frac{(1-\epsilon_{m})}{4}(U_{n-2}^{*}B_{n-1}^{*}-B_{n-2}^{*}U_{n-1}^{*})]+g_{n},\label{eq:Shell-b}\end{eqnarray}
where $f_{n}$ and $g_{n}$ are kinetic and magnetic forcing respectively.
The above equations conserve total energy and cross helicity for any
$\epsilon_{m}$. However, conservation of the third integral imposes
condition on $\epsilon'$s. In 3D, this integral is \[
H_{M}=\sum_{n}(-1)^{n}k_{n}^{-1}|B_{n}|^{2},\]
which is conserved if $\epsilon=1/2$ and $\epsilon_{m}=1/3$. in
2D, the choice of $\epsilon=5/4$ and $\epsilon_{m}=-1/3$ leads to
conservation of \[
a=\sum_{n}k_{n}^{-2}|B_{n}|^{2}.\]

Frick and Sokoloff \cite{Fric} numerically solved Eqs. (\ref{eq:Shell-u},
\ref{eq:Shell-b}) with 30 shells ($-4\le n\le27$). The system was
forced near the $n=0$ shell. The time integration was done using
fourth-order Runge-Kutta method. Frick and Sokoloff studied energy
spectrum, fluxes, and the structure functions. They obtained Kolmogorov-like
energy spectrum (5/3) for nonAlfvénic MHD ($\sigma_{c}\approx0$);
this state is independent of magnetic helicity $H_{M}$. However,
when the magnetic and velocity fields are correlated, Kolmogorov state
is not established, and the result depends on the magnetic helicity.
High level of $H_{M}$ suppresses any cascade of energy, and KID's
spectra was obtained. 

Frick and Sokoloff \cite{Fric} and Basu et al. \cite{Basu:Intermittency}
studied the structure functions of MHD. They found that intermittency
in MHD turbulence is slightly higher than in the hydrodynamic case,
and the level of intermittency for the magnetic field is slightly
higher than the velocity field. Biskamp \cite{Bisk:cascade} has studied
the effect of mean magnetic field using Shell model. For reference,
Gloaguen et al. \cite{Pouq:Shell} constructed one of the first shell
models for MHD turbulence.

Shell models are based on an assumption of local energy transfer.
This assumption appears to be suspect in the light of our results
on shell-to-shell energy transfer described in \ref{sub:Field-theoretic-Calculation-of-shell},
where we showed that there are significant amount of nonlocal energy
transfer in MHD turbulence, specially in presence of magnetic helicity.
This issue requires a closer look.

\subsection{Compressible Turbulence \label{sub:Compressible-Turbulence}}

Terrestrial MHD plasmas are incompressible because plasma velocities
are typically much smaller compared to sound speed or Alfvén speed.
However, astrophysical plasmas are typically compressible. Currently
the energy spectrum of incompressible (infinite sound speed) and fully-compressible
(zero sound speed) turbulence are reasonably well understood. Fully
compressible fluid is described by Burgers equation \[
\frac{\partial\mathbf{u}}{\partial t}+(\mathbf{u}\cdot\nabla)\mathbf{u}=\nu\nabla^{2}\mathbf{u}.\]
for which shocks are exact solution in 1D under $\nu\rightarrow0$
limit. It can be easily shown that $E(k)\sim k^{-2}$ and intermittency
exponents $\zeta_{q}=1$ for $q>1$. Shocks are present in higher
dimensions as well, and the spectral index is expected to be 2. Fully-compressible
MHD turbulence, modelled by generalized Burgers equation \cite{GaltPouq:Compressible},
also show shocks. For properties of shocks, refer to Biskamp \cite{BiskTurb:book}
and Priest \cite{Prie:book}. For the other limiting case, incompressible
fluid turbulence as well as MHD turbulence are well described by Kolmogorov's
theory of turbulence. The difficulty is with finite Mach number.

The velocity in compressible fluids is decomposed into compressible
part $\mathbf{u}^{c}$ and solenoidal part $\mathbf{u}^{s}$. In Fourier
space, $\mathbf{u}^{s}$ is perpendicular to $\mathbf{k}$, and $\mathbf{u}^{c}$
is parallel to $\mathbf{k}$. Corresponding to these fields, we have
solenoidal and compressive velocity spectrum, $E^{s}(k)$ and $E^{c}(k)$.
Porter et al. \cite{Port:PRL} showed that in the supersonic turbulence
($Ma>1$) , $E^{c}(k)\sim k^{-2}$, which is similar to the spectrum
in Burgers turbulence. However for subsonic turbulence ($Ma<1$),
both $E^{c}$ and $E^{s}$ have 5/3 spectral index. 

Pressure spectrum is defined using $\left\langle p^{2}\right\rangle =\int E^{P}(k)dk$.
Assuming 5/3 spectrum for velocity and using $p_{k}\sim\rho_{0}u_{k}^{2}$,
Batchelor \cite{Batc:Pressure}, and Monin and Yaglom \cite{MoniYagl2:book}
obtained\[
\frac{1}{\rho_{0}^{2}}E^{P}(k)\sim\epsilon^{4/3}k^{-7/3}.\]
The above law is expected to be valid for subsonic flows. For polytropic
flows $p\sim\rho^{\gamma}$ , or \[
\delta\rho=C_{s}^{2}\delta\rho,\]
using which we can immediately derive the density spectrum for subsonic
flows\[
N^{\rho}(k)=\epsilon^{4/3}C_{s}^{-4}k^{-7/3}.\]
Note that $\left\langle \rho^{2}\right\rangle =\int N^{\rho}(k)dk$. 

For nearly incompressible MHD turbulence Montgomery et al. \cite{Mont:Density}
argued that $E^{P}(k)\sim N^{\rho}(k)\sim k^{-5/3}$. Their argument
is based on quasi-normal model. For detail refer to Montgomery et
al. \cite{Mont:Density} and Zank and Matthaeus \cite{Zank:Compress_PoF,Zank:Compress_PRL}.
Lithwick and Goldreich \cite{Gold:Compress} also obtained Kolmogorov's
spectrum for the density fluctuations in the ionized interstellar
medium. They calculated the above density spectrum by extending the
theory of incompressible MHD given by Goldreich and Sridhar \cite{Srid1,Srid2}.
Cho and Lazarian \cite{Cho:Compressible} found similar results in
their computer simulation.

It is interesting to note that Burgers equation is local in real space,
contrary to incompressible turbulence which is nonlocal in real space.
Also, {}``mode-to-mode'' energy transfer formulas of Dar et al.
\cite{Dar:flux} cannot be applied to Burgers equation because $\nabla\cdot\mathbf{u}=0$
is not applicable to Burgers equation. We need some kind of generalized
theory which will continuously vary the energy spectrum as we change
the Mach number.

\section{Conclusions and Future Directions \label{sec:Conclusions-and-Future}}

Here we summarize the main results in statistical theory of MHD turbulence.
In this paper, we focussed on the energy spectrum, fluxes, and the
shell-to-shell energy transfers in homogeneous turbulence. When the
mean magnetic field is applied, turbulence is naturally anisotropic.
When the mean magnetic field is much greater that fluctuations (weak
turbulence), the energy cascade is planar, perpendicular to the mean
magnetic field; In this limit Galtiers et al. \cite{Galt:Weak} showed
that \begin{equation}
E_{1,2}(k_{\perp})\sim\left(\Pi B_{0}\right)^{1/2}k_{||}^{1/2}k_{\perp}^{-2}.\end{equation}
When the fluctuations become comparable to the mean magnetic field
(strong turbulence), Goldreich and Sridhar \cite{Srid1,Srid2} showed
that $E(k)\sim k_{\perp}^{-5/3}$, thus establishing Kolmogorov-like
dynamics for MHD turbulence. Verma \cite{MKV:B0_RG} showed that the
nonlinear evolution of Alfvén waves are affected by {}``effective
mean magnetic field'', and showed that Kolmogorov's 5/3 powerlaw
is a valid spectrum for MHD turbulence. The effective mean-magnetic
field turns out to be local ($k$-dependent) field, and can be interpreted
as the field due to the next largest eddy. The above theoretical result
is seen in the numerical simulation of Cho et al. \cite{ChoVish:localB}.
The renormalization group calculations (e. g., Verma \cite{MKV:MHD_RG})
 also favor Kolmogorov's 5/3 energy spectrum for MHD turbulence. All
the above results have been discovered in the last ten years. 

Let us contrast the above conclusions with the earlier results of
Kraichnan \cite{Krai:65} and Iroshnikov \cite{Iros} where effective
time-scale is determined by the mean magnetic field $B_{0}$, and
the energy spectrum is $k^{-3/2}$. Kraichnan's and Iroshnikov's phenomenology
is weak turbulence theory under isotropic situations. This is contradictory
because strong mean magnetic field will create anisotropy. This is
why 3/2 theory is inapplicable to MHD turbulence.

Recently studied energy fluxes and shell-to-shell energy transfers
in MHD turbulence are providing important insights into the energy
exchange between velocity and magnetic fields, and also among various
scales. These calculations have been done using {}``mode-to-mode''
energy transfers in MHD triads. For 3D nonhelical flows ($H_{M}=H_{K}=0$),
all the fluxes $u$-to-$u$, $u$-to-$b$, $b$-to-$u$, $b$-to-$b$
are positive except $b$-to-$u$, which is negative for large Alfvén
ratio. In kinetic-energy dominated regime, kinetic energy flows to
magnetic energy, and the reverse happens in magnetic-energy dominated
regime. Hence, steady-state situation is possible only when $E^{u}\approx E^{b}$;
we believe this to be the reason for the equipartition of kinetic
and magnetic energy in MHD turbulence. The shell-to-shell energy transfer
also suggests that nonhelical transfers are local. The $u$-to-$u$
and $b$-to-$b$ transfers are forward, but $u$-to-$b$ and $b$-to-$u$
are somewhat complex. Helicity induces inverse cascade of magnetic
energy, but their magnitude is smaller than the nonhelical counterparts
for small magnetic and kinetic helicities, which is typical. We also
find a forward cascade for magnetic helicity.

Many of the above analytical work have been motivated by the clues
obtained from numerical simulations, e. g., \cite{Bisk:Kolm1,Bisk:Kolm2,MKV:MHD_Simulation,Dar:flux,Maro:Simulation,Chou:num}.
High resolution simulations, which can test spectrum as well as energy
fluxes, have been made possible by recent powerful computers. In turbulence
research, numerical simulations have become synonymous with experiments.
Similarly, observational results from the solar wind data have been
very useful in understanding the dynamics of MHD turbulence.

Amplification of magnetic field in MHD turbulence, commonly known
as dynamo, has been of interest for almost a century. Earlier theories
were of kinematic origin where given velocity spectrum induces growth
of magnetic field, but the magnetic field cannot affect the velocity
field. In the last ten years, there have been a surge of attempts
to solve the full MHD equation including the back-reaction of the
magnetic field to the velocity field. Pouquet et al. \cite{Pouq:EDQNM}
performed EDQNM calculations and showed that {}``residual helicity''
(difference of kinetic helicity and magnetic helicity) induces growth
of large-scale magnetic field. Some of the recent models are motivated
by the numerical results. Brandenburg \cite{Bran:Alpha} finds that
kinetic helicity induces growth of negative magnetic helicity at large-scales,
which in turn enhances the large-scale magnetic field. Chou \cite{Chou:num}
has shown growth of large-scale magnetic field with small-scale or
large-scale seed magnetic field. Verma's \cite{MKV:MHD_Helical} analytical
findings are in agreement with the the above mentioned numerical results.
Field et al. \cite{Fiel:Dynamo}, Chou \cite{Chou:theo}, Schekochihin
et al. \cite{Maro:DynamoNlin} and Blackman \cite{Blac:Rev_Dynamo}
have constructed theoretical models of dynamics dynamo, and studied
their nonlinear evolution and saturation mechanisms.

Intermittency exponents have been computed numerically by Müller and
Biskamp \cite{Bisk:Kolm2} and others. Generalized She and Leveque's
\cite{SheLeve} theoretical model based on log-Poisson process fits
quite well with the numerical data. Note however that theoretical
calculation of intermittency exponents from the first principles is
still alluding turbulence researchers.

There are many unanswered questions in MHD turbulence. We list some
of them here: 

\begin{enumerate}
\item Goldreich and Sridhar's \cite{Srid2} argument for 5/3 spectral index
for strong MHD turbulence is semi-phenomenological. Generalization
of Verma's field-theoretic calculation for mean magnetic field \cite{MKV:B0_RG}
to anisotropic situations will be very useful. It will help us in
quantifying the effects of mean magnetic field on energy fluxes etc.
\item Effects of magnetic and kinetic helicity on energy spectrum and fluxes
is known only partially through numerical simulations and absolute-equilibrium
theories. 
\item Good understanding of compressible fluid and MHD is lacking. Theoretical
studies of coupling of solenoidal, compressible, pressure modes etc.
will advance our understanding in this area.
\item There are only a couple of large-eddy simulations (LES) of MHD turbulence,
and they are not completely satisfactory. Considering the importance
of LES in modeling large-scale practical systems, e.g., Tokomak flows,
dynamo etc., further investigation of LES of MHD is required.
\item Application of field-theoretic calculation of MHD turbulence to electron
magnetohydrodynamics \cite{Bisk:EMHD}, active scalar \cite{Ruiz:Scalar},
drift wave turbulence \cite{Driftwave} etc. could help us in better
understanding of these models.
\item Role of turbulence in corona heating, accretion disks, and other astrophysical
objects are active area of research. 
\end{enumerate}
With these remarks we conclude our review.

\begin{acknowledgments}
The author gratefully acknowledges the valuable discussions and idea
exchanges he had with his collaborators and friends, V. Eswaran, Gaurav
Dar, J. K. Bhattacharjee, Aaron Roberts, Mel Goldstein, Daniele Carati,
Olivier Debliquy, Arvind Ayyer, Shishir Kumar, Avinash Vijayaraghavan,
Mustansir Barma, Krishna Kumar, Arul Laxminarayan, Agha Afsar Ali,
V. Subrahmanyam, Amit Dutta, Anantha Ramakrishna, R. K. Varma, Supriya
Krishnamurthy, Anurag Sahay, Amar Chandra. He thanks Prof. K. R. Sreenivasan
for the encouragement to write this review, and Prof. I. Procaccia
for useful suggestions. Author is grateful to Amit Dutta and Anantha
Ramakrishna for carefully reading the manuscript. The author also
thanks Open Source community for creating Linux OS and softwares like
GNU, Lyx, FFTW, Gimp, Feynmf, etc. which made writing of this manuscript
much easier. Part of our research work presented here was supported
by a project from Department of Science and Technology, India. 
\end{acknowledgments}
\appendix

\section{Fourier Series vs. Fourier Transform for Turbulent Flows\label{sec:Fourier-Series}}

In statistical theory turbulence we typically assume the flow field
to be homogeneous. Therefore, Fourier transform is not applicable
to these flows in strict sense. However, we can define these quantities
by taking limits carefully. This issue has been discussed by Batchelor
\cite{BatcTurb:book} and McComb \cite{McCo:book}. We briefly discuss
them here because they form the basis of the whole paper.

A periodic function $\mathbf{u}(\mathbf{x})$ in box $L^{d}$ can
be expanded using Fourier series as following: 

\begin{eqnarray*}
\mathbf{u}\left(\mathbf{x}\right) & = & \sum\hat{\mathbf{u}}\left(\mathbf{k}\right)\exp\left(i\mathbf{k\cdot x}\right),\\
\mathbf{\hat{u}(k}) & = & \frac{1}{L^{d}}\int d\mathbf{xu}\left(\mathbf{x}\right)\exp\left(-i\mathbf{k\cdot x}\right),\end{eqnarray*}
where $d$ is the space dimensionality. When we take the limit $L\rightarrow\infty$,
we obtain Fourier transform. Using $\mathbf{u}(\mathbf{k})=\mathbf{\hat{\mathbf{u}}}(\mathbf{k})L^{d}$,
 it can be easily shown that\begin{eqnarray*}
\mathbf{u}\left(\mathbf{x}\right) & = & \int\frac{d\mathbf{k}}{(2\pi)^{d}}\mathbf{u}\left(\mathbf{k}\right)\exp\left(i\mathbf{k\cdot x}\right),\\
\mathbf{u}(\mathbf{k}) & = & \int d\mathbf{xu}\left(\mathbf{x}\right)\exp\left(-i\mathbf{k\cdot x}\right),\end{eqnarray*}
with integrals performed over the whole space. Note however that Fourier
transform (integral converges) makes sense when $u(x)$ vanishes as
$|x|\rightarrow\infty$, which is not the case for homogeneous flows.
However, correlations defined below are sensible quantities. Using
the above equations, we find that\begin{eqnarray}
\left\langle u_{i}(\mathbf{k})u_{j}(\mathbf{k'})\right\rangle  & = & \int d\mathbf{x}d\mathbf{x'}\left\langle u_{i}(\mathbf{x})u_{j}(\mathbf{x'})\right\rangle \exp-i(\mathbf{k}\dot{\cdot\mathbf{x}+\mathbf{k'}\cdot\mathbf{x'})}\nonumber \\
 & = & \int d\mathbf{r}C_{ij}(\mathbf{r})\exp-i\mathbf{k}\cdot\mathbf{r}\int d\mathbf{x}\exp-i(\mathbf{k}\dot{+\mathbf{k')}\cdot\mathbf{x}}\nonumber \\
 & = & C_{ij}(\mathbf{k})(2\pi)^{d}\delta(\mathbf{k+k'})\label{eq:ui_uj}\end{eqnarray}
We have used the fact that $\delta(\mathbf{k})\approx L^{d}/(2\pi)^{d}$.
The above equation holds the key. In experiments we measure correlation
function $C(\mathbf{r})$ which is finite and decays with increasing
$r$, hence spectra $C(\mathbf{k})$ is well defined. Now energy spectrum
as well as total energy can be written in terms of $C(\mathbf{k})$
as the following:

\begin{eqnarray*}
\left\langle u^{2}\right\rangle =\frac{1}{L^{d}}\int d\mathbf{x}u^{2}=\sum_{\mathbf{k}}\left|\mathbf{\hat{u}(k)}\right|^{2} & = & \frac{1}{L^{d}}\int\frac{d\mathbf{k}}{(2\pi)^{d}}\left\langle \left|\mathbf{u(k)}\right|^{2}\right\rangle \\
 & = & (d-1)\int\frac{d\mathbf{k}}{(2\pi)^{d}}C(\mathbf{k)}\end{eqnarray*}
We have used the fact that $\delta(\mathbf{k})\approx L^{d}/(2\pi)^{d}$.
Note that $\left\langle \left|\mathbf{u(k)}\right|^{2}\right\rangle =(d-1)C(\mathbf{k})L^{d}$
{[}see Eq. (\ref{eq:ui_uj}){]} is not well defined in the limit $L\rightarrow\infty$.

In conclusion, the measurable quantity in homogeneous turbulence is
the correlation function, which is finite and decays for large $r$.
Therefore, energy spectra etc. are well defined objects in terms of
Fourier transforms of correlation functions.

We choose a finite box, typically $(2\pi)^{d}$, in spectral simulations
for fluid flows. For these problems we express the equations \emph{(incompressible}
MHD) in terms of Fourier series. We write them below for reference.\begin{eqnarray*}
\left(\frac{\partial}{\partial t}-i\left(\mathbf{B}_{0}\cdot\mathbf{k}\right)+\nu k^{2}\right)\hat{u}_{i}(\mathbf{k},t) & = & -ik_{i}\hat{p}_{tot}\left(\mathbf{k},t\right)-ik_{j}\sum[\hat{u}_{j}(\mathbf{q},t)\hat{u}_{i}(\mathbf{p},t)\\
 &  & +\hat{b}_{j}(\mathbf{q},t)\hat{b}_{i}(\mathbf{p},t)]\\
\left(\frac{\partial}{\partial t}-i\left(\mathbf{B}_{0}\cdot\mathbf{k}\right)+\eta k^{2}\right)\hat{b}_{i}(\mathbf{k},t) & = & -ik_{j}\sum[\hat{u}_{j}(\mathbf{q},t)\hat{b}_{i}(\mathbf{p},t)-\hat{b}_{j}(\mathbf{q,}t)\hat{u}_{i}(\mathbf{p},t)]\end{eqnarray*}

The energy spectrum can be computed using $\hat{u}_{i}(\mathbf{k},t)$:\[
\int E(k)dk=\sum\left|\mathbf{\hat{u}}(\mathbf{k})\right|^{2}/2=\int d\mathbf{n}\left|\mathbf{\hat{u}}(\mathbf{k})\right|^{2}/2=\int d\mathbf{k}\left|\mathbf{\hat{u}}(\mathbf{k})\right|^{2}/2\]
where $\mathbf{n}$ is the lattice vector in $d$-dimensional space.
The above equation implies that\[
E(k)=\frac{\left|\mathbf{\hat{u}}(\mathbf{k})\right|^{2}}{2}S_{d}k^{d-1}.\]

A natural question is why the results of numerical simulations or
experiments done in a finite volume should match with those obtained
for infinite volume. The answer is straight forward. When we go from
size $2\pi$ to $L$, the wavenumbers should be scaled by $(2\pi)/L$.
The velocity and frequency should be should be scaled by $(2\pi)/L$
and $\left[(2\pi)/L\right]^{2}$ to keep dimensionless $\nu$ fixed.
The evolution of the two systems will be identical apart from the
above factors. Hence, numerical simulations in a box of size $2\pi$
can capture the behaviour of a system with $L\rightarrow\infty$,
for which Fourier transform in defined.

\section{ Perturbative Calculation of MHD Equations: ${\bf z}^{\pm}$ Variables}

The MHD equations in terms of ${\bf z^{\pm}}$ can be written as
\begin{eqnarray} 
\left(
\begin{array}{c}
z^+_i(\hat{k}) \\
z^-_i(\hat{k})
\end{array}
\right)
& = & 
\left(
\begin{array}{cc}
G^{++}(\hat{k}) & G^{+-}(\hat{k})  \\
G^{-+}(\hat{k}) & G^{--}(\hat{k}) 
\end{array}
\right) 
\left(
\begin{array}{cc}
-i M_{ijm}({\bf k}) 
\int d\hat{p} [ z^-_j (\hat{p}) z^+_m (\hat{k}-\hat{p})  ]  \\
- i M_{ijm}({\bf k}) 
\int d\hat{p} [ z^+_j (\hat{p}) z^-_m (\hat{k}-\hat{p}) ].
\end{array}
\right)
\end{eqnarray}
The Greens function $G$ is related to self-energy using
\begin{equation}
G^{-1}(k,\omega) = 
\left( 
\begin{array}{cc}
-i\omega - \Sigma^{++}   & \Sigma^{+-}   \\
\Sigma^{-+}  & -i\omega - \Sigma^{--}
\end{array}
\right) .
\end{equation}

We solve the above equation perturbatively keeping the
terms upto the first nonvanishing order.  The
integrals are represented using Feynmann diagrams. To the leading order,
\unitlength=1mm
\begin{fmffile}{fmfz1}
\begin{equation}
\parbox{20mm}
{\begin{fmfgraph*}(20,15) 
  \fmfleft{i} \fmfright{o} \fmf{dbl_dots,label=$z^+_i$}{i,o}
\end{fmfgraph*}} = 
\parbox{20mm}
{\begin{fmfgraph*}(20,15) 
  \fmfleft{i} \fmfright{o1,o2} 
  \fmfv{d.sh=square,d.f=full,d.si=0.1w}{v}
  \fmf{dbl_zigzag,label=$G^{++}$}{i,v} 
  \fmf{dbl_dots,label=$z^+_m$,l.d=5}{v,o1}
  \fmf{dots,label=$z^-_j$,l.d=5}{v,o2}
\end{fmfgraph*}} + 
\parbox{20mm}
{\begin{fmfgraph*}(20,15) 
  \fmfleft{i} \fmfright{o1,o2} 
  \fmfv{d.sh=square,d.f=full,d.si=0.1w}{v}
  \fmf{dbl_zigzag,width=1.5,label=$G^{+-}$}{i,v} 
  \fmf{dots,label=$z^-_m$,l.d=5}{v,o1} 
  \fmf{dbl_dots,label=$z^+_j$,l.d=5}{v,o2}
\end{fmfgraph*}} + 
\end{equation}
\begin{equation}
\parbox{20mm}
{\begin{fmfgraph*}(20,15) 
  \fmfleft{i} \fmfright{o} \fmf{dots,label=$z^-_i$}{i,o}
\end{fmfgraph*}} = 
\parbox{20mm}
{\begin{fmfgraph*}(20,15) 
  \fmfleft{i} \fmfright{o1,o2} 
  \fmfv{d.sh=square,d.f=full,d.si=0.1w}{v}
  \fmf{zigzag,width=1,label=$G^{-+}$}{i,v} 
  \fmf{dbl_dots,label=$z^+_m$,l.d=5}{v,o1}
  \fmf{dots,label=$z^-_j$,l.d=5}{v,o2}
\end{fmfgraph*}} + 
\parbox{20mm}
{\begin{fmfgraph*}(20,15) 
  \fmfleft{i} \fmfright{o1,o2} 
  \fmfv{d.sh=square,d.f=full,d.si=0.1w}{v}
  \fmf{zigzag,label=$G^{--}$}{i,v} 
  \fmf{dots,label=$z^-_m$,l.d=5}{v,o1} 
  \fmf{dbl_dots,label=$z^+_j$,l.d=5}{v,o2}
\end{fmfgraph*}} + 
\end{equation}

\end{fmffile}

The variables $z^+$ and $z^-$ are represented by double-dotted and dotted line 
respectively.  The quantity $G^{++}, G^{+-}, G^{--}, G^{-+}$ are represented
by thick double-zigzag, thin double-zigzag, thick zigzag, and
thin zigzag respectively.  The square represents $-i M_{ijm}$
vertex.  These diagrams appear in renormalization calculations as well
as in energy flux calculation.

\subsection{``Mean Magnetic Field'' Renormalization}

The expansion of $z^+$ in terms of Feynman diagrams are given below:
\unitlength=1mm
\begin{fmffile}{fmfz2}
\begin{eqnarray}
I^+ & = & 
 \parbox{20mm}
{\begin{fmfgraph*}(20,15) 
  \fmfleft{i} \fmfright{o1,o2} 
  \fmfv{d.sh=square,l.a=180,l.d=.1w,d.f=full,d.si=0.1w}{v}
  \fmf{phantom}{i,v}
  \fmf{dots,label=$>$,l.d=5}{v,o2}
  \fmf{dbl_dots,label=$>$,l.d=5}{v,o1}
\end{fmfgraph*}} +
 \parbox{20mm}
{\begin{fmfgraph*}(20,15) 
  \fmfleft{i} \fmfright{o1,o2} 
  \fmfv{d.sh=square,l.a=180,l.d=.1w,d.f=full,d.si=0.1w}{v}
  \fmf{phantom}{i,v}
  \fmf{dots,label=$>$,l.d=5}{v,o2}
  \fmf{dbl_dots,label=$<$,l.d=5}{v,o1}
\end{fmfgraph*}} +
\parbox{20mm}
{\begin{fmfgraph*}(20,15) 
  \fmfleft{i} \fmfright{o1,o2} 
  \fmfv{d.sh=square,l.a=180,l.d=.1w,d.f=full,d.si=0.1w}{v}
  \fmf{phantom}{i,v}
  \fmf{dbl_dots,label=$>$,l.d=5}{v,o1}
  \fmf{dots,label=$<$,l.d=5}{v,o2}
\end{fmfgraph*}} +
 \parbox{20mm}
{\begin{fmfgraph*}(20,15) 
  \fmfleft{i} \fmfright{o1,o2} 
  \fmfv{d.sh=square,l.a=180,l.d=.1w,d.f=full,d.si=0.1w}{v}
  \fmf{phantom}{i,v}
  \fmf{dots,label=$<$,l.d=5}{v,o2}
  \fmf{dbl_dots,label=$<$,l.d=5}{v,o1}
\end{fmfgraph*}}
\end{eqnarray}
We illustrate the expansion of one of the above diagrams:
\begin{eqnarray}
 \parbox{20mm}
{\begin{fmfgraph*}(20,15) 
  \fmfleft{i} \fmfright{o1,o2} 
  \fmfv{d.sh=square,d.f=full,d.si=0.1w}{v}
  \fmf{phantom}{i,v}
  \fmf{dbl_dots,label=$<$,l.d=5}{v,o1}
  \fmf{dots,label=$>$,l.d=5}{v,o2}
\end{fmfgraph*}}
& = &
\parbox{20mm}
{\begin{fmfgraph*}(20,15) 
  \fmfv{d.sh=square,l.a=180,l.d=0.1w,d.f=full,d.si=0.1w}{v1}
  \fmfv{d.sh=square,d.f=full,d.si=0.1w}{v2}
  \fmfleft{i}  
  \fmf{phantom}{i,v1}
  \fmf{dbl_zigzag,tension=0.1,right}{v1,v2}
  \fmf{dots,left,label=$><$}{v1,v2} 
  \fmfright{o}
  \fmf{dbl_dots,label=$<$,l.d=5}{v2,o} 
\end{fmfgraph*}}  +
\parbox{20mm}
{\begin{fmfgraph*}(20,15) 
  \fmfv{d.sh=square,l.a=180,l.d=0.1w,d.f=full,d.si=0.1w}{v1}
  \fmfv{d.sh=square,d.f=full,d.si=0.1w}{v2}
  \fmfleft{i}  
  \fmf{phantom}{i,v1}
  \fmf{dbl_zigzag,tension=0.1,right}{v1,v2}
  \fmf{dots,left,label=$><$}{v1,v2} 
  \fmfright{o}
  \fmf{dbl_dots,label=$>$,l.d=5}{v2,o} 
\end{fmfgraph*}}  +
\parbox{20mm}
{\begin{fmfgraph*}(20,15) 
  \fmfv{d.sh=square,l.a=180,l.d=0.1w,d.f=full,d.si=0.1w}{v1}
  \fmfv{d.sh=square,d.f=full,d.si=0.1w}{v2}
  \fmfleft{i}  
  \fmf{phantom}{i,v1}
  \fmf{dbl_zigzag,tension=0.1,right}{v1,v2}
  \fmf{dots,left,label=$>>$}{v1,v2} 
  \fmfright{o}
  \fmf{dbl_dots,label=$<$,l.d=5}{v2,o} 
\end{fmfgraph*}}  +
\parbox{20mm}
{\begin{fmfgraph*}(20,15) 
  \fmfv{d.sh=square,l.a=180,l.d=0.1w,d.f=full,d.si=0.1w}{v1}
  \fmfv{d.sh=square,d.f=full,d.si=0.1w}{v2}
  \fmfleft{i}  
  \fmf{phantom}{i,v1}
  \fmf{dbl_zigzag,tension=0.1,right}{v1,v2}
  \fmf{dots,left,label=$>>$}{v1,v2} 
  \fmfright{o}
  \fmf{dbl_dots,label=$>$,l.d=5}{v2,o} 
\end{fmfgraph*}} +  \nonumber \\
& & \parbox{20mm}
 {\begin{fmfgraph*}(20,15) 
  \fmfv{d.sh=square,l.a=180,l.d=0.1w,d.f=full,d.si=0.1w}{v1}
  \fmfv{d.sh=square,d.f=full,d.si=0.1w}{v2}
  \fmfleft{i}  
  \fmf{phantom}{i,v1}
  \fmf{dbl_zigzag,tension=0.1,right}{v1,v2}
  \fmf{dots_arrow,left,label=$><$}{v1,v2} 
  \fmfright{o}
  \fmf{dots,label=$<$,l.d=5}{v2,o} 
\end{fmfgraph*}}  +
\parbox{20mm}
{\begin{fmfgraph*}(20,15) 
  \fmfv{d.sh=square,l.a=180,l.d=0.1w,d.f=full,d.si=0.1w}{v1}
  \fmfv{d.sh=square,d.f=full,d.si=0.1w}{v2}
  \fmfleft{i}  
  \fmf{phantom}{i,v1}
  \fmf{dbl_zigzag,tension=0.1,right}{v1,v2}
  \fmf{dots_arrow,left,label=$><$}{v1,v2} 
  \fmfright{o}
  \fmf{dots,label=$>$,l.d=5}{v2,o} 
\end{fmfgraph*}}  +
\parbox{20mm}
{\begin{fmfgraph*}(20,15) 
  \fmfv{d.sh=square,l.a=180,l.d=0.1w,d.f=full,d.si=0.1w}{v1}
  \fmfv{d.sh=square,d.f=full,d.si=0.1w}{v2}
  \fmfleft{i}  
  \fmf{phantom}{i,v1}
  \fmf{dbl_zigzag,tension=0.1,right}{v1,v2}
  \fmf{dots_arrow,left,label=$>>$}{v1,v2} 
  \fmfright{o}
  \fmf{dots,label=$<$,l.d=5}{v2,o} 
\end{fmfgraph*}}  +
\parbox{20mm}
{\begin{fmfgraph*}(20,15) 
  \fmfv{d.sh=square,l.a=180,l.d=0.1w,d.f=full,d.si=0.1w}{v1}
  \fmfv{d.sh=square,d.f=full,d.si=0.1w}{v2}
  \fmfleft{i}  
  \fmf{phantom}{i,v1}
  \fmf{dbl_zigzag,tension=0.1,right}{v1,v2}
  \fmf{dots_arrow,left,label=$>>$}{v1,v2} 
  \fmfright{o}
  \fmf{dots,label=$>$,l.d=5}{v2,o} 
\end{fmfgraph*}} \nonumber \\ 
& & + \mbox{similar diagrams for $G^{+-}$ + higher order terms}
	\label{eq:feyn_I<>}
\end{eqnarray} 
\end{fmffile}   


Equation for $z^-$ can be obtained by interchanging $+$ and $-$.
In the above diagrams, $<z^+({\bf k})z^+({\bf k'})>,
 <z^-({\bf k})z^-({\bf k'})>$, and $<z^\pm({\bf k})z^\mp({\bf k'})>$
are represented
by double-dotted, dotted, and dotted arrow lines respectively.  
All diagrams except
fourth and eighth ones vanish due to gaussian nature of $z^{\pm<}$ variables.
In our calculations, we assume fourth and eighth diagram to vanish. For
its evaluation, refer to Zhou et al. \cite{ZhouVaha88,ZhouVaha93}.
As a consequence, the second term of $I^+$ is zero.  Similar analysis
shows that the third term also vanishes.

The fourth term of $I^+$ is diagramatically represented as

\unitlength=1mm

\begin{fmffile}{fmfz3}
\begin{eqnarray}
I_4^+ & = &
\parbox{20mm}
{\begin{fmfgraph*}(20,15) 
  \fmfleft{i} \fmfright{o1,o2} 
  \fmfv{d.sh=square,d.f=full,d.si=0.1w}{v}
  \fmf{phantom}{i,v}
  \fmf{dots,label=$<$,l.d=5}{v,o2}
  \fmf{dbl_dots,label=$<$,l.d=5}{v,o1}
\end{fmfgraph*}}  \nonumber \\
& = & - \delta \Sigma^{++}(k)   
\parbox{20mm}
{\begin{fmfgraph*}(20,15) 
  \fmfleft{i} \fmfright{o}
  \fmf{dbl_dots,label=$>$,l.s=left}{i,o} 
\end{fmfgraph*}}  - \delta \Sigma^{+-}(k)   
\parbox{20mm}
{\begin{fmfgraph*}(20,15) 
  \fmfleft{i} \fmfright{o}
  \fmf{dots,label=$>$,l.s=left}{i,o} 
\end{fmfgraph*}}  \\
I_4^- & = &
 \parbox{20mm}
{\begin{fmfgraph*}(20,15) 
  \fmfleft{i} \fmfright{o1,o2} 
  \fmfv{d.sh=square,d.f=full,d.si=0.1w}{v}
  \fmf{phantom}{i,v}
  \fmf{dbl_dots,label=$<$,l.d=5}{v,o2}
  \fmf{dots,label=$<$,l.d=5}{v,o1}
\end{fmfgraph*}} \nonumber \\
& = & -\delta  \Sigma^{-+}(k)
\parbox{20mm}
{\begin{fmfgraph*}(20,15) 
  \fmfleft{i}
  \fmfright{o}
  \fmf{dbl_dots,label=$>$,l.s=left}{i,o} 
\end{fmfgraph*}} -  \delta \Sigma^{--}(k)
\parbox{20mm}
{\begin{fmfgraph*}(20,15) 
  \fmfleft{i}
  \fmfright{o}
  \fmf{dots,label=$>$,l.s=left}{i,o} 
\end{fmfgraph*}}
\end{eqnarray}
where
\begin{eqnarray}
-(d-1) \delta \Sigma^{++} & = &
\parbox{20mm}
{\begin{fmfgraph*}(20,15) 
  \fmfv{l.a=180,l.d=.1w,d.sh=square,d.f=full,d.si=0.1w}{v1}
  \fmfv{d.sh=square,d.f=full,d.si=0.1w}{v2}
  \fmfleft{i}  
  \fmf{phantom}{i,v1}
  \fmf{dbl_zigzag,tension=0.3,right}{v1,v2}
  \fmf{dots,tension=0.3,left,label=$<<$}{v1,v2} 
  \fmfright{o}   \fmf{phantom}{v2,o}
\end{fmfgraph*}} +
\parbox{20mm}
{\begin{fmfgraph*}(20,15) 
  \fmfv{l.a=180,l.d=.1w,d.sh=square,d.f=full,d.si=0.1w}{v1}
  \fmfv{d.sh=square,d.f=full,d.si=0.1w}{v2}
  \fmfleft{i}  
  \fmf{phantom}{i,v1}
  \fmf{dbl_zigzag,width=2,tension=0.3,right}{v1,v2}
  \fmf{dots_arrow,tension=0.3,left,label=$<<$}{v1,v2} 
  \fmfright{o}   \fmf{phantom}{v2,o}
\end{fmfgraph*}} +
\parbox{20mm}
{\begin{fmfgraph*}(20,15) 
  \fmfv{l.a=180,l.d=.1w,d.sh=square,d.f=full,d.si=0.1w}{v1}
  \fmfv{d.sh=square,d.f=full,d.si=0.1w}{v2}
  \fmfleft{i}  
  \fmf{phantom}{i,v1}
  \fmf{zigzag,width=1,tension=0.3,left}{v1,v2}
  \fmf{dots_arrow,tension=0.3,right,label=$<<$}{v1,v2} 
  \fmfright{o}   \fmf{phantom}{v2,o} 
\end{fmfgraph*}} +
\parbox{20mm}
{\begin{fmfgraph*}(20,15) 
  \fmfv{l.a=180,l.d=.1w,d.sh=square,d.f=full,d.si=0.1w}{v1}
  \fmfv{d.sh=square,d.f=full,d.si=0.1w}{v2}
  \fmfleft{i}  
  \fmf{phantom}{i,v1}
  \fmf{zigzag,tension=0.3,left}{v1,v2}
  \fmf{dots_arrow,tension=0.3,right,label=$<<$}{v1,v2} 
  \fmfright{o}   \fmf{phantom}{v2,o} 
\end{fmfgraph*}}  \label{eq:feyn_Sigma++} \\
(d-1) \delta \Sigma^{+-} & = &
\parbox{20mm}
{\begin{fmfgraph*}(20,15) 
  \fmfv{l.a=180,l.d=.1w,d.sh=square,d.f=full,d.si=0.1w}{v1}
  \fmfv{d.sh=square,d.f=full,d.si=0.1w}{v2}
  \fmfleft{i}  
  \fmf{phantom}{i,v1}
  \fmf{dbl_zigzag,width=2,tension=0.3,right}{v1,v2}
  \fmf{dots_arrow,tension=0.3,left,label=$<<$}{v1,v2} 
  \fmfright{o}   \fmf{phantom}{v2,o}
\end{fmfgraph*}} +
\parbox{20mm}
{\begin{fmfgraph*}(20,15) 
  \fmfv{l.a=180,l.d=.1w,d.sh=square,d.f=full,d.si=0.1w}{v1}
  \fmfv{d.sh=square,d.f=full,d.si=0.1w}{v2}
  \fmfleft{i}  
  \fmf{phantom}{i,v1}
  \fmf{dbl_zigzag,tension=0.3,right}{v1,v2}
  \fmf{dots_arrow,tension=0.3,left,label=$<<$}{v1,v2} 
  \fmfright{o}   \fmf{phantom}{v2,o}
\end{fmfgraph*}} +
\parbox{20mm}
{\begin{fmfgraph*}(20,15) 
  \fmfv{l.a=180,l.d=.1w,d.sh=square,d.f=full,d.si=0.1w}{v1}
  \fmfv{d.sh=square,d.f=full,d.si=0.1w}{v2}
  \fmfleft{i}  
  \fmf{phantom}{i,v1}
  \fmf{zigzag,tension=0.3,left}{v1,v2}
  \fmf{dbl_dots,tension=0.3,right,label=$<<$}{v1,v2} 
  \fmfright{o}   \fmf{phantom}{v2,o} 
\end{fmfgraph*}} +
\parbox{20mm}
{\begin{fmfgraph*}(20,15) 
  \fmfv{l.a=180,l.d=.1w,d.sh=square,d.f=full,d.si=0.1w}{v1}
  \fmfv{d.sh=square,d.f=full,d.si=0.1w}{v2}
  \fmfleft{i}  
  \fmf{phantom}{i,v1}
  \fmf{zigzag,width=1,tension=0.3,left}{v1,v2}
  \fmf{dbl_dots,tension=0.3,right,label=$<<$}{v1,v2} 
  \fmfright{o}   \fmf{phantom}{v2,o} 
\end{fmfgraph*}} \\
-(d-1) \delta \Sigma^{-+} & = &
\parbox{20mm}
{\begin{fmfgraph*}(20,15) 
  \fmfv{l.a=180,l.d=.1w,d.sh=square,d.f=full,d.si=0.1w}{v1}
  \fmfv{d.sh=square,d.f=full,d.si=0.1w}{v2}
  \fmfleft{i}  
  \fmf{phantom}{i,v1}
  \fmf{zigzag,width=1,tension=0.3,right}{v1,v2}
  \fmf{dots_arrow,tension=0.3,left,label=$<<$}{v1,v2} 
  \fmfright{o}   \fmf{phantom}{v2,o}
\end{fmfgraph*}} +
\parbox{20mm}
{\begin{fmfgraph*}(20,15) 
  \fmfv{l.a=180,l.d=.1w,d.sh=square,d.f=full,d.si=0.1w}{v1}
  \fmfv{d.sh=square,d.f=full,d.si=0.1w}{v2}
  \fmfleft{i}  
  \fmf{phantom}{i,v1}
  \fmf{zigzag,tension=0.3,right}{v1,v2}
  \fmf{dots_arrow,tension=0.3,left,label=$<<$}{v1,v2} 
  \fmfright{o}   \fmf{phantom}{v2,o}
\end{fmfgraph*}} +
\parbox{20mm}
{\begin{fmfgraph*}(20,15) 
  \fmfv{l.a=180,l.d=.1w,d.sh=square,d.f=full,d.si=0.1w}{v1}
  \fmfv{d.sh=square,d.f=full,d.si=0.1w}{v2}
  \fmfleft{i}  
  \fmf{phantom}{i,v1}
  \fmf{dbl_zigzag,tension=0.3,left}{v1,v2}
  \fmf{dots,tension=0.3,right,label=$<<$}{v1,v2} 
  \fmfright{o}   \fmf{phantom}{v2,o} 
\end{fmfgraph*}} +
\parbox{20mm}
{\begin{fmfgraph*}(20,15) 
  \fmfv{l.a=180,l.d=.1w,d.sh=square,d.f=full,d.si=0.1w}{v1}
  \fmfv{d.sh=square,d.f=full,d.si=0.1w}{v2}
  \fmfleft{i}  
  \fmf{phantom}{i,v1}
  \fmf{dbl_zigzag,width=2,tension=0.3,left}{v1,v2}
  \fmf{dots,tension=0.3,right,label=$<<$}{v1,v2} 
  \fmfright{o}   \fmf{phantom}{v2,o} 
\end{fmfgraph*}}  \\
-(d-1) \delta \Sigma^{--} & = &
\parbox{20mm}
{\begin{fmfgraph*}(20,15) 
  \fmfv{l.a=180,l.d=.1w,d.sh=square,d.f=full,d.si=0.1w}{v1}
  \fmfv{d.sh=square,d.f=full,d.si=0.1w}{v2}
  \fmfleft{i}  
  \fmf{phantom}{i,v1}
  \fmf{zigzag,tension=0.3,right}{v1,v2}
  \fmf{dbl_dots,tension=0.3,left,label=$<<$}{v1,v2} 
  \fmfright{o}   \fmf{phantom}{v2,o}
\end{fmfgraph*}} +
\parbox{20mm}
{\begin{fmfgraph*}(20,15) 
  \fmfv{l.a=180,l.d=.1w,d.sh=square,d.f=full,d.si=0.1w}{v1}
  \fmfv{d.sh=square,d.f=full,d.si=0.1w}{v2}
  \fmfleft{i}  
  \fmf{phantom}{i,v1}
  \fmf{zigzag,width=1,tension=0.3,right}{v1,v2}
  \fmf{dbl_dots,tension=0.3,left,label=$<<$}{v1,v2} 
  \fmfright{o}   \fmf{phantom}{v2,o}
\end{fmfgraph*}} +
\parbox{20mm}
{\begin{fmfgraph*}(20,15) 
  \fmfv{l.a=180,l.d=.1w,d.sh=square,d.f=full,d.si=0.1w}{v1}
  \fmfv{d.sh=square,d.f=full,d.si=0.1w}{v2}
  \fmfleft{i}  
  \fmf{phantom}{i,v1}
  \fmf{dbl_zigzag,width=2,tension=0.3,left}{v1,v2}
  \fmf{dots_arrow,tension=0.3,right,label=$<<$}{v1,v2} 
  \fmfright{o}   \fmf{phantom}{v2,o} 
\end{fmfgraph*}} +
\parbox{20mm}
{\begin{fmfgraph*}(20,15) 
  \fmfv{l.a=180,l.d=.1w,d.sh=square,d.f=full,d.si=0.1w}{v1}
  \fmfv{d.sh=square,d.f=full,d.si=0.1w}{v2}
  \fmfleft{i}  
  \fmf{phantom}{i,v1}
  \fmf{dbl_zigzag,tension=0.3,left}{v1,v2}
  \fmf{dots_arrow,tension=0.3,right,label=$<<$}{v1,v2} 
  \fmfright{o}   \fmf{phantom}{v2,o} 
\end{fmfgraph*}} \label{eq:feyn_Sigma--} 
\end{eqnarray}
\end{fmffile}


In Eqs.~(\ref{eq:feyn_Sigma++}-\ref{eq:feyn_Sigma--}) we have omitted all the 
vanishing diagrams (similar
to those appearing in Eq.~[\ref{eq:feyn_I<>}]).  
These terms contribute to $\Sigma$s.

The algebraic expressions for the above diagrams are given in Section
\ref{sub:Mean-Magnetic-Field-RG}.
These expressions have Green's functions, correlation functions, and
algebraic factors resulting from the contraction of tensors.
The algebraic factors for $-(d-1) \delta \Sigma^{++}$, denoted
by $S_{(1-4)}(k,p,q)$, are given below. 
\begin{eqnarray}
S_1(k,p,q) & = & M_{bjm}(k) M_{mab}(p) P_{ja}(q) 
            =  kp(d-2+z^2)(z+xy) \nonumber \\
S_2(k,p,q) & = & M_{ajm}(k) M_{mab}(p) P_{jb}(q) 
            =  kp(-z+z^3+y^2 z+ x y z^2) \nonumber \\
S_3(k,p,q) & = & M_{bjm}(k) M_{jab}(p) P_{ma}(q) 
            =  kp(-z+z^3+x^2 z+ x y z^2) \nonumber \\
S_4(k,p,q) & = & M_{ajm}(k) M_{jab}(p) P_{mb}(q) 
            =  kp(-z+z^3+x y+x^2 z+y^2 x+x y z^2)  \nonumber
\end{eqnarray}
Here,  $x,y,z$ are direction cosines defined as
\begin{equation}
{\bf p \cdot q} = -pqx; \hspace{1cm} {\bf q \cdot k}= qky; 
\hspace{1cm}{\bf p \cdot k}=pkz.
\end{equation}
\subsection{Renormalization of Dissipative Parameters}

The Feynman diagrams for renormalization of $\nu_{\pm\pm}, \nu_{\pm\mp}$
are identical to the given above except that in renormalization of
dissipative parameters, $>$ modes are averaged instead of $<$ modes.

\subsection{Mode-to-Mode Energy Transfer in MHD Turbulence}

In Section 3, we studied the ``mode-to-mode'' energy transfer
$S^{\pm}({\bf k'|p|q})$  from
${\bf z}^\pm({\bf p})$ to  ${\bf z}^\pm({\bf k'})$ with the mediation
of ${\bf z}^\mp({\bf q})$.  The expression for this transfer is
\begin{equation}
S^{\pm}({\bf k'|p|q}) = -\Im ({\bf [k' \cdot z^{\mp}(q)]
			[z^{\pm}(k') \cdot z^{\pm}(p)]}) 
\end{equation}

In perturbative calculation of $S$ we assume the field variables
${\bf z}^\pm$ to be quasi-gaussian.  Hence, $S$ vanish to zeroth
order.  To first order, $S^+$ is
\unitlength=1mm
\begin{fmffile}{fmfz4}
\begin{eqnarray}
S^{+}({\bf k'|p|q}) & = &
\parbox{30mm}
{\begin{fmfgraph*}(30,20) 
  \fmfv{d.sh=square,d.f=full,d.si=0.01w}{v1}
  \fmfv{d.sh=square,d.f=full,d.si=0.1w}{v2}
  \fmfleft{i}  
  \fmf{phantom}{i,v1}
  \fmf{dots_arrow,tension=0.1,right,label=$q$}{v1,v2}
  \fmf{dots_arrow,tension=0.1,label=$p$}{v1,v2}
  \fmf{dbl_zigzag,tension=0.1,left,label=$k$}{v1,v2} 
  \fmfright{o}   \fmf{phantom}{v2,o}
\end{fmfgraph*}} +
\parbox{30mm}
{\begin{fmfgraph*}(30,20) 
  \fmfv{d.sh=square,d.f=full,d.si=0.01w}{v1}
  \fmfv{d.sh=square,d.f=full,d.si=0.1w}{v2}
  \fmfleft{i}  
  \fmf{phantom}{i,v1}
  \fmf{dots_arrow,tension=0.1,right,label=$q$}{v1,v2}
  \fmf{dbl_zigzag,tension=0.1,label=$p$}{v1,v2}
  \fmf{dots_arrow,tension=0.1,left,label=$k$}{v1,v2} 
  \fmfright{o}   \fmf{phantom}{v2,o}
\end{fmfgraph*}} +
\parbox{30mm}
{\begin{fmfgraph*}(30,20) 
  \fmfv{d.sh=square,d.f=full,d.si=0.01w}{v1}
  \fmfv{d.sh=square,d.f=full,d.si=0.1w}{v2}
  \fmfleft{i}  
  \fmf{phantom}{i,v1}
  \fmf{zigzag,width=1,tension=0.1,right,label=$q$}{v1,v2}
  \fmf{dbl_dots,tension=0.1,label=$p$}{v1,v2}
  \fmf{dots_arrow,tension=0.1,left,label=$k$}{v1,v2} 
  \fmfright{o}   \fmf{phantom}{v2,o}
\end{fmfgraph*}} \nonumber \\ 
\nonumber \\ \nonumber \\ \nonumber \\
 &  & +
\parbox{30mm}
{\begin{fmfgraph*}(30,20) 
  \fmfv{d.sh=square,d.f=full,d.si=0.01w}{v1}
  \fmfv{d.sh=square,d.f=full,d.si=0.1w}{v2}
  \fmfleft{i}  
  \fmf{phantom}{i,v1}
  \fmf{dots,tension=0.1,right,label=$q$}{v1,v2}
  \fmf{dbl_dots,tension=0.1,label=$p$}{v1,v2}
  \fmf{dbl_zigzag,tension=0.1,left,label=$k$}{v1,v2} 
  \fmfright{o}   \fmf{phantom}{v2,o}
\end{fmfgraph*}} +
\parbox{30mm}
{\begin{fmfgraph*}(30,20) 
  \fmfv{d.sh=square,d.f=full,d.si=0.01w}{v1}
  \fmfv{d.sh=square,d.f=full,d.si=0.1w}{v2}
  \fmfleft{i}  
  \fmf{phantom}{i,v1}
  \fmf{dots,tension=0.1,right,label=$q$}{v1,v2}
  \fmf{dbl_zigzag,tension=0.1,label=$p$}{v1,v2}
  \fmf{dbl_dots,tension=0.1,left,label=$k$}{v1,v2} 
  \fmfright{o}   \fmf{phantom}{v2,o}
\end{fmfgraph*}} +
\parbox{30mm}
{\begin{fmfgraph*}(30,20) 
  \fmfv{d.sh=square,d.f=full,d.si=0.01w}{v1}
  \fmfv{d.sh=square,d.f=full,d.si=0.1w}{v2}
  \fmfleft{i}  
  \fmf{phantom}{i,v1}
  \fmf{zigzag,width=1,tension=0.1,right,label=$q$}{v1,v2}
  \fmf{dots_arrow,tension=0.1,label=$p$}{v1,v2}
  \fmf{dbl_dots,tension=0.1,left,label=$k$}{v1,v2} 
  \fmfright{o}   \fmf{phantom}{v2,o}
\end{fmfgraph*}}  \nonumber \\ 
\nonumber \\ \nonumber \\ \nonumber \\
&  & +
\parbox{30mm}
{\begin{fmfgraph*}(30,20) 
  \fmfv{d.sh=square,d.f=full,d.si=0.01w}{v1}
  \fmfv{d.sh=square,d.f=full,d.si=0.1w}{v2}
  \fmfleft{i}  
  \fmf{phantom}{i,v1}
  \fmf{dots,tension=0.1,right,label=$q$}{v1,v2}
  \fmf{dbl_dots,tension=0.1,label=$p$}{v1,v2}
  \fmf{dbl_zigzag,width=2,tension=0.1,left,label=$k$}{v1,v2} 
  \fmfright{o}   \fmf{phantom}{v2,o}
\end{fmfgraph*}} +
\parbox{30mm}
{\begin{fmfgraph*}(30,20) 
  \fmfv{d.sh=square,d.f=full,d.si=0.01w}{v1}
  \fmfv{d.sh=square,d.f=full,d.si=0.1w}{v2}
  \fmfleft{i}  
  \fmf{phantom}{i,v1}
  \fmf{dots,tension=0.1,right,label=$q$}{v1,v2}
  \fmf{dbl_zigzag,width=2,tension=0.1,label=$p$}{v1,v2}
  \fmf{dbl_dots,tension=0.1,left,label=$k$}{v1,v2} 
  \fmfright{o}   \fmf{phantom}{v2,o}
\end{fmfgraph*}} +
\parbox{30mm}
{\begin{fmfgraph*}(30,20) 
  \fmfv{d.sh=square,d.f=full,d.si=0.01w}{v1}
  \fmfv{d.sh=square,d.f=full,d.si=0.1w}{v2}
  \fmfleft{i}  
  \fmf{phantom}{i,v1}
  \fmf{zigzag,tension=0.1,right,label=$q$}{v1,v2}
  \fmf{dots_arrow,tension=0.1,label=$p$}{v1,v2}
  \fmf{dbl_dots,tension=0.1,left,label=$k$}{v1,v2} 
  \fmfright{o}   \fmf{phantom}{v2,o}
\end{fmfgraph*}} \nonumber \\ 
\nonumber \\ \nonumber \\ \nonumber \\
&  & +
\parbox{30mm}
{\begin{fmfgraph*}(30,20) 
  \fmfv{d.sh=square,d.f=full,d.si=0.01w}{v1}
  \fmfv{d.sh=square,d.f=full,d.si=0.1w}{v2}
  \fmfleft{i}  
  \fmf{phantom}{i,v1}
  \fmf{dots_arrow,tension=0.1,right,label=$q$}{v1,v2}
  \fmf{dots_arrow,tension=0.1,label=$p$}{v1,v2}
  \fmf{dbl_zigzag,width=2,tension=0.1,left,label=$k$}{v1,v2} 
  \fmfright{o}   \fmf{phantom}{v2,o}
\end{fmfgraph*}} +
\parbox{30mm}
{\begin{fmfgraph*}(30,20) 
  \fmfv{d.sh=square,d.f=full,d.si=0.01w}{v1}
  \fmfv{d.sh=square,d.f=full,d.si=0.1w}{v2}
  \fmfleft{i}  
  \fmf{phantom}{i,v1}
  \fmf{dots_arrow,tension=0.1,right,label=$q$}{v1,v2}
  \fmf{dbl_zigzag,width=2,tension=0.1,label=$p$}{v1,v2}
  \fmf{dots_arrow,tension=0.1,left,label=$k$}{v1,v2} 
  \fmfright{o}   \fmf{phantom}{v2,o}
\end{fmfgraph*}} +
\parbox{30mm}
{\begin{fmfgraph*}(30,20) 
  \fmfv{d.sh=square,d.f=full,d.si=0.01w}{v1}
  \fmfv{d.sh=square,d.f=full,d.si=0.1w}{v2}
  \fmfleft{i}  
  \fmf{phantom}{i,v1}
  \fmf{zigzag,tension=0.1,right,label=$q$}{v1,v2}
  \fmf{dbl_dots,tension=0.1,label=$p$}{v1,v2}
  \fmf{dots_arrow,tension=0.1,left,label=$k$}{v1,v2} 
  \fmfright{o}   \fmf{phantom}{v2,o}
\end{fmfgraph*}} 
\end{eqnarray}
\end{fmffile}

where the left vertex denotes $k_i$, and the right vertex (square) represents
$-i M_{ijm}$.  The diagrams for $S^-$ can be obtained by interchanging 
$+$ and $-$.  Some of the diagrams may vanish depending on the form of 
correlation function.

The corresponding expressions to each diagram  would involve two 
correlation functions,  one Green's function, and an algebraic factor.
For isotropic flows, these factors, denoted by $T_{(13-24)}(k,p,q)$,
are given by
\begin{eqnarray}
T_{13,15}(k,p,q) & = & k_i M_{jab}(k') P_{ja}(p) P_{ib}(q)
	           =  - k p  y z (y+x z)  \nonumber \\
T_{14,16}(k,p,q) & = & k_i M_{jab}(k') P_{jb}(p) P_{ia}(q)
                   =  k^2 (1-y^2) (d-2+z^2) \nonumber\\
T_{17,19}(k,p,q) & = & k_i M_{jab}(p) P_{ja}(k) P_{ib}(q)
	           =  k p  x z (x+y z) \nonumber  \\
T_{18,20}(k,p,q) & = & -T_{14}(k,p,q) \nonumber\\
T_{21,23}(k,p,q) & = & k_i M_{iab}(q) P_{ja}(k) P_{jb}(p)
	           =  -k p x y (1-z^2)\nonumber  \\
T_{22,24}(k,p,q) & = & -T_{13}(k,p,q) \nonumber
\end{eqnarray}

\section{ Perturbative Calculation of MHD Equations: ${\bf u,b}$ Variables}

The MHD equations can be written as
\begin{eqnarray} 
\left(
\begin{array}{c}
u_i(\hat{k}) \\
b_i(\hat{k})
\end{array}
\right)
& = & 
\left(
\begin{array}{cc}
G^{uu}(\hat{k}) & G^{ub}(\hat{k})  \\
G^{bu}(\hat{k}) & G^{bb}(\hat{k}) 
\end{array}
\right) 
\left(
\begin{array}{cc}
-\frac{i}{2} P^+_{ijm}({\bf k}) 
\int d\hat{p} [ u_j (\hat{p}) u_m (\hat{k}-\hat{p}) -
   b_j (\hat{p}) b_m (\hat{k}-\hat{p}) ]  \\
- i P^-_{ijm}({\bf k}) 
\int d\hat{p} [ u_j (\hat{p}) b_m (\hat{k}-\hat{p}) ]
\end{array}
\right)
\end{eqnarray}
where the Greens function $G$ 
can be obtained from $G^{-1}(\hat{k})$ 
\begin{equation}
G^{-1}(k,\omega) = 
\left( 
\begin{array}{cc}
-i\omega - \Sigma^{uu}   & \Sigma^{ub}   \\
\Sigma^{bu}  & -i\omega - \Sigma^{bb}
\end{array}
\right) .
\end{equation}

We solve the above equation perturbatively keeping the
terms upto the first nonvanishing order.  Feynmann diagrams
representing various terms are

\unitlength=1mm
\begin{fmffile}{fmfub1}
\begin{equation}
\parbox{20mm}
{\begin{fmfgraph*}(20,15) 
  \fmfleft{i} \fmfright{o} \fmf{plain,label=$u_i$}{i,o}
\end{fmfgraph*}} = 
\parbox{20mm}
{\begin{fmfgraph*}(20,15) 
  \fmfleft{i} \fmfright{o1,o2} 
  \fmfv{d.sh=circle,d.f=full,d.si=0.1w}{v}
  \fmf{photon,label=$G^{uu}$}{i,v} 
  \fmf{plain,label=$u_m$,l.d=5}{v,o1}
  \fmf{plain,label=$u_j$,l.d=5}{v,o2}
\end{fmfgraph*}} - 
\parbox{20mm}
{\begin{fmfgraph*}(20,15) 
  \fmfleft{i} \fmfright{o1,o2} 
  \fmfv{d.sh=circle,d.f=full,d.si=0.1w}{v}
  \fmf{photon,label=$G^{uu}$}{i,v} 
  \fmf{dashes,label=$b_m$,l.d=5}{v,o1} 
  \fmf{dashes,label=$b_j$,l.d=5}{v,o2}
\end{fmfgraph*}} + 
\parbox{20mm}
{\begin{fmfgraph*}(20,15) 
  \fmfleft{i} \fmfright{o1,o2} 
  \fmfv{d.sh=circle,d.f=empty,d.si=0.1w}{v}
  \fmf{photon,label=$G^{ub}$,width=1}{i,v} 
  \fmf{dashes,label=$b_m$,l.d=5}{v,o1} 
  \fmf{plain,label=$u_j$,l.d=5}{v,o2}
\end{fmfgraph*}} + \ldots
\end{equation}
\begin{equation}
\parbox{20mm}
{\begin{fmfgraph*}(20,15) 
  \fmfleft{i} \fmfright{o} \fmf{dashes,label=$b_i$}{i,o}
\end{fmfgraph*}} =  
\parbox{20mm}
{\begin{fmfgraph*}(20,15) 
  \fmfleft{i} \fmfright{o1,o2} 
  \fmfv{d.sh=circle,d.f=full,d.si=0.1w}{v}
  \fmf{gluon,label=$G^{bu}$,width=1}{i,v} 
  \fmf{plain,label=$u_j$,l.d=5}{v,o2} 
  \fmf{plain,label=$u_m$,l.d=5}{v,o1}
\end{fmfgraph*}} - 
\parbox{20mm}
{\begin{fmfgraph*}(20,15) 
  \fmfleft{i} \fmfright{o1,o2} 
  \fmfv{d.sh=circle,d.f=full,d.si=0.1w}{v}
  \fmf{gluon,label=$G^{bu}$,width=1}{i,v} 
  \fmf{dashes,label=$b_j$,l.d=5}{v,o2} 
  \fmf{dashes,label=$b_m$,l.d=5}{v,o1}
\end{fmfgraph*}} + 
\parbox{20mm}
{\begin{fmfgraph*}(20,15) 
  \fmfleft{i} \fmfright{o1,o2} 
  \fmfv{d.sh=circle,d.f=empty,d.si=0.1w}{v}
  \fmf{gluon,label=$G^{bb}$}{i,v} 
  \fmf{plain,label=$u_j$,l.d=5}{v,o2} 
  \fmf{dashes,label=$b_m$,l.d=5}{v,o1}
\end{fmfgraph*}} +  \ldots
\end{equation}

\end{fmffile}

Solid and dashed lines represent fields $u$ and $b$ 
respectively. Thick wiggly (photon), thin wiggly, thick curly (gluon),
and thin curly lines denote $G^{uu}, G^{ub}, G^{bb}$, and
$G^{bu}$ respectively.  The filled circle denotes  $-(i/2) P^+_{ijm}$
vertex, while the empty circle denotes $-i P^-_{ijm}$ vertex.
  These diagrams appear in renormalization calculations as well
as in energy flux calculations.

\subsection{Viscosity and Resistivity Renormalization}

The expansion of $u,b$ in terms of Feynman diagrams are given below:
\unitlength=1mm
\begin{fmffile}{fmfub2}
\begin{eqnarray}
I^u & = & 
 \parbox{20mm}
{\begin{fmfgraph*}(20,15) 
  \fmfleft{i} \fmfright{o1,o2} 
  \fmfv{d.sh=circle,l.a=180,l.d=.1w,d.f=full,d.si=0.1w}{v}
  \fmf{phantom}{i,v}
  \fmf{plain,label=$<$,l.d=5}{v,o2}
  \fmf{plain,label=$<$,l.d=5}{v,o1}
\end{fmfgraph*}} - 
 \parbox{20mm}
{\begin{fmfgraph*}(20,15) 
  \fmfleft{i} \fmfright{o1,o2} 
  \fmfv{d.sh=circle,l.a=180,l.d=.1w,d.f=full,d.si=0.1w}{v}
  \fmf{phantom}{i,v}
  \fmf{dashes,label=$<$,l.d=5}{v,o2}
  \fmf{dashes,label=$<$,l.d=5}{v,o1}
\end{fmfgraph*}} +
 \parbox{20mm}
{\begin{fmfgraph*}(20,15) 
  \fmfleft{i} \fmfright{o1,o2} 
  \fmfv{d.sh=circle,l=2,l.a=180,l.d=.1w,d.f=full,d.si=0.1w}{v}
  \fmf{phantom}{i,v}
  \fmf{plain,label=$>$,l.d=5}{v,o2}
  \fmf{plain,label=$<$,l.d=5}{v,o1}
\end{fmfgraph*}} - 
 \parbox{20mm}
{\begin{fmfgraph*}(20,15) 
  \fmfleft{i} \fmfright{o1,o2} 
  \fmfv{d.sh=circle,l=2,l.a=180,l.d=.1w,d.f=full,d.si=0.1w}{v}
  \fmf{phantom}{i,v}
  \fmf{dashes,label=$>$,l.d=5}{v,o2}
  \fmf{dashes,label=$<$,l.d=5}{v,o1}
\end{fmfgraph*}}  +
 \parbox{20mm}
{\begin{fmfgraph*}(20,15) 
  \fmfleft{i} \fmfright{o1,o2} 
  \fmfv{d.sh=circle,l.a=180,l.d=.1w,d.f=full,d.si=0.1w}{v}
  \fmf{phantom}{i,v}
  \fmf{plain,label=$>$,l.d=5}{v,o2}
  \fmf{plain,label=$>$,l.d=5}{v,o1}
\end{fmfgraph*}} - 
 \parbox{20mm}
{\begin{fmfgraph*}(20,15) 
  \fmfleft{i} \fmfright{o1,o2} 
  \fmfv{d.sh=circle,l.a=180,l.d=.1w,d.f=full,d.si=0.1w}{v}
  \fmf{phantom}{i,v}
  \fmf{dashes,label=$>$,l.d=5}{v,o2}
  \fmf{dashes,label=$>$,l.d=5}{v,o1}
\end{fmfgraph*}} \\
I^b & = &
\parbox{20mm}
{\begin{fmfgraph*}(20,15) 
  \fmfleft{i} \fmfright{o1,o2} 
  \fmfv{d.sh=circle,d.f=empty,d.si=0.1w}{v}
  \fmf{phantom}{i,v}
  \fmf{plain,label=$<$,l.d=5}{v,o2}
  \fmf{dashes,label=$<$,l.d=5}{v,o1}
\end{fmfgraph*}} +
 \parbox{20mm}
{\begin{fmfgraph*}(20,15) 
  \fmfleft{i} \fmfright{o1,o2} 
  \fmfv{d.sh=circle,d.f=empty,d.si=0.1w}{v}
  \fmf{phantom}{i,v}
  \fmf{plain,label=$>$,l.d=5}{v,o2}
  \fmf{dashes,label=$<$,l.d=5}{v,o1}
\end{fmfgraph*}} +
 \parbox{20mm}
{\begin{fmfgraph*}(20,15) 
  \fmfleft{i} \fmfright{o1,o2} 
  \fmfv{d.sh=circle,d.f=empty,d.si=0.1w}{v}
  \fmf{phantom}{i,v}
  \fmf{plain,label=$<$,l.d=5}{v,o2}
  \fmf{dashes,label=$>$,l.d=5}{v,o1}
\end{fmfgraph*}} +
\parbox{20mm}
{\begin{fmfgraph*}(20,15) 
  \fmfleft{i} \fmfright{o1,o2} 
  \fmfv{d.sh=circle,d.f=empty,d.si=0.1w}{v}
  \fmf{phantom}{i,v}
  \fmf{plain,label=$>$,l.d=5}{v,o2}
  \fmf{dashes,label=$>$,l.d=5}{v,o1}
\end{fmfgraph*}} 
\end{eqnarray}
\end{fmffile}

Factor of 2 appears in $I^u$ because of $<>$ symmetry in the corresponding
term. To zeroth order, the terms with $<>$ are zero because of quasi-gaussian
nature of $>$ modes.  To the next order in perturbation, the third term
of $I^u$ is

\unitlength=1mm
\begin{fmffile}{fmfub3}
\begin{eqnarray}
 \parbox{20mm}
{\begin{fmfgraph*}(20,15) 
  \fmfleft{i} \fmfright{o1,o2} 
  \fmfv{d.sh=circle,d.f=full,d.si=0.1w}{v}
  \fmf{phantom}{i,v}
  \fmf{plain,label=$>$,l.d=5}{v,o2}
  \fmf{plain,label=$<$,l.d=5}{v,o1}
\end{fmfgraph*}}
& = &
\parbox{20mm}
{\begin{fmfgraph*}(20,15) 
  \fmfv{d.sh=circle,l=2,l.a=180,l.d=0.1w,d.f=full,d.si=0.1w}{v1}
  \fmfv{d.sh=circle,d.f=full,d.si=0.1w}{v2}
  \fmfleft{i}  
  \fmf{phantom}{i,v1}
  \fmf{photon,tension=0.1,left}{v1,v2}
  \fmf{plain,right,label=$<>$}{v1,v2} 
  \fmfright{o}
  \fmf{plain,label=$>$,l.d=5}{v2,o} 
\end{fmfgraph*}}  +
\parbox{20mm}
{\begin{fmfgraph*}(20,15) 
  \fmfv{d.sh=circle,d.f=full,d.si=0.1w}{v1}
  \fmfv{d.sh=circle,d.f=full,d.si=0.1w}{v2}
  \fmfleft{i}  
  \fmf{phantom}{i,v1}
  \fmf{photon,tension=0.1,left}{v1,v2}
  \fmf{plain,right,label=$<>$}{v1,v2} 
  \fmfright{o}
  \fmf{plain,label=$<$,l.d=5}{v2,o} 
\end{fmfgraph*}}  +
\parbox{20mm}
{\begin{fmfgraph*}(20,15) 
  \fmfv{d.sh=circle,d.f=full,d.si=0.1w}{v1}
  \fmfv{d.sh=circle,d.f=full,d.si=0.1w}{v2}
  \fmfleft{i}  
  \fmf{phantom}{i,v1}
  \fmf{photon,tension=0.1,left}{v1,v2}
  \fmf{plain,right,label=$<<$}{v1,v2} 
  \fmfright{o}
  \fmf{plain,label=$>$,l.d=5}{v2,o} 
\end{fmfgraph*}}  +
\parbox{20mm}
{\begin{fmfgraph*}(20,15) 
  \fmfv{d.sh=circle,l=2,l.a=180,l.d=.1w,d.f=full,d.si=0.1w}{v1}
  \fmfv{d.sh=circle,d.f=full,d.si=0.1w}{v2}
  \fmfleft{i}  
  \fmf{phantom}{i,v1}
  \fmf{photon,tension=0.1,left}{v1,v2}
  \fmf{plain,right,label=$<<$}{v1,v2} 
  \fmfright{o}
  \fmf{plain,label=$<$,l.d=5}{v2,o} 
\end{fmfgraph*}} -  \nonumber \\
& & 
\parbox{20mm}
{\begin{fmfgraph*}(20,15) 
  \fmfv{d.sh=circle,label=2,l.a=180,l.d=.1w,d.f=full,d.si=0.1w}{v1}
  \fmfv{d.sh=circle,d.f=full,d.si=0.1w}{v2}
  \fmfleft{i}  
  \fmf{phantom}{i,v1}
  \fmf{photon,tension=0.1,left}{v1,v2}
  \fmf{dots,right,label=$<>$}{v1,v2} 
  \fmfright{o}
  \fmf{dashes,label=$>$,l.d=5}{v2,o} 
\end{fmfgraph*}}  -
\parbox{20mm}
{\begin{fmfgraph*}(20,15) 
  \fmfv{d.sh=circle,d.f=full,d.si=0.1w}{v1}
  \fmfv{d.sh=circle,d.f=full,d.si=0.1w}{v2}
  \fmfleft{i}  
  \fmf{phantom}{i,v1}
  \fmf{photon,tension=0.1,left}{v1,v2}
  \fmf{dots,right,label=$<>$}{v1,v2} 
  \fmfright{o}
  \fmf{dashes,label=$<$,l.d=5}{v2,o} 
\end{fmfgraph*}}  -
\parbox{20mm}
{\begin{fmfgraph*}(20,15) 
  \fmfv{d.sh=circle,d.f=full,d.si=0.1w}{v1}
  \fmfv{d.sh=circle,d.f=full,d.si=0.1w}{v2}
  \fmfleft{i}  
  \fmf{phantom}{i,v1}
  \fmf{photon,tension=0.1,left}{v1,v2}
  \fmf{dots,right,label=$<<$}{v1,v2} 
  \fmfright{o}
  \fmf{dashes,label=$>$,l.d=5}{v2,o} 
\end{fmfgraph*}}  -
\parbox{20mm}
{\begin{fmfgraph*}(20,15) 
  \fmfv{d.sh=circle,label=2,l.a=180,l.d=.1w,d.f=full,d.si=0.1w}{v1}
  \fmfv{d.sh=circle,d.f=full,d.si=0.1w}{v2}
  \fmfleft{i}  
  \fmf{phantom}{i,v1}
  \fmf{photon,tension=0.1,left}{v1,v2}
  \fmf{dots,right,label=$<<$}{v1,v2} 
  \fmfright{o}
  \fmf{dashes,label=$<$,l.d=5}{v2,o} 
\end{fmfgraph*}} + \nonumber \\
& & 
\parbox{20mm}
{\begin{fmfgraph*}(20,15) 
  \fmfv{d.sh=circle,d.f=full,d.si=0.1w}{v1}
  \fmfv{d.sh=circle,d.f=empty,d.si=0.1w}{v2}
  \fmfleft{i}  
  \fmf{phantom}{i,v1}
  \fmf{photon,width=1,tension=0.1,left}{v1,v2}
  \fmf{plain,right,label=$<>$}{v1,v2} 
  \fmfright{o}
  \fmf{dashes,label=$>$,l.d=5}{v2,o} 
\end{fmfgraph*}} +
\parbox{20mm}
{\begin{fmfgraph*}(20,15) 
  \fmfv{d.sh=circle,d.f=full,d.si=0.1w}{v1}
  \fmfv{d.sh=circle,d.f=empty,d.si=0.1w}{v2}
  \fmfleft{i}  
  \fmf{phantom}{i,v1}
  \fmf{photon,width=1,tension=0.1,left}{v1,v2}
  \fmf{plain,right,label=$<>$}{v1,v2} 
  \fmfright{o}
  \fmf{dashes,label=$<$,l.d=5}{v2,o} 
\end{fmfgraph*}} +
\parbox{20mm}
{\begin{fmfgraph*}(20,15) 
  \fmfv{d.sh=circle,d.f=full,d.si=0.1w}{v1}
  \fmfv{d.sh=circle,d.f=empty,d.si=0.1w}{v2}
  \fmfleft{i}  
  \fmf{phantom}{i,v1}
  \fmf{photon,width=1,tension=0.1,left}{v1,v2}
  \fmf{plain,right,label=$<<$}{v1,v2} 
  \fmfright{o}
  \fmf{dashes,label=$>$,l.d=5}{v2,o} 
\end{fmfgraph*}} + 
\parbox{20mm}
{\begin{fmfgraph*}(20,15) 
  \fmfv{d.sh=circle,d.f=full,d.si=0.1w}{v1}
  \fmfv{d.sh=circle,d.f=empty,d.si=0.1w}{v2}
  \fmfleft{i}  
  \fmf{phantom}{i,v1}
  \fmf{photon,width=1,tension=0.1,left}{v1,v2}
  \fmf{plain,right,label=$<<$}{v1,v2} 
  \fmfright{o}
  \fmf{dashes,label=$<$,l.d=5}{v2,o} 
\end{fmfgraph*}} +  \nonumber \\
& & 
\parbox{20mm}
{\begin{fmfgraph*}(20,15) 
  \fmfv{d.sh=circle,d.f=full,d.si=0.1w}{v1}
  \fmfv{d.sh=circle,d.f=empty,d.si=0.1w}{v2}
  \fmfleft{i}  
  \fmf{phantom}{i,v1}
  \fmf{photon,width=1,tension=0.1,left}{v1,v2}
  \fmf{dots,right,label=$<>$}{v1,v2} 
  \fmfright{o}
  \fmf{plain,label=$>$,l.d=5}{v2,o} 
\end{fmfgraph*}} +
\parbox{20mm}
{\begin{fmfgraph*}(20,15) 
  \fmfv{d.sh=circle,d.f=full,d.si=0.1w}{v1}
  \fmfv{d.sh=circle,d.f=empty,d.si=0.1w}{v2}
  \fmfleft{i}  
  \fmf{phantom}{i,v1}
  \fmf{photon,width=1,tension=0.1,left}{v1,v2}
  \fmf{dots,right,label=$<>$}{v1,v2} 
  \fmfright{o}
  \fmf{plain,label=$<$,l.d=5}{v2,o} 
\end{fmfgraph*}} +
\parbox{20mm}
{\begin{fmfgraph*}(20,15) 
  \fmfv{d.sh=circle,d.f=full,d.si=0.1w}{v1}
  \fmfv{d.sh=circle,d.f=empty,d.si=0.1w}{v2}
  \fmfleft{i}  
  \fmf{phantom}{i,v1}
  \fmf{photon,width=1,tension=0.1,left}{v1,v2}
  \fmf{dots,right,label=$<<$}{v1,v2} 
  \fmfright{o}
  \fmf{plain,label=$>$,l.d=5}{v2,o} 
\end{fmfgraph*}} +
\parbox{20mm}
{\begin{fmfgraph*}(20,15) 
  \fmfv{d.sh=circle,d.f=full,d.si=0.1w}{v1}
  \fmfv{d.sh=circle,d.f=empty,d.si=0.1w}{v2}
  \fmfleft{i}  
  \fmf{phantom}{i,v1}
  \fmf{photon,width=1,tension=0.1,left}{v1,v2}
  \fmf{dots,right,label=$<<$}{v1,v2} 
  \fmfright{o}
  \fmf{plain,label=$<$,l.d=5}{v2,o} 
\end{fmfgraph*}} \label{eq:feyn_I3ub} \nonumber \\
& & + \mbox{higer oder diagrams}
\end{eqnarray} 
\end{fmffile}   


In the above diagrams solid lines denote $<u({\bf k}) u({\bf k'})>$,
and the dashed-arrow lines denote $<u({\bf k}) b({\bf k'})>$.  The
correlation function $<u({\bf k}) u({\bf k'})>$ is denoted by dashed
line.  As mentioned earlier, the wiggly and curly lines denote various
Green's functions.  All the diagrams except 4,8,12, and 16th can be
shown to be trivially zero using
Eqs.~(\ref{eqn:avgbegin}---\ref{eqn:avgend}).  We assume that 4,8,12,
and 16th diagrams are also zero, as usually done in turbulence RG
calculations \cite{YakhOrsz,McCo:book,ZhouVaha88,ZhouVaha93}.
Hence, the term is zero. Following the similar
procedure we can show that the 4th term of $I^u$, and the 
2nd and 3rd terms of $I^b$ are
zero to first order. Now we are left with $>>$ terms 
(5th and 6th of $I^u$, and 4th term of $I^b$), which are

\unitlength=1mm
\begin{fmffile}{fmfub4}
\begin{eqnarray}
I_3^u & = &
\parbox{20mm}
{\begin{fmfgraph*}(20,15) 
  \fmfleft{i} \fmfright{o1,o2} 
  \fmfv{d.sh=circle,d.f=full,d.si=0.1w}{v}
  \fmf{phantom}{i,v}
  \fmf{plain,label=$>$,l.d=5}{v,o2}
  \fmf{plain,label=$>$,l.d=5}{v,o1}
\end{fmfgraph*}} -
\parbox{20mm}
{\begin{fmfgraph*}(20,15) 
  \fmfleft{i} \fmfright{o1,o2} 
  \fmfv{d.sh=circle,d.f=full,d.si=0.1w}{v}
  \fmf{phantom}{i,v}
  \fmf{dashes,label=$>$,l.d=5}{v,o2}
  \fmf{dashes,label=$>$,l.d=5}{v,o1}
\end{fmfgraph*}} \nonumber \\
& = & - \delta \Sigma^{uu}(k)   
\parbox{20mm}
{\begin{fmfgraph*}(20,15) 
  \fmfleft{i} \fmfright{o}
  \fmf{plain,label=$<$,l.s=left}{i,o} 
\end{fmfgraph*}}  - \delta \Sigma^{ub}(k)   
\parbox{20mm}
{\begin{fmfgraph*}(20,15) 
  \fmfleft{i} \fmfright{o}
  \fmf{dashes,label=$<$,l.s=left}{i,o} 
\end{fmfgraph*}}  \\
I_3^b & = &
 \parbox{20mm}
{\begin{fmfgraph*}(20,15) 
  \fmfleft{i} \fmfright{o1,o2} 
  \fmfv{d.sh=circle,d.f=empty,d.si=0.1w}{v}
  \fmf{phantom}{i,v}
  \fmf{plain,label=$>$,l.d=5}{v,o2}
  \fmf{dashes,label=$>$,l.d=5}{v,o1}
\end{fmfgraph*}} \nonumber \\
& = & -\delta  \Sigma^{bu}(k)
\parbox{20mm}
{\begin{fmfgraph*}(20,15) 
  \fmfleft{i}
  \fmfright{o}
  \fmf{plain,label=$<$,l.s=left}{i,o} 
\end{fmfgraph*}} - \delta \Sigma^{bb}(k)
\parbox{20mm}
{\begin{fmfgraph*}(20,15) 
  \fmfleft{i}
  \fmfright{o}
  \fmf{dashes,label=$<$,l.s=left}{i,o} 
\end{fmfgraph*}}
\end{eqnarray}
where
\begin{eqnarray}
-(d-1) \delta \Sigma^{uu} & = &
\parbox{20mm}
{\begin{fmfgraph*}(20,15) 
  \fmfv{l=4,l.a=180,l.d=.1w,d.sh=circle,d.f=full,d.si=0.1w}{v1}
  \fmfv{d.sh=circle,d.f=full,d.si=0.1w}{v2}
  \fmfleft{i}  
  \fmf{phantom}{i,v1}
  \fmf{photon,tension=0.3,right}{v1,v2}
  \fmf{plain,tension=0.3,left,label=$>>$}{v1,v2} 
  \fmfright{o}   \fmf{phantom}{v2,o}
\end{fmfgraph*}} -
\parbox{20mm}
{\begin{fmfgraph*}(20,15) 
  \fmfv{l=2,l.a=180,l.d=.1w,d.sh=circle,d.f=full,d.si=0.1w}{v1}
  \fmfv{d.sh=circle,d.f=empty,d.si=0.1w}{v2}
  \fmfleft{i}  
  \fmf{phantom}{i,v1}
  \fmf{gluon,tension=0.3,right}{v1,v2}
  \fmf{dashes,tension=0.3,left,label=$>>$}{v1,v2} 
  \fmfright{o}   \fmf{phantom}{v2,o}
\end{fmfgraph*}} +
\parbox{20mm}
{\begin{fmfgraph*}(20,15) 
  \fmfv{l=2,l.a=180,l.d=.1w,d.sh=circle,d.f=full,d.si=0.1w}{v1}
  \fmfv{d.sh=circle,d.f=empty,d.si=0.1w}{v2}
  \fmfleft{i}  
  \fmf{phantom}{i,v1}
  \fmf{photon,width=1,tension=0.3,right}{v1,v2}
  \fmf{dashes_arrow,tension=0.3,left,label=$>>$}{v1,v2} 
  \fmfright{o}   \fmf{phantom}{v2,o} 
\end{fmfgraph*}} -
\parbox{20mm}
{\begin{fmfgraph*}(20,15) 
  \fmfv{l=4,l.a=180,l.d=.1w,d.sh=circle,d.f=full,d.si=0.1w}{v1}
  \fmfv{d.sh=circle,d.f=full,d.si=0.1w}{v2}
  \fmfleft{i}  
  \fmf{phantom}{i,v1}
  \fmf{gluon,width=1,tension=0.3,right}{v1,v2}
  \fmf{dashes_arrow,tension=0.3,left,label=$>>$}{v1,v2} 
  \fmfright{o}   \fmf{phantom}{v2,o} 
\end{fmfgraph*}}  \label{eq:feyn_Sigmauu} \\
-(d-1) \delta \Sigma^{ub} & = & -
\parbox{20mm}
{\begin{fmfgraph*}(20,15) 
  \fmfv{l=4,l.a=180,l.d=.1w,d.sh=circle,d.f=full,d.si=0.1w}{v1}
  \fmfv{d.sh=circle,d.f=full,d.si=0.1w}{v2}
  \fmfleft{i}  
  \fmf{phantom}{i,v1}
  \fmf{photon,tension=0.3,right}{v1,v2}
  \fmf{dashes_arrow,tension=0.3,left,label=$>>$}{v1,v2} 
  \fmfright{o}   \fmf{phantom}{v2,o}
\end{fmfgraph*}} +
\parbox{20mm}
{\begin{fmfgraph*}(20,15) 
  \fmfv{l=2,l.a=180,l.d=.1w,d.sh=circle,d.f=full,d.si=0.1w}{v1}
  \fmfv{d.sh=circle,d.f=empty,d.si=0.1w}{v2}
  \fmfleft{i}  
  \fmf{phantom}{i,v1}
  \fmf{photon,width=1,tension=0.3,right}{v1,v2}
  \fmf{plain,tension=0.3,left,label=$>>$}{v1,v2} 
  \fmfright{o}   \fmf{phantom}{v2,o}
\end{fmfgraph*}} +
\parbox{20mm}
{\begin{fmfgraph*}(20,15) 
  \fmfv{l=4,l.a=180,l.d=.1w,d.sh=circle,d.f=full,d.si=0.1w}{v1}
  \fmfv{d.sh=circle,d.f=full,d.si=0.1w}{v2}
  \fmfleft{i}  
  \fmf{phantom}{i,v1}
  \fmf{gluon,width=1,tension=0.3,right}{v1,v2}
  \fmf{dashes_arrow,tension=0.3,left,label=$>>$}{v1,v2} 
  \fmfright{o}   \fmf{phantom}{v2,o} 
\end{fmfgraph*}} -
\parbox{20mm}
{\begin{fmfgraph*}(20,15) 
  \fmfv{l=2,l.a=180,l.d=.1w,d.sh=circle,d.f=full,d.si=0.1w}{v1}
  \fmfv{d.sh=circle,d.f=empty,d.si=0.1w}{v2}
  \fmfleft{i}  
  \fmf{phantom}{i,v1}
  \fmf{gluon,tension=0.3,right}{v1,v2}
  \fmf{dashes_arrow,tension=0.3,left,label=$>>$}{v1,v2} 
  \fmfright{o}   \fmf{phantom}{v2,o} 
\end{fmfgraph*}} \\
-(d-1) \delta \Sigma^{bu} & = & 
\parbox{20mm}
{\begin{fmfgraph*}(20,15) 
  \fmfv{l=2,l.a=180,l.d=.1w,d.sh=circle,d.f=empty,d.si=0.1w}{v1}
  \fmfv{d.sh=circle,d.f=full,d.si=0.1w}{v2}
  \fmfleft{i}  
  \fmf{phantom}{i,v1}
  \fmf{photon,tension=0.3,left}{v1,v2}
  \fmf{dashes_arrow,tension=0.3,right,label=$>>$}{v1,v2} 
  \fmfright{o}   \fmf{phantom}{v2,o}
\end{fmfgraph*}} +
\parbox{20mm}
{\begin{fmfgraph*}(20,15) 
  \fmfv{d.sh=circle,d.f=empty,d.si=0.1w}{v1}
  \fmfv{d.sh=circle,d.f=empty,d.si=0.1w}{v2}
  \fmfleft{i}  
  \fmf{phantom}{i,v1}
  \fmf{gluon,width=1,tension=0.3,right}{v1,v2}
  \fmf{dashes_arrow,tension=0.3,left,label=$>>$}{v1,v2} 
  \fmfright{o}   \fmf{phantom}{v2,o}
\end{fmfgraph*}} +
\parbox{20mm}
{\begin{fmfgraph*}(20,15) 
  \fmfv{d.sh=circle,d.f=empty,d.si=0.1w}{v1}
  \fmfv{d.sh=circle,d.f=empty,d.si=0.1w}{v2}
  \fmfleft{i}  
  \fmf{phantom}{i,v1}
  \fmf{photon,width=1,tension=0.3,left}{v1,v2}
  \fmf{dashes,tension=0.3,right,label=$>>$}{v1,v2} 
  \fmfright{o}   \fmf{phantom}{v2,o} 
\end{fmfgraph*}} -
\parbox{20mm}
{\begin{fmfgraph*}(20,15) 
  \fmfv{l=2,l.a=180,l.d=.1w,d.sh=circle,d.f=empty,d.si=0.1w}{v1}
  \fmfv{d.sh=circle,d.f=full,d.si=0.1w}{v2}
  \fmfleft{i}  
  \fmf{phantom}{i,v1}
  \fmf{gluon,width=1,tension=0.3,right}{v1,v2}
  \fmf{plain,tension=0.3,left,label=$>>$}{v1,v2} 
  \fmfright{o}   \fmf{phantom}{v2,o} 
\end{fmfgraph*}} \\
-(d-1) \delta \Sigma^{bb} & = & 
\parbox{20mm}
{\begin{fmfgraph*}(20,15) 
  \fmfv{l=2,l.a=180,l.d=.1w,d.sh=circle,d.f=empty,d.si=0.1w}{v1}
  \fmfv{d.sh=circle,d.f=full,d.si=0.1w}{v2}
  \fmfleft{i}  
  \fmf{phantom}{i,v1}
  \fmf{photon,tension=0.3,left}{v1,v2}
  \fmf{dashes,tension=0.3,right,label=$>>$}{v1,v2} 
  \fmfright{o}   \fmf{phantom}{v2,o}
\end{fmfgraph*}} +
\parbox{20mm}
{\begin{fmfgraph*}(20,15) 
  \fmfv{d.sh=circle,d.f=empty,d.si=0.1w}{v1}
  \fmfv{d.sh=circle,d.f=empty,d.si=0.1w}{v2}
  \fmfleft{i}  
  \fmf{phantom}{i,v1}
  \fmf{gluon,tension=0.3,right}{v1,v2}
  \fmf{plain,tension=0.3,left,label=$>>$}{v1,v2} 
  \fmfright{o}   \fmf{phantom}{v2,o}
\end{fmfgraph*}} +
\parbox{20mm}
{\begin{fmfgraph*}(20,15) 
  \fmfv{d.sh=circle,d.f=empty,d.si=0.1w}{v1}
  \fmfv{d.sh=circle,d.f=empty,d.si=0.1w}{v2}
  \fmfleft{i}  
  \fmf{phantom}{i,v1}
  \fmf{photon,width=1,tension=0.3,left}{v1,v2}
  \fmf{dashes_arrow,tension=0.3,right,label=$>>$}{v1,v2} 
  \fmfright{o}   \fmf{phantom}{v2,o} 
\end{fmfgraph*}} -
\parbox{20mm}
{\begin{fmfgraph*}(20,15) 
  \fmfv{d.sh=circle,d.f=empty,d.si=0.1w}{v1}
  \fmfv{d.sh=circle,d.f=empty,d.si=0.1w}{v2}
  \fmfleft{i}  
  \fmf{phantom}{i,v1}
  \fmf{gluon,width=1,tension=0.3,right}{v1,v2}
  \fmf{dashes_arrow,tension=0.3,left,label=$>>$}{v1,v2} 
  \fmfright{o}   \fmf{phantom}{v2,o} 
\end{fmfgraph*}}  \label{eq:feyn_Sigmabb}
\end{eqnarray}
\end{fmffile}


In Eqs.~(\ref{eq:feyn_Sigmauu}-\ref{eq:feyn_Sigmabb}) we have omitted all the 
vanishing diagrams (similar
to those appearing in Eq.~[\ref{eq:feyn_I3ub}]).  
These terms contribute to $\Sigma$s.

The algebraic expressions for the above diagrams are given in Section 7.
For isotropic flows, the algebraic factors $S_i(k,p,q)$ resulting from tensor 
contractions are given below. The 
factors for the diagrams are
$S, S_6, S_6, S, S, S_5, S, S_5, S_8, S_{10}, S_{12}, S_7,
S_8, S_9, S_{11}, S_9$ in sequential order.
\begin{eqnarray}
S(k,p,q)   & = & P^{+}_{bjm}(k) P^{+}_{mab}(p) P_{ja}(q) 
            =  kp \left( (d-3)z+2 z^3+(d-1)xy \right) \nonumber \\ 
S_5(k,p,q) & = & P^{+}_{bjm}(k) P^{-}_{mab}(p) P_{ja}(q) 
            =  kp \left( (d-1)z+(d-3)xy-2 y^2 z \right)\nonumber  \\
S_6(k,p,q) & = & P^{+}_{ajm}(k) P^{-}_{mba}(p) P_{jb}(q) 
            =  -S_5(k,p,q) \nonumber \\
S_7(k,p,q) & = & P^{-}_{ijm}(k) P^{+}_{mab}(p) P_{ja}(q) P_{ib}(k) 
            =  S_5(p,k,q) \nonumber \\
S_8(k,p,q) & = & P^{-}_{ijm}(k) P^{+}_{jab}(p) P_{ma}(q) P_{ib}(k)
            =  -S_5(p,k,q) \nonumber \\
S_9(k,p,q)    & = & P^{-}_{ijm}(k) P^{-}_{mab}(p) P_{ja}(q) P_{ib}(k) 
               =  kp(d-1)(z+xy) \nonumber\\
S_{10}(k,p,q) & = & P^{-}_{ijm}(k) P^{-}_{mab}(p) P_{jb}(q) P_{ia}(k) 
                                =  -S_9(k,p,q) \nonumber \\
S_{11}(k,p,q) & = & P^{-}_{ijm}(k) P^{-}_{jab}(p) P_{ma}(q) P_{ib}(k) 
               =  -S_9(k,p,q) \nonumber \\
S_{12}(k,p,q) & = & P^{-}_{ijm}(k) P^{-}_{jab}(p) P_{mb}(q) P_{ia}(k) 
                    =  S_9(k,p,q) \nonumber
\end{eqnarray}

\subsection{Mode-to-Mode Energy Transfer in MHD Turbulence}

In Section 3, we studied the ``mode-to-mode'' energy transfer
$S^{YX}({\bf k'|p|q})$  from mode {\bf p} of field X to
mode {\bf k'} of field Y, with mode {\bf q} acting as a mediator.
The perturbative calculation of $S$ involves many terms.
However when cross helicity is zero, then many of them vanish.and yield
\unitlength=1mm
\begin{fmffile}{fmfub5}
\begin{eqnarray}
\left\langle S^{uu}(k'|p|q) \right\rangle & = &
\parbox{30mm}
{\begin{fmfgraph*}(30,20) 
  \fmfv{d.sh=circle,d.f=full,d.si=0.01w}{v1}
  \fmfv{d.sh=circle,d.f=full,d.si=0.1w}{v2}
  \fmfleft{i}  
  \fmf{phantom}{i,v1}
  \fmf{plain,tension=0.1,right,label=$q$}{v1,v2}
  \fmf{plain,tension=0.1,label=$p$}{v1,v2}
  \fmf{photon,tension=0.1,left,label=$k$}{v1,v2} 
  \fmfright{o}   \fmf{phantom}{v2,o}
\end{fmfgraph*}} +
\parbox{30mm}
{\begin{fmfgraph*}(30,20) 
  \fmfv{d.sh=circle,d.f=full,d.si=0.01w}{v1}
  \fmfv{d.sh=circle,d.f=full,d.si=0.1w}{v2}
  \fmfleft{i}  
  \fmf{phantom}{i,v1}
  \fmf{plain,tension=0.1,right,label=$q$}{v1,v2}
  \fmf{photon,tension=0.1,label=$p$}{v1,v2}
  \fmf{plain,tension=0.1,left,label=$k$}{v1,v2} 
  \fmfright{o}   \fmf{phantom}{v2,o}
\end{fmfgraph*}} +
\parbox{30mm}
{\begin{fmfgraph*}(30,20) 
  \fmfv{d.sh=circle,d.f=full,d.si=0.01w}{v1}
  \fmfv{d.sh=circle,d.f=full,d.si=0.1w}{v2}
  \fmfleft{i}  
  \fmf{phantom}{i,v1}
  \fmf{photon,tension=0.1,right,label=$q$}{v1,v2}
  \fmf{plain,tension=0.1,label=$p$}{v1,v2}
  \fmf{plain,tension=0.1,left,label=$k$}{v1,v2} 
  \fmfright{o}   \fmf{phantom}{v2,o}
\end{fmfgraph*}} ; \\ 
\nonumber \\ \nonumber \\ \nonumber \\
-\left\langle S^{ub}(k'|p|q) \right\rangle & = & -
\parbox{30mm}
{\begin{fmfgraph*}(30,20) 
  \fmfv{d.sh=circle,d.f=full,d.si=0.01w}{v1}
  \fmfv{d.sh=circle,d.f=full,d.si=0.1w}{v2}
  \fmfleft{i}  
  \fmf{phantom}{i,v1}
  \fmf{dashes,tension=0.1,right,label=$q$}{v1,v2}
  \fmf{dashes,tension=0.1,label=$p$}{v1,v2}
  \fmf{photon,tension=0.1,left,label=$k$}{v1,v2} 
  \fmfright{o}   \fmf{phantom}{v2,o}
\end{fmfgraph*}} +
\parbox{30mm}
{\begin{fmfgraph*}(30,20) 
  \fmfv{d.sh=circle,d.f=full,d.si=0.01w}{v1}
  \fmfv{d.sh=circle,d.f=empty,d.si=0.1w}{v2}
  \fmfleft{i}  
  \fmf{phantom}{i,v1}
  \fmf{dashes,tension=0.1,right,label=$q$}{v1,v2}
  \fmf{gluon,tension=0.1,label=$p$}{v1,v2}
  \fmf{plain,tension=0.1,left,label=$k$}{v1,v2} 
  \fmfright{o}   \fmf{phantom}{v2,o}
\end{fmfgraph*}} +
\parbox{30mm}
{\begin{fmfgraph*}(30,20) 
  \fmfv{d.sh=circle,d.f=full,d.si=0.01w}{v1}
  \fmfv{d.sh=circle,d.f=empty,d.si=0.1w}{v2}
  \fmfleft{i}  
  \fmf{phantom}{i,v1}
  \fmf{gluon,tension=0.1,right,label=$q$}{v1,v2}
  \fmf{dashes,tension=0.1,label=$p$}{v1,v2}
  \fmf{plain,tension=0.1,left,label=$k$}{v1,v2} 
  \fmfright{o}   \fmf{phantom}{v2,o}
\end{fmfgraph*}} ; \\ 
\nonumber \\ \nonumber \\ \nonumber \\
-\left\langle S^{bu}(k'|p|q) \right\rangle & = &
\parbox{30mm}
{\begin{fmfgraph*}(30,20) 
  \fmfv{d.sh=circle,d.f=full,d.si=0.01w}{v1}
  \fmfv{d.sh=circle,d.f=empty,d.si=0.1w}{v2}
  \fmfleft{i}  
  \fmf{phantom}{i,v1}
  \fmf{dashes,tension=0.1,right,label=$q$}{v1,v2}
  \fmf{plain,tension=0.1,label=$p$}{v1,v2}
  \fmf{gluon,tension=0.1,left,label=$k$}{v1,v2} 
  \fmfright{o}   \fmf{phantom}{v2,o}
\end{fmfgraph*}} -
\parbox{30mm}
{\begin{fmfgraph*}(30,20) 
  \fmfv{d.sh=circle,d.f=full,d.si=0.01w}{v1}
  \fmfv{d.sh=circle,d.f=full,d.si=0.1w}{v2}
  \fmfleft{i}  
  \fmf{phantom}{i,v1}
  \fmf{dashes,tension=0.1,right,label=$q$}{v1,v2}
  \fmf{photon,tension=0.1,label=$p$}{v1,v2}
  \fmf{dashes,tension=0.1,left,label=$k$}{v1,v2} 
  \fmfright{o}   \fmf{phantom}{v2,o}
\end{fmfgraph*}} +
\parbox{30mm}
{\begin{fmfgraph*}(30,20) 
  \fmfv{d.sh=circle,d.f=full,d.si=0.01w}{v1}
  \fmfv{d.sh=circle,d.f=empty,d.si=0.1w}{v2}
  \fmfleft{i}  
  \fmf{phantom}{i,v1}
  \fmf{gluon,tension=0.1,right,label=$q$}{v1,v2}
  \fmf{plain,tension=0.1,label=$p$}{v1,v2}
  \fmf{dashes,tension=0.1,left,label=$k$}{v1,v2} 
  \fmfright{o}   \fmf{phantom}{v2,o}
\end{fmfgraph*}} ; \\ 
\nonumber \\ \nonumber \\ \nonumber \\
\left\langle S^{bb}(k'|p|q) \right\rangle & = & 
\parbox{30mm}
{\begin{fmfgraph*}(30,20) 
  \fmfv{d.sh=circle,d.f=full,d.si=0.01w}{v1}
  \fmfv{d.sh=circle,d.f=empty,d.si=0.1w}{v2}
  \fmfleft{i}  
  \fmf{phantom}{i,v1}
  \fmf{plain,tension=0.1,right,label=$q$}{v1,v2}
  \fmf{dashes,tension=0.1,label=$p$}{v1,v2}
  \fmf{gluon,tension=0.1,left,label=$k$}{v1,v2} 
  \fmfright{o}   \fmf{phantom}{v2,o}
\end{fmfgraph*}} +
\parbox{30mm}
{\begin{fmfgraph*}(30,20) 
  \fmfv{d.sh=circle,d.f=full,d.si=0.01w}{v1}
  \fmfv{d.sh=circle,d.f=empty,d.si=0.1w}{v2}
  \fmfleft{i}  
  \fmf{phantom}{i,v1}
  \fmf{plain,tension=0.1,right,label=$q$}{v1,v2}
  \fmf{gluon,tension=0.1,label=$p$}{v1,v2}
  \fmf{dashes,tension=0.1,left,label=$k$}{v1,v2} 
  \fmfright{o}   \fmf{phantom}{v2,o}
\end{fmfgraph*}} -
\parbox{30mm}
{\begin{fmfgraph*}(30,20) 
  \fmfv{d.sh=circle,d.f=full,d.si=0.01w}{v1}
  \fmfv{d.sh=circle,d.f=full,d.si=0.1w}{v2}
  \fmfleft{i}  
  \fmf{phantom}{i,v1}
  \fmf{photon,tension=0.1,right,label=$q$}{v1,v2}
  \fmf{dashes,tension=0.1,label=$p$}{v1,v2}
  \fmf{dashes,tension=0.1,left,label=$k$}{v1,v2} 
  \fmfright{o}   \fmf{phantom}{v2,o}
\end{fmfgraph*}} 
\end{eqnarray}
\end{fmffile}
\vspace{1cm}

In all the diagrams, the left vertex denotes
$k_i$, while the filled circle and the empty circles of right vertex
represent $(-i/2) P^+_{ijm}$ and $-i P^-_{ijm}$ respectively.
For isotropic nonhelical flows, the algebraic factors are given
below. The factors for the diagrams are
$T_1, T_5, T_9, T_2, T_6, T_{10}, T_3, T_7, T_{11},
T_4, T_8, T_{12}$ in sequential order.
\begin{eqnarray}
T_1(k,p,q) & = & k_i P^{+}_{jab}(k) P_{ja}(p) P_{ib}(q)
            =  kp\left( (d-3)z + (d-2)xy +2 z^3+ 2 x y z^2
			  + x^2 z \right) \nonumber \\
T_3(k,p,q) & = & k_i P^{-}_{jab}(k) P_{ja}(p) P_{ib}(q)
             =  -k^2 \left( (d-2)(1-y^2) + z^2 +x y z \right) \nonumber \\
T_5(k,p,q) & = & -k_i P^{+}_{jab}(p) P_{ja}(k) P_{ib}(q)
             =  -kp\left( (d-3)z + (d-2)xy +2 z^3+ 2 x y z^2
			  + y^2 z \right) \nonumber \\
T_7(k,p,q) & = & -k_i P^{-}_{jab}(p) P_{ja}(k) P_{ib}(q)
            =  -kp\left( (2-d)x y + (1-d)z +y^2 z \right) \nonumber \\
T_9(k,p,q) & = & -k_i P^{+}_{iab}(q) P_{ja}(k) P_{jb}(p)
            =  -k q \left(x z - 2 x y^2 z - y z^2 \right) \nonumber \\
T_{11}(k,p,q) & = & -k_i P^{-}_{iab}(q) P_{ja}(k) P_{jb}(p)
                =  -k q z \left(x + y z \right) \nonumber \\
T_{2 n}(k,p,q) & = & -T_{2 n -1 }(k,p,q) 
			\hspace{1cm} \mbox{for $n=1..6$} \nonumber
\end{eqnarray}

For helical flows, we get additional terms involving helicities.
We are skipping those terms due to lack of space.

Following the similar procedure, we can obtain Feynman diagrams for
 mode-to-mode magnetic-helicity transfer, which is

\unitlength=1mm
\begin{fmffile}{fmfub6}
\begin{eqnarray}
\left\langle S^{H_M}(k'|p|q) \right\rangle & = &
\parbox{30mm}
{\begin{fmfgraph*}(30,20) 
  \fmfv{d.sh=triangle,d.f=empty,d.si=0.1w}{v1}
  \fmfv{d.sh=circle,d.f=empty,d.si=0.1w}{v2}
  \fmfleft{i}  
  \fmf{phantom}{i,v1}
  \fmf{dashes,tension=0.1,right,label=$q$}{v1,v2}
  \fmf{plain,tension=0.1,label=$p$}{v1,v2}
  \fmf{gluon,tension=0.1,left,label=$k$}{v1,v2} 
  \fmfright{o}   \fmf{phantom}{v2,o}
\end{fmfgraph*}} -
\parbox{30mm}
{\begin{fmfgraph*}(30,20) 
  \fmfv{d.sh=triangle,d.f=empty,d.si=0.1w}{v1}
  \fmfv{d.sh=circle,d.f=full,d.si=0.1w}{v2}
  \fmfleft{i}  
  \fmf{phantom}{i,v1}
  \fmf{dashes,tension=0.1,right,label=$q$}{v1,v2}
  \fmf{photon,tension=0.1,label=$p$}{v1,v2}
  \fmf{dashes,tension=0.1,left,label=$k$}{v1,v2} 
  \fmfright{o}   \fmf{phantom}{v2,o}
\end{fmfgraph*}} +
\parbox{30mm}
{\begin{fmfgraph*}(30,20) 
  \fmfv{d.sh=triangle,d.f=empty,d.si=0.1w}{v1}
  \fmfv{d.sh=circle,d.f=empty,d.si=0.1w}{v2}
  \fmfleft{i}  
  \fmf{phantom}{i,v1}
  \fmf{gluon,tension=0.1,right,label=$q$}{v1,v2}
  \fmf{plain,tension=0.1,label=$p$}{v1,v2}
  \fmf{dashes,tension=0.1,left,label=$k$}{v1,v2} 
  \fmfright{o}   \fmf{phantom}{v2,o}
\end{fmfgraph*}}  +
\nonumber \\ \nonumber \\ \nonumber \\  \nonumber \\
&  & 
\parbox{30mm}
{\begin{fmfgraph*}(30,20) 
  \fmfv{d.sh=triangle,d.f=shaded,d.si=0.1w}{v1}
  \fmfv{d.sh=circle,d.f=empty,d.si=0.1w}{v2}
  \fmfleft{i}  
  \fmf{phantom}{i,v1}
  \fmf{plain,tension=0.1,right,label=$q$}{v1,v2}
  \fmf{dashes,tension=0.1,label=$p$}{v1,v2}
  \fmf{gluon,tension=0.1,left,label=$k$}{v1,v2} 
  \fmfright{o}   \fmf{phantom}{v2,o}
\end{fmfgraph*}} +
\parbox{30mm}
{\begin{fmfgraph*}(30,20) 
  \fmfv{d.sh=triangle,d.f=shaded,d.si=0.1w}{v1}
  \fmfv{d.sh=circle,d.f=empty,d.si=0.1w}{v2}
  \fmfleft{i}  
  \fmf{phantom}{i,v1}
  \fmf{plain,tension=0.1,right,label=$q$}{v1,v2}
  \fmf{gluon,tension=0.1,label=$p$}{v1,v2}
  \fmf{dashes,tension=0.1,left,label=$k$}{v1,v2} 
  \fmfright{o}   \fmf{phantom}{v2,o}
\end{fmfgraph*}} -
\parbox{30mm}
{\begin{fmfgraph*}(30,20) 
  \fmfv{d.sh=triangle,d.f=shaded,d.si=0.1w}{v1}
  \fmfv{d.sh=circle,d.f=full,d.si=0.1w}{v2}
  \fmfleft{i}  
  \fmf{phantom}{i,v1}
  \fmf{photon,tension=0.1,right,label=$q$}{v1,v2}
  \fmf{dashes,tension=0.1,label=$p$}{v1,v2}
  \fmf{dashes,tension=0.1,left,label=$k$}{v1,v2} 
  \fmfright{o}   \fmf{phantom}{v2,o}
\end{fmfgraph*}}  +
\nonumber \\ \nonumber \\ \nonumber \\ \nonumber \\
 &  &
\parbox{30mm}
{\begin{fmfgraph*}(30,20) 
  \fmfv{d.sh=triangle,d.f=full,d.si=0.1w}{v1}
  \fmfv{d.sh=circle,d.f=empty,d.si=0.1w}{v2}
  \fmfleft{i}  
  \fmf{phantom}{i,v1}
  \fmf{dashes,tension=0.1,right,label=$q$}{v1,v2}
  \fmf{plain,tension=0.1,label=$p$}{v1,v2}
  \fmf{gluon,tension=0.1,left,label=$k$}{v1,v2} 
  \fmfright{o}   \fmf{phantom}{v2,o}
\end{fmfgraph*}} -
\parbox{30mm}
{\begin{fmfgraph*}(30,20) 
  \fmfv{d.sh=triangle,d.f=full,d.si=0.1w}{v1}
  \fmfv{d.sh=circle,d.f=full,d.si=0.1w}{v2}
  \fmfleft{i}  
  \fmf{phantom}{i,v1}
  \fmf{dashes,tension=0.1,right,label=$q$}{v1,v2}
  \fmf{photon,tension=0.1,label=$p$}{v1,v2}
  \fmf{dashes,tension=0.1,left,label=$k$}{v1,v2} 
  \fmfright{o}   \fmf{phantom}{v2,o}
\end{fmfgraph*}} +
\parbox{30mm}
{\begin{fmfgraph*}(30,20) 
  \fmfv{d.sh=triangle,d.f=full,d.si=0.1w}{v1}
  \fmfv{d.sh=circle,d.f=empty,d.si=0.1w}{v2}
  \fmfleft{i}  
  \fmf{phantom}{i,v1}
  \fmf{gluon,tension=0.1,right,label=$q$}{v1,v2}
  \fmf{plain,tension=0.1,label=$p$}{v1,v2}
  \fmf{dashes,tension=0.1,left,label=$k$}{v1,v2} 
  \fmfright{o}   \fmf{phantom}{v2,o}
\end{fmfgraph*}}
\end{eqnarray}
\end{fmffile}

where empty, shaded, and filled triangles (vertices) represent
$\epsilon_{ijm}, - \epsilon_{ijm} k_i k_l /k^2$ and $\epsilon_{ijm}
k_i k_l/k^2$ respectively.  The algebraic factors
can be easily computed for these diagrams.

\section{Digression to Fluid Turbulence \label{sec:Digression-to-Fluid}}

Many of the MHD turbulence work have been motivated by theories of
fluid turbulence. Therefore, we briefly sketch some of the main results
are statistical theory of fluid turbulence. 

\begin{enumerate}
\item McComb and coworkers \cite{McCo:book,McCoWatt,ZhouMcCo:RGrev} have
successfully applied self-consistent renormalization group theory
to \emph{3D} fluid turbulence. The RG procedure has been described
in Section \ref{sub:Nonhelical-nonAlfvenic-MHD}. They showed that
\begin{eqnarray}
E(k) & = & K_{Ko}\Pi^{2/3}k^{-5/3},\label{eq:fluid-Ek}\\
\nu(k=k_{n}k') & = & K_{Ko}^{1/2}\Pi^{1/3}k_{n}^{-4/3}\nu^{*}(k'),\label{eq:fluid-nu}\end{eqnarray}
is a consistent solution of renormalization group equation. Here,
$K_{Ko}$ is Kolmogorov's constant, $\Pi$ is the energy flux, and
$\nu^{*}(k')$ is a universal function that is a constant ($\approx0.38$)
as $k'\rightarrow0$. See Fig. \ref{Fig:Fluid-RG} for an illustration.
\begin{figure}
\includegraphics[bb=85 300 530 500]{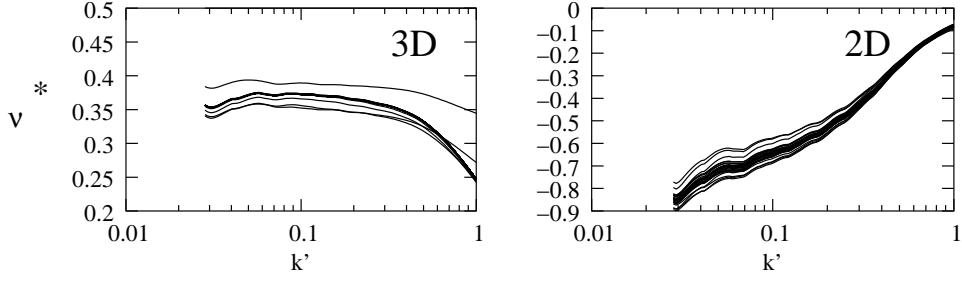}

\caption{\label{Fig:Fluid-RG} Plot of $\nu^{*}(k')$ vs. $k'$ for 2D and
3D fluid turbulence. In 2D, $\nu^{*}$ is negative.}
\end{figure}

\item Energy flux for 3D fluid turbulence can be computed using field-theoretic
technique described in Section \ref{sub:Nonhelical-nonAlfvenic-MHD-flux}.
This technique is same as Direct Interaction Approximation of Kraichnan.
The computation yields Kolmogorov's constant $K$ to be close to 1.58. 
\item The above analysis can be extended to 2D fluid turbulence. We find
that Eqs. (\ref{eq:fluid-Ek}, \ref{eq:fluid-nu}) are the solution
of RG equations, but $\nu^{*}(k')$ is negative as shown in Fig. \ref{Fig:Fluid-RG}.
The function $\nu^{*}$ is not very well behaved as $k'\rightarrow0$.
Still, negative renormalized viscosity is consistent with negative
eddy viscosity obtained using Test Field Model \cite{Krai:71} and
EDQNM calculations. We estimate $\nu^{*}\approx-0.60$ . The energy
flux calculation yields $K_{Ko}^{2D}\approx6.3$.
\item Incompressible fluid turbulence is nonlocal in real space due to incompressibility
condition. Field-theoretic calculation also reveals that mode-to-mode
transfer $S(k|p|q)$ is large when $p\ll k$, but small for $k\sim p\sim q$,
hence Navier-Stokes equation is nonlocal in Fourier space too. However,
in 3D shell-to-shell energy transfer rate $T_{nm}^{YX}$ is forward
and most significant to the next-neighbouring shell. Hence, shell-to-shell
energy transfer rate is local even though the interactions appear
to be nonlocal in both real and Fourier space. Refer to Zhou \cite{Zhou:Local},
Domaradzki and Rogallo \cite{Doma:Local2}, and Verma et al. \cite{Ayye}.
\item In 2D fluid turbulence, energy transfer to the next neighbouring shell
is forward, but the transfer is backward for the more distant shells
(see Fig. \ref{Fig:Fluid-T}). %
\begin{figure}
\includegraphics[bb =85 300 530 500 ]{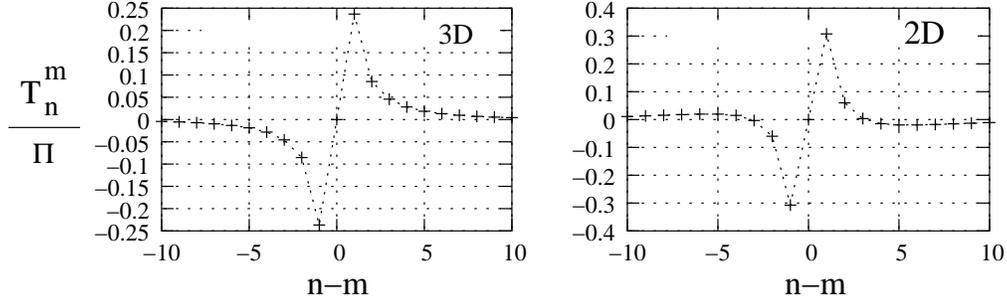}

\caption{\label{Fig:Fluid-T} Plots of shell-to-shell energy transfer rates
$T_{nm}^{YX}/\Pi$ vs. $n-m$ for 3D and 2D fluid turbulence. In 3D
energy transfer is forward and local. In 2D energy transfer for energy
transfer is forward for the nearest neighbours, but the is backward
for fourth neighbour onward; these backward transfers are one of the
major factors in the inverse cascade of energy.}
\end{figure}
The sum of all these transfers is negative energy flux, consistent
with the inverse cascade result of Kraichnan \cite{Krai:71}. For
details refer to Verma et al. \cite{Ayye}.
\item Kinetic helicity suppress the energy flux. Field-theoretic calculation
discussed in Section \ref{sub:Helical-nonAlfvenic-MHD-flux} yield
\[
\Pi=K^{3/2}\Pi(0.53-0.28r_{K}^{2}),\]
where $r_{K}=H_{K}(k)/(kE(k))$ (see the entry of $\Pi_{u>}^{u<}$
in Table \ref{table:hm_rA1_5000}). 
\item All the above conclusions are for large Reynolds number or $\nu\rightarrow0$
limit. The behaviour of Navier-Stokes equation for viscosity $\nu=0$
(inviscid) is very different, and has been analyzed using absolute
equilibrium theory (see Section \ref{sub:Absolute-Equilibrium-States}).
It can be shown using this theory that under steady state, energy
is equipartitioned among all the modes, resulting in $C(k)=const$
\cite{Orsz:Rev}. Using this result we can compute mode-to-mode energy
transfer rates $\left\langle S^{uu}(k|p|q)\right\rangle $ to first
order in perturbation theory (Eq. {[}\ref{eqn:Suu-nonhelical}{]}),
which yields\[
\left\langle S^{uu}(k|p|q)\right\rangle \propto\int\frac{\left(T_{1}(k,p,q)+T_{5}(k,p,q)+T_{9}(k,p,q)\right)Const}{\nu(k)k^{2}+\nu(p)p^{2}+\nu(q)q^{2}}=0\]
because $T_{1}(k,p,q)+T_{5}(k,p,q)+T_{9}(k,p,q)=0$. Hence, under
steady-state, their is no energy transfer among Fourier modes in inviscid
Navier-Stokes. In other words {}``principle of detailed balance''
holds here. Note that the above result holds for all space dimensions.
Contrast this result with the turbulence situation when energy preferentially
gets transferred from smaller wavenumber to larger wavenumber. This
example contrasts equilibrium and nonequilibrium systems.\end{enumerate}


\end{document}